\documentclass[letterpaper,11pt]{article}
\usepackage[usenames,dvipsnames,svgnames,table]{xcolor}
\usepackage[utf8]{inputenc}
\usepackage{amsmath}
\usepackage{amssymb, bbm}
\usepackage{wasysym,relsize}
\usepackage{amsfonts}
\usepackage{amsthm}
\usepackage{graphicx}
\usepackage{color}
\usepackage{hyperref}
\usepackage{cite}
\usepackage{tikz}
\usepackage{subfigure}
\usepackage{xspace}
\usepackage{needspace}
\usepackage{url}
\usepackage{enumitem}
\usepackage{tikz}
\usetikzlibrary{calc,trees,positioning,arrows,chains,shapes.geometric,%
    decorations.pathreplacing,decorations.pathmorphing,shapes,%
    matrix,shapes.symbols,positioning,patterns}
\usepackage[top=1.3in, bottom=1.2in, left=1.1in, right=1.1in]{geometry}
\usetikzlibrary{decorations.pathmorphing}
\usetikzlibrary{shapes}
\newtheorem{theorem}{Theorem}[section]

\newtheorem{lemma}[theorem]{Lemma}
\newtheorem{corollary}[theorem]{Corollary}

\theoremstyle{definition}
\newtheorem{definition}[theorem]{Definition}
\newtheorem{fact}[theorem]{Fact}
\newtheorem{convention}[theorem]{Convention}

\newtheorem{algorithm}[theorem]{Algorithm}
\let\oldalgorithm\algorithm
\renewcommand{\algorithm}{\oldalgorithm\normalfont}

\newcommand{\termasm}[1]{\mathcal{A}_{\Box}[{#1}]}
\newcommand{\prodasm}[1]{\mathcal{A}[{#1}]}

\newcommand{\dom}[1]{{\rm dom}(#1)}

\newcommand{\glue}{\mathcal{G}}

\newcommand\pos{\mathrm{pos}}
\newcommand\type{\mathrm{type}}

\newcommand{\wind}{W}

\newcommand{\traj}{D}
\newcommand{\boundv}{\mathcal{B}^v}
\newcommand{\boundf}{\mathcal{B}^-}
\newcommand{\domi}{d}
\newcommand{\fresetstep}{f^s}

\newcommand{\ind}[1]{\mathbf{#1}}
\newcommand{\vsets}[1]{\mathcal{V}_{\seed}^{#1}}
\newcommand{\vset}[1]{\mathcal{V}^{#1}}
\newcommand{\pstakem}{F}
\newcommand{\omeplus}[1]{#1^{+\omega}}
\newcommand{\omepluse}[1]{#1^{*+\omega}}
\newcommand{\omemoi}[1]{#1^{-\omega}}
\newcommand{\ome}[1]{#1^{\omega}}
\newcommand{\omemoinse}[1]{#1^{*-\omega}}
\newcommand{\tile}[2]{\draw[fill=white] (#1.12,#2.12) rectangle (#1.88,#2.88);}
\newcommand{\tileg}[2]{\draw[fill=gray] (#1.12,#2.12) rectangle (#1.88,#2.88);}
\newcommand{\tileh}[2]{\draw[fill=gray!60] (#1.12,#2.12) rectangle (#1.88,#2.88);}
\newcommand{\tiles}[2]{\draw[fill=black] (#1.12,#2.12) rectangle (#1.88,#2.88);}

\newcommand{\free}{\mathcal{F}}
\newcommand{\indm}{s^-}
\newcommand{\indp}{s^+}
\newcommand{\cutp}{C^+}
\newcommand{\cutm}{C^-}
\newcommand{\lplus}{U}
\newcommand{\lmin}{D}
\newcommand{\lplusk}[1]{U^{{#1}}}
\newcommand{\lmink}[1]{D^{{#1}}}
\newcommand{\gint}{G^I}
\newcommand{\gext}{G^E}
\newcommand{\per}{p}
\newcommand{\movie}{\mathcal{M}}
\newcommand{\stakem}{\mathcal{S}^-}
\newcommand{\stakep}{\mathcal{S}^+}
\newcommand{\inds}{s}
\newcommand{\hunt}{\mathcal{H}}
\newcommand{\border}{\mathcal{B}}
\newcommand{\seed}{\sigma}%rectangle seed
\newcommand{\rseed}{\mathcal{R}^\seed}
\newcommand{\rect}{\mathcal{R}}
\newcommand{\fbound}{\mathcal{B}^f}
\newcommand{\freset}{f^r}
\newcommand{\fjail}{f^j}
\newcommand{\ebound}{\mathcal{B}^e}
\newcommand{\seedbound}{\mathcal{B}^\seed}
\newcommand{\sbound}{\mathcal{B}^s}
\newcommand{\fband}{f^b}

\newcommand{\grid}{G}

\newcommand{\lastp}{|P|}

\newcommand{\vu}{\ensuremath{\overrightarrow{v}}}
\newcommand{\wu}{\ensuremath{\overrightarrow{w}}}

\newcommand{\tiling}{\mathcal{T}}

\newcommand{\vect}[1]{{\protect\overrightarrow{#1}}}

\usepackage{ titling}

\newcommand{\ignore}[1]{}

\newcommand{\zs}{\mathbb{Z}^2}

\title{A pumping lemma for non-cooperative self-assembly}
\author{Pierre-Étienne Meunier\\
  Aalto University \\
\href{mailto:pierre-etienne.meunier@aalto.fi}{pierre-etienne.meunier@aalto.fi}
\thanks{Aalto University, Helsinki, Finland and Aix Marseille Université, CNRS, LIF UMR 7279, 13288, Marseille, France. Supported in part by National Science Foundation Grant CCF-1219274.}
\and
Damien Regnault\\
Université d'Évry Val-d'Essonne\\
\href{mailto:damien.regnault@ibisc.fr}{damien.regnault@ibisc.fr}
\thanks{Université d'Évry Val-d'Essonne, IBISC  EA 4526, 91037, Évry, France. Supported in part by ANR project Quasicool (ANR-12-JS02-011-01)}}

\date{}
\begin{document}

\maketitle

\let\oldlabel=\label
\let\oldref=\ref
\renewcommand\label[1]{\oldlabel{ext-#1}}
\renewcommand\ref[1]{\oldref{ext-#1}}
\let\oldcite=\cite
\renewcommand\cite[1]{\oldcite{ext-#1}}

\newcommand\var[1]{\ocwlowerid{#1}}
\renewcommand\label[1]{\oldlabel{#1}}
\renewcommand\ref[1]{\oldref{#1}}
\renewcommand\cite[1]{\oldcite{#1}}
\newcommand\prog{\href{http://users.ics.aalto.fi/meunier/pumpability.html}{\tt http://users.ics.aalto.fi/meunier/pumpability.html}}

\begin{abstract}
We prove here a result which strongly hints at the computational weakness of a model of tile assembly that has so far resisted many attempts of formal analysis or positive constructions. Specifically, we prove that, in Winfree's abstract Tile Assembly Model, when restricted to use only noncooperative bindings, any long enough path starting from the seed that can grow in all terminal assemblies is \emph{pumpable}, meaning that this path can be extended into an infinite, ultimately periodic path.

This result can be seen as a geometric generalization of the pumping lemma of finite state automata, and is a great step to solve the question of what can be computed deterministically in this model.  Moreover, this question has motivated the development of a new method called \emph{visible glues}. We believe that this method can also be used to tackle other long-standing problems in computational geometry, in relation for instance with self-avoiding paths.

Tile assembly (including non-cooperative tile assembly) was originally introduced by Winfree and Rothemund in STOC 2000 to understand how to \emph{program shapes}.
The \emph{non-cooperative} variant, also known as \emph{temperature 1 tile assembly}, is the model where tiles are allowed to bind as soon as they match on one side, whereas in cooperative tile assembly, some tiles need to match on several sides in order to bind.
Previously, exactly one known result (SODA 2014) showed a restriction on the assemblies general non-cooperative self-assembly could achieve, without any implication on its computational expressiveness.
With non-square tiles (like polyominos, SODA 2015), other recent works have shown that the model quickly becomes computationally powerful.

%As a corollary, we prove a conjecture by Doty, Patitz and Summers, published in 2009 in DNA~15. Moreover, since most known constructions using non-cooperative bindings to compute non-trivial algorithms have the property that we use, it gives an indication that two-dimensional, non-cooperative tile assembly may not be Turing universal, or at least that new constructions are needed.

\end{abstract}
\clearpage
\section{Introduction}

A possible approach to natural sciences is to try and write programs using the same kind of programming language as we think nature uses. If we can implement our theoretical algorithms in the actual natural systems, we will know the theory is meaningful to the systems studied. Through this process, we can learn from theorems \emph{reasons why} something is true, yielding insights beyond the \emph{modeling} of observed phenomena. This approach contrasts with other approaches where natural scientists test hypotheses against experiments to understand \emph{what happens}.

Although present since Turing~\cite{Turing1936}'s and Von Neumann's~\cite{vonNeumann1966} works, this idea has really been able to develop and extend into physical realizations only in recent years.
One of these realizations is the first implementation of Shor's algorithm in 2001~\cite{vandersypen2001}, providing a precious link between techniques for programming qubits devised by computer scientists, and the bricks actually used by nature.
Another achievement is the connection observed by Winfree in 1998~\cite{Winf98} between core concepts from theoretical computer science (computing machines, tilings) and a kind of building bricks devised by Seeman~\cite{Seem82} using DNA. One of the main models used in this connection, called the \emph{abstract Tile Assembly Model}, has yielded an impressive number of experimental demonstrations~\cite{WinLiuWenSee98,yurke2000dna,RothOrigami}.

Although they use different concepts and tools, these works use the same approach: trying to write programs using the language of nature (of physics in the former case, of chemistry in the latter), and confront these programs to the physical world by implementing them.

In this work, we study a long-standing open problem from the second approach by showing that a simple version of the programming language of tile assembly, although ubiquitous in many systems, is almost surely not able to perform computation. More precisely, in the abstract Tile Assembly Model, we are interested in the interactions and bindings of grounds of matter represented by square tiles, with glues of a certain \emph{color} and integer \emph{strength} on each of their four borders.
The dynamics start from an initial assembly called the \emph{seed}, and proceeds asynchronously and nondeterministically, one tile at a time, according to the following rule: a tile may attach to the current assembly if the sum of glue strengths on its sides that match the colors of adjacent tiles sum up to at least a parameter of the model called the temperature $\tau=1,2,3\ldots$. In particular, this means that unlike in Wang tilings (one inspiration of this model), adjacent tiles may have a \emph{mismatch}, i.e. disagree on the glue types of their common border.

This model is an abstraction of a simple chemical fact: when the temperature of a solution is increased, so is molecular agitation; for a tile to stay stably attached to an assembly, it needs then either stronger bonds to that assembly, or bonds to a larger neighborhood.

\paragraph{Temperature 1} This work will exclusively focus on \emph{temperature 1} tile assembly, also called \emph{non-cooperative} self-assembly. At higher temperatures, fewer assemblies are stable, allowing more control over producible assemblies: indeed, temperature~2 self-assembly is able to simulate arbitrary Turing machines \cite{Winf98,RotWin00,jCCSA}, and produce arbitrary connected shapes with a number of tile types within a log factor of their Kolmogorov complexity \cite{SolWin07}.  More surprisingly, this model has even been shown intrinsically universal \cite{IUSA}, meaning that there is a single tileset capable of simulating arbitrary tile assembly systems, modulo rescaling. In generalizations of this model, a single tile can even be sufficient to simulate all tile assembly systems and therefore all Turing machines~\cite{Demaine2014}.

In all known generalizations of the model, non-cooperative self-assembly is capable of arbitrary Turing computation: 3D cubic tiles~\cite{Cook-2011}, stochastic assembly sequences~\cite{Cook-2011}, hierarchical self-assembly~\cite{Versus}, polyominoes~\cite{Fekete2014}, duples~\cite{Hendricks-2014}, tiles with signals~\cite{Jonoska2014}, geometric tiles~\cite{geotiles} or negative glues~\cite{Patitz-2011}. Moreover, the synchronous version of this model can simulate arbitrary cellular automata.

The framework of intrinsic simulations (i.e. simulations up to rescaling) has recently yielded the first proof of a qualitative (and indeed geometric) difference between non-cooperative tile assembly and the more general model~\cite{Meunier-2014}. However, that result had no computational implications: indeed, it also holds in the three-dimensional generalization of temperature~1, known to be Turing-universal~\cite{Cook-2011}.

Therefore, an absolute requirement to understand this model seems to be an intuition on the role of planarity, and the shape of tiles.
Here, we introduce a new framework to study how information can be communicated in a planar space, via geometric interactions.
This framework will then (in the end of our proof) allow us to abstract geometric considerations away and reason on large boxes in which paths are forced to grow. This is a significant progress in the field, since the ``low-level geometry'' of paths producible at temperature~1 has been notoriously difficult to understand.

\paragraph{Relation to other works}
Doty, Patitz and Summers conjectured~\cite{Doty-2011}: \emph{for all temperature 1 directed tile assembly systems, there is a constant $c$ such that all paths longer than $c$ producible by the system are pumpable}. They also have shown that if this conjecture is true then the set of producible assemblies of temperature~1 directed tile assembly systems is therefore semi-linear, and hence computationally simple.

%In another recent work~\cite{t1notiu}, Meunier and Woods used the visible glues method to prove that there is no single tileset that can simulate all temperature~1 tile assembly systems at temperature 1. Exploiting stronger geometry characteristics of their problem, that work handles the full complexity of blocking many different paths in the same assembly, whereas ours deals with the case of a single path, but geometrically totally unrestricted.

Moreover, counting and sampling self-avoiding walks in the plane is an old problem at the intersection polymer chemistry and computer science, introduced by Flory~\cite{Flory53}; an early attempt to solve it was made by Knuth~\cite{knuth:math}, and this field has remained active today~\cite{Bousquet2010}. Another interpretation of our questions is the following problem, related to these works: starting from any point in $\zs$, start a self-avoiding walk deterministically (with memory). How far can you go without ever entering a cycle, if you only have $n$ bits of memory?

% \paragraph{Towards a synthesis?}
% The tools and problems studied in the various works discussed above, and in particular the probabilistic model~\cite{Cook-2011}, along with the other known use of the visible glues method~\cite{t1notiu} and the present work, certainly call for a synthesis; in particular, unifying our Lemma~\oldref{lem:uturns} and Lemma~5.6 of~\cite{t1notiu} could yield strong new insights on these problems.
% Moreover, an adaptation of this method to a probabilistic setting could certainly shed a new algorithmic light on the geometric counting and sampling problems described above.

\subsection{Main result}

Our result can be seen as a two-dimensional equivalent of the pumping lemma on deterministic finite automata \cite{Sipser}: we prove that if a non-cooperative tile assembly system can \emph{always} grow assemblies over a certain size (depending only on the size of their seed and on the number of tile types used), then these paths can be extended into ultimately periodic paths.

However, remark that non-cooperative systems can grow at least the same assemblies as cooperative ones: intuitively, their growth is ``harder to control'', resulting in more possible assemblies.  This is why our result is specific to patterns that can grow \emph{in all assemblies} producible by the system:

\begin{theorem}
\label{def:main:graph}
  Let $\mathcal{T}=(T,\sigma,1)$ be a tile assembly system such that the seed assembly $\sigma$ is finite and connected.  There is a constant $c(|T|,|\dom\sigma|)$ such that any path $P$, that can grow in all assemblies of $\mathcal{T}$ and reaches a point at a distance more than $c(|T|,\sigma)$ from $\sigma$, is \emph{pumpable}.
\end{theorem}

This result is different from the conjecture of Doty, Patitz and Summers since our paths starts from the seed. Nevertheless, using the tools developed in this paper, we think that our result is sufficient to prove that the set of producible assemblies of temperature~1 directed tile assembly systems is semi-linear. We will provide this result within a few months.

% \begin{theorem}
% \label{def:main:graph}
% Let $\mathcal{T}=(T,\sigma,1)$ be a tile assembly system, and $P$ be a path assembly producible by $\mathcal{T}$. If $P$ has at least one tile more than \pemm{TODO} tiles away from the seed (in Manhattan distance), then $P$ is pumpable or fragile (or both).
% \end{theorem}

The term \emph{pumpable} will be defined in Section \ref{def:main}. Intuitively, we say that $P$ is pumpable if one of its subpaths $P_{i,i+1,\ldots,j}$ can be repeated infinitely many times immediately after $P_{1,2,\ldots,j-1}$, without \emph{conflicting} with $\sigma$ nor with $P_{1,2,\ldots,j-1]}$, nor with any other repetition.

\label{sec:intro}

\section{Definitions and basic properties}
\label{def:main}
In this section, we give the definitions necessary to formalize the main result and some fundamental properties which will be used all along the article. First, we remind the classical notations on the $2D$ grid. Then, we introduce non-cooperative tile assembly and assemblies using standard formalism. Finally, we define path assemblies, fragility and pumpability. All the figures of this section are in appendix \ref{app:def}.

\subsection{Graphs and paths}
\label{def:graph}

\noindent \textbf{Graphs.} Let $\grid$ be the grid graph of $\mathbb{Z}^2$, i.e. the undirected graph whose vertices are the points of $\mathbb{Z}^2$, and for any two points $A,B\in\mathbb{Z}^2$, there is an edge between $A$ and $B$ if and only if $||AB||=\max(|x_B-x_A|,|y_B-y_A|)=1$. Note that along the article, we will exclusively use the maximum norm, \emph{i.e.} $||(x,y)||=\max(|x|,|y|)$. An element of $\mathbb{Z}^2$ is called a \emph{position}. For any graph $G'$, we will denote by $V(G')$ the set of its vertices and by $E(G')$ the set of its edges (for example, $V(\grid)=\mathbb{Z}^2$). Consider two graphs $G^1$ and $G^2$, $G^1$ is a \emph{subgraph} of $G^2$ if and only if $V(G^1)\subset V(G^2)$ and $E(G^1)\subset E(G^2)$. All the graphs considered in this article are subgraph of the grid graph $\grid$. The union of $G^1$ and $G^2$ is the graph $(G^1\cup G^2)=(V(G^1)\cup V(G^2), E(G^1)\cup E(G^2))$ (see Figure \ref{fig:def:graph}). All graphs considered in this article are connected and when the union of two graphs is done, the resulting graph will still be connected. For any position $A \in V(\grid)$, we denote $x_A$ its abscissa and by $y_A$ its ordinate, \emph{i.e.} $A=(x_A,y_A)$. For a graph $G'$, we denote by $x_{G'}=\min\{x_A: A \in V(G')\}$ if this value is finite and $X_{G'}=\max\{x_A: A \in V(G')\}$ if this value is finite (see Figure \ref{fig:def:graph}). The notations $y_{G'}$ and $Y_{G'}$ are defined similarly. Consider a vector $\vu \in \mathbb{Z}^2$, the translation of $G'$ by $\vu$ is the graph $G'+\vu$ where $V(G'+\vu)=\{A+\vu=A \in V(G)\}$ and $E(G'+\vu)=\{(x+\vu,y+\vu):(x,y) \in E(G')\}$. If the graph $G'$ is finite then its size is $|G'|=|V(G')|$.

\vspace{+0.5em}

\noindent \textbf{Paths.}  Since any path $P$ is also a graph, then all paths considered in this article are a subgraph of the grid graph $\grid$ and all the previous notations can also be used to study paths. A \emph{simple} path is a path with no loop. We will often represent a path as a sequence of vertices: if $P$ is finite we denote it as $(P_i)_{1\leq i \leq |P|}$ where for all $1\leq i \leq |P|$, $P_i \in V(P)$, for all $1\leq i \leq |P|-1$, $(P_i,P_{i+1}) \in E(P)$ and for all $2\leq i \leq \lastp-1$, $P_{i-1} \neq P_{i+1}$. We call such a representation of $P$, an \emph{indexing} of $P$. Remark that since $P$ is a non-oriented path, then if $V(P)$ is finite there exists two possible indexing of $P$ and along the article, we will precise which indexing is considered if necessary. %Also, when a path is defined, its indexing is always implicitly defined. 
When $V(P)$ is infinite, defining an indexing is more complex and two main cases may occur. In the first case, there exists a vertex of $V(P)$ with only one neighbor in $P$, then we say that $P$ is \emph{infinite} and there exists an indexing $(P_i)_{1\leq i}$ of $P$ where $P_1$ is the vertex with one neighbors in $P$ (see Figure \ref{fig:def:path}a). In the second case, all vertices of $P$ possess two neighbors in $P$ and then we say that $P$ is \emph{bi-infinite} and an indexing $(P_i)_{i \in \mathbb{Z}}$ of $P$ (see Figure \ref{fig:def:path}b) is defined as follow: one position of $V(P)$ is labeled with the index $0$ and one of the two possible orientations of the path has to be chosen. In this case, there exists an infinity of possible indexing.

Some proofs will require to consider a lot of different indices of the path, thus we introduce a specific notation to write them, let $1\leq \ind{1} \leq \lastp$ to denote a specific index $\ind{1}$ of the indexing of $P$. Consider a path $P$ and two indices $1\leq \ind{1} \leq \ind{2} \leq \lastp$ of $P$, we write $[\ind{1},\ind{2}]=\{\ind{1},\ind{1}+1,\ldots,\ind{2}\}$ and the notation $P_{[\ind{1},\ind{2}]}$ designs the path $Q$ where $|Q|=\ind{2}-\ind{1}+1$ and for all $1\leq \ind{3} \leq |Q|$ we have $Q_{\ind{3}}=Q_{\ind{1}+\ind{3} -1}$. Note that the path $Q$ is a subgraph of $P$, we say that $Q$ is a \emph{segment} of $P$. Consider an index $1\leq \ind{1} \leq \lastp$, such that $P_{\ind{1}}=(x,y)$ then we denote by $x_{P_{\ind{1}}}=x$ and by $y_{P_{\ind{1}}}=y$. A path $P$ is simple if and only if for all $1\leq \ind{1} <\ind{2} \leq \lastp$, we have $P_{\ind{1}} \neq P_{\ind{2}}$. 

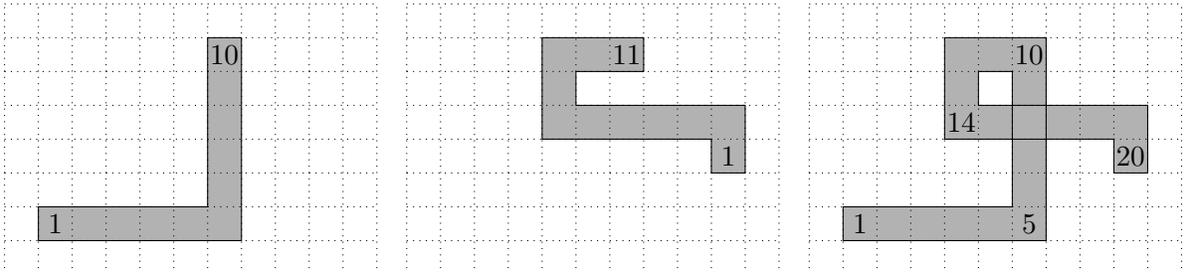
Concerning the union of two paths $P$ and $Q$, remark that $P \cup Q$ is not necessary a path. To avoid this problem, we introduce a specific operation to replace the union of two paths. Consider two paths $P$ and $Q$ such that $P_{\lastp}=Q_{1}$ (resp. $P_{\lastp}=Q_{|Q|}$), then the path $R=P\cdot Q$, called the \emph{concatenation} of $P$ and $Q$, is the path of length $\lastp+|Q|-1$ such that for all $1\leq \ind{1} \leq \lastp$ we have $R_{\ind{1}}=P_{\ind{1}}$ and for all $\lastp\leq \ind{1} \leq \lastp+|Q|-1$, we have $R_{\ind{1}}=Q_{\ind{1}-\lastp+1}$ (resp. $R_{\ind{1}}=Q_{|Q|+\lastp-\ind{1}}$). In this case, the concatenation of two paths is still a path. Nevertheless, remark that the concatenation of two simple paths is not necessary a simple path (see Figure \ref{fig:def:concat:path}).

For a finite path $P$, the \emph{direction} of $P$ is $\vu=\overrightarrow{P_{1}P_{\lastp}}$. Along the article, we will often need to extend a path $P$ of direction $\vu$ into an infinite or a bi-infinite path. To achieve this goal remark that making the concatenation $P\cdot (P+\vu)$ is possible. Then, we can introduce the following notations (see Figure \ref{fig:def:extension}): the paths $\omeplus{P}=\cdot_{0\leq i} (P+i\vu)$, $\omepluse{P}=\cdot_{1 \leq i} (P+i\vu)$, $\omemoi{P}=\cdot_{0\leq i} (P-i\vu)$ and $\omemoinse{P}=\cdot_{1 \leq i} (P-i\vu)$ are infinite paths and the path $\ome{P}=\cdot_{i \in \mathbb{Z}} (P+i\vu)$ is a bi-infinite path. These five paths are called the \emph{extensions} of $P$. Moreover, we generally need for these paths to be simple. Fortunately, there exists an easy test to determine if extending a finite path will generate a simple path: if $P$ is simple and $V(P)\cap V(P+\vu)=P_{\lastp}$ then all its extensions are simple. This result was proven in \cite{Demaine2014}. We formalize this result in the following definition and lemma. 

\begin{definition}
\label{def:good:cond}
Consider a finite path $P$ of direction $\vu$, we say that $P$ is a \emph{good} path if and only if $P$ is simple, finite and $V(P) \cap V(P+\vu)=P_{\lastp}$.
\end{definition}

\begin{lemma}
\label{lem:good:cond}
Consider a good path $P$ then $\omeplus{P}$, $\omepluse{P}$, $\omemoi{P}$, $\omemoinse{P}$ and $\ome{P}$ are simple paths.
\end{lemma}

Remark that, the extensions of a finite path $P$ are periodic. We will always consider an indexing of an extension of $P$ with the following property:

\begin{fact}
\label{fact:path:good}
Consider a finite path $P$ of direction $\vu$ and let $\ell=|P|-1$, then for all $\ind{1} \in \mathbb{Z}$, we have $\ome{P}_{\ind{1}+\ell}=\ome{P}_{\ind{1}}+\vu$. 
\end{fact}

Finally, consider a path $P$ and two indices $1\leq \ind{1} \leq \ind{2} \leq \lastp$ then the notation $\omepluse{P_{[\ind{1},\ind{2}]}}$ designs $\omepluse{(P_{[\ind{1},\ind{2}]})}$, \emph{i.e.} the extension of the segment $P_{[\ind{1},\ind{2}]}$ of path $P$.

\subsection{The abstract Tile Assembly Model.}

\vspace{+0.5em}
A \emph{tile type} is a unit square with four sides, each consisting of a \emph{glue label} and a nonnegative integer strength (see figure \ref{fig:tile}a). Formally, a tile $t=(n,e,s,o)$ is an element  of $(\glue\times\mathbb{N})^4$, where $\glue$ is a finite set of glue labels. Moreover, $n$ is called its \emph{north glue}, $e$ its \emph{east glue}, $s$ its \emph{south glue} and $w$ its \emph{west glue}.

Let $T$ be a finite set of tile types (see figure \ref{fig:tile:system}). An \emph{assembly over $T$} is a partial function of $\mathbb{Z}^2\dashrightarrow T$, whose domain is a connected component of $\mathbb{Z}^2$. Intuitively, an assembly is a positioning of tile types at some positions in the plane (see figure \ref{fig:assembly:exemple}a).

We say that two neighboring tiles of an assembly \emph{interact} if the glue labels on their abutting side are equal, and have positive strength (see figure \ref{fig:tile}b). An assembly $\alpha$ induces a weighted \emph{binding graph} $G_\alpha=(V_\alpha,E_\alpha)$, where $V_\alpha=\dom\alpha$, and there is an edge $(a,b)\in E_{\alpha}$ if and only if tiles $a$ and $b$ interact. An assembly $\alpha$ is said to be \emph{$\tau$-stable} if any cut of $G_\alpha$ has weight at least~$\tau$.

A \emph{tile assembly system} is a triple $\mathcal{T}=(T,\sigma,\tau)$, where $T$ is a finite tile set, $\sigma$ is a $\tau$-stable assembly called the seed, and $\tau\in\mathbb{N}$ is the temperature.
In this paper, $\tau$ will always be equal to~$1$.

Given two $\tau$-stable assemblies $\alpha$ and $\beta$, we say that $\alpha$ is a \emph{subassembly} of $\beta$, and write $\alpha\sqsubseteq\beta$ if $\dom\alpha\subseteq\dom\beta$, and for all position $p\in\dom\alpha$, $\alpha(p)=\beta(p)$. We also write $\alpha\rightarrow_1^{\mathcal T}\beta$ if $\alpha\sqsubseteq\beta$ and $|\dom\beta\setminus\dom\alpha|=1$ (i.e. if we can obtain $\beta$ from $\alpha$ by a single tile attachment).

We say that $\beta$ is \emph{producible} from $\alpha$, and write $\alpha\rightarrow^{\mathcal T}\beta$ (or simply $\alpha\rightarrow\beta$ if there is no ambiguity), if there is a (possibly empty) sequence $\alpha=\alpha_0\rightarrow_1^{\mathcal T}\alpha_1\rightarrow_1^{\mathcal T}\ldots\rightarrow_1^{\mathcal T}\alpha_{n-1}=\beta$.
The set of \emph{productions} of a tile assembly system $\mathcal{T}=(T,\sigma,\tau)$ is $\prodasm {\mathcal T}=\{\alpha | \sigma\rightarrow^{\mathcal T}\alpha\}$. Moreover, an assembly $\alpha$ is called \emph{terminal} if there is no $\beta$ such that $\alpha\rightarrow_1^{\mathcal{T}}\beta$, and the set of productions of a tile assembly system $\mathcal T$, that are terminal assemblies, is written $\termasm{\mathcal T}$. A tile assembly system is \emph{deterministic} if it has exactly one (potentially infinite) terminal assembly.

Note that the notations introduced for graph in section \ref{def:graph} can be adapted to assembly. For example, consider an assembly $\alpha$, then $X_\alpha=\max\{x_A:A \in V(G_\alpha)\}$. Only the union of two assemblies involves a subtle technical difficulty, that can be easily dealt with using proper vocabulary. When assemblies overlap, two different things can happen: either the assemblies disagree on the tile types they place at their common positions, or they agree. Consider two assemblies $\alpha$ and $\beta$ such that there is a position $A \in\dom\alpha\cap\dom\beta$, we say that $\alpha$ and $\beta$ intersect at position $A$. If $\alpha(A)\neq \beta(A)$, we say that $\alpha$ and $\beta$ \emph{conflict} at position $A$. On the other hand, we say that $\alpha$ and $\beta$ \emph{agree} at position $A$ if $\alpha(A)=\beta(A)$. If two assemblies $\alpha$ and $\beta$ agree on all their common positions, i.e. if there is an assembly $\gamma$ such that $\alpha\sqsubseteq\gamma$, $\beta\sqsubseteq\gamma$ and $\dom{\gamma}=\dom\alpha\cup\dom\beta$ is connected, we write $\alpha\cup\beta$ for this assembly $\gamma$ (see figure \ref{fig:union}).

\subsection{Path assemblies}

An important point about temperature $1$ tile assembly, is that any path in the binding graph of an assembly can start to grow, independent from anything else. More precisely, if $\mathcal{T}=(T,\sigma,1)$ is a tile assembly system, then for any $\alpha\in\prodasm{\mathcal{T}}$, and any path $P$ in the binding graph of $\alpha$ such that $P_1$ is in $\sigma$, an immediate induction on the length of $P$ shows that the restriction of $\alpha$ to $(\dom{\sigma} \cup P)$ is in $\prodasm{\mathcal{T}}$. %Note that to simplify notations, all finite sequences $S$ in the article are indexed from $1$ to $|S|$. For example, the path $P$ starts in $P_1$ and ends in $P_{|P|}$, where $|P|$ means ``the length of $P$''. We will denote such a sequence by $S=(S_i)_{1\leq i \leq |S|}$.

Since assemblies following paths are particularly important in our proof, we define them now using \emph{sequences} instead of the more general formalism of \emph{assemblies}: first, for any element $a=(p,t)\in\mathbb{Z}^2\times T$, we call $p$ the \emph{position} of $a$, written $\pos(a)$, and $t$ the \emph{type} of $a$, written as $\type(a)$. We also write the position of $a$ as $(x_a,y_a)$. Let then $P=(P_{i})_{1\leq i \leq |P|}$ be any sequence of $\mathbb{Z}^2\times T$. If for all $1\leq \ind{1} \leq \ind{2} \leq \lastp$, we have $\pos(P_{\ind{1}})=\pos(P_{\ind{2}}) \Rightarrow\type(P_{\ind{1}})=\type(P_{\ind{2}})$ and $\dom{P}=\bigcup_{1\leq i \leq |P|} (\pos(P_i))$ is connected, we define the assembly \emph{induced} by a sequence $P$ as the assembly  $\alpha_P$ such that $\dom{\alpha_P}=\dom{P}$ and for all $i$, $\alpha(\pos(P_i))=\type(P_i)$. We call the sequence of positions $(\pos(P_i))_{i\in\{1,2,\ldots,|P|\}}$ the \emph{underlying path} of $P$. Moreover, if the underlying path of $P$ is a path and if $P$ induces an assembly $\alpha_P$ such that for all $i\in\{1,2,\ldots,|P|-1\}$, the tiles at positions $(x_i,y_i)$ and $(x_{i+1},y_{i+1})$ in $\alpha_P$ interact, we call $P$ a path assembly
%the sequence of positions of $P$, i.e. $(\pos(P_i))_{i\in\{1,2,\ldots,|P|\}}$, is a path of $\grid$, and if $P$ induces an assembly $\alpha_P$ such that for all $i\in\{1,2,\ldots,|P|-1\}$, the tiles at positions $(x_i,y_i)$ and $(x_{i+1},y_{i+1})$ in $\alpha_P$ interact, we call $P$ a \emph{path assembly} 
(even though not formally an assembly, since a path assembly is a sequence of $\mathbb{Z}^2\times T$, and an assembly is a function of $\mathbb{Z}^2\rightarrow T$), see Figure \ref{fig:path}. The binding graph of $P$ is the binding of $\alpha_P$. A path assembly $P$ is \emph{simple} if and only if its underlying path is a simple path. Abusively, we denote the assembly $\alpha_P \cup \seed$ by $P\cup\seed$. Then, a path assembly $P$ is producible by a tiling system $\mathcal{T}=(T,\sigma,1)$ if only if there is no conflict between $P$ and $\seed$ and the assembly $P \cup \seed$ is producible by $\mathcal{T}$. The distance from a path assembly to the seed is defined by $\max \{||AB||: A \in \dom{P} \text{ and } B \in \dom{\seed}\}$.

Also, notations used for paths and assemblies have a counterpart for path assemblies. For example, we denote by $Y_P=\max_{1 \leq i \leq \lastp}\{y_{P_i}\}$ and for any two indices $1\leq \ind{1}\leq \ind{2}\leq \lastp$, we denote by $Q=P_{[\ind{1},\ind{2}]}$ the path assembly of length $\ind{2}-\ind{1}+1$ such that for all $1\leq i \leq \ind{2}-\ind{1}+1$, we have $Q_i=P_{\ind{1}+i}$. The path assembly $P_{[\ind{1},\ind{2}]}$ is called a \emph{segment} of the path assembly $P$ and the assembly induced by $P_{[\ind{1},\ind{2}]}$ is a subassembly of the assembly induced by $P$. Consider a path assembly $P$ producible by a tiling system  $\mathcal{T}=(T,\sigma,1)$, if $\pos(P_1) \in \dom{\seed}$ then for all $1\leq \ind{1} \leq \lastp$, the path assembly $P_{[1,\ind{1}]}$ is also producible by this tiling system. This is not necessary the case of $P_{[\ind{1},\ind{2}]}$ with $1<\ind{1} \leq \ind{2} \leq \lastp$. Consider two path assemblies $P$ and $P'$ such that $P_{\lastp}=P'_1$, then the concatenation of $P$ and $P'$ is defined as the sequence $Q=P\cdot P'$  such that $|Q|=|P|+|P'|-1$ and for all $1\leq i \leq \lastp$, we have $Q_i=P_i$ and for all $1 \leq i \leq |P'|$, we have $Q_{i+\lastp-1}=P'_i$. Note that the sequence $Q$ is not a path assembly if there is a conflict between $P$ and $P'$. If all intersections between $P$ and $P'$ are agreements then $Q$ is also a path assembly. Also note that $P$ can be producible by a tiling system $\tiling$ but not the path assembly $Q$, this event occurs when $P'$ collides with the seed. Finally, remark that the concatenation of two simple path assemblies which are in agreement is not necessarily simple. 

Extending a finite path assembly into an infinite or bi-infinite sequence is more complex than extending a path. First, the \emph{direction} of a finite path assembly is $\vu=\overrightarrow{\pos(P_1)\pos(P_{\lastp})}$ (in the rest of the article, we will abusively denote by $\overrightarrow{P_iP_j}$ the vector $\overrightarrow{\pos(P_i)\pos(P_j)}$). Secondly, a path assembly $P$ of direction $\vu$ is a \emph{candidate} if and only if $P$ is finite, its direction $\vu$ is not null and $\type(P_1)=\type(P_{\lastp})$. Consider a candidate path assembly $P$, then we introduce the extensions of $P$  as $\omeplus{P}=\cdot_{0\leq i} (P+i\vu)$, $\omepluse{P}=\cdot_{1 \leq i} (P+i\vu)$, $\omemoi{P}=\cdot_{0\leq i} (P-i\vu)$ and $\omemoinse{P}=\cdot_{1\leq i} (P-i\vu)$ which are infinite sequences and $\ome{P}=\cdot_{i \in \mathbb{Z}} (P+i\vu)$ which is a bi-infinite sequence. These extensions are correctly defined if and only if $P$ is a candidate path assembly. Similarly to what has been done to path, we introduce good path assemblies:

\begin{definition}
\label{def:good:cond:assem}
Consider a path assembly $P$ of direction $\vu$, we say that $P$ is a \emph{good} path assembly if and only if $P$ is simple, finite, $\dom{P} \cap \dom{P+\vu}=\pos(P_{\lastp})$ and a candidate path assembly.
\end{definition}

\begin{lemma}
\label{lem:good:cond:assem}
Consider a good path assembly $P$ then $\omeplus{P}$, $\omepluse{P}$, $\omemoi{P}$, $\omemoinse{P}$ and $\ome{P}$ are simple path assemblies.
\end{lemma}

Moreover, the periodicity of the extensions of $P$ are summarized in the following fact:

\begin{fact}
\label{fact:ext:per}
Consider a candidate path assembly $P$ of direction $\vu$ and let $\ell=|P|-1$, then for all $\ind{1} \in \mathbb{Z}$, we have $\ome{P}_{\ind{1}+\ell}=\ome{P}_{\ind{1}}+\vu$. \end{fact}

\subsection{Fragility and pumpability}
\label{sec:def:fragility}

We now define the two properties of path assemblies that we will study along the article and we discuss some fundamental facts. First a path assembly is fragile if it is does not appear in all terminal assembly.

\begin{definition}[Fragility]
\label{def:fragile}
Let $\mathcal{T}=(T,\sigma,1)$ be a tile assembly system. We say that a path assembly $P$ is \emph{fragile} when there is at least one terminal assembly $\alpha\in\termasm{\mathcal{T}}$ of which the assembly induced by $P$ is not a subassembly. 
\end{definition}

According to this definition, if an assembly admits a non-fragile path assembly, then it can always be produced from any assembly. An efficient and practical way to prove the fragility of a path assembly $P$ is to find another path assembly $P'$ which is in conflict with $P$ (see figure \ref{fig:fragile}).

\begin{fact}
\label{fact:fragile}
Consider a path assembly $P$ producible by a tiling system $\tiling=(T,\sigma,1)$, if there exists another path assembly $P'$ producible by $\tiling$ such that $P$ and $P'$ are in conflict then $P$ is fragile.
\end{fact}

Note that in this case $P'$ is also fragile and the tiling system is not deterministic. Now, the segment of a path assembly is pumpable if it is possible to extend it infinitely (see figure \ref{fig:def:pump}a). %In the following definition and in the rest of the article, we will abusively denote by $\overrightarrow{P_iP_j}$ the vector $\overrightarrow{\pos(P_i)\pos(P_j)}$.

\begin{definition}[Pumpability]
\label{def:pump:seg}
Let $\mathcal{T}=(T,\sigma,1)$ be a temperature $1$ tile assembly system, and $P$ be a path assembly producible by $\mathcal{T}$, of length at least $2$ and two indices $1\leq \ind{1} <\ind{2} \leq \lastp$. The segment $P_{[\ind{1},\ind{2}]}$ of $P$ is pumpable if and only if it is a candidate segment and either the sequence $P_{[1,\ind{1}]}\cdot \omeplus{P_{[\ind{1},\ind{2}]}}$ is a path assembly producible by $\mathcal{T}$ or the sequence $P_{[1,\ind{1}]}\cdot \omemoinse{P_{[\ind{1},\ind{2}]}}$ is a path assembly producible by $\mathcal{T}$.
%following sequence $Q$ of $\mathbb{Z}^2\times T$ is a path assembly producible by $\tiling$: $Q$ is defined for all integer $k\in\mathbb{N}$ by $Q_k=P_k$ if $k<i$, and $Q_k=P_{i+((k-i)\mod (j-i))} + \left\lfloor\frac{k-i}{j-i}\right\rfloor\vect{P_iP_j}$. 
%The path assembly $P$ is \emph{pumpable} if and only if there exists a pumpable segment in $P$. 
\end{definition}

Since our representation of a path assembly implicitly implies an orientation of its underlying path, we have two distinguish two cases for pumpability. Consider a segment of direction $\vu$ of a path assembly $P$, then the segment can be pumped in direction $\vu$ (see Figure \ref{fig:def:pump}a) or in direction $\vu$ (see Figure \ref{fig:def:pump}b). %Remark that we distinguish two cases of pumpability. Consider the path of figure , the segment $P_{[2,4]}$ is not pumpable (figure \ref{fig:reverse}b) whereas it can be pumped in the direction $\overrightarrow{P_4P_2}$ (see figure \ref{fig:reverse}c). 
In the second case, we call this operation \emph{reverse pumping}. Reserve pumping will be used only one time in the proof of lemma \ref{hunt:peszr}. Otherwise, we will always pump a segment of a path assembly according to the direction of the path assembly. Also, in this definition, we require that $\ind{1}<\ind{2}$ whereas in some cases, it is possible to pump segment of length $1$. Nevertheless, we will never consider such segments along the proof.

There is a subtlety to define the pumpability of a path assembly or an assembly. Indeed, if a path assembly contains a pumpable segment then it is not necessary "pumpable": it could be fragile. For example the path assembly of Figure~\ref{fig:def:pump}a is in conflict with the assembly $P$ of figure~\ref{fig:path}. The following definition take this case into account. 

\begin{definition}[Pumpability or fragility]
\label{def:pump:til}
Let $\mathcal{T}=(T,\sigma,1)$ be a temperature 1 tile assembly system, and $P$ a path assembly producible by $\tiling$. The path assembly $P$ is \emph{fragile or pumpable} if and only if $P$ is fragile or there exists $1\leq \ind{1} \leq \ind{2} \leq \lastp$ such that $P_{[\ind{1},\ind{2}]}$ is a pumpable candidate segment. Moreover, consider $\beta$ an assembly producible by $\tiling$, we say that $\beta$ is \emph{fragile or pumpable} if and only if $\beta$ is fragile or if there exists a path assembly $P$ such that $\alpha_P \sqsubseteq \beta$ and $P$ is fragile or pumpable.
\end{definition}

%Similarly we defined an assembly fragile or pumpable with reversed pumping allowed.

%\begin{figure}[th]
%\centerline{
%\includegraphics[width=5cm]{./photo/imagesintro/pumpability.jpeg}
%}
%\caption{The segment $P_{[2,7]}$ of path assembly $P$ (figure \ref{fig:path}) is pumpable. The assembly $\beta$ of figure \ref{fig:union}c admits $\alpha_P$ as a subassembly, leading to the fragility or pumpability of $\beta$. In fact, $\beta$ is fragile since it conflicts with the pumping of $P$.}
%\label{fig:pump}
%\end{figure}

%Consider a candidate segment $P_{[i,j]}$ of $P$, the \emph{pumping} of the candidate segment $P_{[i,j]}$ is the infinite sequence $P^\infty=\cdot_{k \geq 1} (P_{[i,j]}+k\overrightarrow{P_iP_j})$. The segment $P_{[i,j]}$ is pumpable if and only if $P_{[1,j]} \cdot P^\infty$ is a path assembly producible by $\tiling$. 

Consider a path assembly $P$ and a candidate segment $P_{[\ind{1},\ind{2}]}$ of $P$. In the best case, this segment is pumpable and we have obtained the desired result. Otherwise, remark that $Q=P_{[1,\ind{1}]} \cdot \ome{P_{[\ind{1},\ind{2}]}}$ is a sequence where two consecutive tiles interact, then if $Q$ is not a path assembly producible by $\tiling$ then $\ome{P_{[\ind{1},\ind{2}]}}$ creates at least one conflict. We need to know the exact position of this conflict. If several conflicts exist, we are interested in the first one (see Figure \ref{fig:pumpAcc}).

\begin{definition}
Consider a path assembly $P$ producible by a tiling system $\tiling=(T,\seed,1)$ and two indices $1\leq \ind{1} < \ind{2} \leq \lastp$ such that $P_{[\ind{1},\ind{2}]}$ is a candidate segment of $P$ which is not pumpable. Let $Q=\omepluse{P_{[\ind{1},\ind{2}]}}$, then the first conflict of $P_{[\ind{1},\ind{2}]}$ is the index $1 \leq \ind{3}$ such that $\ind{3}=\min\{i: P_{[1,\ind{2}]} \cdot Q_{[1,i]} \text{ is not producible by } \tiling \}$.
\end{definition} 

Note that this definition implies that $\ind{3}>1$. Also, we do not consider reverse pumping for this definition since reverse pumping is marginal. Two kinds of obstacle may prevent the candidate segment to be pumpable, either the sequence $Q$ is not a path assembly ($Q$ conflicts with itself) or $Q$ is a path assembly which conflicts with $\sigma \cup P_{[1,\ind{2}]}$ . We now show that whatever happens the first conflict always occurs in the domain of $\sigma \cup P_{[1,\ind{2}]}$. This result is due to the periodicity of $Q$. %if there is a conflict between $Q_{\ind{4}}$ and $Q_{\ind{5}}$ then there is a conflict between $Q_{\ind{4}}-\vu$ and $Q_{\ind{5}}-\vu$.

\begin{lemma}
\label{lem:good:week}
Consider a path assembly $P$ producible by a tiling system $\tiling=(T,\sigma,1)$ and two indices $1\leq \ind{1} \leq \ind{2} \leq \lastp$ such that $P_{[\ind{1},\ind{2}]}$ is a candidate segment of $P$ which is not pumpable. Let $Q=\omepluse{P_{[\ind{1},\ind{2}]}}$ and $\ind{3}$ be the first conflict of $P_{[\ind{1},\ind{2}]}$, then $\pos(Q_{\ind{3}}) \in \dom{\seed \cup P_{[1,\ind{2}]}}$.
%if $P^\infty$ is not a path assembly and if $P^\infty$ does not conflict with $P_{[0,i-1]}\cup \seed$ then the first conflict occurs in $\dom{P_{[i,j]}}$, \emph{i.e.} let $Q$ be the pumping of $P^\infty$ until the first conflict then $\pos(Q_{|Q|}) \in \dom{P_{[i,j]}}$.
\end{lemma}

\begin{proof}
If $P_{[\ind{1},\ind{2}]}$ is not pumpable then either $Q$ is a path assembly or $Q$ is not a path assembly. If $Q$ is a path assembly, since $P_{[\ind{1},\ind{2}]}$ is not pumpable then there exists a conflict between $Q$ and $\seed \cup P_{[1,\ind{2}]}$. In this case, the lemma is true. Otherwise if $Q$ is not a path assembly, for the sake of contradiction suppose that $Q_{\ind{3}} \notin \dom{\seed \cup P_{[1,\ind{2}]}}$. Then there exists $1 \leq \ind{4} < \ind{3}$ such that $\pos(Q_{\ind{3}})=\pos(Q_{\ind{4}})$ and $\type(Q_{\ind{3}}) \neq \type(Q_{\ind{4}})$. Let $\ell=|P_{[\ind{1},\ind{2}]}|-1$ and $\vu$ be the direction of $P_{[\ind{1},\ind{2}]}$. If $\ell<\ind{4}$ then by fact \ref{fact:ext:per}, we have $Q_{\ind{3}-\ell}=Q_\ind{3}-\vu$ and $Q_{\ind{4}-\ell}=Q_\ind{4}-\vu$. Then there exists a conflict between $Q_{\ind{3}-\ell}$ and $Q_{\ind{4}-\ell}$ which contradicts the definition of $\ind{3}$. If $\ind{4} \leq \ell$ then by fact \ref{fact:ext:per} and by definition of $Q$, there exists $\ind{1} \leq \ind{5} \leq \ind{2}$ such that $P_{\ind{5}}=Q_\ind{4}-\vu$. Now, if $\ell<\ind{3}$ then $Q_{\ind{3}-\ell}=Q_{\ind{3}}-\vu$ and there exists a conflict between $Q_{\ind{3}-\ell}$ and $P_{\ind{5}}$ which contradicts the definition of $\ind{3}$. Finally if $\ind{3} \leq \ell$ then by fact \ref{fact:ext:per} and by definition of $Q$, there exists $\ind{1} \leq \ind{6} \leq \ind{2}$ such that $P_{\ind{6}}=Q_\ind{3}-\vu$. Thus there is a conflict between $P_{\ind{5}}$ and $P_{\ind{6}}$ and $P$ is not a path assembly which is a contradiction. All cases lead to a contradiction and the lemma is true.
%Then there exists a conflict between $Q_{\ind{3}-\ell}$ and $Q_{\ind{4}-\ell}$ which contradicts the definition of $\ind{3}$.
%
%By hypothesis, the first conflict does not belong to $\dom{\seed \cup P_{[1,i-1]}}$. Thus, the only possibility is that $Q$ creates a conflict with itself: $Q$ is not a path assembly. Then, we have $Q_{|Q|} \in \dom{Q_{[2,|Q|-1]}}$. Thus, there exists $1< k < |Q|$ such that $Q_k$ and $Q_{|Q|}$ are in conflict. If $|Q|\leq \ell$ then $Q_k$ and $Q_{|Q|}$ both belong to $P_{[i,j]}$ (see fact \ref{fact:pumping}) and this would contradict the fact that $P$ is a path assembly. If $|Q|> \ell$ then there exist a conflict between $Q_{|Q|-\ell}$ and $Q_{k}-\overrightarrow{P_iP_j}$. Note that $Q_{k}-\overrightarrow{P_iP_j}$ either belongs to $P_{[i,j]}$ or $Q_{[1,|Q|-\ell]}$ (see fact \ref{fact:pumping}) and this would contradict the definition of $Q$ which should have ended at index $|Q|-\ell$.  
%If $k \leq \ell$ then the lemma is true. For the sake of contradiction, assume that $k>\ell$ and $\pos(Q_k)\notin \dom{P_{i,j}}$. Then by the periodicity of the pumping (fact \ref{fact:pumping}), $\pos(P_{k-\ell})=\pos(P_{k})-\overrightarrow{P_iP_j}=\pos(P_{|Q|})-\overrightarrow{P_iP_j}=\pos(P_{|Q|-\ell})$ and $\type(P_{k-\ell})=\type(P_{k})\neq \type(P_{Q-\ell})$. This implies that $P_{k-\ell}$ and $P_{|Q|-\ell}$ are in conflict and that $Q_{[1,|Q|-\ell]}$ is not a path assembly. This fact contradicts the definition of $Q$ as the pumping of $P_{[i,j]}$ until the first collision.
\end{proof}

%As a corollary of this result, the first conflict always occurs in $\dom{\seed} \cup \dom{P_{[1,j]}}$. Now, we will give a sufficient condition for a candidate segment to avoid conflict in its pumping. This condition is simple, the candidate segment does not collide with its first translation. 
%
%\begin{definition}
%\label{def:good:cond}
%Consider a candidate segment $P_{[i,j]}$ of a path assembly $P$. We say that $P_{[i,j]}$ is a \emph{good} candidate segment if and only if $P_{[i,j]}$ and $P_{[i,j]}+\overrightarrow{P_iP_j}$ intersect only in $\pos(P_j)$, \emph{i.e.} $\dom{P_{[i,j]}}\cap \dom{P_{[i,j]}+\overrightarrow{P_iP_j}}=\pos(P_j)$.
%\end{definition}
%
%\begin{lemma}
%\label{lem:good:cond}
%Consider a candidate segment $P_{[i,j]}$ of a simple path assembly $P$. If $P_{[i,j]}$ is a good candidate segment then its pumping is a simple path assembly.
%\end{lemma}

Note that, if we consider a good candidate segment $P_{[\ind{1},\ind{2}]}$ of a path assembly $P$ then its first collision cannot occurs in $\dom{P_{[\ind{1},{\ind{2}}]}}$. 

\begin{corollary}
\label{cor:good:week}
Consider a path assembly $P$ producible by a tiling system $\tiling=(T,\sigma,1)$ and two indices $1\leq \ind{1} \leq \ind{2} \leq \lastp$ such that $P_{[\ind{1},\ind{2}]}$ is a good segment of $P$ which is not pumpable. Let $Q=\omepluse{P_{[\ind{1},\ind{2}]}}$ and $\ind{3}$ be the first conflict of $P_{[\ind{1},\ind{2}]}$, then $\pos(Q_{\ind{3}}) \in \dom{\seed \cup P_{[1,\ind{1}-1]}}$.
%if $P^\infty$ is not a path assembly and if $P^\infty$ does not conflict with $P_{[0,i-1]}\cup \seed$ then the first conflict occurs in $\dom{P_{[i,j]}}$, \emph{i.e.} let $Q$ be the pumping of $P^\infty$ until the first conflict then $\pos(Q_{|Q|}) \in \dom{P_{[i,j]}}$.
\end{corollary}

\section{Roadmap}
\label{sec:roadmap2}

%\textcolor{red}{twelve} different parts corresponding to the \textcolor{red}{twelve} major steps of the main proof. The document is divided in \textcolor{red}{three} sections and these sections contains \textcolor{red}{twelve} subsections, one for each of these sets of lemmas and three to introduce the tools and the notations needed in the sections. Unfortunately, these twelve parts do not depend from each other in a sequential order. In fact, the dependencies between them are rather labyrinthic as shown in Figure \ref{fig:dependancies}. Note that these dependencies do not introduce a cycle, five parts have no prerequisite (\emph{Discrete Geometry Toolbox}, \emph{Imperfect Algorithm for building stakes}, \emph{Jailed path assembly}, \emph{Stripe lemma} and \emph{Visible glue toolbox}) and the seven remaining ones can be proven in the following order: \emph{U-turn}, \emph{Seed near the origin}, \emph{Reset lemma}, \emph{Black box and stakes}, \emph{Hunt}, \emph{Eight "easy" cases} and finally \emph{Firing at the black box}. This complexity comes from our need to develop a toolbox of four macroscopic lemmas. %In fact, we prove four lemmas which lead to the final result.  

\begin{figure}
\begin{tikzpicture}[node distance=0.45cm, auto]  
\tikzset{
    mynode/.style={rectangle,rounded corners,draw=black, very thick, inner sep=0.5em, minimum size=0.5em, text centered},
    myarrow/.style={->, >=latex', shorten >=1pt, thick},
    mylabel/.style={text width=3em, text centered} 
}  
%noeud
\node[mynode, text width=2.4cm] (DGS) {\scriptsize Two Dimensional Discrete Toolbox\\ \scriptsize Section \ref{sec:dgt}};  
\node[mynode,below=1cm of DGS,text width=2.4cm] (WML) {\scriptsize Window Movie Lemma\\ \scriptsize See \cite{Meunier-2014}};
\node[mynode,right=3.8cm of WML.south,anchor=south, text width=1.9cm] (MWL) {\scriptsize Modified Movie Lemma\\ \scriptsize Lemma \ref{lem:uwl:final}};
\node[mynode,right=2.8cm of MWL.south, anchor=south, text width=1.8cm] (TPA) {\scriptsize Thin Path Assembly\\ \scriptsize Lemma \ref{lem:thin}};
\node[mynode,right=2.8cm of TPA.south, anchor=south, text width=1.7cm] (Fork) {\scriptsize Fork\\ \scriptsize Lemma \ref{lem:fork}};
\node[mynode,right=2.8cm of Fork.south, anchor=south, text width=1.7cm] (U-turn) {\scriptsize U-turn\\ \scriptsize Lemma \ref{lem:uturn}};
\path (Fork.center) -- (U-turn.center) node[midway,anchor=center] (tempS7) {};
\node[mynode,above=1.5cm of tempS7,text width=4.4cm] (VTT) {\scriptsize Visible Tile Toolbox\\ \scriptsize Section \ref{sec:sub:vis}};
\node[mynode,below=2cm of MWL,text width=1.7cm] (Jail) {\scriptsize Jail Lemma\\ \scriptsize Lemma \ref{macro:seed}};
\node[mynode,right=2.8cm of Jail.south, anchor=south, text width=1.9cm] (Stakes) {\scriptsize Stakes Lemma\\ \scriptsize Lemma \ref{macro:reset}};
\node[mynode,right=2.8cm of Stakes.south, anchor=south, text width=1.7cm] (Reset) {\scriptsize Reset Lemma\\ \scriptsize Lemma \ref{macro:stakes}};
\node[mynode,right=2.8cm of Reset.south,anchor=south, text width=1.7cm] (Seed) {\scriptsize Seed Lemma\\ \scriptsize Lemma \ref{macro:jailed}};
\node[mynode,below=2cm of Stakes,text width=5.6cm] (MacroL) {\scriptsize Macroscopic Reasoning and Main Result\\ \scriptsize Section \ref{sec:macro} and Theorem \ref{theorem:final:reverse}};

%cadre
\path (MWL.west) -- (TPA.east) node[midway,anchor=center, mynode, text width=5.2cm,minimum height=1.7cm] (S6) {};
\node[text centered, above=0.15cm of S6,text width=2cm] (S6cap) {Section \ref{sec:uwl}};
\path (tempS7.center) -- (VTT.center) node[midway,anchor=center, mynode, text width=5cm,minimum height=3.6cm] (S7) {};
\node[text centered, above=0.15cm of S7,text width=2cm] (S7cap) {Section \ref{sec:micro}};
\path (MWL.west) -- (U-turn.east) node[midway,anchor=center] (MicroTemp) {};
\node[text centered, below=0.84cm of MicroTemp,anchor=south, mynode, text width=11.2cm,minimum height=4.64cm] (Micro) {};
\node[text centered, above=0.15cm of Micro,text width=4cm] (MicroCap) {Microscopic Reasoning};
\path (Jail.west) -- (Seed.east) node[midway,anchor=center] (S8temp) {};
\node[text centered, below=0.75cm of S8temp,anchor=south, mynode, text width=11.2cm,minimum height=2.1cm] (Macro) {};
\node[left=1.93cm of Macro,anchor=center] (Macrotemp) {};
\node[above=1cm of Macrotemp,anchor=north,text width=3.9cm] (MicroCap) {Macroscopic Toolbox \\ \vspace{+0.5em} \scriptsize Statements are in Section \ref{sec:macro} \\ \vspace{+0.25em} \scriptsize Proof of Jailed lemma is in  Section \ref{sec:uwl} \\ \vspace{+0.25em} \scriptsize Proof of remaining lemmas are in Section \ref{sec:lastpart} };

\draw[myarrow] (WML) -- (MWL);	
\draw[myarrow] (DGS) -- (Micro);	
\draw[myarrow] (MWL) -- (TPA);	
\draw[myarrow] (MWL) -- (Jail);	
\draw[myarrow] (Jail) -- (Stakes);	
\draw[myarrow] (Reset) -- (Stakes);	
\draw[myarrow] (Jail) -- (MacroL);	
\draw[myarrow] (Reset) -- (MacroL);	
\draw[myarrow] (Stakes) -- (MacroL);	
\draw[myarrow] (Fork) -- (U-turn);	
\draw[myarrow] (VTT) -- (Fork);	
\draw[myarrow] (VTT) -- (U-turn);	
\draw[myarrow] (Seed) -- (Reset);	
\draw[myarrow] (U-turn) -- (Seed);	
\draw[myarrow] (U-turn) -- (Reset);	
\draw[myarrow] (Fork) -- (Stakes);	
\draw[myarrow] (TPA) -- (Stakes);	
\draw[myarrow] (Seed.south)  -- ++(0,-0.4cm) -- ++(0,-0.4cm) -|   (Stakes.310);	
\draw[myarrow] (VTT.east) -- ++(0.4cm,0) -- ++(0.4cm,0) |- (Seed.east);	

%\draw[myarrow] (manufacturer.south) -- ++(-.5,0) -- ++(0,-1) -|  (retailer1.north);	

\end{tikzpicture} 
\caption{The dependencies between the different parts of the proof.}
\label{fig:dependancies}
\end{figure}
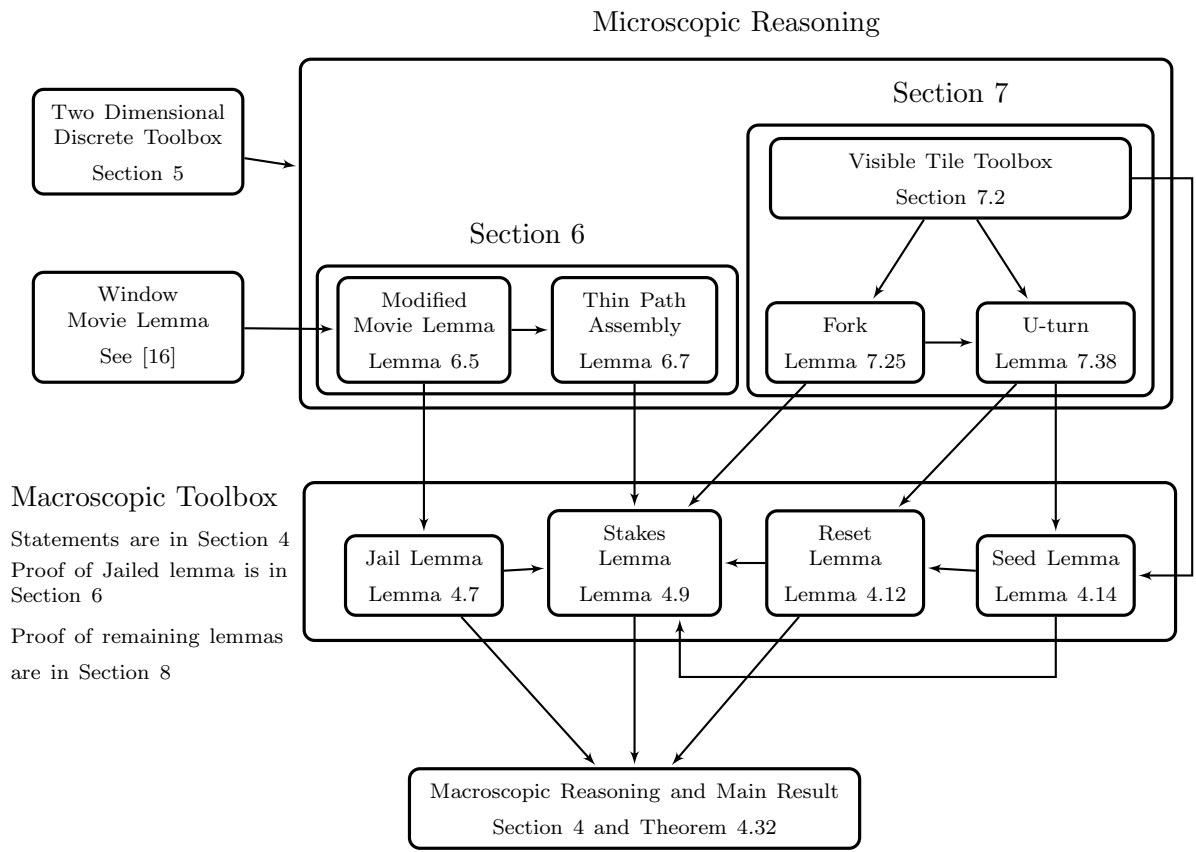

The proof of the main result is long, difficult and requires a lot of transitional lemmas. These lemmas are gathered into different parts as shown in Figure \ref{fig:dependancies} (these dependencies do not introduce a cycle). Note that, %except for the \emph{Discrete Geometry Toolbox} which contains some results about discrete $2D$ geometry which are not specific to tile assembly system, 
all results of this paper can be categorized in two categories: microscopic and macroscopic ones. Microscopic results focus on a local part of an assembly. A simple typical example is "a segment of $|T|+1$ tiles of an assembly contains at least two identical tiles". On the one hand, these results requires no prerequisites but on the other hand, they deal with very specific conditions (\emph{i.e.} a long list of hypothesis), they are not very useful alone and their proofs are technical and could lead to an explosion of cases to study if not done carefully. Thus microscopic reasoning is lengthy, tedious, technical and not intuitive. Macroscopic results zoom out and focus on the whole assembly. A typical macroscopic results is formulated as "an assembly growing at distance $d$ of the seed should respect property $\mathcal{P}$", the main result of this article is a macroscopic one. Macroscopic results are powerful and intuitive. Their only weakness  is that their proofs rely on lengthy microscopic reasoning and thus their proofs are extremely complex. %To our knowledge the only macroscopic results currently known is the window movie lemma \cite{}. 

Thus, the main difficulty is to get through microscopic reasonings to reach macroscopic ones. To achieve this aim, we develop a toolbox made of four macroscopic lemmas. The combination of these lemmas leads to the final result. To present all of our results we divide the rest of the article in five main sections. In section \ref{sec:macro}, we present the four lemmas of our macroscopic toolbox and we show that this toolbox is enough to solve the conjecture. This approach is more pedagogical, the reader starts with the easiest part of the proof and the aim of the following sections have become clear. Section~\ref{sec:dgt} is dedicated to a $2D$ discrete toolbox. Here we present results some  tools and results which are fundamental to all microscopic reasonings. It aims to develop an effective way to cut the grid into different zones. In section \ref{sec:uwl}, we prove one lemma of our macroscopic toolbox whose proof is independent from the other ones (see jail lemma in Figure \ref{fig:dependancies}). In the following section \ref{sec:micro}, we develop a powerful tool about U-turn, this part is the longest and hardest of the proof and a prerequisite for all the remaining macroscopic lemma. In the last section \ref{sec:lastpart}, we conclude by proving the three remaining macroscopic lemmas. Their proofs are consecutive: the reset lemma is required for the stakes lemma and the seed lemma is required for the reset lemma (see Figure \ref{fig:dependancies}). With this approach, the difficulty increases along the article. Also, this presentation allow us to introduce the different tools one by one and to illustrate them in "simple" cases before moving to the hardest ones. %Moreover, when reaching the section dedicated to microscopic reasoning, aims have become clear. 

%The demonstrations of three of these lemmas are heavily linked and 

%If this document followed the logical order, then it would start by microscopic reasoning to prove one of this tool which would be saved for later and we would start again microscopic reasoning to prove another part of our toolbox. When all proofs of our toolbox would have been proven, then we would move on to solve the conjecture. Such a presentation forces the reader to deal with the hardest and more labyrinthic parts of the proof first without giving any intuition of the aims to be achieved. Moreover, we would have to start with a lengthy section of definitions since all the different tools introduced in the article are used in the microscopic reasoning. Thus, we chose a more pedagogical approach. %In section \ref{sec:macro}, we start by presenting the \emph{Pumping Toolbox} which contains some simple but fundamental results, conventions and notations which are used all along the proof. 

\section{Macroscopic reasoning}
\label{sec:macro}
This section deals with all the macroscopic reasoning part of the proof. We start by giving conventions and definitions which are necessary to this part. Then, we describe our macroscopic toolbox developed in sections \ref{sec:uwl} and \ref{sec:lastpart}. The conclusion of the proof contains the last step of figure \ref{fig:dependancies}. All the figures of this section are in appendix \ref{app:macro}.

\subsection{Conventions and macroscopic tools}

First, in this section we will introduce several bounds ($\border$, $\seedbound$, $\sbound$, $\ebound$, $\fbound$) and two functions $\freset$ and $\fjail$, these bounds and functions will depend on the size of the seed ($|\seed|$) and the size of the tiles set ($|T|$). Their exact values are not trivial to compute and not very useful since they quickly get over exponential. We are more interested into their existence. To avoid lengthy definitions, the exact values of these bounds and functions are all regrouped in appendix \ref{app:bound}. Conventions introduced here are summarized in figure \ref{fig:convention}. 

\vspace{+0.5em}
\noindent \textbf{Conventions.} Consider a path assembly $P$ producible by $\mathcal{T}=(T,\sigma,1)$, we would like to work only with positive coordinates. This could easily be achieved by choosing the convention that the southernmost (resp. westernmost) tile of the seed or $P$ has an ordinate (resp. abscissa) of~$0$. Unfortunately, one of our macroscopic lemma (lemma \ref{macro:stakes}) requires to construct a path assembly which goes below the southernmost tile of the seed and $P$ by a distance of at most $\border$ (see appendix \ref{app:bound} for the exact value of $\border$). In order to keep this new path assembly in the first quadrant of the plane, we will use the following convention.

\begin{convention}
\label{convention:quadrant}
Consider a path assembly $P$ producible by $\mathcal{T}=(T,\sigma,1)$, we say that $P$ satisfies the \emph{first quadrant convention} if $\min(y_P,y_\seed)=\border$ and $\min(x_P,x_\seed)=\border$. 
\end{convention}

The statements of the different lemmas  will always consider a path assembly $P$ which satisfies this convention along this section. Nevertheless, we may construct other path assemblies which does not satisfy this convention along their proofs. Also if a path assembly $P$ producible by a tiling system $\tiling=(T,\seed,1)$ does not satisfied the first quadrant convention, then there exists a vector $\vu$ such that $P+\vu$ is producible by $(\tiling+\vu)=(T,\seed+\vu,1)$, $P+\vu$ satisfies the first quadrant convention and $P+\vu$ is pumpable or fragile if and only if $P$ is pumpable or fragile. The second convention deals with the position of the seed. We want the seed to be near the origin. 

\begin{convention}
\label{convention:seed}
Consider a path assembly $P$ producible by $\mathcal{T}=(T,\sigma,1)$, we say that $P$ satisfies the \emph{axis conventions} if $y_{\seed} - y_P \leq Y_P-Y_{\seed}$ and $x_{\seed} - x_P \leq X_P-X_{\seed}$. 
\end{convention}

Consider a path assembly $P$ producible by $\tiling=(T,\seed,1)$, then if $P$ does not satisfy the axis conventions then it is possible to find another path assembly $P'$ producible by another tiling system $\tiling'=(T,\seed',1)$ with $|T'|=|T|$, $|\seed'|=|\seed|$ and such that $P'$ satisfies the axis conventions and $P$ is pumpable or fragile if and only if $P'$ is pumpable or fragile. This result is achieved by a symmetry of the $x$ axis or/and of the $y$ axis. For the tile set, this operation is equivalent to switch the north and south glues of each tile (or west and east glues). 

\vspace{+0.5em}
\noindent \textbf{Squares.} The following definition is dedicated to squares which are an useful tool to delimit a part of the grid graph. A square with sides of length $\ell$ is the subset of the grid graph $G$ containing vertices of the first quadrant at distance less than $\ell$ of the origin.

\begin{definition}
\label{def:rect}
Consider $\ell \in \mathbb{N}$, the square $\rect(\ell)$ is a subset of  $\mathbb{Z}^2$ defined by $\rect(\ell)=\{(x,y): 0\leq x\leq \ell \text{ and } 0\leq y \leq \ell\}$. 
\end{definition}

Squares will come handy to classify the different parts of the path assembly according to their distance to the origin of the plan. Note that squares are $2D$ objects defined by only one parameter: the length of their sides. This is the main reason why the macroscopic part of the proof is easier than the microscopic one. We will show that the more a path assembly grows away from the origin the more it is constrained. This allows us to consider only one parameter instead of the two dimensions. 

%\begin{figure}[th]
%\centerline{
%\includegraphics[width=5cm]{./photo/conventions/square.JPG}
%}
%\caption{The squares $\rect(3)$ and $\rect(7)$, note that $\rect(3) \subset \rect(7)$.}
%\label{fig:def:square}
%\end{figure}

Along the article, we will need to describe precisely some patterns of the path assembly. In this model, describing a pattern is unfortunately rather complex. Proofs will rely on mathematical descriptions which are precise but not intuitive. Nevertheless during preliminary presentations of our results we will use a more intuitive but informal language. For example, ``the path assembly $P$ reaches an ordinate of $L$ and then enters square $\rect(\ell)$" means that there exist two indices $1\leq \ind{1} \leq \ind{2} \leq \lastp$ such that $y_{P_{\ind{1}}}>L$ and $\pos(P_{\ind{2}}) \in \rect(\ell)$. For another example, consider a path assembly $P$ and a square $\rect(\ell)$ such that $\pos(P_1) \in \rect(\ell)$ then ``the first time the path assembly $P$ leaves the square $\rect(\ell)$ is by its east side" means that $X_P>\ell$ and $\ind{1}=\min\{j:j\notin \rect(\ell)\}$ is such that $x_{P_{\ind{1}}}=\ell+1$. Finally, consider a position $A \in V(\grid)$, ``$P$ ends in $A$" means that $\pos(P_{\lastp})=A$.

\vspace{+0.5em}
\noindent \textbf{Path assembly with macroscopic initial conditions.} Path assemblies that we will study in this section will have six constraints. We start by showing that these constraints are not restrictive. We say that a path assembly which satisfies these properties, satisfies \emph{macroscopic initial conditions} (see appendix \ref{app:bound} for the value of $\fbound$):

\begin{definition}
Consider a tiling system $\tiling=(T,\seed,1)$, the path assembly $P$ satisfies macroscopic initial conditions if:
\begin{itemize}
\item $P$ is finite;
\item $\dom{P} \cap \dom{\seed}= \pos(P_1)$; 
\item $y_{(P_{\lastp})}>\fbound$;
\item $P$ is simple.
\item $P$ satisfies the first quadrant convention and the axis convention.
\end{itemize}
\end{definition}

Note that the third item implies that $\pos(P_{\lastp}) \notin \rect(\fbound)$.

\begin{lemma}
\label{lem:goodMacroCond}
Consider a path assembly $P$ producible by $\tiling=(T,\seed,1)$ such that the distance from $P$ to the seed is strictly greater than $\fbound$ then there exists a path assembly $P'$ producible by a tiling system $\tiling'=(T',\seed',1)$ with $|T'|=|T|$, $|\seed'|=|\seed|$ and such that $P'$ satisfies macroscopic initial conditions and such that if $P'$ is pumpable or fragile then $P$ is pumpable or fragile.
\end{lemma}

\begin{proof}
In a first time, we will focus on the four first properties, we will deal with the two conventions at the end. By hypothesis there exists an index $1\leq \ind{1} \leq \lastp$ such that the distance between $\pos(P_{\ind{1}})$ and the seed is at least $\fbound+1$. Moreover, since $P$ is producible by $\tiling=(T,\seed,1)$ then there exists an index $1\leq \ind{2} \leq \lastp$ such that $\pos(P_{\ind{2}}) \in \dom{\seed}$. Suppose that $\ind{1}<\ind{2}$ (the other case is symmetric and simpler), see Figure \ref{fig:goodMacro}a for graphical representation of the following notations, then let $\ind{3}= \min_{\ind{1} \leq i \leq \ind{2}}\{i :\pos(P_{i}) \in \dom{\seed}\}$ and $\ind{4}=\max_{\ind{1} \leq i \leq \ind{3}}\{i :\pos(P_{i}) \text { is at distance $\fbound+1$ of the seed}\}$. Now, let $P'$ be the path assembly such that for all $1\leq i \leq \ind{4}-\ind{3}+1$, we have $P'_{i}=P_{\ind{4}-i+1}$ ($P'$ is obtained by switching the indexing of the segment $P_{[\ind{4},\ind{3}]}$, see Figure \ref{fig:goodMacro}b). Now, if $P'$ is not simple then there exist two indices $1< \ind{5} < \ind{6} < |P'|$ such that $P_{\ind{5}}=P_{\ind{6}}$ then let $P''=P_{[1,\ind{5}]} \cdot P_{[\ind{6},|P'|]}$ (see Figure \ref{fig:goodMacro2}a). W. l. o.g, we suppose that the path assembly $P''$ is simple (other cycles could be removed the same way). Then $\alpha_{P''} \sqsubseteq \alpha_{P}$, and $\dom{P''} \cap \dom{\seed}= \pos(P'_1)$ and then $P''$ is producible by $\tiling$. This means that if $P''$ is pumpable or fragile then $P$ is pumpable or fragile. Moreover $P''$ is finite ($|P''| \leq |P'| \leq \ind{4}-\ind{3}+1$), simple and $P''_{|P''|}$ is the only tile of $P''$ which is at distance more than $\fbound+1$ of the seed. 

Now, up to some symmetries and a translation, there exists a path assembly $Q$ producible by a tiling system $\tiling'=(T',\seed',1)$ with $|T'|=|T|$, $|\seed'|=|\seed|$ and such that $Q$ satisfies both conventions and $Q$ is pumpable or fragile if and only if $P''$ is fragile or pumpable (see Figure \ref{fig:goodMacro2}b). The symmetries and translation do not modify the previous properties of $P''$. The path assembly $Q$ is finite, simple and such that $\dom{Q} \cap \dom{\seed'}= \pos(Q_1)$. Moreover the last tile of $Q$ is the only one at distance $\fbound$ of the seed. This means that either $|y_{Q_{|Q|}}-Y_{\seed}|>\fbound$ or $|x_{Q_{|Q|}}-X_{\seed}|>\fbound$. If $|x_{Q_{|Q|}}-X_{\seed}|>\fbound$ then by a symmetry of axis $x=y$ (we switch the $x$ and $y$ axis) we obtain that $|y_{Q_{|Q|}}-Y_{\seed}|>\fbound$. This symmetry does not change the previous properties. Then we have a path $Q$ such that $|y_{Q_{|Q|}}-Y_{\seed}|>\fbound$. Since $Q$ satisfies the axis convention and $Q_{|Q|}$ is the only tile of $Q$ at distance $\fbound+1$ of seed, we have $y_{Q_{|Q|}}>Y_{\seed}+\fbound$. %Since by the first quadrant convention $Y_{\seed}>0$, we have $Q_{|Q|} \notin \rect(\fbound)$. 
The path assembly $Q$ with the tiling system $\tiling'$ satisfies the required properties.
\end{proof}

This result allows us to restrict our study to path assemblies satisfying macroscopic initial conditions. Our final aim goal is to show that such path assemblies are pumpable or fragile. This result will imply that all assemblies growing at distance more than $\fbound$ of the seed are pumpable or fragile. %Here is the alternative version of this lemma for reverse pumping.

%\begin{lemma}
%\label{lem:goodMacroCond:reverse}
%Consider a path assembly $P$ producible by $\tiling=(T,\seed,1)$ such that there exists $k$ such that $\min_{A\in \seed}\{||\overrightarrow{AP_k}||\}>\fbound$ then there exists a path assembly $P'$ producible $\tiling=(T',\seed',1)$ such that $P'$ satisfies macroscopic initial conditions and such that $\alpha_{P'}$ is fragile or pumpable with reverse pumping allowed if and only if $\alpha_{P}$ is fragile or pumpable with reverse pumping allowed.
%\end{lemma}

\subsection{The macroscopic toolbox}

The key idea of our proof is that the more a path assembly grows away from the seed, the more constraints appear. If one of them is not fulfilled by the path assembly then it is fragile or pumpable. Eventually, all the constraints could not be all fulfilled at the same time and this will conclude the main proof. In this subsection, we present the four main constraints and lemmas of our macroscopic toolbox. Constraints \ref{constraint:seed}, \ref{constraint:stakes} and \ref{constraint:jailed} are the main constraints which have to be fulfilled while constraint \ref{constraint:reset} is a useful tool to allow independent studies of the constraints. Note that these constraints admit several interpretations, their formulations use square but during their introductions we will discuss about more intuitive interpretations. The first two lemmas are illustrated in Figure \ref{fig:reset} while Figure \ref{fig:macro4} is dedicated to the last two ones. 

The first macroscopic constraint shows that a path assembly has to choose between growing north or south (resp. west or east) of the seed. More formally, consider a tile assembly system $\mathcal{T}=(T,\sigma,1)$ then for any path assembly $P$, if $y_\seed-y_P>4|T|+\sigma$ and $Y_P-Y_\seed>4|T|-\sigma$ then $P$ is fragile or pumpable. With the chosen conventions, it means that if the seed is at distance more than $\seedbound$ from the origin of the plane then it is pumpable or fragile. In terms of squares, this result is stated as follow. For the definition of bound $\seedbound$ used in the following definition, see appendix \ref{app:bound}.

\begin{definition}[Seed constraint]
\label{constraint:seed}
Let $P$ be a path assembly producible by $\mathcal{T}=(T,\sigma,1)$ satisfying macroscopic initial conditions, $P$ satisfies the \emph{seed constraint} if and only if $\dom{\seed} \subset \rect(\seedbound)$.
\end{definition}

\begin{lemma}[Macro 1: seed lemma]
\label{macro:seed}
Let $P$ be a path assembly producible by $\mathcal{T}=(T,\sigma,1)$ satisfying macroscopic initial conditions, if $P$ does not satisfies the seed constraint then it is fragile or pumpable.
\end{lemma}

This lemma allows us to consider that the path assembly starts near the origin of the plane, up to constant $\seedbound$ which will quickly become negligible compared to the following ones. Also, this result shows the convenience of the chosen conventions and the rectangle notation: consider $\ell \geq \seedbound$, then the square $\rect(\ell)$ contains the seed (otherwise the path assembly is fragile or pumpable by lemma \ref{macro:seed}) and if a tile of a path assembly does not belong to the square $\rect(\ell)$ then it is at least at a distance $\ell-\seedbound$ of the seed. 

Now comes the second macroscopic constraint which shows that the further a path assembly grows away from the origin, the further it has to stay away from it. This constraint is very useful to show that the path assembly has to avoid artifacts created near the origin of the plane. After stating this lemma, we illustrate how to use it with a quick example. For the definition of function $\freset:\mathbb{N}\rightarrow \mathbb{N}$ used in the following definition, see appendix \ref{app:bound}.

\begin{definition}[Reset constraint]
\label{constraint:reset}
Let $P$ be a path assembly producible by $\mathcal{T}=(T,\sigma,1)$ satisfying macroscopic initial conditions and $\ell \in \mathbb{N}$, then $P$ does not satisfy the \emph{reset constraint} for $\rect(\ell)$ if and only if there exists $1\leq \ind{1} \leq \ind{2} \leq \lastp$ such that $\pos(P_{\ind{1}}) \notin\rect(\freset(\ell))$ and $\pos(P_{\ind{2}}) \in \rect(\ell)$. 
\end{definition}

\begin{lemma}[Macro 2: reset lemma]
\label{macro:reset}
Let $P$ be a path assembly producible by $\mathcal{T}=(T,\sigma,1)$ satisfying macroscopic initial conditions and $\ell \in \mathbb{N}$ such that $\ell>\seedbound$. If $P$ does not satisfy the reset constraint for square $\rect(\ell)$ then $P$ is fragile or pumpable.
\end{lemma}

The reset lemma could be interpreted as follow: if the path assembly reaches a distance more than $\freset(\ell)$ of the origin then it must stay away from the origin by at least a distance $\ell$. Figure~\ref{fig:reset} illustrates a quick example of how to use the first two macro lemmas: consider a path assembly $P$ satisfying macroscopic initial conditions and $d\in \mathbb{N}$. Then by the seed constraint, either the path assembly $P$ is fragile or pumpable or the seed $\seed$ belongs to the square $\rect(\seedbound)$ and thus any tiles outside of $\rect(\seedbound+d)$ is at distance $d$ of $\seed$. Let $D=\freset(\seedbound+d)$. By the reset lemma after exiting the square $\rect(D)$, the path assembly $P$ has to stay away from the seed by at least a distance $d$ or it is fragile or pumpable.

The third constraint is the key to the final result. When a path assembly $P$ grows far away from the seed, it is possible to shift its end by a non null vector $\vu$ (see Figure \ref{fig:macro4}). More precisely, there exists an index $1\leq \inds \leq \lastp$ such that it is possible to build two other assemblies where $P_{[\inds,\lastp]}-\vu$ (resp. $P_{[\inds,\lastp]}+\vu$) appears in the first (resp. second) one. For the definition of the bounds $\sbound$ and $\fbound$ used in the following definition, see appendix \ref{app:bound}. A graphical representation of this lemma is available, see Figure \ref{fig:macro4}.

\begin{definition} [Stakes constraint I]
\label{constraint:stakes}
Let $P$ be a simple path assembly producible by $\mathcal{T}=(T,\sigma,1)$ satisfying macroscopic initial conditions. Consider an index $1\leq \inds \leq \lastp$, a vector $\vu$ of $\mathbb{Z}^2$ and two path assemblies $\stakem$ and $\stakep$ producible by $\mathcal{T}$. The path assembly $P$ satisfies the \emph{stakes constraint} at index $\inds$ of direction $\vu$ with path assemblies $\stakem$ and $\stakep$ if and only if :
\begin{itemize}
\item $y_{\vu}\geq 1$ and $||\vu|| \leq \border$;
\item $\stakem \cdot (P_{[s,\lastp]}-\vu)$ and $\stakep \cdot (P_{[s,\lastp]}+\vu)$ are path assemblies producible by $\mathcal{T}$; 
\item $\seed, \stakep, \stakem$ and $P_{[1,s]}$ belong to $\rect(\sbound)$.
\end{itemize}
The square $\rect(\sbound)$ is called the \emph{stakes zone} and $\stakem$, $\stakep$ are called stakes.
\end{definition}

\begin{definition} [Stakes constraint II]
\label{notation:stake}
Let $P$ be a path assembly producible by $\mathcal{T}=(T,\sigma,1)$ satisfying macroscopic initial conditions, $P$ satisfies the stakes constraint if there exists $\inds$, $\vu$, $\stakem$ and $\stakep$ such that $P$ satisfies the stake constraint of direction $\vu$ at index $\inds$ with path assemblies $\stakem$ and $\stakep$ .
\end{definition}

\begin{lemma} [Macro 3: stakes lemma]
\label{macro:stakes}
Let $P$ be a path assembly producible by $\mathcal{T}=(T,\sigma,1)$ satisfying macroscopic initial conditions. If $P$ does not satisfy the stake constraint then it is fragile or pumpable.
\end{lemma}

Note that the proof of the stakes lemma requires lot of microscopic reasoning and is the last and most technical part of the article. Also note that the seed constraint \ref{macro:seed} seems to be a sub-case of the stakes constraint. Nevertheless, the seed constraint \ref{macro:seed} is still useful because it is needed to prove the stakes lemma and the bound ($\seedbound-\border$) is the only linear bound in our result. Additionally, since path assemblies satisfying macroscopic initial conditions also satisfy the first quadrant convention, the translations of $P$ by $\vu$ and $-\vu$ also belong to the first quadrant of the plane but they may not satisfy the first quadrant convention. This lemma is the reason why the border $\border$ is needed in the definition of the seed constraint.

Finally, the last macroscopic constraint is more specific but crucial to conclude the proof. After stating the stakes constraint, it is natural to wonder how the end of the path assembly $P_{[s,\lastp]}$ interacts with its translations $(P_{[s,\lastp]}-\vu)$ and $(P_{[s,\lastp]}+\vu)$. Listing all the possible interactions is the aim of the final part of the proof but we partially answer here this question: if $P_{[s,\lastp]}$ and $(P_{[s,\lastp]}-\vu)$ does not intersect for a long time then $P$ is fragile or pumpable. For the definition of function $\fjail:\mathbb{N}\rightarrow \mathbb{N}$ used in the following definition, see appendix \ref{app:bound}.

\begin{definition} [Macro 4: jail constraint]
\label{constraint:jailed}
Let $P$ be a path assembly producible by $\mathcal{T}=(T,\sigma,1)$ satisfying macroscopic initial conditions. Consider an index $1\leq \ind{1} \leq \lastp$ and a vector $\vu$. The path assembly $P$ is jailed at index $\ind{1}$ according to $\vu$ if and only if there exists $\ell \in \mathbb{N}$ and an index $\ind{1}\leq \ind{2} \leq \lastp$ such that:
\begin{itemize}
\item $\dom{P_{[1,\ind{1}]}} \subset \rect(\ell)$,  $\dom{\seed} \subset \rect(\ell)$ and $\pos(P_{\ind{2}}) \notin\rect(\fjail(\ell))$;
\item $y_{\vu}\geq 1$ and $||\vu|| \leq \border$;
\item $P_{[\ind{1},\ind{2}]}$ and $(P_{[\ind{1},\ind{2}]}-\vu)$ does not intersect.
\end{itemize}
\end{definition}

\begin{lemma} [Macro 4: jail lemma]
\label{macro:jailed}
Let $P$ be a path assembly producible by $\mathcal{T}=(T,\sigma,1)$ satisfying macroscopic initial conditions and $1\leq \ind{1} \leq \lastp$. If the path assembly $P$ is jailed at index $\ind{1}$ then it is fragile or pumpable.
\end{lemma}

Now in the next section, we show that the combination of these four macroscopic results is enough to solve the conjecture.

\subsection{Macroscopic reasoning}

Here is a quick roadmap of the end of the proof. %First, consequences of the stakes lemma (lemma \ref{macro:stakes}) are studied. 
Consider a path assembly $P$ producible by  $\mathcal{T}=(T,\sigma,1)$ which respects the stakes constraint at index $\inds$ of direction $\vu$ then how does $P_{[s,\lastp]}$ and $P_{[s,\lastp]}-\vu$ interacts? We give a list of seven possibilities and for each of them, we show that they imply directly the fragility or pumpability of the path assembly. To conclude, we show that $P_{[s,\lastp]}$ and $P_{[s,\lastp]}-\vu$ have to match at least one of these seven cases.

Since we need to consider the different intersections between $P_{[s,\lastp]}$ and $P_{[s,\lastp]}-\vu$, we need an efficient tool to find them. We will introduce the following notation $\hunt(\ind{1})$ which for index $s\leq \ind{1} \leq \lastp$ returns the position of the next intersection between $P_{[\ind{1},\lastp]}$ and $P_{[\ind{1},\lastp]}-\vu$. The notion of the "next" intersection implies some subtleties. We are interesting in growing $P_{[\ind{1},\lastp]}$ and $P_{[\ind{1},\lastp]}-\vu$ in parallel and stop when the first intersection occurs. This definition is not equivalent to the first time that $P_{[\ind{1},\lastp]}$ intersects with $P_{[\ind{1}, \lastp]}-\vu$ (see Figure \ref{fig:hunt}).

\begin{definition}
\label{def:hunt}
Consider a path assembly $P$ producible by  $\mathcal{T}=(T,\sigma,1)$ satisfying the macroscopic initial conditions and the stakes constraint at index $\inds$ according to vector $\vu$. For any index $\inds \leq \ind{1} \leq \lastp$, we define $\hunt(\ind{1})$ as follow, $\hunt(\ind{1})=\lastp$ if $P_{[\ind{1},\lastp]}$ and $P_{[\ind{1},\lastp]}-\vu$ does not intersect otherwise $\hunt(\ind{1})=\min_{\ind{1}\leq i \leq \lastp}\{i: \dom{P_{[\ind{1},i]}} \cap \dom{P_{[\ind{1},i]}-\vu} \neq \emptyset\}$. 
\end{definition}

Note that the function $\hunt$ should also take as argument $P$ and $\vu$ but there will have no ambiguities along the article so to reduce the amount of notation, we will write only $\hunt(\ind{1})$. Also note that, $\hunt(\ind{1})>\ind{1}$ since $\vu$ is a non null vector. Now, consider the index $\inds \leq \ind{1} \leq \lastp$, we define a list of seven cases that can be fulfilled by index $\ind{1}$.  At the end of the section, we conclude that index $\ind{1}$ has to match at least one of these cases. Sometimes, several cases can be true for the same index.
 
The first final case occurs when the end of the path assembly does not intersect with its translation. The fourth macroscopic lemma (jailed path assembly) was designed to deal with this case. For the value $\ebound$ used in the following lemma, see appendix \ref{app:bound}.

\begin{definition}[Final case $1$: Jail]
Consider a path assembly $P$ producible by  $\mathcal{T}=(T,\sigma,1)$ satisfying the macroscopic initial conditions and the stakes constraint of direction $\vu$ at index~$\inds$. For any index $\inds \leq \ind{1} \leq \lastp$, the index $\ind{1}$ matches the \emph{jail} final case if $P_{[\ind{1},\lastp]}$ and $P_{[\ind{1},\lastp]}-\vu$ does not intersect.
\end{definition}

\begin{lemma}
\label{hunt:jail}
Consider a path assembly $P$ producible by $\mathcal{T}=(T,\sigma,1)$ satisfying the macroscopic initial conditions and the stakes constraint of direction $\vu$ at index $\inds$. If there exists an index $\inds \leq \ind{1} \leq \lastp$ which matches the \emph{jail} hunt case and such that $\dom{P_{[1,\ind{1}]}}  \subset \rect(\ebound+1)$ then path assembly $P$ is fragile or pumpable.
\end{lemma}

\begin{proof}
By the definition of stakes constraint (definition \ref{constraint:jailed}), we have $y_{\vu}\geq1$ and $||\vu||\leq \border$. Since $P$ satisfies the macroscopic initial conditions, it does not fit in $\rect(\fbound)$ and $\fbound=\fjail(\ebound+1)+1$ (see appendix \ref{app:bound}). Then there exists an index $\ind{1}<\ind{2}\leq \lastp$ such that $\pos(P_{\ind{2}}) \notin \rect(\fjail(\ebound+1))$. Thus, the path assembly $P$ is jailed at index $\ind{1}$ of direction $\vu$ and by the jail lemma (lemma \ref{macro:jailed}), $P$ is fragile or pumpable.
\end{proof}

The second final case occurs when the end of the path assembly is in conflict with its translation. By the definition of the stakes constraint, the fragility of the path assembly is direct.

\begin{definition}[Final case $2$: Conflict]
Consider a path assembly $P$ producible by  $\mathcal{T}=(T,\sigma,1)$ satisfying the macroscopic initial conditions and the stakes constraint of direction $\vu$ at index $\inds$. For any index $\inds \leq \ind{1} \leq \lastp$, index $\ind{1}$ matches the \emph{conflict} final case if $P_{[\ind{1},\lastp]}$ and $P_{[\ind{1},\lastp]}-\vu$ are in conflict.
\end{definition}

\begin{lemma}
\label{hunt:conf}
Consider a path assembly $P$ producible by $\mathcal{T}=(T,\sigma,1)$ satisfying the macroscopic initial conditions and the stakes constraint of direction $\vu$ at index $\inds$. If there exists $\inds \leq \ind{1} \leq \lastp$ which matches the \emph{conflict} hunt case then path assembly $P$ is fragile.
\end{lemma}

\begin{proof}
By the definition of stakes constraint (definition \ref{constraint:jailed}), the path assembly $P'=\stakem \cdot (P_{[\inds,\lastp]}-\vu)$ is producible by $\tiling$. Then, since $P$ and $P'$ are in conflict and producible by $\tiling$, they are both fragile.
\end{proof}

If the end of the path assembly intersects with its translation and if the first intersection is an agreement, then we will show that it is possible to find a good candidate segment in $P_{[\ind{1},\lastp]}$. The third final case occurs when this good candidate segment is pumpable which directly implies the pumpability or fragility of $P$. 

\begin{definition}[Final case $3$: Pumpable]
Consider a path assembly $P$ producible by  $\mathcal{T}=(T,\sigma,1)$ satisfying the macroscopic initial conditions and the stakes constraint  of direction $\vu$ at index $\inds$. For any index $\inds \leq \ind{1} \leq \lastp$, index $\ind{1}$ matches the \emph{pumpable} final case if there exists $\ind{1} \leq \ind{2} < \hunt(\ind{1})$ such that $P_{[\ind{2},\hunt(\ind{1})]}$ is a good pumpable candidate segment.
\end{definition}

\begin{fact}
\label{fact:pump}
Consider a path assembly $P$ producible by $\mathcal{T}=(T,\sigma,1)$ satisfying the macroscopic initial conditions and the stakes constraint at index $\inds$ according to vector $\vu$ (then $\vu \neq \overrightarrow{0}$). If there exists $\inds \leq \ind{1} \leq \lastp$ which matches the \emph{pumpable} final case then path assembly $P$ is pumpable or fragile.
\end{fact}

As the previous case, the intersection between the end of the path assembly and its translation creates a good segment in $P_{[\ind{1},\lastp]}$. The fourth final case occurs when this good segment is of direction $\vu$ and conflicts with $P$ after index $\inds$ (see Figure \ref{fig:case4}a). Using $\stakep$, we will be able to prove the fragility of $P$ (see Figure \ref{fig:case4}b). As a side note, this case is the reason why the condition ``$\stakep \cdot (P_{[\inds,\lastp]}+\vu)$ is producible by $\tiling$" is required in the stakes condition.

\begin{definition}[Final case $4$: Conflict after stake]
Consider a path assembly $P$ producible by  $\mathcal{T}=(T,\sigma,1)$ satisfying the macroscopic initial conditions and the stakes constraint of direction $\vu$ at index $\inds$. Consider an index $\inds \leq \ind{1} \leq \lastp$ and let $\ind{3}=\hunt(\ind{1})$. The index $\ind{1}$ matches the \emph{Conflict after stake} final case if there exists $\ind{1} \leq \ind{2} < \ind{3}$ such that $\overrightarrow{P_{\ind{2}}P_{\ind{3}}} =\vu$ and the segment $P_{[\ind{2},\ind{3}]}$ is a good segment which is not pumpable. Moreover, let $Q=\omepluse{P_{[\ind{2},\ind{3}]}}$, then the first conflict $\ind{4}$ of $P_{[\ind{2},\ind{3}]}$ should also occurs in $\dom{P_{[s,\ind{2}-1]}}$ (\emph{i.e.} $\pos(Q_{\ind{4}}) \in \dom{P_{[s,\ind{2}-1]}}$) and should not enter the stakes zone (\emph{i.e.} $\dom{Q_{[1,\ind{4}]}}\cap\rect(\sbound)=\emptyset$).
 \end{definition}

\begin{lemma}
\label{hunt:fcac}
Consider a path assembly $P$ producible by $\mathcal{T}=(T,\sigma,1)$ satisfying the macroscopic initial conditions and the stakes constraint of direction $\vu$ at index $\inds$. If there exists an index $\inds \leq \ind{1} \leq \lastp$ which matches the \emph{Conflict after stake} final case then path assembly $P$ is fragile.
\end{lemma}

\begin{proof}
Consider a path assembly $P$ producible by $\mathcal{T}=(T,\sigma,1)$ satisfying the macroscopic initial conditions and the stakes constraint of direction $\vu$ at index $\inds$ with path assembly $\stakep$. Suppose that there exists $\inds \leq \ind{1} \leq \lastp$ which matches the \emph{Conflict after stake} final case. Then, let $\ind{3}=\hunt(\ind{1})$ and let $s\leq \ind{2} < \ind{3}$ such that $P_{[\ind{2},\ind{3}]}$ is a good candidate segment with $\overrightarrow{P_{\ind{2}}P_{\ind{3}}}=\vu$. By the definition of the stakes constraint, $\stakep \cdot (P_{[s,\ind{2}]}+\vu)$ is a path assembly producible by $\tiling$ and this path assembly ends at position $\pos(P_{\ind{2}}+\vu)=\pos(P_{\ind{3}})$. Now, let $Q=\omepluse{P_{[\ind{2},\ind{3}]}}$ and let $\ind{4}$ be the first conflict of $P_{[\ind{2},\ind{3}]}$ in $P$. Since $P_{[\ind{2},\ind{3}]}$ is a good candidate segment then $Q$ is a path assembly by lemma \ref{lem:good:cond}. Since index $\ind{1}$ matches the \emph{Conflict after stake} final case then $\dom{Q_{[1,\ind{4}]}} \cap \rect(\sbound)=\emptyset$ and by definition of the stakes constraint, $\dom{\seed \cup \stakep} \subset \rect(\sbound)$ then $Q_{[1,\ind{4}]}$ and $\seed \cup \stakep$ cannot intersect. If there exists a conflict between $Q_{[1,\ind{4}]}$ and $P_{[s,\ind{2}]}+\vu$ then let $1\leq \ind{5} \leq \ind{4}$ be the index such that this conflict occurs in $Q_{\ind{5}}$. By the periodicity of $Q$ (fact~\ref{fact:ext:per}) then $Q_{\ind{5}}-\vu$ either belongs to $Q$ or to $P_{[\ind{2},\ind{3}]}$. If $Q_{\ind{5}}-\vu$ belongs to $P_{[\ind{2},\ind{3}]}$ since there is a conflict between $Q_{\ind{5}}$ and $P_{[s,\ind{2}]}+\vu$ then there is a conflict between $(P_{[s,\ind{2}]}+\vu)-\vu=P_{[s,\ind{2}]}$ and $P_{[\ind{2},\ind{3}]}$ which is a contradiction. Otherwise if $Q_{\ind{5}}-\vu$ belongs to $Q$ then by the periodicity of $Q$ (fact~\ref{fact:ext:per}), there exists $1\leq \ind{6} < \ind{5}$ such that $Q_{\ind{6}}=Q_{\ind{5}}-\vu$. Then, since there is a conflict between $Q_{\ind{5}}$ and $P_{[s,\ind{2}]}+\vu$, there exists a conflict between $Q_{\ind{6}}$ and $(P_{[s,\ind{2}]}+\vu)-\vu=P_{[s,\ind{2}]}$ which would contradict the definition of index $\ind{4}$. Thus, there is no conflict between $(P_{[s,\ind{2}]}+\vu)$ and $Q_{[1,\ind{4}]}$. To conclude, the path assembly $\stakep \cdot (P_{[s,\ind{2}]}+\vu) \cdot Q_{[1,\ind{4}]}$ is producible by $\tiling$. By definition of $Q$ and $\ind{4}$, $P$ has a conflict with $Q_{[1,\ind{4}]}$ at position $\pos(Q_{\ind{4}})$. All these results lead to the fragility of $P$. 
\end{proof}

The fifth final case is similar to the previous one except that the good candidate segment is of direction $-\vu$ (see Figure \ref{fig:case5}a). This time using $\stakem$, we will be able to prove the fragility of $P$ in a similar way of the previous lemma (see Figure \ref{fig:case5}b). 

\begin{definition}[Final case $5$: Conflict after stake (reverse)]
Consider a path assembly $P$ producible by  $\mathcal{T}=(T,\sigma,1)$ satisfying the macroscopic initial conditions and the stakes constraint of direction $\vu$ at index $\inds$. Consider an index $\inds \leq \ind{1} \leq \lastp$ and let $\ind{3}=\hunt(\ind{1})$. The index $\ind{1}$ matches the \emph{Conflict after stake (reverse)} final case if there exists $\ind{1} \leq \ind{2} < \ind{3}$ such that $\overrightarrow{P_{\ind{2}}P_{\ind{3}}} =-\vu$ and the segment $P_{[\ind{2},\ind{3}]}$ is a good candidate segment which is not pumpable. Moreover, let $Q=\omepluse{P_{[\ind{2},\ind{3}]}}$, then the first conflict $\ind{4}$ of $P_{[\ind{2},\ind{3}]}$ should also occurs in $\dom{P_{[s,\ind{2}-1]}}$ (\emph{i.e.} $\pos(Q_{\ind{4}}) \in \dom{P_{[s,\ind{2}-1]}}$) and should not enter the stakes zone (\emph{i.e.} $\dom{Q_{[1,\ind{3}]}}\cap\rect(\sbound)=\emptyset$).
%For any index $\inds \leq i \leq \lastp$, the index $i$ matches the \emph{First conflict after stake (reverse)} hunt case if there exists $i\leq j < \hunt(i)$ such that $\overrightarrow{P_jP_{\hunt(i)}} =-\vu$ and the segment $P_{[j,\hunt(i)]}$ is a good candidate segment whose pumping until first conflict $Q$ ends in $\dom{P_{[s,j-1]}}$ ($\pos(Q_{|Q|}) \in \dom{P_{[s,j-1]}}$) and does not enter the stakes zone ($\dom{Q}\cap\rect(\sbound)=\emptyset$).
\end{definition}

\begin{lemma}
\label{hunt:fcacr}
Consider a path assembly $P$ producible by $\mathcal{T}=(T,\sigma,1)$ satisfying the macroscopic initial conditions and the stakes constraint at index $\inds$ according to vector $\vu$. If there exists $\inds \leq \ind{1} \leq \lastp$ which matches the \emph{Conflict after stake (reverse)} final case then path assembly $P$ is fragile.
\end{lemma}

\begin{proof}
Consider a path assembly $P$ producible by $\mathcal{T}=(T,\sigma,1)$ satisfying the macroscopic initial conditions and the stakes constraint of direction $\vu$ at index $\inds$ with path assembly $\stakem$. Suppose that there exists $\inds \leq \ind{1} \leq \lastp$ which matches the \emph{Conflict after stake (reverse)} final case. Then, let $\ind{3}=\hunt(\ind{1})$ and let $s\leq \ind{2} < \ind{3}$ such that $P_{[\ind{2},\ind{3}]}$ is a good candidate segment with $\overrightarrow{P_{\ind{2}}P_{\ind{3}}}=-\vu$. By the definition of the stakes constraint, $\stakem \cdot (P_{[s,\ind{2}]}-\vu)$ is a path assembly producible by $\tiling$ and this path assembly ends at position $\pos(P_{\ind{2}}-\vu)=\pos(P_{\ind{3}})$. %Now, if $\stakem \cdot (P_{[s,\ind{2}]}-\vu) \cdot P_{[\ind{2},\ind{3}]}$ is not a path assembly producible by $\tiling$ then $P$ is fragile. 
Otherwise, let $Q=\omepluse{P_{[\ind{2},\ind{3}]}}$ and let $\ind{4}$ be the first conflict of $P_{[\ind{2},\ind{3}]}$ in $P$. Since $P_{[\ind{2},\ind{3}]}$ is a good candidate segment then $Q$ is a path assembly by lemma \ref{lem:good:cond}. Since index $\ind{1}$ matches the \emph{Conflict after stake (reverse)} final case then $\dom{Q_{[1,\ind{4}]}} \cap \rect(\sbound)=\emptyset$ and by definition of the stakes constraint, $\dom{\seed \cup \stakem} \subset \rect(\sbound)$ then $Q_{[1,\ind{4}]}$ and $\seed \cup \stakem$ cannot intersect. If there exists a conflict between $Q_{[1,\ind{4}]}$ and $P_{[s,\ind{2}]}-\vu$ then let $1\leq \ind{5} \leq \ind{4}$ be the index such that this conflict occurs in $Q_{\ind{5}}$. By the periodicity of $Q$ (fact~\ref{fact:ext:per}) then $Q_{\ind{5}}+\vu$ either belongs to $Q$ or to $P_{[\ind{2},\ind{3}]}$. If $Q_{\ind{5}}+\vu$ belongs to $P_{[\ind{2},\ind{3}]}$ since there is a conflict between $Q_{\ind{5}}$ and $P_{[s,\ind{2}]}-\vu$ then there is a conflict between $(P_{[s,\ind{2}]}-\vu)+\vu=P_{[s,\ind{2}]}$ and $P_{[\ind{2},\ind{3}]}$ which is a contradiction. Otherwise if $Q_{\ind{5}}+\vu$ belongs to $Q$ then by the periodicity of $Q$ (fact~\ref{fact:ext:per}), there exists $1\leq \ind{6} < \ind{5}$ such that $Q_{\ind{6}}=Q_{\ind{5}}+\vu$. Then, since there is a conflict between $Q_{\ind{5}}$ and $P_{[s,\ind{2}]}-\vu$, there exists a conflict between $Q_{\ind{6}}$ and $(P_{[s,\ind{2}]}-\vu)+\vu=P_{[s,\ind{2}]}$ which would contradict the definition of index $\ind{4}$. Thus, there is no conflict between $(P_{[s,\ind{2}]}-\vu)$ and $Q_{[1,\ind{4}]}$. Also, since $Q$ is a good candidate segment there is no conflict between $P_{[\ind{2},\ind{3}]}$ and $Q_{[1,\ind{4}]}$. To conclude, the path assembly $\stakem \cdot (P_{[s,\ind{2}]}-\vu) \cdot Q_{[1,\ind{4}]}$ is producible by $\tiling$. By definition of $Q$ and $\ind{4}$, $P$ has a conflict with $Q_{[1,\ind{4}]}$ at position $\pos(Q_{\ind{4}})$. All these results lead to the fragility of $P$. 

\end{proof}

The sixth final case occurs if the good candidate segment is of direction $\vu$ and its pumping enters the stakes zone. This time using the reset lemma, we will be able to prove the fragility or pumpability of $P$. Note that the bound $\ebound$ used in the following lemma is defined in appendix~\ref{app:bound}.

\begin{definition}[Final case $6$: Pumping enters stake zone]
Consider a path assembly $P$ producible by  $\mathcal{T}=(T,\sigma,1)$ satisfying the macroscopic initial conditions and stakes constraint of direction $\vu$ at index $\inds$. Consider an index $\inds \leq \ind{1} \leq \lastp$ and let $\ind{3}=\hunt(\ind{1})$. The index $\ind{1}$ matches the \emph{Pumping enters stake zone} final case if there exists $\ind{1} \leq \ind{2} < \ind{3}$ such that $\overrightarrow{P_{\ind{2}}P_{\ind{3}}} =\vu$ and the segment $P_{[\ind{2},\ind{3}]}$ is a good candidate segment whose pumping intersects the stakes zone, \emph{i.e.} $\dom{\omepluse{P_{[\ind{2},\ind{3}]}}} \cap \rect(\sbound) \neq \emptyset$.
\end{definition}

\begin{lemma}
\label{hunt:pesz}
Consider a path assembly $P$ producible by $\mathcal{T}=(T,\sigma,1)$ satisfying the macroscopic initial conditions and stakes constraint at index $\inds$. If there exists $\inds \leq \ind{1} \leq \lastp$ which matches the \emph{Pumping enters stakes zone} final case and such that $\dom{P_{[1,\ind{1}]}} \not \subset \rect(\ebound)$ then path assembly $P$ is fragile or pumpable.
\end{lemma}

\begin{proof}
See Figure \ref{fig:case6} for an illustration of this proof. By definition of the \emph{Pumping enters stake zone} final case, let $\ind{3}=\hunt(\ind{1})$ and let $\ind{1} \leq \ind{2} < \ind{3}$ such that $\overrightarrow{P_{\ind{2}}P_{\ind{3}}} =\vu$. Moreover, let $Q=\omepluse{P_{[\ind{2},\ind{3}]}}$ and then, there exists $(a,b)\in V(\grid)$ such $(a,b) \in \rect(\sbound)$ and $(a,b) \in \dom{Q}$. By the periodicity of $Q$ (fact \ref{fact:ext:per}), there exists $\ind{2} \leq \ind{4} \leq \ind{3}$ and $\ell \in \mathbb{N}$ such that $\pos(P_{\ind{4}})+\ell \vu=(a,b)$. By definition of the stakes constraint (definition \ref{constraint:stakes}), we have $y_{\vu} \geq 1$ and by the first quadrant convention (convention \ref{convention:quadrant}), we have $y_{P_{\ind{4}}}\geq 0$. Since $y_{P_{\ind{4}}}=b-\ell y_{\vu}$, we have $b=y_{P_{\ind{4}}}+\ell y_{\vu}$ and it follows that $\ell \leq b$. Moreover, since $(a,b)\in \rect(\sbound)$, we $b\leq \sbound$ and then $\ell \leq \sbound$. Now, since $(a,b) \in \rect(\sbound)$, we have $a\leq \sbound$ and $x_{P_{\ind{4}}} = a-\ell x_{\vu}  \leq a+\ell ||\vu|| \leq a+\sbound||\vu||\leq \sbound (||\vu||+1)\leq \sbound(\border +1)$. Finally, $y_{P_{\ind{4}}}=b-\ell y_{\vu} \leq b \leq \sbound \leq \sbound(\border +1)$ and then $\pos(P_{\ind{4}})$ belongs to $\rect(\sbound(\border +1))$. By hypothesis, $\dom{P_{[1,\ind{1}]}} \not \subset \rect(\ebound)$ then there exists $1\leq \ind{5} \leq \ind{1}$ such that $\pos(P_{\ind{5}}) \notin \rect(\ebound)$. Since $\freset(\sbound(\border +1))\leq \ebound$ (see appendix \ref{app:bound}), it follows that $P$ does not satisfy the reset constraint for $\rect(\sbound(\border +1))$ and thus by the reset lemma (lemma \ref{macro:reset}), $P$ is fragile or pumpable.
\end{proof}

The seventh final case occurs if the good segment is of direction $-\vu$ and its pumping enters the stakes zone. This is the only case where reverse pumping is necessary.%, it is possible to prove the fragility or pumpability of $P$ as in the previous lemmas. If reverse pumping is not allowed, then section \ref{sec:last} is dedicated to deal with this case. 

\begin{definition}[Final case $7$: Pumping enters stake zone (reverse)]
Consider a path assembly $P$ producible by $\mathcal{T}=(T,\sigma,1)$ satisfying the macroscopic initial conditions and the stakes constraint of direction $\vu$ at index $\inds$. Consider an index $\inds \leq \ind{1} \leq \lastp$ and let $\ind{3}=\hunt(\ind{1})$. The index $\ind{1}$ matches the \emph{Pumping enters stake zone} final case if there exists $\ind{1} \leq \ind{2} < \ind{3}$ such that $\overrightarrow{P_{\ind{2}}P_{\ind{3}}} =-\vu$ and the segment $P_{[\ind{2},\ind{3}]}$ is a good segment whose pumping intersects the stakes zone, \emph{i.e.} $\dom{\omepluse{P_{[\ind{2},\ind{3}]}}} \cap \rect(\sbound) \neq \emptyset$.

%For any index $\inds \leq \ind{1} \leq \lastp$, the index $\ind{1}$ matches the \emph{Pumping enters stake zone (reverse)} final case if there exists $\ind{1} \leq \ind{2} < \ind{3}$ such that $\overrightarrow{P_jP_{\hunt(i)}} =-\vu$ and the segment $P_{[j,\hunt(i)]}$ is a good candidate segment whose pumping intersects the stakes zone.
\end{definition}

\begin{lemma}
\label{hunt:peszr}
Consider a path assembly $P$ producible by $\mathcal{T}=(T,\sigma,1)$ satisfying the macroscopic initial conditions and stakes constraint at index $\inds$. If there exists $\inds \leq \ind{1} \leq \lastp$ which matches the \emph{Pumping enters stakes zone (reverse)} final case and such that $\pos(P_{\ind{1}}) \notin \rect(\ebound)$ then $P$ is fragile or pumpable.
\end{lemma}

\begin{proof}
Consider a path assembly $P$ producible by $\mathcal{T}=(T,\sigma,1)$ satisfying the macroscopic initial conditions and stakes constraint of direction $\vu$ at index $\inds$ with path assembly $\stakep$. If there exists $\inds \leq \ind{1} \leq \lastp$ which matches the \emph{Pumping enters stakes zone (reverse)} final case and such that $\pos(P_{\ind{1}}) \notin \rect(\ebound)$. Then let $\ind{3}=\hunt(\ind{1})$ and $\ind{1} \leq \ind{2} < \ind{3}$ such that $P_{[\ind{2},\ind{3}]}$ is a good segment and $\overrightarrow{P_{\ind{2}}P_{\ind{3}}}=-\vu$. Let $Q=\omemoinse{P_{[\ind{2},\ind{3}]}}$ and if $P_{[1,\ind{2}]} \cdot Q$ is path assembly producible by $\tiling$ then $P_{[\ind{2},\ind{3}]}$ is pumpable (we use reverse pumping in this case). Otherwise if $\dom{Q} \cap \rect(\sbound)$ is not empty then by a reasoning similar to the one of lemma \ref{hunt:pesz}, we can show that $P$ does not satisfy the reset constraint for $\rect(\sbound(\border +1))$ and thus by the reset lemma (lemma \ref{macro:reset}), $P$ is fragile or pumpable.  Otherwise, let $\ind{4}$ be the first index such that $P_{[1,\ind{2}]} \cdot Q_{[1,\ind{4}]}$ is not a path assembly producible by $\tiling$ and let $\ell=|P_{[\ind{2},\ind{3}]}|=\ind{3}-\ind{2}+1$. If $\ind{4} \leq \ell$ then $P$ conflicts with $P_{[\ind{2},\ind{3}]}+\vu$ and then $P$ is fragile by the stakes lemma \ref{macro:stakes} (see Figure \ref{fig:hunt:proof}). Otherwise, by a reasoning similar to lemma \ref{hunt:fcac}, $Q_{[\ell,\ind{4}]}$ and $P_{[s,\ind{2}]}+\vu$ have no conflict and then $\stakep \cdot (P_{[s,\ind{2}]}+\vu) \cdot Q_{[\ell,\ind{4}]}$ is producible by $\tiling$ and $P$ is fragile. 
  
\end{proof}

To conclude this section, we show that for any $\inds \leq \ind{1} \leq \lastp$ at least one of these seven cases cases is matched.

\begin{lemma}
\label{theorem:hunt}
Consider a path assembly $P$ producible by $\mathcal{T}=(T,\sigma,1)$ satisfying the macroscopic initial conditions and the stakes constraint at index $\inds$ and an index $\inds \leq \ind{1} \leq \lastp$. Then, index $\ind{1}$ has to match at least one of the seven final cases.
\end{lemma}

\begin{proof}
Figure \ref{fig:hunt:proof} illustrates the reasoning made along this proof. Consider a path assembly $P$ producible by  $\mathcal{T}=(T,\sigma,1)$ satisfying the macroscopic initial conditions and the stakes constraint at index $\inds$ according to vector $\vu$ and an index $\inds \leq \ind{1} \leq \lastp$. First if $P_{[\ind{1},\lastp]}$ and $P_{[\ind{1},\lastp]}-\vu$ does not intersect then index $\ind{1}$ matches the \emph{jail} final case. Otherwise if $P_{[\ind{1},\lastp]}$ and $P_{[\ind{1},\lastp]}-\vu$ intersect then either this intersection is a conflict (index $\ind{1}$ matches the \emph{conflict} final case) or there is an agreement between $P_{[\ind{1},\lastp]}$ and $P_{[\ind{1},\lastp]}-\vu$. If this intersection is an agreement let $\ind{3}=\hunt(\ind{1})$ and by definition of $\ind{3}$, two cases can occurs either $\pos(P_{\ind{3}}-\vu) \in \dom{P_{[\ind{1},\ind{3}]}}$ or $\pos(P_{\ind{3}}) \in \dom{P_{[\ind{1},\ind{3}]}-\vu}$. In the first (resp. second) case, there exists $\ind{1} \leq \ind{2} < \ind{3}$ such that $P_{\ind{2}}=P_{\ind{3}}-\vu$ (resp. $P_{\ind{2}}-\vu=P_{\ind{3}}$) and then the segment $P_{[\ind{2},\ind{3}]}$ is a candidate segment and $\overrightarrow{P_{\ind{2}}P_{\ind{3}}} =\vu$ (resp. $\overrightarrow{P_{\ind{2}}P_{\ind{3}}}=-\vu$). Moreover, since $P_{[\ind{1},\ind{3}]}$ and $P_{[\ind{1},\ind{3}]}-\vu$ intersect only at position $\pos(P_{\ind{2}})$ (resp. $\pos(P_{\ind{3}})$) and since $\ind{1}\leq \ind{2}$ then the segment $P_{[\ind{2},\ind{3}]}$ is also a good segment of $P$. Consider the case $\overrightarrow{P_{\ind{2}}P_{\ind{3}}} =\vu$ (the other case is symmetric) then if the good segment $P_{[\ind{2},\ind{3}]}$ is pumpable then index $\ind{1}$ matches the \emph{pumpable} final case. Otherwise, let $Q=\omepluse{P_{[\ind{2},\ind{3}]}}$ and let $\ind{4}$ be the first conflict of the good segment $P_{[2,3]}$ of $P$. By lemma \ref{lem:good:cond}, we have $\pos(Q_{\ind{4}}) \in \dom{\seed \cup P_{[1,\ind{2}-1]}}$. We consider the different cases:
\begin{itemize}
\item $\pos(Q_{\ind{4}}) \in \dom{P_{[\inds,\ind{2}-1]}}$, if $\dom{Q}\cap \rect(\sbound)\neq \emptyset$ then index $\ind{1}$ matches the \emph{Pumping enters stakes zone} final case otherwise  index $\ind{1}$ matches the \emph{Conflict after stake} final case;
\item $\pos(Q_{\ind{4}}) \in \dom{\seed \cup P_{[1,\inds]}}$: by definition of the stakes constraint, $\dom{\seed \cup P_{[1,\inds]}}$ is a subset of the stakes zone and thus index $\ind{1}$ matches the \emph{Pumping enters stakes zone} final case.
\end{itemize}
\end{proof}

Then, all the lemmas of this subsection can be summarized in the following corollary. 

\begin{corollary}
\label{cor:hunt:reverse}
Consider a path assembly $P$ producible by $\mathcal{T}=(T,\sigma,1)$ satisfying the macroscopic initial conditions and the stakes constraint. Then path assembly $P$ is fragile or pumpable.
\end{corollary}

\begin{proof}
Consider a path assembly $P$ producible by $\mathcal{T}=(T,\sigma,1)$ satisfying the macroscopic initial conditions and the stakes constraint at index $\inds$. Then consider the first index $\ind{1}$ such that $s\geq \ind{1}$ and $P_{[1,\ind{1}]} \notin \rect(\ebound)$ ($\ind{1}$ exists since $\sbound < \ebound < \fbound$). By lemma \ref{theorem:hunt}, index $\ind{1}$ has to match one of the seven final cases. Remark that $\dom{P_{[1,\ind{1}]}} \subset \rect(\ebound+1)$ and whatever final case $\ind{1}$ matches we can always conclude that $P$ is always fragile or pumpable by lemmas or facts \ref{hunt:jail}, \ref{hunt:conf}, \ref{fact:pump}, \ref{hunt:fcac}, \ref{hunt:fcacr}, \ref{hunt:pesz} or \ref{hunt:peszr}. 
\end{proof}

Finally we can state the main theorem and prove it.

\begin{theorem}[Main theorem]
\label{theorem:final:reverse}
Consider a path assembly $P$ producible by $\mathcal{T}=(T,\sigma,1)$, if the distance between $P$ and the seed is more than $\fbound$ then $P$ is fragile or pumpable.
\end{theorem}

\begin{proof}
Consider a path assembly $P$ producible by $\mathcal{T}=(T,\sigma,1)$ such that its distance from the seed is strictly more than $\fbound$ then by lemma~\ref{lem:goodMacroCond}, there exists a path assembly $P'$ producible by a tiling system $\mathcal{T'}=(T',\sigma',1)$ (with $|\sigma|=|\sigma'|$, $|T'|=|T|$) such that $P'$ satisfies macroscopic initial conditions and such that if $P'$ is fragile or pumpable then $P$ is fragile or pumpable. By lemma \ref{macro:stakes}, if $P'$ does not satisfy the stakes constraint, it is fragile or pumpable. By lemma \ref{cor:hunt:reverse}, if $P'$ satisfies the stakes constraint then $P'$ is fragile or pumpable. Thus in all cases, $P'$ is fragile or pumpable.
\end{proof}

\section{Two dimensional discrete toolbox}
\label{sec:dgt}
Along the proofs, we will need to isolate some specific parts of the $2D$ grid. Achieving this result is far from trivial. Firstly, we start by defining the kind of cut we will use along the articles. This definition is quite general and we list the specific settings which will appear along the article. Secondly, we study how a path interact with our cuts. Thirdly, we introduce the decomposition of a path according to a cut: we divide a path into segments such that each of these segments belongs only to one part of the cut. Although these notions seems intuitive, some special cases require precise definitions. Finally, we end this section by an application of our results to a specific case which will be important later. The figures of this section are in appendix \ref{app:tool}.

\subsection{Definition of a cut of the grid}
\label{sec:sub:cut}

To disconnect a graph, there exist two classical methods: cuts where a subset of edges are removed and vertex separators where a subset of vertices are removed. In this article, these methods are not efficient because they may cut the grid into more than two parts. Instead, we will use an hybrid approach to cut the $2D$ discrete grid $\grid$. We will use a simple path, called a \emph{window}, which is either a bi-infinite path or a cycle to divide the grid into two parts. One part corresponds to the left side of the window; the other one corresponds to the right side (see Figure \ref{fig:cut:example} for an example) of the window. The definition of our cut is done in several steps. A \emph{cut} $(G_1,G_2)$ and its \emph{window} $\wind$ are three graphs such that: 
\begin{itemize}
\item $V(G_1) \cup V(G_2)=V(\grid)$ and $E(G_1) \cup E(G_2)=E(\grid)$;
\item $G_1$ and $G_2$ are connected graphs;
\item $V(G_1)\cap V(G_2)= V(\wind)$;
\item $E(G_1) \cap E(G_2) =E(\wind)$;
\item $\wind$ is either a simple cycle or a simple bi-infinite path;
%\item straight edge common out of $\wind$ belong to $V_1$ or $V_2$;
%\item corner edge common out of $\wind$ belong to $V_1$ or $V_2$; besoin de turn left or turn right 
\end{itemize}
Intuitively, the union of graphs $G_1$ and $G_2$ is the grid graph $\grid$. Nevertheless, these two graphs shares some vertices and edges in common. These vertices and edges create a simple path $\wind$ (bi-infinite or a cycle) which is the window of the cut. The graphs $G_1$ and $G_2$ will correspond to the left and right side of the window and thus to the two zones of our cut.  Remark that with our definition, there exists no edge with an extremity in $V_1\setminus V(\wind)$ and the other one in $V_2\setminus V(\wind)$. Thus any path $P$ starting in $V_1\setminus V(\wind)$ (resp. $V_2\setminus V(\wind)$) and ending in $V_2\setminus V(\wind)$ (resp $V_1\setminus V(\wind)$) has to intersect with the window $\wind$. These intersections between $P$ and $W$ will allow us to decompose $P$ into subpaths, called \emph{arcs}, which belongs exclusively to $G_1$ or exclusively to $G_2$. Before this, we still have to deal with one obstacle to define properly our cut.
% We will later proves some useful property on the window but 
%Intuitively, the window $W$ has two sides, we would like that one side correspond to $G_1$ and the other one to $G_2$ (see figure \ref{}). 
Indeed, the current definition allow us to define $G_1=G$ and $G_2=\wind$ which is not useful for our study. To solve this problem we have to define the right and left side of the window, to achieve this aim we need an indexing of the window $W$.

We remind that the \emph{indexing}  (see section \ref{def:graph}) of the window $W$ is a representation of this path as a bi-infinite sequence of vertices $(W_i)_{i \in \mathbb{Z}}$ such that for any $i \in \mathbb{Z}$, $W_i$ is a vertex of $V(\wind)$, the edge $(W_i,W_{i+1})$ belongs to $E(W)$ and $W_{i-1}\neq W_{i+1}$. Moreover, if $\wind$ is a cycle then for any $i\in \mathbb{Z}$, we have $W_{i+ |V(\wind)|}=W_i$. Note that for a given window $\wind$, an indexing of this window is not unique and depends on two parameters: the \emph{origin} and the \emph{orientation}. Indeed, one position of $W$ has to be chosen as the origin $W_0$ of the index. If $\wind$ is a cycle there exists $|\wind|$ possibilities, otherwise there exists an infinity of possibilities. After choosing an origin $W_0$, one of the two neighbors of $W_0$ has to be chosen as $W_1$, and the other one as $W_{-1}$ (see figure \ref{fig:cut:index}). Here, two possibilities arise. Consider a window $W$ and two indexing $(W_i)_{i \in\mathbb{Z}}$ and  $(W'_i)_{i \in \mathbb{Z}}$ of $W$, we say that these two indexing \emph{have the same orientation} if and only if for all $i,j \in \mathbb{Z}$ such that $W_i=W'_j$, we have $W_{i+1}=W'_{j+1}$ otherwise they do not have the same orientation. Along the article, the choice of $W_0$ and of the orientation is either not relevant or specified when necessary.% and can be done arbitrarily while the orientation will depend on the type of cut we will use. We will generally specify the origin and orientation when these criteria are relevant. 

Now that indexing have been defined, we can analyze the structure of $W$ more precisely. Consider a window $\wind$ and an indexing of $\wind$, then   the window $\wind$ is made of two kinds of vertices: vertex $W_i$ is \emph{straight} if $x_{W_{i-1}}=x_{W_{i+1}}$ or $y_{W_{i-1}}=y_{W_{i+1}}$ otherwise it is a \emph{corner} (see figure \ref{fig:cut:example}). Note that being straight or being a corner does not depend on the origin or orientation of the index. Now remark, that any position $W_i$ of $V(W)$ is the extremities of four different edges of $E(\grid)$. Among these four edges, two of them belongs to $V(\wind)$ since $\wind$ is a bi-infinite path or a cycle. Among the two others edges, we impose new constraints: if $W_i$ is a corner then the two other edges both belong to the graph $G_j$ with $j \in \{1,2\}$, otherwise if $W_i$ is straight then one of the edge belongs to $G_1$ and the other one belongs to $G_2$. To finish our definition of a cut of the grid, we need one last more constraint. Consider a vertex $A \in V(G)$ which is a neighbor of $W_i$ and such that the edge $(A,W_i)$ belongs to $E(\grid) \setminus E(\wind)$ then if the angle $\widehat{W_{i-1}W_iA}=\frac{\pi}{2}$ (resp. $\widehat{W_{i-1}W_iA}=-\frac{\pi}{2}$) then the edge $(W_i,A)$ belongs to $G_1$ (resp. $G_2$), we say that that $A$ is in the left side (resp. right side) of $W_i$. Note that the definition of left and right depends on the orientation of the indexing, switching the orientation will switch the left and right side  (see figure \ref{fig:cut:index}). Now all constraints of our cut have been specified. We conjecture that that for any window $W$ which is a simple bi-infinite path or a simple cycle there exists a unique couple of graphs $(G_1,G_2)$ such that $G_1$, $G_2$, $\wind$ satisfies all these properties. Hopefully, we will only consider four kinds of cut where these properties could be achieved "easily". Before defining these cuts we recap the different kind of vertices and edges which will be encountered along the proof (see figure \ref{fig:cut:type}).

There exists three kinds of vertices: the ones in $V(W)$ (and thus in both $V(G_1)$ and $V(G_2)$), the ones in $V(G_1) \setminus V(W)$ and the ones in $V(G_2) \setminus V(W)$. Things are more complicated for edges since there exists five kinds of edges (see figure \ref{fig:cut:type}): 
\begin{itemize}
\item the ones where at least one extremity is in $V(G_1) \setminus V(W)$ and which belongs to $E(G_1)$; 
\item the ones where at least one extremity is in $V(G_2) \setminus V(W)$ and which belongs to $E(G_2)$;
\item the ones which belongs of $E(\wind)$ and thus to both $E(G_1)$ and $E(G_2)$;
\item the ones where both extremities are in $V(W)$ and belongs to $E(G_1) \setminus E(\wind)$;
\item the ones where both extremities are in $V(W)$ and belongs to $E(G_2) \setminus E(\wind)$;
\end{itemize}
Edges of the fourth and fifth kinds requires that there there exists $i,j \in \mathbb{Z}$ such that $W_i \notin \{W_{j-1},W_j, W_{j+1}\}$ and $W_i$ is a neighbors of $W_j$ in $\grid$. Finally, consider an edge $(u,v) \in E(\grid)$ and and index $\ind{1}$ such that $u= W_{\ind{1}}$ then to determine to which part of the cut of the grid this edge belongs, we only need to know $W_{[\ind{1}-1,\ind{1}+1]}$. In this case, we say that $(u,v)$ is \emph{strictly on the left side} (res. right side) of $W_{[\ind{1}-1,\ind{1}+1]}$ if and only if $(u,v) \notin E(W)$ and $(u,v)$ is on the left side of the cut of the grid (resp. right side), see Figure \ref{fig:cut:type}. Since this notion relies only on local argument, it could be extended to finite paths.

\vspace{+1em}

\noindent \textbf{Cutting the grid with a simple cycle.}
%\vspace{+0.5em}
%\noindent \textbf{Cutting the plane} 
%Along the proof, we will cut the grid $G$ into different parts. These parts could be finite or infinite. We use two kinds of cut: cuts by cycle and cuts by infinite line. We also make combinations of cuts. Determining to which component a position belongs is a tricky question, we introduce tools to answer it. We start by the cut of the plane by a finite simple cycle. 
Consider a simple cycle $C$ then it is always possible to find a cut $(\gint,\gext)$ such that its window is $C$. Moreover, one of the two graphs is infinite while the other one is finite, see Figure \ref{fig:cut:cycle}. We will consider that $\gint$, called the \emph{interior} of $C$, is the finite one and $\gext$, called the \emph{exterior} of $C$, is the infinite one. This result is due to Jordan curve theorem. This theorem requires a planar embedding of the grid graph, as well as ``non-integer half-square-tiles''. Although our drawings use a planar embedding, our proofs do not. Jordan curve theorem stipulates that any continuous simple cycle of the planar graph cuts $\mathbb{R}$ in two components: a finite one and an infinite one. To apply our result in our setting, we just have to consider the continuous counterpart of our cycle $C$. A vertex $A$ is in $\gint$ if the point in the continuous grid is inside the finite component of $C$. Same reasoning is done for edges especially for the ones linking two vertices of the window. For the indexing, its origin and orientation will be specified when required.

\vspace{+1em}

\noindent \textbf{Cutting the grid with lines.}
Now, we define a cut of the grid graph $G$ using discrete lines, which typically live in $\mathbb{R}^2$. Consider a vector $\vu$ of $\mathbb{Z}^2$ and a line of slope $\vu$, we would like to cut the grid in two parts, one part with vertices which are over the line and the other part with vertices which are under the line. Fortunately, discrete approximation of continuous line with rational slope is a well known topic: in language theory they are represented as Christoffel word (Sturmian words represented discrete approximation of continuous lines with irrational slopes). The following tools are folklore and easy to check. The \emph{height function} of direction $\vu$ is the function $h:V(G)\rightarrow \mathbb{Z}$ where the \emph{height} of vertex $(x,y) \in V(G)$ is $h(x,y)=x_{\vu}y-xy_{\vu}$. Remark that for any vertex $X \in V(G)$, we have $h(X)=h(X+\vu)$. The \emph{cut of the grid by the line of direction $\vect{v}$ of height $h$} is the following two infinite graphs $\lplus$ and $\lmin$ (see Figure \ref{fig:cut:line}a):

\begin{eqnarray*}
  V(\lplus)&=&\{X\in V(G) | h(X)\geq h\}\\
  E(\lplus)&=&\{(X,Y) \in E(G) | X\in V(\lplus), Y\in V(\lplus)\}\\
  V(\lmin)&=&\{X \in V(G) | h(X) < h+x_{\vu}+y_{\vu}\}\\
  E(\lmin)&=&\{(X,Y) \in E(G) | X\in V(\lmin), Y\in V(\lmin)\}\\
\end{eqnarray*}

The graph $\lplus$ is called the upper part of the cut and the graph $\lmin$ is the lower part of the cut. The window $L$ is $V(L)=V(\lplus)\cap V(\lmin)=\{X\in\mathbb{Z}^2 | h\geq h(X) > h -x_{\vu}-y_{\vu}\}$ and $E(L)= \{(X,Y) \in E(G) | X\in V(W), Y\in V(W) \text{ and } (X,Y) \in E\}$. This window is a simple bi-infinite path which is stable under translation of direction $\vu$: $L+\vu=L$. Moreover, the window is made by a periodic pattern of length $\per=x_{\vu}+y_{\vu}$ (see figure \ref{fig:cut:line}). This remark allow us to always choose an orientation of the indexing such that for all indices $i \in \mathbb{Z}$, we have $L_{i+\per-1}=L_i+\vu$. Such an indexing is called \emph{coherent} with the direction $\vu$. With this convention, the zone $\lplus$ is the left side and the zone $\lmin$ is the right side of the cut of the grid. Also, for this kind of cut there exists no edge of the four or fifth kind: if we consider an edge $(u,v) \in E(\grid)$ such that $u \in V(W)$ and $v\in V(W)$ then $(u,v) \in E(W)$. Along the article, the vector $\vu$ of $\mathbb{Z}^2$ will always be such that either $y_{\vu} \geq 1$ or $y=0$ and $x>0$. Finally, no confusion on the direction $\vu$ will occur along the article: if different cuts of the grid using discrete lines are considered at the same time in the article, then these lines will all have the same direction but with different heights. The considered height function will be the same for all these cuts of the grid.

\vspace{+1em}

\noindent \textbf{Cutting the grid with the extension of a good path.} This cut is a generalization of the previous case. Consider a good path $P$ and let $\vu=\overrightarrow{P_1P_{\lastp}}$ be its direction. Let $W=\ome{P}$. Since $P$ is a good path, then by lemma \ref{lem:good:cond} the path $W$ is a simple bi-infinite path. By the periodicity of $W$ (see fact \ref{fact:path:good}), there exists an indexing of $W$ such that for all $i\in \mathbb{Z}$, we have $W_{i+\lastp-1}=W_i+\vu$. Such an indexing of $W$ is \emph{coherent} with $P$ if $W_0=P_1$. Consider the height function $h$ of direction $\vu$ and let $H= \max\{h(A): A \in P\}$ and $h= \min\{h(A): A \in P\}$, then for any $W_i$, we have $H \geq h(W_i) \geq h$. Then, there exists a cut $(\lplus,\lmin)$ of the grid such that $W$ is its window (see Figure \ref{fig:cut:goodPath}) and such that the graph $\lplus$ contains all vertices of height greater than $H$ and is the left side of $W$ (if the indexing of $W$ is coherent) while the graph $\lmin$ contains all vertices of height less than $h$ and is the right side of $W$ (if the indexing of $W$ is coherent). Using Jordan's curve theorem, we can determine to which component the vertices and edges of height between $h$ and $H$ belong. This result is possible from the periodicity of $W$, we only have to consider a finite part of the window to determine if the vertices or the edges belong to $\lplus$ or $\lmin$. Morever if a position $A\in V(G)$ or an edge $u \in E(G)$ belongs to $\lplus$ (resp. $\lmin$) then $A+\vu \in V(\lplus)$ (resp. $A+\vu \in V(\lmin)$) or $u+\vu \in E(\lplus)$  (resp. $u+\vu \in E(\lmin)$). Note that cutting the grid with a line is a special case of cutting the grid with a good path since the discretization of a continuous line of rational slope is obtained from a periodic finite pattern. Nevertheless, we made a distinction between these two kinds of cut because they will be used in different settings along the proof. Moreover, cutting the grid with a line is more intuitive: all properties can be checked by computing the height of vertices. %More details will be given to these kinds of cut when we will define stripes and give an alternate proof of lemma \ref{} at the end of this section.

\vspace{+1em}

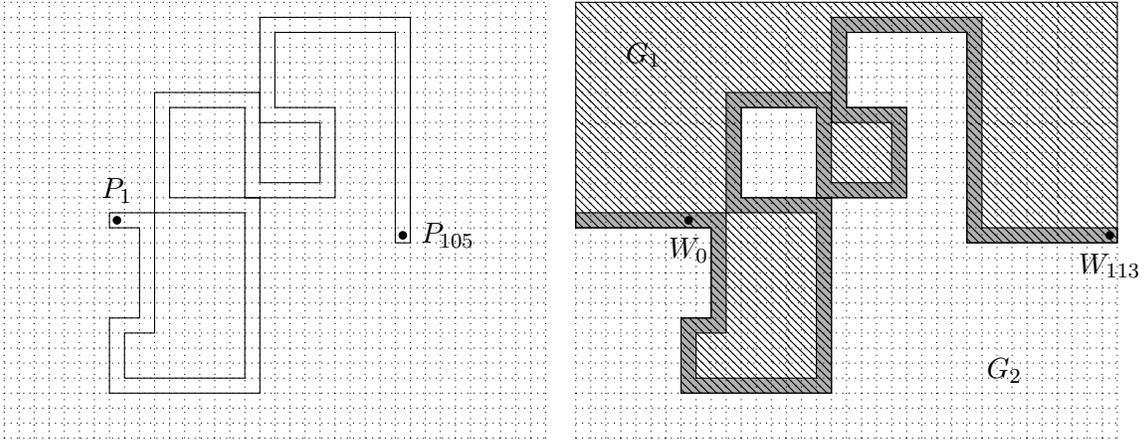
\noindent \textbf{Cutting the grid with a simple path with visible extremities.} Consider a finite simple path $P$, we say that its extremities $P_1$ (resp. $P_{\lastp}$) is \emph{visible the from west} if and only if there exists no $1\leq i \leq \lastp$ such that $x_{P_i}<x_{P_1}$ (resp. $x_{P_i}<x_{P_{\lastp}}$) and $y_{P_i}=y_{P_{1}}$ (resp. $y_{P_i}=y_{P_{\lastp}}$).
% its extremities are \emph{visible} if and only if there exists no $1\leq i \leq \lastp$ such that $x_{P_i}<x_{P_1}$ and $y_{P_i}=y_{P_1}$ and there exists no $1\leq i \leq \lastp$ such that $x_{P_i}>x_{P_{\lastp}}$ and $y_{P_i}=y_{P_{\lastp}}$ (see figure \ref{fig:cut:visible}a). 
Intuitively, $P$ does not pass by a position west of its starting  (resp. ending) point. Similarly, we define the visibility from the east, north and south (see Figure \ref{fig:cut:visible} for an example). The path $P$ has \emph{visible extremities} if and only if $P_1$ is visible (from the west in our example ) and $P_{\lastp}$ is visible (from the east in our example) and the ray starting in their extremities does not intersect (in our example there is no position $A \in V(G)$ such that $x_A\leq x_{P_1}$, $y_A= y_{P_1}$, $x_A> x_{P_{\lastp}}$ and $y_A= y_{P_{\lastp}}$). Such a path could be completed into a bi-infinite path by adding horizontal ray in its ending and starting position (see Figure \ref{fig:cut:visible}b). More formally, let $\wind$ be the bi-infinite path such that for all $0 \leq i \leq \lastp-1$, we have $\wind_{i}=P_{i+1}$, for all $i< 0$, we have $W_i=W_{i+1}+(-1,0)$ (if $P_1$ is visible from the west) and for all $i\geq \lastp$, we have $W_i=W_{i-1}+(1,0)$ (if $P_{\lastp}$ is visible from the east). Then the path $W$ is a simple bi-infinite path since both extremities of $P$ are visible. Moreover, there exists a cut of the grid $(\cutp,\cutm)$ such that $W$ is the window of this cut. This result is a combination of the previous kinds of cut of the grid: $W_{[-\infty,0]}$ and $W_{[\lastp-1,+\infty]}$ are infinite lines while $W_{[0,\lastp-1]}=P$ is a finite path which belongs to a finite rectangle of size $(X_P-x_P+2) \times (Y_P-y_P+2)$. This finite zone could be partitioned using Jordan curve theorem. Note that when we have defined $W$, we also have defined an indexing of $W$. We will always consider this indexing of $W$ which is called \emph{coherent} with $P$. Also this kind of cut could be generalize as previously by extending the extremities of $P$ by any periodic pattern instead of simple lines.%Also, we consider that $\cutp$ is the left side of the cut and $\cutm$ is the right side of the cut.

\vspace{+1em}

\noindent \textbf{Splitting a cut in two.} We have defined the four kinds of cut of the grid that will be used along the article. Afterwards, in the statement or in the proof of a lemma, we will mainly consider only one cut of the grid at a time but sometimes we need to consider two cuts of the grid at the same time. Here lies a subtlety. The aim of a cut of the grid is to divide the grid into two connected components. Thus making two consecutive cuts of the grid should divide the grid into four connected components. The intuitive approach to deal with two cuts of the grid $(G_1,G_2)$ and $(G_3,G_4)$ is to consider the four following graphs: $G_1\cap G_3$, $G_1 \cap G_4$, $G_2 \cap G_3$ and $G_2\cap G_4$. Unfortunately, this approach is not correct because in some cases these graphs are not connected (see Figure \ref{fig:split:tikz}). We now explain how to achieve this goal in a very specific setting which is the one used in section \ref{sec:Uturn}, see Figure \ref{fig:split:tikz:corr}. 

Consider a finite simple path $P$ with $P_1$ visible from the west and $P_{\lastp}$ visible from the east and consider the cut of grid $(\cutp,\cutm)$ according to the finite simple path $P$ with visible extremities. Let $W$ be the window of this cut of the grid and consider a coherent indexing of $W$. Now consider a good path $Q$ such that $\omeplus{Q}$ is a subgraph of $\cutp$ and such that there exists an index $\ind{1} \in \mathbb{Z}$ such that $\dom{Q} \cap V(W)=Q_1=W_{\ind{1}}$. We say that the extension of $Q$ splits $\cutp$ at position $W_\ind{1}$ into $(A^+,A^-)$ where $A^+$ and $A^-$ is a cut of the graph $\cutp$, \emph{i.e.} $A^+$ and $A^-$ are subgraphs of $\cutp$, $A^+\cup A^-=\cutp$, $A^+\cap A^-=\omeplus{Q}$ and $A^+$ contains $W_{[\ind{1},+\infty]}$ and $A^-$ contains $W_{[-\infty,\ind{1}]}$.

\subsection{Interactions between a path and a window}

Now, that we have an efficient way to decompose the grid in different zones, we analyze how a path interacts with these zones. In this section we consider a cut $(G_1,G_2)$, its window $\wind$ and an indexing of this window. Until the end of the section, we will always assume that this window is a bi-infinite path. Now, our final aim is to decompose a path $P$ into a sequence of subpaths such that all these subpaths are either subgraphs of $G_1$ or subgraphs of $G_2$. Finding such a decomposition is easy but there exists several of them and to find an interesting one, we have to be careful and precise in our definitions. First, we introduce some definitions to classify different kinds of path according to their interactions with $W$ (see Figure \ref{fig:cut:arc}).

In this section, we consider a simple finite path $P$ such that $P_{1} \in V(W)$ and $P_{\lastp} \in V(W)$. From now on, be careful that we deal with two indexings, the one of the window $W$ and the one of the path $P$. Let $\ind{1}=\min\{i: W_i \in V(P)\}$ and $\ind{2}=\max\{i: W_i \in V(P)\}$, then we say that $P$ starts in $W_{\ind{1}}$ and ends in $W_{\ind{2}}$. Remark that $\ind{2} \geq \ind{1}$ and the \emph{width} of path $P$ is $w(P)=\ind{2}-\ind{1} \geq 0$. The \emph{segment of the window associated} to $P$ is $W_{[\ind{1},\ind{2}]}$. Note that $V(P) \cap V(\wind) \subset V(\wind_{[\ind{1},\ind{2}]})$. The path $P$ is \emph{extremum} if and only if $\{P_1, P_{\lastp}\}=\{W_{\ind{1}},W_{\ind{2}}\}$ (we will mainly consider extremum paths in the article).  A path $P$ is \emph{basic} if there exists two indices $\ind{3},\ind{4}$ such that $W_{\ind{3}}=P_1$, $W_{\ind{4}}=P_{\lastp}$ and if $\ind{3} \leq \ind{4}$ then $V(W_{[\ind{3},\ind{4}]}) \cap V(P)=\{W_{\ind{3}},W_{\ind{4}}\}$ or $V(W_{[\ind{4},\ind{3}]}) \cap V(P)=\{W_{\ind{3}},W_{\ind{4}}\}$ otherwise. Consider $i \in \{1,2\}$, then the path $P$ is an \emph{arc} of $G_i$ if and only if $P$ is a subgraph of $G_i$. Note that the hypothesis $E(P)\subset E(G_i)$ implies that $V(P) \subset V(G_i)$ (and thus that $P$ is a subgraph of $G_i$) whereas the reciprocity is false, see the path $P$ in Figure \ref{fig:cut:arc}. Similarly to path, an arc could be basic or extremum. Finally an arc is \emph{elementary} if and only if $V(A) \cap V(\wind) = \{A_1, A_{|A|}\}$, \emph{i.e.} the arc intersects the window only at its extremities. Being an elementary arc is equivalent to being a basic and extremum path. %Along the proof, basic arcs are only needed during the proofs of lemmas 

Now, consider a path $P$ such that $P$ starts in $W_{\ind{1}}$ and ends in $W_{\ind{2}}$ and $w(P)>0$, then there exists $1 \leq \ind{3} \leq \lastp$ (resp. $1 \leq \ind{4} \leq \lastp$) such that $P_{\ind{3}}=W_{\ind{1}}$ (resp. $P_{\ind{4}}=W_{\ind{2}}$). Since $w(P)>0$ then $\ind{3} \neq \ind{4}$. If $\ind{3} < \ind{4}$ then we say that $P$ is a \emph{positive} path otherwise $P$ is \emph{negative}. If $P$ is positive (resp. negative) then $P_{[\ind{3},\ind{4}]}$ (resp. $P_{[\ind{4},\ind{3}]}$) is an extremum path and a subgraph of $P$, this path is called \emph{the extremum path} extracted from $P$ (see Figure \ref{fig:cut:extremum}). Note that we have defined the orientation of path of width at least $1$. The definition of the orientation of path of width $0$ is postpone to the next subsection.

When paths interact with a window, they delimit different zones of the grid. Firstly, consider a basic path $P$ and $\ind{1},\ind{2}$ such that $W_{\ind{1}}=P_1$ and $W_{\ind{2}}=P_{\lastp}$. We suppose that $\ind{1} \leq \ind{2}$, the other case is symmetric, then $P$ and $W_{[\ind{1},\ind{2}]}$ creates a simple cycle $C$. Now, consider the cut $(\gint,\gext)$ of $\grid$ defined by $C$, then the finite graph $\gint$ is called the \emph{restrained interior} of the basic path $P$ (see Figure \ref{fig:def:basic}). This graph is not necessarily a subgraph of $G_1$ or a subgraph of $G_2$.

%\begin{figure}[th]
%\centering
%\begin{tikzpicture}[x=0.3cm,y=0.3cm]
%
%%restrained interior P1
%\path [draw, fill=gray!60] (16,7) -| (24,19) -| (34,8) -| (31,14) -|(29,-1) -| (12,4) -| (16,7);
%\path [draw, fill= white] (17,6) -| (25,18) -| (33,9) -| (32,15) -|(28,0) -| (13,3) -| (17,6);
%\path [pattern=north west lines] (16,7) -| (24,19) -| (34,8) -| (31,14) -|(29,-1) -| (12,4) -| (16,7);
%
%%grid
%\path [dotted, draw, thin] (0,-2) grid[step=0.3cm] (36,27);
%
%%window W 
%\draw (7.5,14) node {$W_0$};
%\draw[fill] (7.5,12.5) circle (0.25) ;
%%\draw (37.8,11.5) node {$W_{121}$};
%\draw (35.5,13) node {$W_{111}$};
%\draw[fill] (33.5,11.5) circle (0.25) ;
%
%\path [draw] (0,12) -| (9,6) -| (7,1) -| (17,14) -| (11,20) -| (16,14) -| (22,20) -| (18,25) -| (26,11) -| (36,12) -| (27,26) -| (17,19) -| (21,15) -| (17,21) -| (10,13) -| (16,2) -| (8,5) -| (10,13) -| (0,12);
%
%%chemin P1 basic démarre en 20 (28) fin en 111 (119)
%
%\draw (18.5,3.5) node {$W_{26}$};
%\draw[fill] (16.5,3.5) circle (0.25) ;
%
%\draw (14.5,6.5) node {$W_{29}$};
%\draw[fill] (16.5,6.5) circle (0.25) ;
%
%%\path [draw, pattern= north west lines] (16,7) -| (24,19) -| (34,8) -| (31,14) -|(29,-1) -| (12,4) -| (17,3) -| (13,0) -| (28,15) -| (32,9) -| (33,18) -| (25,6) -| (16,7);
%
%\end{tikzpicture}
%\caption{The window of the restrained interior of path $P^1$ of figure \ref{fig:cut:arc} is represented in grey, it is defined has $W_{[26,29]} \cdot P^1$. Note that the restrained interior is neither a subgraph of $G_1$ or $G_2$. The original window is in white.}
%\label{fig:def:basic}
%\end{figure}

Secondly, consider a path $P$ with starts in $W_{\ind{1}}$ and ends in $W_{\ind{2}}$, let $Q$ be the extremum path extracted from $P$. Then $W'=W_{[-\infty,\ind{1}]} \cdot Q \cdot W_{[\ind{2},+\infty]}$ is a simple bi-infinite path which can be used to delimit a cut $(G_3,G_4)$ of the grid. This cut is called the \emph{cut associated} to $P$ and $W'$ is the \emph{window associated} to $P$ (see Figure \ref{fig:cut:fund:ext}a for a simple example where $P$ is an arc and see Figures \ref{fig:cut:defP} and \ref{fig:decompo:extremum} for a more complex example). Moreover, if $P$ is an arc of $G_1$, if $G_1$ is the left side of $W$ and if $G_3$ is the left side of $W'$ then $G_3$ is a subgraph of $G_1$ and $G_2$ is a subgraph of $G_4$. In this case, we define the \emph{interior} of the extremum arc $P$ as the graph $\gint$ where $V(\gint)=\{A: A\in (V(G_4) \cap V(G_1)) \}$ and $E(\gint)=\{(u,v): u \in V(\gint), v \in V(\gint) \text{ and } (u,v) \in V(G_4)\}$ (see Figure \ref{fig:cut:fund:ext}b). Note that $\gint$ is always a finite graph and let $S$ be the segment of $W$ associated to $P$ then  $P$ and $S$ are both subgraph of $\gint$. If $P$ is a basic arc then $P \cdot S$ is a simple cycle and $\gint$ is the interior of this cycle. If $P$ is not basic then we cannot define a window for the graph $\gint$. In every cases, note that the only way to enter in the interior of an extremum arc of $G_i$ is either by intersecting the arc $P$ or by passing by an edge of $G_{2}$. 

Now, consider fours indices $\ind{1}, \ind{2}, \ind{3}, \ind{4}$ and two arcs $A$ and $B$ such that $A$ (resp. $B$) starts in $W_{\ind{1}}$ (resp. $W_{\ind{3}}$) and ends in $W_{\ind{2}}$ (resp. $W_{\ind{4}}$), if $\ind{1} \leq \ind{3} \leq \ind{2} < \ind{4}$ then $B$ starts in the interior of $A$ and has to leave it, this means that $A$ and $B$ intersect (see Figure \ref{fig:cut:fund}). This remark is a fundamental property of arc and window:

\begin{fact}
\label{fact:arc}
Consider a cut $(G_1,G_2)$ of the grid, its window $\wind$, $k \in \{1,2\}$ and two arcs $A$ and $B$ of $G_k$ such that $A$ (resp. $B$) starts in $W_{\ind{1}}$ and ends in $W_{\ind{2}}$ (resp. starts in $W_{\ind{3}}$ and ends in $W_{\ind{4}}$), then if $\ind{1}< \ind{3} < \ind{2} < \ind{4}$, then $A$ and $B$ intersect, \emph{i.e.} $V(A)\cap V(B)\neq \emptyset$. %Moreover $A'$ contains an edge which is on the west side of $A$ and one edge on the right side of $A$.
\end{fact}

First, this fact has some consequences for the cuts of the grid by the extension of a good path (see Figure \ref{fig:cut:trans}).

\begin{lemma}
\label{lem:per:arc}
Consider a good path $P$ of direction $\vu$ and let ($U$,$D$) be the cut of the grid by the extension of $P$. Consider an arc $A$ of $U$ or $D$ of width greater than $\lastp$ then $A$ and $A-\vu$ intersect.
\end{lemma}

\begin{proof}
Let $W=\ome{P}$ be the window of this cut of the grid. Without loss of generality, we suppose that $A$ is an arc of $U$. Let $m \geq \lastp$ be the width of $A$ and consider an indexing of $W$ such that $A$ starts in $W_0$ and ends in $W_m$. Since we consider a cut of the grid by the extension of a good path of direction $\vu$ then $A-\vu$ is also an arc of $U$. Moreover, the arc $A-\vu$ starts in $W_{-\lastp}$ and ends in $W_{m-\lastp}$. Since by hypothesis $-\lastp<0\leq m-\lastp <m$, then by fact \ref{fact:arc} the arcs $A$ and $A-\vu$ intersect.
\end{proof}

\begin{corollary}
\label{lem:jail:arc}
Consider a vector $\vu$ of $\mathbb{Z}^2$ and a cut $(U,D)$ of the grid by a line of direction $\vu$ and its window $L$. Consider an arc $A$ of $U$ or $D$ of width greater than $x_{\vu}+y_{\vu}$ then $A$ and $A-\vu$ intersect.
\end{corollary}

Also, fact \ref{fact:arc} is important to categorize the different possible interactions between several arcs. Let $k \in \{1,2\}$ and consider two arcs $A$ and $B$ of $G_k$ such that $A$ (resp. $B$) starts in $W_{\ind{1}}$ and ends in $W_{\ind{2}}$ (resp. starts in $W_{\ind{3}}$ and ends in $W_{\ind{4}}$). Then, there exists three kinds of possible interactions between these two arcs:
\begin{itemize}
\item if $\ind{3} \leq \ind{4}$ then we say that $B$ is \emph{consective} to $A$ (if $\ind{3}<\ind{4}$ then $B$ is \emph{stricly consecutive} to $A$). %Moreover, the \emph{segment of $W$ between $A$ and $A'$} is $W_{[\ind{2},\ind{3}]}$;
\item if $\ind{1}\leq \ind{3}\leq \ind{4} \leq \ind{2}$ then either $B$ is a subgraph of the interior of $A$ or $A$ and $B$ intersect. In this case we say that $A$ \emph{dominates} $B$;%Minside the cycle delimited by the arc $A$ and $W_{[\ind{1},\ind{2}]}$, in this case we say that $A$ \emph{dominates} $A'$;
\item if $\ind{1} < \ind{3} < \ind{2} < \ind{4}$ then in this case $A$ intersects with $B$ by fact \ref{fact:arc}.%, this notion is stronger than the notion of intersection $A$ contains an edge which is on the right side of $A'$ and on edge on the left side of $A'$ (and vice versa).
\end{itemize}

%\begin{fact}
%Let $k \in \{1,2\}$, consider a simple cycle $C$ which is a subgraph of $G_k$ and consider the partition $(\gint,\gext)$ admitting $C$ as a window, then $\gint$ is a subgraph of $G_k$.
%\end{fact}

%\begin{fact}
%Let $k \in \{1,2\}$, consider a window $\wind$ and two simple arcs $A$ and $A'$ of $G_k$ such that $A$ (resp. $A'$) starts in $i$ and ends in $j$ (resp. starts in $i'$ and ends in $j'$) according to the index of $W$, then if $i<i'<j$ and $j<j'$, then $A$ and $A'$ intersect.
%\end{fact}

Remark that if $\ind{3}=\ind{1}$ and $\ind{4}=\ind{2}$ then arc $A$ dominates $B$ and $B$ dominates $A$ but in this case $A$ and $B$ intersect.  Thus if $A$ and $B$ does not intersect then only two cases are possible: either one arc dominates the other one or one arc is strictly consecutive to the other one. %Otherwise these three cases are exclusive: two arcs $A$ and $B$ cannot matches several of this cases at the same times. 
Moreover if $A$ dominates $B$ and does not intersect with $B$ then $B$ is a subgraph of the interior of $A$. Also note that the property of being consecutive is transitive: consider three arcs $A$, $B$ and $C$, if  $A$ is consecutive to $B$ and $B$ is consecutive to $C$ then $A$ is consecutive to $C$. Sequences of consecutive arcs will play a major role in the rest of the article.
%Our aim is to obtain a decomposition of an extremum arc or of a simple path into a sequence of arcs. Thus, the different arcs will not intersect and then the third case cannot occur. The arcs which are dominated are in fact useless in our future study and can be removed. Thus, we will only consider arcs of the first kind which can be ordered according to their starting and ending point in $W$.  

\subsection{Decomposition of a path into extremum arcs}
\label{sub:sec:decompo}

The aim of this part is to decompose a path $P$ into a sequence of strictly consecutive arcs. This decomposition will remove all unnecessary information in $P$ and will help studying it: all positions of $P$ which belong to the other zone of the grid are removed and we keep only extremum and dominating arcs. Consider a cut $(G_1,G_2)$ of the grid, its window $W$ and a simple finite path $P$ such that $P_1 \in V(W)$ and $P_{\lastp} \in V(W)$ (see Figure \ref{fig:cut:defP}). The \emph{decomposition of $P$ into extremum arcs} in $G_1$ is defined in three steps. First consider the finite set of arcs $\mathcal{A}$ of $G_1$ (see Figure \ref{fig:cut:setA}a) such that :

\begin{itemize}
\item for all $A \in \mathcal{A}$, $A$ is an arc of $G_1$;
\item for all $A \in \mathcal{A}$, there exists $1\leq \ind{1} \leq \ind{2} \leq \lastp$ such that $A = P_{[\ind{1},\ind{2}]}$ and if $\ind{1}>1$ (resp. $\ind{2}<\lastp$) then the edge $(P_{\ind{1}-1},P_\ind{1})$ (resp. $(P_{\ind{2}},P_{\ind{2}}+1)$) does not belongs to $G_1$;
\item for all $1\leq i \leq \lastp$ such that $P_i \in V(G_1)$, there exists $A \in \mathcal{A}$ such that $P_i \in V(A)$.
\end{itemize}
Now, we define the finite set of arcs $\mathcal{A}'$ of $G_1$ by replacing each arc of $\mathcal{A}$ by the extremum arc extracted from it, \emph{i.e.} an arc $A'$ belongs to $\mathcal{A'}$ if and only if there exists an arc $A$ of $\mathcal{A}$ such that $A'$ is the extremum arc extracted from $A$ (see Figure \ref{fig:cut:setA}b). Finally the decomposition of $P$ into extremum arcs is the sequence $(A^i)_{1\leq i \leq \ell}$ such that (see Figure \ref{fig:cut:max}): 
\begin{itemize}
\item for all $1\leq i \leq \ell$, there exists $A' \in \mathcal{A'}$ such that $A^i=A'$;
\item for all $A'\in \mathcal{A'}$, there exists $1\leq i\leq \ell$ such that $A^{i}=A'$ if and only if there exists no $A'' \in \mathcal{A'}$ such that $A'$ is dominated by $A''$.
\item for all $1\leq i <\ell$, the arc $A^{i+1}$ is strictly consecutive to $A^{i}$.
\end{itemize}

Remark that all simple paths admit a unique decomposition in extremum arcs and that all arcs of this decomposition are extremum ones. Moreover since $P$ is simple, for all $1\leq i < j \leq \ell$, the arc $A^i$ does not intersect with the arc $A^j$. One important property of a decomposition is that it divides the window in several parts (see Figure \ref{fig:decompo:win}). The \emph{decomposition} of $W$ associated to $(A^i)_{1\leq i \leq \ell}$ is the finite sequence $(S^i)_{1\leq i \leq \ell}$ such that $S^i$ is the segment of $W$ associated to the arc $A^i$. The \emph{dual decomposition} of $\wind$ according to $(A^i)_{1\leq i \leq \ell}$ is the finite sequence $(D^i)_{0\leq i \leq \ell}$ defined as follow:
\begin{itemize}
\item for all $1\leq i \leq \ell-1$, let $\ind{1},\ind{2} \in \mathbb{Z}$ such that $A^i$ ends in $W_{\ind{1}}$ and $A^{i+1}$ starts in $W_{\ind{2}}$ then $D^i=W_{[\ind{1},\ind{2}]}$;
\item let $\ind{1} \in \mathbb{Z}$ such that $A^1$ starts in $W_{\ind{1}}$ (resp. $A^{\ell}$ ends in $W_{\ind{1}}$) then $D^0$ (resp. $D^{\ell}$) is the infinite path $W_{[-\infty,\ind{1}]}$ (resp. $W_{[\ind{1},+\infty]}$).
\end{itemize}
Remark that for all $1\leq i \leq \ell$, if $|A^i|=1$ then $|S^i|=1$ but for all $0\leq i \leq \ell$, we have $|D^i|>1$. Also, remark that $W=D^0 \cdot_{1\leq  i \leq \ell} (S^i \cup D^i)$.

Now, we study the properties of these definitions. The main aim of this subsection is lemma \ref{lem:conc:prel} which states that if a positive simple path $P$ is extremum then its decomposition in extremum arcs is positive. To achieve this result, a lengthy sequence of lemmas is required. Their aims is to characterized the different interactions between the different parts of the grid which are defined by a decomposition. Remark that most of these lemmas will also be used in the next sections and are an important part of our toolbox. Now, we start by proving the main properties of a decomposition into extremum arcs.

\begin{lemma}
\label{lem:decompo:path}
Consider a cut $(G_1,G_2)$ of the grid and its bi-infinite window $W$. Consider a simple finite path $P$ such that $P_1 \in V(W)$ and $P_{\lastp} \in V(W)$ and consider the decomposition $(A^i)_{1\leq i \leq \ell}$  of $P$ in extremum arcs in $G_1$, its decomposition $(S^i)_{1\leq i \leq \ell}$ of $W$ and its dual decomposition $(D^i)_{0\leq i \leq \ell}$ of $W$. Then for all $1\leq \ind{1} \leq \lastp$ such that $P_\ind{1} \in V(G^1)$, there exists $1\leq i \leq \ell$ such that $P_{\ind{1}} \in V(\gint)$ where $\gint$ is the interior of $A^i$, for all $0\leq \ind{1} \leq \ell$, $V(D^i) \cap V(P) \subset \{D^i_1,D^i_{|D^i|}\}$ and the edge $(D^i_1,D^i_2)$ is not an edge of $E(P)$. 
\end{lemma}

\begin{proof}
Let $\mathcal{A}, \mathcal{A'}$ be the two sets used in the construction of the decomposition of $P$ in extremum arcs and consider $1\leq \ind{1} \leq \lastp$ such that $P_{\ind{1}} \in V(G^1)$. Then there exists an arc $A \in \mathcal{A}$ such that $P_{\ind{1}} \in V(A)$. Then there exists an arc $A' \in \mathcal{A}'$ such that $P_{\ind{1}}$ belongs to the interior of $A'$. Now there exists an index $1\leq i \leq \ell$ such that either $A^i=A'$ or $A^i$ dominates $A'$. In both cases $P_{\ind{1}}$ belongs to the interior of $A^i$. Then for all $1\leq \ind{1} \leq \lastp$ such that $P_\ind{1} \in V(G_1)$, there exists $1\leq i \leq \ell$ such that $P_{\ind{1}} \in \gint$ where $\gint$ is the interior of $A^i$. Since $W=D^0 \cdot_{1\leq  i \leq \ell} (S^i \cup D^i)$, then for all $0\leq \ind{1} \leq \ell$, $V(D^i) \cap V(W) \subset \{D^i_1,D^i_{D^i}\}$. Moreover, if there exists $0 \leq i \leq \ell$ such that the edge $(D^i_1,D^i_2)$ is an edge of $E(P)$ then by the previous remark we have $1\leq i \leq \ell-1$ and $|D^i|=2$ but in this case $A^i \cup D^i \cup A^{i+1}$ is an arc of $P$ which dominates both $A^i$ and $A^{i+1}$ and this is a contradiction.
\end{proof}

Now, we can extend the definition of orientation to dual segments and to arcs of length $1$ (see Figure \ref{fig:def:forbid}). Consider $1\leq i \leq \ell-1$ and two indices $\ind{1}$ and $\ind{2}$ such that $P_{\ind{1}}=D^i_1$ and $P_{\ind{2}}=D^i_{|D^i|}$ then the dual segment $D^i$ is \emph{positive} (resp. \emph{negative}) if and only if $\ind{1}< \ind{2}$ (resp. $\ind{1}>\ind{2}$). For an arc of length $1$, we give the following definition. 

\begin{definition}
\label{def:forbidden}
Consider a cut $(G_1,G_2)$ of the grid, its window $W$, a finite path $P$ such that $P_1 \in V(W)$ and $P_{\lastp} \in V(W)$ and the decomposition $(A^i)_{1\leq i \leq \ell}$ of $P$ into extremum arcs in $G_1$. Consider $1\leq i \leq \ell$ such that $|A^i|=1$ and let $1\leq \ind{1} \leq \lastp$ and $\ind{2} \in \mathbb{Z}$ such that $P_{\ind{1}}=W_{\ind{2}}$. If $\ind{1}=1$ (resp. $\ind{1}=\lastp$) then let $\ind{3}=\min\{i>1:P_i \in V(W)\}$ (resp.  $\ind{3}=\max\{i<\lastp:P_i \in V(W)\}$) and let $\ind{4}$ such that $W_{\ind{4}}=P_\ind{3}$. If $\ind{4}>\ind{2}$ then $A^i$ is positive (resp. negative) otherwise $A^i$ is negative (resp. positive). Now if $1<\ind{1}<\lastp$, the arc $A^i$ is a negative arc if and only if up to some rotation $W_{\ind{2}+1}=W_{\ind{2}}+(0,1)$, $W_{\ind{2}-1}=W_{\ind{2}}-(1,0)$, $P_{\ind{1}+1}=P_{\ind{1}}-(0,1)$ and $P_{\ind{1}-1}=P_{\ind{1}}+(1,0)$ otherwise it is positive.
\end{definition}

A decomposition implies the existence of many different finite zones of the grid. Each arc $A^i$ defines a different zone of the grid with its interior. Using all these finite zones we define a new cut of the grid by considering the bi-infinite path $W'=D^0 \cdot_{1\leq  i \leq \ell} (A^i \cup D^i)$ which is called the \emph{window associated} to the decomposition $(A^i)_{1\leq i \leq \ell}$. Then $W'$ is the window of a cut of the grid $(G^3,G^4)$ such that $G^3$ is a subgraph of $G^1$ and $G^{2}$ is a subgraph of $G^4$, see Figure \ref{fig:decompo:newwindow}. Moreover if $G^1$ is the left (resp. right) side of $W$ then $G^3$ is the left (resp. right) side of $W'$. A corollary of lemma \ref{lem:decompo:path} is that $(V(G^3) \setminus V(W))\cap V(P)= \emptyset$: the zone of the grid $G^3$ does not contains any position of $P$ except in its window. 

\begin{corollary}
\label{fact:decompo:newwind}
Consider a cut $(G^1,G^2)$ of the grid and its window $W$. Consider a simple finite path $P$ such that $P_1 \in V(W)$ and $P_{\lastp} \in V(W)$ and the decomposition $(A^i)_{1\leq i \leq \ell}$ of $P$ into extremum arcs in $G^1$. Let $W'$ and $(G^3,G^4)$ be the window and the cut of the grid associated to the decomposition $(A^i)_{1\leq i \leq \ell}$ such that $G^3$ is a subgraph of $G^1$, then $(V(P)\cap V(G^3))\setminus V(W')=\emptyset$.
\end{corollary}

Now, consider $1\leq i \leq \ell-1$, then there exists $1\leq \ind{1} \leq \lastp$ (resp. $1\leq \ind{2} \leq \lastp$) such that $P_{\ind{1}}=D^i_1$ (resp. $P_{\ind{2}}=D^i_{|D^i|}$). If $\ind{1} < \ind{2}$ (resp. $\ind{1} >\ind{2}$) then by lemma \ref{lem:decompo:path}, $P_{[\ind{1},\ind{2}]}$ (resp.$P_{[\ind{2},\ind{1}]}$ ) is a basic path of $W$ and we can define the \emph{interior} of $D^i$ as the restrained interior of $P_{[\ind{2},\ind{1}]}$ (resp. $P_{[\ind{2},\ind{1}]}$), see Figure \ref{fig:decompo:windowdual}. Remark that the interior of an arc is a subgraph of $G_i$ but the interior of dual is not necessarily a subgraph of $G_1$ or $G_2$. Consider the cut $(G_5,G_6)$ associated to $P$ and its windows $W''$, suppose that $G_1$ is the left side of $W''$ and $G_5$ is the left side of $W'$ then the interior of $D^0$ and $D^\ell$ is the graph $G_6$ (see Figure \ref{fig:decompo:extremum}).

To summarize, if $P$ is an extremum path, these three cuts of the grid interact as follow (see Figure \ref{fig:deco:path:ext}):
\begin{itemize}
%\item the cut of the grid $(G_1,G_2)$ and its window $W$;
\item %the cut of the grid $(G_3,G_4)$ associated to the decomposition of $P$ in extremum arcs in $G_1$. 
The zone $G_3$ is a subgraph of $G_1$ (intuitively $G_3$ is obtained by removing the interiors of all arcs $A^i$ from $G_1$);
\item %the cut of the grid $(G_5,G_6)$ associated to the extremum path $P$. 
The zone $G_3$ is a subgraph of $G_5$ (intuitively $G_5$ is obtained by adding to $G_3$ the interiors of all segments $D^i$).
\end{itemize} 

We are mainly interested in the decomposition of extremum paths but we will prove more general results. We starts by two properties about the decomposition of an extremum path into extremum arcs. These results are useful to show lemma \ref{lem:dual:empty} which states that the intersection between the interior of two dual segments is a subset of $V(P)$. The first result is that the first and last arcs of a decomposition of an extremum path $P$ into extremum arcs are positive if $P$ is positive (see Figure \ref{fig:deco:path:ext}b). In fact, we will later prove that all arcs and dual segments of the decomposition of an extremum path are positive but we need this preliminary result first.

\begin{lemma}
\label{lem:firstlastarc:prel}
Consider a cut $(G_1,G_2)$ of the grid, its bi-infinite window $W$ and a simple path $P$ such that $P_1 \in V(W)$ and $P_{\lastp} \in V(W)$. Consider the decomposition $(A^i)_{1\leq i \leq \ell}$ of $P$ in extremum arcs in $G_1$. Let $\ind{1}$ such that $P$ starts in $W_{\ind{1}}$ (resp. ends in $W_{\ind{1}}$) and suppose that $P_1=W_{\ind{1}}$ (resp. $P_{\lastp}=W_{\ind{1}}$) then $A^1$ is positive (resp. $A^\ell$ is positive). Moreover, $P_1=A^1_1$ (resp. $P_{\lastp}=A^\ell_{|A^\ell|}$).
\end{lemma}

\begin{proof}
Suppose that there exists $\ind{1}$ such that $P$ starts in $W_{\ind{1}}$ and $P_{1}=W_{\ind{1}}$, the other case is symmetric. If $P_{1} \neq A^1_1$ then $A^1$ starts in $W_{\ind{2}}$ with $\ind{2}<\ind{1}$ which contradicts the fact that $P$ starts in $W_{\ind{1}}$. Now, if $|A^1|>1$ then $A^1$ ends in $W_{\ind{3}}$ with $\ind{3}>\ind{1}$ and thus $A$ is positive. Otherwise let $\ind{4}=\min\{i>1:P_{1} \in V(W)\}$ and let $\ind{3}$ such that $W_{\ind{3}}=P_{\ind{4}}$ then since $P$ starts in $W_{\ind{1}}$, we have $\ind{3}>\ind{2}$ and thus $A^1$ is positive. 
\end{proof}

\begin{corollary}
\label{lem:firstlastarc}
Consider a cut $(G_1,G_2)$ of the grid, its bi-infinite window $W$, a simple finite extremum path $P$ and the decomposition $(A^i)_{1\leq i \leq \ell}$ of $P$ in extremum arcs in $G_1$. If $P$ is positive (resp. negative) then $A^1$ and $A^\ell$ are both positive (resp. negative). Moreover, $P_1=A^1_1$ (resp. $P_1=A^\ell_{|A^\ell|}$) and $P_{\lastp}=A^\ell_{A^\ell}$ (resp. $P_{\lastp}=A^1_1$).
\end{corollary}

%\begin{proof}
%Suppose that $P$ is positive, the other case is symmetric. Let $\ind{2}$ such that $P_{1}=W_{\ind{2}}$, if $P_{1} \neq A^1_1$ then $A^1$ starts in $W_{\ind{1}}$ with $\ind{1}<1$ which contradicts the fact that $P$ is an extremum positive path. Now, if $|A^1|>1$ then $A^1$ ends in $W_{\ind{3}}$ with $\ind{3}>\ind{2}$ and thus $A$ is positive. Otherwise let $\ind{4}=\min\{i>1:P_{1} \in V(W)\}$ and let $\ind{3}$ such that $W_{\ind{3}}=P_{\ind{4}}$ then since $P$ is an extremum positive path, we have $\ind{3}>\ind{2}$ and thus $A^1$ is positive. Same reasoning could be done for $P_{\lastp}$.
%\end{proof}

Now, consider an extremum path $P$ and a decomposition of this path in the left side of the grid, then all the dual segments belongs to the left side of the cut of the grid associated to $P$ (see Figures \ref{fig:decompo:decompoQ}b and \ref{fig:deco:path:ext}d).

\begin{lemma}
\label{lem:dual:bon:cote}
Consider a cut $(G_1,G_2)$ of the grid, its window $W$ such that $G_1$ is the left side of $W$. Consider a simple finite extremum path $P$ and its decomposition $(A^i)_{1\leq i \leq \ell}$ in extremum arcs in $G_1$. Consider the dual decomposition $(D^i)_{0\leq i \leq \ell}$ of $W$ according to $(A^i)_{1\leq i \leq \ell}$ and consider the cut of the grid $(G_5,G_6)$ according to the extremum path $P$. If $G_5$ is the left side of the cut, then for all $0\leq i \leq \ell$, the dual segment $D^i$ is a subgraph of $G_5$. 
\end{lemma}

\begin{proof}
Consider $(G_3,G_4)$ the cut of the grid associated associated to the decomposition $(A^i)_{1\leq i \leq \ell}$, let $W'$ be the window of this cut and consider that $G_3$ is the left side of this cut. For all $0\leq i \leq \ell$ the dual segment  $D^i$ is a subpath of $W'$ and thus a subgraph of $G_3$. Now let $W''$ be the window of $(G_5,G_6)$. Since $G_5$ is the left side of $W''$ and since $G_3$ is the left side of $W'$ then $G_3$ is a subgraph of $G_5$. Thus for all $0\leq i \leq \ell$, we have $D^i$ is a subgraph of $G_5$. 
\end{proof}

%A corollary of this result and of lemma \ref{lem:decompo:path} is that the first edge of a dual segment is on the left side of $P$.
%
%\begin{corollary}
%\label{cor:dual:firstedge:prel}
%Consider a cut $(G_1,G_2)$ of the grid, its window $W$ such that $G_1$ is the left side of $W$. Consider a simple finite extremum path $P$ and its decomposition $(A^i)_{1\leq i \leq \ell}$ in extremum arcs in $G_1$. Consider the dual decomposition $(D^i)_{0\leq i \leq \ell}$ of $W$ according to $(A^i)_{1\leq i \leq \ell}$ then for all $1\leq i \leq \ell-1$, there $1< \ind{1} < \lastp$ such that we have the edge $(D^i_1,D^i_2)$ is on the left side of $P_{[\ind{1}-1,\ind{1}+1]}$. 
%\end{corollary}

Now, we prove a technical lemma: consider $1\leq i <j \leq \ell$ then the intersection between the interior of $D^i$ and the interior of $D^j$ is a subset of $V(P)$. This result is needed to prove two powerful tools, Lemma \ref{lem:conc:toolbox} and Lemma \ref{lem:extact:ext:path}.

\begin{lemma}
\label{lem:dual:empty}
Consider a cut $(G_1,G_2)$ of the grid, its bi-infinite window $W$ and a simple path $P$. Consider the decomposition $(A^i)_{1\leq i \leq \ell}$ of $P$ into extremum arcs according to $G_1$  and its dual decomposition $(D^i)_{0\leq i \leq \ell}$ of the window $W$. Consider $0\leq i <j \leq \ell$ such that either $i\neq 0$ or $j \neq \ell$ and let $G_7$ be the interior of $D^i$ and let $G_8$ be the interior of $D^j$, then $V(G_7) \cap V(G_8) \subset V(P)$.% its window. If there exists $1\leq j \leq \ell-1$ and $1\leq k \leq |D^j|$ such that $D^j_{k} \in V(\gint)$ then $D^j_k\in V(C)$.
\end{lemma}

\begin{proof}
Without loss of generality, we suppose that $G_1$ is the left side of $W$. Let $(G_3,G_4)$ be the cut of the grid associated to the decomposition $(A^i)_{1\leq i \leq \ell}$ and let $W'$ be its window. We suppose that $G_3$ is the left side of $W'$. Then by lemma \ref{fact:decompo:newwind}, the graph $G_7$ and $G_8$ are both subgraphs of $G_4$ (see Figure \ref{fig:decompo:windowdual}). We consider two cases: either $i=0$ or $0 < i <j < \ell$. Note that the case $j=\ell$ is equivalent to $i=0$.

\vspace{+0.5em}

\noindent \textbf{Case $i=0$}: for an illustration of this proof, see Figure \ref{fig:decompo:proof:intdual}a. In this case consider $\ind{1}$ and $\ind{2}$ such that $P_{[\ind{1},\ind{2}]}$ is the extremum path extracted from $P$ and consider the cut of the grid $(G_5,G_6)$ associated to $P$. Suppose that $G_6$ is the right side of this cut of the grid, then $G_7=G_6$ and let $(B^i)_{1\leq i \leq \ell'}$ be the decomposition associated to $P_{[\ind{1},\ind{2}]}$ and $(S^i)_{1\leq i \leq \ell'}$ (resp. $(E^i)_{0\leq i \leq \ell'}$) the decomposition (resp. dual decomposition) of $W$ associated to $(B^i)_{1\leq i \leq \ell'}$. Since $G_5$ is the left side of the cut of the grid associated to $P$ then by lemma \ref{lem:dual:bon:cote}, for all $0\leq i \leq \ell'$, $E^i$ is a subpath of $G_5$. Now by lemma \ref{lem:decompo:path}, $V(D^j) \cap V(P) \subset \{D^j_1,D^j_{D^{j}}\}$. Thus either there exists $1\leq k \leq \ell'$ such that $D^j$ is a subset of $S^k$ or there exists $1\leq k \leq \ell'-1$ such that $D^j$ is a subset of $E^k$. In the first case either the arc $A^j$ is dominated by $B^k$ (which contradicts the definition of a decomposition in extremum arcs) or $A^j$ intersects with $B^k$ (which contradict the fact that $P$ is a simple path). In the second case $D^j$ is a subgraph of $G_5$ and since $P$ is simple then $G_8$ is a subgraph of $G_5$ and thus $V(G_7) \cap V(G_8) \subset V(P)$.

\vspace{+0.5em}

\noindent \textbf{Case $0 < i <j < \ell$}: for an illustration of this proof, see Figure \ref{fig:decompo:proof:intdual}b. In this case, consider the cut of the grid $(G_3,G_4)$ associated to the decomposition $(A^i)_{1\leq i \leq \ell}$ and let $W'$ be its window. We suppose that $G_3$ is the left side of $W'$. Then $G_7$ and $G_8$ are subgraph of $G_4$. Now, let $\ind{1}$ and $\ind{2}$ such that $P_{\ind{1}}=D^i_1$ and $P_{\ind{2}}=D^i_{|D^i|}$; let $\ind{3}$ and $\ind{4}$ such that $P_{\ind{3}}=D^j_1$ and $P_{\ind{4}}=D^j_{|D^j|}$; let $\ind{5}$ and $\ind{6}$ such that $P_{\ind{1}}=W'_{\ind{5}}$ and $P_{\ind{2}}=W'_{\ind{6}}$ and let $\ind{7}$ and $\ind{8}$ such that $P_{\ind{3}}=W'_{\ind{7}}$ and $P_{\ind{4}}=W'_{\ind{8}}$. We have $\ind{5}< \ind{6} \leq \ind{7} < \ind{8}$. Let $C$ be the window of $G_7$ and let $R$ such that $R=P_{[\ind{2},\ind{1}]}$ if $\ind{2} \leq \ind{1}$ or $R$ is equal to $P_{[\ind{1},\ind{2}]}$ where the indexing has been inverted if $\ind{1}<\ind{2}$. Then in both cases, we have $C=D^i \cdot R$, $R_1=W_{\ind{6}}$ and $R_{|R|}=W_{\ind{5}}$. Let $C'$ be the window of $G_8$ and let $R'$ such that $R'=P_{[\ind{4},\ind{3}]}$ if $\ind{4} \leq \ind{3}$ or $R'$ is equal to $P_{[\ind{3},\ind{4}]}$ where the indexing has been inverted if $\ind{3}<\ind{4}$. In both cases, we have $C'=D^j \cdot R'$, $R'_1=W_{\ind{8}}$ and $R'_{|R|}=W_{\ind{7}}$. If $R'$ is not a subgraph of $R$ and if an edge of $R'$ belongs to $E(G_7) \setminus E(R)$ then $P$ is not simple. If $R'$ is a subgraph of $R$ then for the sake of contradiction suppose that $G_8$ is a subgraph of $G_7$. Remark that $G_7$ is the right side of $C$ and $G_8$ is the right side of $C'$. Then there exists $1 \leq \ind{9} < \ind{10} \leq |R|$ such that $R_{\ind{9}}=R'_1=W_{\ind{8}}$ and  $R_{\ind{10}}=R'_{|R'|}=W_{\ind{7}}$. Then $R_{[1,\ind{9}]}$ is an arc of $G_4$ such that $R_1=W_{\ind{6}}$, $R_{\ind{9}}=W_{\ind{8}}$ and $R_{[\ind{10},|R|]}$ is an arc of $G_4$ such that $R_{\ind{10}}=W_{\ind{7}}$, $R_{|R|}=W_{\ind{5}}$. Since $\ind{5}\leq \ind{6} < \ind{7}\leq \ind{8}$ then $R_{[1,\ind{9}]}$ intersects with  $R_{[\ind{10},|R|]}$ and then $R$ is not simple. Thus $P$ is not simple which is an contradiction.

\end{proof}

Now, we study the interactions of two non intersecting paths

\begin{lemma}
\label{lem:conc:toolbox}
Consider a cut $(G_1,G_2)$ of the grid, its bi-infinite window $W$ and a simple path $P$ which starts in $W_{\ind{1}}$ and ends in $W_{\ind{2}}$. Consider the decomposition $(A^i)_{1\leq i \leq \ell}$ of $P$ into extremum arcs according to $G_1$ (which is the left side of $W$) and its dual decomposition $(D^i)_{0\leq i \leq \ell}$ of the window $W$. Let $(G_3,G_4)$ be the cut of the grid associated to the decomposition $(A^i)_{1\leq i \leq \ell}$ and let $W'$ be its window. Suppose that $G^3$ is the left side of $W'$. Now, consider a simple path $Q$ such that:
\begin{itemize}
\item $V(Q) \cap V(P) =\emptyset$;
\item there exists no arc $A$ of $G^4$ such that $A$ is a subgraph of $Q$ and $A$ starts in $W_{\ind{3}}$ with $\ind{3} \leq \ind{1}$ and ends in $W_{\ind{4}}$ with $\ind{4} \geq \ind{2}$);
\item there exists $0\leq c \leq d \leq \ell$ such that $Q_1 \in V(G_5)$ and  $Q_{|Q|} \in V(G_6)$ where $G_5$ is the interior of $D^c$ and $G_6$ is the interior of $D^d$;
\item if $c=0$ (resp. $d=\ell$) then $d<\ell$ (resp. $c>0$) and $V(Q) \cap V(D^{\ell})=\emptyset$  (resp. $V(Q) \cap V(D^{0})=\emptyset$).
\end{itemize}
Let $(B_i)_{1\leq i \leq \ell'}$ be the decomposition of $Q$ into extremum arc in $G_1$ then for all $c+1\leq j \leq d$ there exists $1\leq i \leq \ell'$ such that the arc $A^j$ is dominated by $B^i$.
\end{lemma}

\begin{proof}
See figure \ref{fig:decom:pathQ} for a graphical representation of the proof. For the sake of contradiction suppose that there exists $c+1\leq j \leq d$ such that for all $1\leq i \leq \ell'$ the arc $A^j$ is not dominated by $B^i$. Since either $c>0$ or $d<\ell$, then by lemma \ref{lem:dual:empty}, all positions which belongs to both the interior of $D^c$ and the interior $D^d$ are also positions of $V(P)$ and then since $Q$ and $P$ does not intersect then there exists $1\leq \ind{1} \leq \ind{4}\leq  |Q|$ such that $Q_{\ind{1}} \in V(D^c)$ and $Q_{\ind{4}} \in V(D^d)$. Let $W'$ be the window associated to the decomposition $(A^i)_{1\leq i \leq \ell}$, then $Q_{\ind{1}} \in V(W')$ and $Q_{\ind{2}} \in V(W')$.
Without loss of generality, we suppose that $A^j$ is positive and let $\ind{6} \leq \ind{7}$ such that $W'_{\ind{6}}=A^j_{1}$ and $W'_{\ind{7}}=A^j_{|A^j|}$. Let $\ind{1} \leq \ind{3} \leq \ind{4}$ such that $\ind{3}=\min\{i : \text{ there exists $k\geq \ind{6}$ such that } Q_{i} = W'_k\}$ (note that this index exists since $Q_{\ind{6}} \in V(D^d)$, $D^d$ is a subgraph of $W'$ and $d\geq j$). Let $\ind{8}$ such that $W'_{\ind{8}}=Q_{\ind{3}}$, by definition of $\ind{3}$ and since $P$ and $Q$ does not intersect then $\ind{8}>\ind{7}$. Now, let $\ind{1} \leq \ind{2} < \ind{3}$ such that $\ind{2}=\max\{i <\ind{3}: \text{there exists $k\leq \ind{3}$ such that } Q_{i} = W'_k\}$  (note that this index exists since $Q_{\ind{1}} \in V(D^c)$, $D^c$ is a subgraph of $W'$ and $c< j$). Then, let $\ind{5}$ such that $W'_{\ind{5}}=Q_{\ind{2}}$. By definition of $\ind{2}$ and since $P$ and $Q$ does not intersect then $\ind{5}<\ind{6}$. Then, we have $\ind{5} < \ind{6} \leq \ind{7}< \ind{8}$. Now since $P$ and $Q$ does not intersect and by the definition of $W'$, then there exists $c'<j$ and $d'\geq j$ such that $W'_{\ind{5}} \in V(D^{c'})$ and $W'_{\ind{8}} \in V(D^{d'})$. 
%If $\ind{3}=\ind{1}$ then $Q_{\ind{1}} \in V(D^j)$, $D^c$ is a subgraph of $W'$  $Q_{\ind{1}} \in V(D^{j-1})$.
%This means that $A^j$ is of size $1$ (see figure \ref{fig:decom:pathQ}c) and that $Q_{\ind{1}}=A^j_1$ which contradict the hypothesis that $P$ and $Q$ do no intersect. 
%Otherwise, 
Now, by definition of $\ind{2}$ and $\ind{3}$, we have $V(Q_{[\ind{2},\ind{3}]}) \cap V(W')=\{Q_{\ind{2}},Q_{\ind{3}}\}$. Then the path $Q_{[\ind{2},\ind{3}]}$ is either an elementary arc of $G_3$ or an elementary arc of $G_4$. In the first case, since $G_3$ is a subgraph of $G_1$ then $Q_{[\ind{2},\ind{3}]}$ is also an arc of $G_1$ and this arc dominates $A^j$ which is a contradiction (see Figure \ref{fig:decom:pathQ}). If $Q_{[\ind{2},\ind{3}]}$ is an elementary arc of $G_4$ and if $c'=0$ and $d'=\ell$, then the second hypothesis of our lemma is contradicted. Finally if $Q_{[\ind{2},\ind{3}]}$ is an elementary arc of $G_4$ and if $c'>0$ (the case $d'<\ell$ is symmetric to this one) then %$W'_{\ind{6}} \in D^{c'}\setminus \{D^i_1,D^i_{|D^i|}\}$ (see Figure \ref{fig:decom:pathQ}b). 
let $\gint$ be the interior of $D^{c'}$ and let $C$ be its window. Remark that since $Q_{[\ind{2},\ind{3}]}$ is an arc of $G_4$ (which is the right side of $W'$) then the edge $(Q_{\ind{2}},Q_{\ind{2}+1})\in V(\gint)$ (which is the right side of $C$). By definition of $\ind{2}$ and since $P$ is simple then $V(Q_{[\ind{2},\ind{3}]}) \cap V(C)=\{Q_1\}$. %Since $P$ and $Q$ is simple then $V(Q_{[\ind{2},\ind{3}]}) \cap V(C)=\{Q_1\}$. 
Then $Q_{[\ind{2},\ind{3}]}$ is a subgraph of $\gint$ and $Q_{\ind{3}} \in V(\gint)$. Since $Q_{\ind{3}}$ is also in the interior of $D^{d'}$, by lemma \ref{lem:dual:empty}, we have $Q_{\ind{3}} \in V(P)$ which contradicts the fact that $P$ and $Q$ does not intersect.
%Since $0<i<\ell$ there exists $1\leq \ind{5} \leq |P|$ (resp. $1\leq \ind{6} \leq |P|$) be the such that $P_{\ind{5}}=D^i_1$ (resp. $P_{\ind{6}}=D^i_{|D^i|}$). Without loss of generality we suppose that $\ind{5} \leq \ind{6}$ and we define $C$ as the simple cycle $P_{[\ind{5},\ind{6}]} \cup D^e$ and we define $\gint$ as its interior. The graph $\gint$ is the interior of $D^i$ and then a subset of $G_4$. We consider an indexing of $C$ such that $C_0=D^i_1$ and $C_{|D^i|-1}=D^i_{D^i}$. By definition $G^3$ is the left side of $W'$ and $G^4$ is the right side of $W'$. Since $W'_{\ind{3}} \in D^i\setminus \{D^i_1,D^i_{|D^i|}\}$ then the edge $Q_{[\ind{1},\ind{1}+1]}$ is on the right side of $D^i$ and then this edge belong to $\gint$.  By lemma \ref{lem:dual:empty}, either $Q_{\ind{2}} \notin \gint$ or $Q_{\ind{2}} \in C$. In the first case, $Q$ has to intersect $C$ and since $Q$ is a simple arc of $G^4$ then $Q$ intersect $P_{[\ind{5},\ind{6}]}$. In the second case, since $Q_{\ind{2}} \in V(D^j)$ and $j>i$ then we have $Q_{\ind{2}} \in P_{[\ind{5},\ind{6}]}$. Then in both cases there exists $1<\ind{7}$ such that $Q_{\ind{7}}\in V(P_{[\ind{5},\ind{6}}])$ which is a contradiction. 
\end{proof}

Now, we show that in some case there exists an index $\ind{2}$ such that $P_{[\ind{2},\lastp]}$ contain no relevant information.

\begin{lemma}
\label{lem:extact:ext:path}
Consider a cut $(G_1,G_2)$ of the grid, its bi-infinite window $W$, a simple path $P$ such that $P_1 \in V(W)$ and an index $1\leq \ind{2} \leq \lastp$ such that $P_{\ind{2}} \in V(W)$. Consider the decomposition $(A^i)_{1\leq i \leq \ell}$ of $P_{[1,\ind{2}]}$ in $G_1$ and let $(D^i)_{1\leq i \leq \ell}$ be its dual decomposition of $W$. Consider $0\leq k \leq \ell$ and let $\gint$ be the interior of $D^k$ and $W'$ be its window. Then if $P_{[\ind{2},\lastp]}$ is a subgraph of $\gint$ and if $V(P) \cap V(W') \subset \{P_{\ind{2}}\}$ then the decomposition of $P$ in extremum arcs in $G_1$ is $(A^i)_{1\leq i \leq \ell}$.
%and let $\ind{3}$ and $\ind{4}$ such that $P_{[\ind{1},\ind{2}]}$ starts in $W_{\ind{3}}$ and ends in $W_{\ind{4}}$. Let $(G_3,G_4)$ be the cut of the grid associated to $P_{[\ind{1}, \ind{2}]}$ and $W'$ its window such that $G_3$ is the left side of $W'$. Then if $P_{[1,\ind{1}]}$ and $P_{[\ind{2},\lastp]}$ are subgraph of $G_4$ and if $V(P_{[1,\ind{1}]}) \cap V(W')=\{P_{\ind{1}}\}$ and $V(P_{[\ind{2},\lastp]}) \cap V(W')=\{P_{\ind{2}}\}$ then the decomposition of $P$ in extremum arc in $G^1$ is the same decomposition as the decomposition of $P_{[\ind{1},\ind{2}]}$ in extremum arc in $G^1$.
\end{lemma}

\begin{proof}
See Figure \ref{fig:decom:ext:caseB} for an illustration of this proof. Let $(S^i)_{1\leq i \leq \ell}$ be the decomposition of $W$ according to $(A^i)_{1\leq i \leq \ell}$. Consider an arc $A$ of $G_1$ which is a subgraph of $P_{[\ind{2},\lastp]}$ then $A_1 \in V(W)$ and $A_{|A|} \in V(W)$. If there exists $0 \leq j \leq \ell$ such that $A_1 \in V(D^j)$ then by lemma \ref{lem:decompo:path}, since $P$ is simple and since by hypothesis $V(P) \cap V(W') \subset \{P_{\ind{2}}\}$, then we have $A_1=P_{\ind{2}}$. Moreover by lemma \ref{lem:decompo:path}, either $P_{\ind{2}}=D^k_1=S^k_{|S^k|}$ or $P_{\ind{2}}=D^k_{|D^k|}=S^{k+1}_{1}$. Same reasoning could be done with $A_{|A|}$ and then there exists $1\leq i \leq j \leq \ell$ such that $A_1 \in V(S^i)$ and $A_{|A|} \in V(S^j)$. If $i<j$ then $A$ intersects with both $A^i$ and $A^j$ which is a contradiction and if $i=j$ then this arc is dominated by $A^i$ and does not appear in the decomposition of $P$ in extremum arcs in $G_1$. 
\end{proof}

%%%%%%%%%%%%%%new version
Now we want to study in what order the different arcs of the decomposition appear in $P$. We say that a decomposition is positive if all the arcs of the decomposition appears according to the indexing of $P$.

\begin{definition}
\label{def:decomp:position}
Consider a cut $(G_1,G_2)$ of the grid, its bi-infinite window $W$ and a simple path $P$ such that $P_1 \in V(W)$ and $P_{\lastp} \in V(W)$. Consider the decomposition $(A^i)_{1\leq i \leq \ell}$ of $P$ in extremum arcs  in $G_1$. The decomposition is positive (resp. negative) if and only if for all $1\leq i \leq \ell$ and $1\leq j \leq \ell$, for all indices $1\leq \ind{1} \leq |A^i|$ and $1\leq \ind{2} \leq |A^j|$ and for all indices $1\leq \ind{3} \leq \lastp$ and $1\leq \ind{4} \leq \lastp$ such that $P_{\ind{3}}=A^i_{\ind{1}}$ and $P_{\ind{4}}=A^j_{\ind{2}}$, we have:
\begin{itemize}
\item if $i<j$ then $\ind{3}< \ind{4}$ (resp. $\ind{3} > \ind{4}$);
\item if $i=j$ and $\ind{1}< \ind{2}$ then $\ind{3} < \ind{4}$ (resp. $\ind{3} > \ind{4}$).
\end{itemize}
\end{definition}

A decomposition in extremum arcs is not necessarily positive or negative, see a counter-example in Figure \ref{fig:cut:max}. Nevertheless, one of the final lemma of this section is that the decomposition in extremum arcs of an extremum positive path is always positive (see Lemma \ref{lem:conc:prel}). Note that a similar property is always true for the dual: consider two indices $\ind{1}, \ind{2} \in \mathbb{Z}$ then if there exists $0 \leq i < j \leq \lastp$ such that $W_{\ind{1}} \in V(D^i)$ and $W_{\ind{2}} \in V(D^j)$ then $\ind{1} < \ind{2}$. Also, if there exists $1\leq \ind{3}<\ind{4} \leq |D^i|$ such that $W_{\ind{1}}= D^i_{\ind{3}}$ and $W_{\ind{2}}=D^i_{\ind{4}}$ then $\ind{1} < \ind{2}$. 
%\begin{figure}[th]
%\centerline{
%\subfigure[A path $P$ which is not extremum.]{
%\includegraphics[width=6cm]{./photoFinal/photodecompositionfinal/PnotExt.JPG}}
%\subfigure[The decomposition of $P$ into extremum arcs is neither positive nor negative.]{
%\includegraphics[width=6cm]{./photoFinal/photodecompositionfinal/PnotExtDec.JPG}
%}}
%\caption{The decomposition of a simple path $P$ into extremum arcs is not always ordered and oriented.}
%\label{fig:deco:path:notex}
%\end{figure}
We state a sufficient condition for a path to be positive.

\begin{fact}
\label{fact:sufficient:positive}
Consider a cut $(G_1,G_2)$ of the grid, its bi-infinite window $W$, a finite path $P$, its decomposition $(A^i)_{1\leq i \leq \ell}$ in extremum arcs in $G_1$ and its dual decomposition $(D^i)_{0\leq i \leq \ell}$ of $W$. If for all $1\leq i \leq \ell$, the arc $A^i$ is positive (resp. negative) and for all $1\leq i \leq \ell-1$, the dual segment $D^i$ is positive (resp. negative) then the decomposition $(A^i)_{1\leq i \leq \ell}$ is positive (resp. negative).
\end{fact}

Now, our aim is to prove that for a decomposition of a positive simple path there exists two values $1\leq j \leq k \leq \ell$ such that for all $i < j$ or $i> k$,  the arc $A^i$ is negative and for all $j \leq i \leq k$ the arc $A^i$ is positive (a similar property hold for the orientation of the dual segments). This result and Lemma \ref{lem:firstlastarc} will imply that if $P$ is extremum then its decomposition is positive. This result is obtained through a sequence of technical lemmas. 

\begin{lemma}
\label{lem:exit:notfree}
Consider a cut $(G_1,G_2)$ of the grid, its bi-infinite window $W$, a finite path $P$ such that $P_1 \in V(W)$ and $P_{\lastp} \in V(W)$ and the decomposition $(A^i)_{1\leq i \leq \ell}$ of $P$ into extremum arcs in $G_1$. Consider $1\leq i \leq \ell$ and $1\leq \ind{1} < \lastp$ such that $P_{\ind{1}}=A^i_{|A^i|}$. Consider the decomposition $(B^i)_{1\leq i \leq \ell'}$ of $P_{[1,\ind{1}]}$ into extremum arcs in $G_1$ and let $W'$ be the window associated to the decomposition $(B^i)_{1\leq i \leq \ell'}$ and let $(G_3,G_4)$ be the cut defined by $W'$. We suppose that $G_3$ is a subgraph of $G_1$. Then the edge $(P_{\ind{1}},P_{\ind{1}}+1)$ does not belong to $E(G_3)$.
\end{lemma}

\begin{proof}
%Without loss of generality, we suppose that $A^i$ is positive, the other case is symmetric. 
See Figures \ref{lem:exit:notfree:fig} and \ref{lem:exit:notfree:fig2} for an illustration of this lemma. Let $m=w(A^i)$ be the width of arc $A^i$ and consider an indexing of $W$ such that $W_0=A^i_1$ and $W_{m}=A^i_{|A^i|}$. For the sake of contradiction, suppose that the edge $(P_{\ind{1}},P_{\ind{1}}+1)$ belongs to $G_3$. By hypothesis, $G_3$ is a subgraph of $G_1$ and then $G_2$ is a subgraph of $G_4$. Thus $W$ is a subgraph of $G_4$ and since  $P_{\lastp} \in V(W)$ and $(P_{\ind{1}},P_{\ind{1}}+1) \in E(G^3)$ then there exists $\ind{2}=\min\{i >\ind{1}: P_i \in V(W')\}$. Moreover, $P_{[\ind{1},\ind{2}]}$ is an elementary arc of $G_3$. Now either $P_{\ind{2}} \in V(P_{[1,\ind{1}]})$ or $P_{\ind{2}} \in V(W)$. In the first case, the path $P$ is not simple which is a contradiction. In the second case, since by hypothesis $G_3$ is a subgraph of $G_1$, then $P_{[\ind{1},\ind{2}]}$ is an elementary arc of $G_1$ and there exists $\ind{3}$ such that $W_{\ind{3}}=P_{\ind{2}}$. If $\ind{3} < 0$ then the arc $P_{[\ind{1},\ind{2}]}$ dominates the arc $A^i$ which contradicts the definition of $A^i$. If $0 \leq \ind{3} \leq m$ then $P$ is not simple which is a contradiction. If $\ind{3} >m$ then let $1\leq \ind{0} \leq \ind{1}$ such that $P_{\ind{0}}=A^i_1$ and thus the arc $P_{[\ind{0},\ind{1}]}$ dominates $A^i$ which is a contradiction.

\end{proof}

\begin{lemma}
\label{lem:exit:intoint:prel}
Consider a cut $(G_1,G_2)$ of the grid, its bi-infinite window $W$, a finite path $P$ such that $P_1 \in V(W)$ and an index $1<\ind{3}\leq \lastp$ such $P_{\ind{3}} \in V(W)$. Consider the  decomposition $(A^i)_{1\leq i \leq \ell}$ of $P_{[1,\ind{3}]}$ into extremum arcs in $G_1$. Consider $1\leq i \leq \ell$ and $1\leq \ind{1} < \ind{3}$ such that $P_{\ind{1}}=A^i_{|A^i|}$. Now consider the decomposition $(B^i)_{1\leq i \leq \ell'}$ of $P_{[1,\ind{1}]}$ and its dual decomposition $(D^i)_{0\leq i \leq \ell'}$ of the window $W$. Let $(G_3,G_4)$ be the cut associated to the decomposition $(B^i)_{1\leq i \leq \ell'}$, we suppose that $G_3$ is a subgraph of $G_1$. Then there exists $1\leq j \leq \ell'$ such that $A^i=B^j$.  Moreover if $A^i$ is positive (resp. negative), let $\gint$ be the interior of $D^j$ (resp. $D^{j-1}$) and then the edge $(P_{\ind{1}},P_{\ind{1}}+1)$ belongs either to $E(\gint)$ or to $E(G_3)$.
\end{lemma}

\begin{proof}
This proof is made by a local reasoning along tile $P_{\ind{1}}$. First, consider the special case where $\ind{1}=1$ then $i=1$, $|A^1|=1$. Moreover the window associated to $A^1$ is $W$ and then $\gint=G_2$ and $G_3=G_1$ and the edge $(P_{1},P_{2})$ belongs either to $G_3$ or to $\gint$. Otherwise, we supposer that $A^i$ is positive (the other case is symmetric) since $P_{\ind{1}}=A^i_{|A^i|}$ then there exists $\ind{2}$ such that $P_{\ind{1}}=W_{\ind{2}}$. Now, among the neighbors of $P_{\ind{1}}$, there are $P_{\ind{1}-1}$, $P_{\ind{1}+1}$, $W_{\ind{2}-1}$ and $W_{\ind{2}+1}$. Since $W$ is simple then $W_{\ind{1}-1} \neq W_{\ind{1}+1}$ and since the arc $A^i$ is positive then it ends in $W_{\ind{2}}$ and thus $W_{\ind{2}+1} \neq P_{\ind{1}-1}$ and $W_{\ind{2}+1} \neq P_{\ind{1}+1}$. Without loss of generality we can assume that $W_{\ind{2}+1}=W_{\ind{2}}+(0,1)$. Since $P_{\ind{1}-1} \neq P_{\ind{1}+1}$ we are left with eighteen cases to study. This cases are classified in three categories:

\noindent \textbf{Edge $(P_{\ind{1}},P_{\ind{1}+1})$ belongs to $E(\gint)$:} see Figure \ref{fig:local:interior}. A typical example of this case occurs when $W_{\ind{2}-1}=W_{\ind{2}}-(1,0)$, $P_{\ind{1}+1}=P_{\ind{1}}+(1,0)$ and $P_{\ind{1}-1}=P_{\ind{1}}-(0,1)$. In this case, consider the window $W'$ of $\gint$. Either $W'$ is a finite cycle if $j<\ell'$ or $W'$ is a the window associated to $P_{[1,\ind{1}]}$ if $i=\ell'$. In both cases, there exists $\ind{3}$ such that $W'_{\ind{3}}=W_{\ind{2}}$, $W'_{\ind{3}+1}=W_{\ind{2}+1}$ and $W'_{\ind{3}-1}=P_{\ind{1}-1}$ and $\gint$ is the right side of $W'$. In this case, the edge $(P_{\ind{1}},P_{\ind{1}+1})$ is on the right side of $W'_{[\ind{3}-1,\ind{3}+1]}$ and thus this edge belong to $\gint$. The lemma is true in this case.

\vspace{+0.5em}

\noindent \textbf{Edge $(P_{\ind{1}},P_{\ind{1}+1})$ belongs to $E(G_3)$:} see Figure \ref{fig:local:interior2}. A typical example of this case occurs when $W_{\ind{2}-1}=W_{\ind{2}}+(1,0)$, $P_{\ind{1}+1}=P_{\ind{1}}-(1,0)$ and $P_{\ind{1}-1}=P_{\ind{1}}-(0,1)$. In this case, consider the window $W''$ associated to the decomposition $(A^i)_{1\leq i \leq \ell}$, we suppose that $G_3$ is the left side of $W''$ and thus $G_3$ is a subgraph of $G_1$. With these conventions, there exists $\ind{4}$ such that $W''_{\ind{4}}=W_{\ind{2}}$, $W''_{\ind{4}+1}=W_{\ind{2}+1}$ and either $W''_{\ind{4}-1}=P_{\ind{1}-1}$ if the edge $(P_{\ind{1}}-1,P_{\ind{1}})$ belongs to $E(G_1)$ or $W''_{\ind{4}-1}=W_{\ind{2}-1}$ otherwise (in this case $|A^i|=1$). In this case, the edge $(P_{\ind{1}},P_{\ind{1}+1})$ is on the left side of $W''_{[\ind{4}-1,\ind{4}+1]}$ and thus this edge belong to $G_3$. The lemma is true in this case.
%Since $A^i$ is positive, this result is in contradiction with lemma \ref{lem:exit:notfree}.

\vspace{+0.5em}

\noindent \textbf{Arc $A^i$ is a negative arc of length $1$:} see the combination of the two dotted neighborhood of Figure \ref{fig:local:interior2}. This case occurs only time when $W_{\ind{2}-1}=W_{\ind{2}}-(1,0)$, $P_{\ind{1}+1}=P_{\ind{1}}-(0,1)$ and $P_{\ind{1}-1}=P_{\ind{1}}+(1,0)$. In this case, by definition \ref{def:forbidden}, the arc $A^i$ is a negative arc of length $1$ which contradicts the hypothesis that $A^i$ is positive.

\end{proof}

For Lemma \ref{lem:exit:intoint:prel}, when $\ind{3}=\lastp$ where are in case where the edge $(P_{\ind{1}},P_{\ind{1}}+1)$ cannot belongs to $G_3$ by lemma \ref{def:decomp:position}. Thus we have the following corollary (see Figure \ref{lem:exit:notfree:fig2}b).

\begin{corollary}
\label{lem:exit:intoint}
Consider a cut $(G_1,G_2)$ of the grid, its bi-infinite window $W$, a finite path $P$ such that $P_1 \in V(W)$ and $P_{\lastp} \in V(W)$. Consider the decomposition $(A^i)_{1\leq i \leq \ell}$ of $P$ into extremum arcs in $G_1$. Consider $1\leq i \leq \ell$ and $1\leq \ind{1} \leq \lastp$ such that $P_{\ind{1}}=A^i_{|A^i|}$. Suppose that $\ind{1} < \lastp$ and consider the decomposition $(B^i)_{1\leq i \leq \ell'}$ of $P_{[1,\ind{1}]}$ and its dual decomposition $(D^i)_{0\leq i \leq \ell'}$ of the window $W$ then there exists $1\leq j \leq \ell'$ such that $A^i=B^j$. Moreover if $A^i$ is positive (resp. negative), let $\gint$ be the interior of $D^j$ (resp. $D^{j-1}$) and then the edge $(P_{\ind{1}},P_{\ind{1}}+1)$ belongs to $\gint$.
\end{corollary}

Now, that we have dealt with the technical lemmas, we start by studying the orientation of the dual segments. This first lemma studies the orientation of the dual segment between two positives arcs. 

\begin{lemma}
\label{lem:deal:dual}
Consider a cut $(G_1,G_2)$ of the grid, its bi-infinite window $W$, a finite path $P$ such that $P_1 \in V(W)$ and $P_{\lastp} \in V(W)$ and the decomposition $(A^i)_{1\leq i \leq \ell}$ of $P$ into extremum arcs in $G_1$. Consider $1\leq i < j \leq \ell$ such that $A^i$ and $A^j$ are both positive (resp. negative) arcs and let $\ind{1}$ such that $P_{\ind{1}}=A^i_{|A^i|}$ and $\ind{2}$ such that $P_{\ind{2}}=A^j_{|A^j|}$. If $P$ is positive (resp. negative) then $\ind{1}< \ind{2}$ (resp. $\ind{1}>\ind{2}$). 
\end{lemma}

\begin{proof}
Let $1\leq \ind{3} \leq \ind{4} \leq \lastp$ such that $P_{[\ind{3},\ind{4}]}$ is the extremum arcs extracted from $P$. We suppose that $P_{[\ind{3},\ind{4}]}$ is positive (the other case is symmetric) then $A^i$ and $A^j$ are both positive. By contradiction suppose that $\ind{1}>\ind{2}$.  See Figure \ref{fig:deal:dual}a for an illustration of this setting. We distinguish three cases, $\ind{3} \leq \ind{2}$, $\ind{2} < \ind{1}  \leq \ind{3}$ and $\ind{2} \leq \ind{3} \leq \ind{1}$. The last case generates two sub-cases. We now show that they all lead to a contradiction.

\vspace{+0.5em}

\noindent \textbf{Case $\ind{3} \leq \ind{2}$}: see Figure \ref{fig:deal:dual}b for an illustration of this case. In this case, consider the decomposition $(B^i)_{1\leq i \leq \ell'}$ of $P_{[\ind{3},\ind{2}]}$ and let $(D^i)_{0\leq i \leq \ell'}$ be the dual decomposition of $W$ associated to $(B^i)_{1\leq i \leq \ell'}$. Note that, there exists $1\leq k \leq \ell'$ such that $B^k=A^j$. Let $\gint$ be the interior of $D^{k}$. Since $\ind{2}<\ind{1}$ then $\ind{2}<\lastp$  and since $B^k$ is positive, then by lemma \ref{lem:exit:intoint} the edge $(P_{\ind{2}},P_{\ind{2}+1})$ belongs to $E(\gint)$. Moreover, since $i<j$ there exists $k'<k$ such that $P_{\ind{1}} \in D^{k'}$. Moreover since $\ind{3}<\ind{1}$ then $k' \geq 1$. By definition of $P_{\ind{3}}$, we have $V(D^0) \cap V(P_{[\ind{2}+1,\ind{1}]})=\emptyset$ and then there exists no arc which starts in $D^0$ and ends in $D^{\ell'}$ and which is a subgraph of $P_{[\ind{2}+1,\ind{1}]}$. Moreover, since $P$ is simple then $P_{[\ind{2}+1,\ind{1}]}$ does not intersect with $P_{[\ind{3},\ind{2}]}$. Then by applying lemma \ref{lem:conc:toolbox} on $(B^i)_{1\leq i \leq \ell'}$ and $P_{[\ind{2}+1,\ind{1}]}$, there exists an arc $A$ which dominates $B^k=A^j$ and which is a subgraph of $P_{[\ind{2}+1,\ind{1}]}$ which is a contradiction.

\vspace{+0.5em}

\noindent \textbf{Case $\ind{2} < \ind{1}  \leq \ind{3}$}: see Figure \ref{fig:deal:dual}c for an illustration of this case. In this case, let $1 \leq \ind{5} \leq \lastp$ be such that $P_{\ind{5}}=A^j_1$ then $\ind{5}\leq \ind{2}$ and consider the decomposition $(B^i)_{1\leq i \leq \ell'}$ of $P_{[\ind{5},\ind{1}]}$ and let $(D^i)_{0\leq i \leq \ell'}$ be the dual decomposition of $W$ associated to $(B^i)_{1\leq i \leq \ell'}$. Note that, there exists $1\leq i' \leq j' \leq \ell'$ such that $B^{i'}=A^i$ and $B^{j'}=A^j$. Let $\gint$ be the interior of $D^{i'}$. Since $\ind{1}<\ind{4}$ then $\ind{1}<\lastp$  and since $B^{i'}$ is positive, then by lemma \ref{lem:exit:intoint} the edge $(P_{\ind{1}},P_{\ind{1}+1})$ belongs to $E(\gint)$. Moreover, by definition of $\ind{4}$, we have $P_{\ind{4}} \in D^{\ell'}$ and then there exists $\ind{6}=\min\{i>\ind{5}:P_i \in D^0 \text{ or } D^{\ell'}\}$. By definition of $\ind{6}$, there exists no arc which starts in $D^0$ and ends in $D^{\ell'}$ and which is a subgraph of $P_{[\ind{1}+1,\ind{6}]}$. Moreover, since $P$ is simple then $P_{[\ind{1}+1,\ind{6}]}$ does not intersect with $P_{[\ind{5},\ind{1}]}$. Then by applying lemma \ref{lem:conc:toolbox} on $(B^i)_{1\leq i \leq \ell'}$ and $P_{[\ind{1}+1,\ind{6}]}$, if $P_{\ind{6}} \in V(D^0)$ (resp. $V(D^{\ell'})$) then there exists an arc $A$ which dominates $B^{i'}=A^i$ (resp. $B^{j'}=A^j$) and which is a subgraph of $P_{[\ind{2}+1,\ind{6}]}$ which is a contradiction.

\vspace{+0.5em}

\noindent \textbf{Case $\ind{2} \leq \ind{3} \leq \ind{1}$}: in this case, let $1 \leq \ind{5} \leq \lastp$ be such that $P_{\ind{5}}=A^j_1$, then $\ind{5}\leq \ind{2}$ and let $1\leq \ind{6} \leq \lastp$ such that $P_{\ind{6}}=A^1_{|A^1|}$. By definition of $\ind{3}$ we have either $A^1_1=P_{\ind{3}}$ (if $A^1$ is positive) or $A^1_{|A^1|}=P_{\ind{3}}$ (if $A^1$ is negative), in both cases we have $\ind{6}\geq \ind{3}$. We have $\ind{6}<\ind{4}$ otherwise $A^1$ would dominate $A^i$ and $A^j$ and then $\ind{6} < \lastp$. Now, consider the decomposition $(B^i)_{1\leq i \leq \ell'}$ of $P_{[\ind{5},\ind{6}]}$ and let $(D^i)_{0\leq i \leq \ell'}$ be its dual decomposition of $W$. Note that $A^1=B^1$ and that there exists $1\leq k \leq \ell'$ such that $B^k=A^j$. Now, we distinguish two subcases either $B^1$ is positive or negative.

\vspace{+0.5em}

\textbf{Sub-case $B^1$ is positive}: see Figure \ref{fig:deal:dual}d for an illustration of this case. Let $\gint$ be the interior of $D^{1}$. Since $B^1$ is positive, then by lemma \ref{lem:exit:intoint} the edge $(P_{\ind{6}},P_{\ind{6}+1})$ belongs to $E(\gint)$. Moreover by definition of $\ind{3}$, we have $V(D^0) \cap V(P_{[\ind{6}+1,\ind{4}]})=\emptyset$ and then there exists no arc which starts in $D^0$ and ends in $D^{\ell'}$ and which is a subgraph of $P_{[\ind{6}+1,\ind{4}]}$. Moreover, since $P$ is simple then $P_{[\ind{6}+1,\ind{4}]}$ does not intersect with $P_{[\ind{5},\ind{6}]}$. Also by definition of $\ind{4}$, we have $P_{\ind{4}} \in V(D^{\ell'})$. Then by applying lemma \ref{lem:conc:toolbox} on $(B^i)_{1\leq i \leq \ell'}$ and $P_{[\ind{6}+1,\ind{4}]}$, there exists an arc $A$ which dominates $B^k=A^j$ and which is a subgraph of $P_{[\ind{6}+1,\ind{4}]}$ which is a contradiction.

% we have $P_{\ind{4}} \in D^{\ell'}$. Finally, remark that by definition of $P_{\ind{3}}$, we have $V(D^0) \cap P_{[\ind{6}+1,\lastp]}=\emptyset$ and then by lemma \ref{}, there exists an arc $A$ which dominates $A^j$ and which is a subgraph of $P_{[\ind{6}+1,\lastp]}$ which is a contradiction.

\vspace{+0.5em}

\textbf{Sub-case $B^1$ is negative}: see Figure \ref{fig:deal:dual}e for an illustration of this case. Let $\gint$ be the interior of $D^{0}$. Since $B^1$ is negative, then by lemma \ref{lem:exit:intoint}, the edge $(P_{\ind{6}},P_{\ind{6}+1})$ belongs to $E(\gint)$. Moreover by definition of $\ind{4}$, we have $P_{\ind{4}} \in V(D^{\ell'})$, then there exists $\ind{7} = \min \{ i >\ind{6} : P_{\ind{7}} \in V(D^{\ell'})\}$. Note that since $P$ is a simple path, then $P_{[\ind{6},\ind{7}]}$ is a subgraph of $\gint$. Moreover, since $i<j$ there exists $k'<k$ such that $P_{\ind{1}} \in D^{k'}$ and since $\ind{3}<\ind{1}$ then $k' \geq 1$. Thus by lemma \ref{lem:dual:empty}, we have $\ind{7}<\ind{1}$.  By definition of $P_{\ind{3}}$, we have $V(D^0) \cap V(P_{[\ind{7},\ind{1}]})=\emptyset$ and then there exists no arc which starts in $D^0$ and ends in $D^{\ell'}$ and which is a subgraph of $P_{[\ind{7},\ind{1}]}$. Moreover, since $P$ is a simple path then $P_{[\ind{7},\ind{1}]}$ does not intersect with $P_{[\ind{5},\ind{6}]}$. Then by applying lemma \ref{lem:conc:toolbox} on $(B^i)_{1\leq i \leq \ell'}$ and $P_{[\ind{7},\ind{1}]}$, there exists an arc $A$ which dominates $B^k=A^j$ and which is a subgraph of $P_{[\ind{7},\ind{1}]}$ which is a contradiction.
\end{proof}

This second lemma studies the case where the dual segment and its following arc have an opposite orientation.

\begin{lemma}
\label{lem:deal:dual:oppo}
Consider a cut $(G_1,G_2)$ of the grid, its bi-infinite window $W$, a finite path $P$ such that $P_1 \in V(W)$ and $P_{\lastp} \in V(W)$, the decomposition $(A^i)_{1\leq i \leq \ell}$ of $P$ into extremum arcs in $G_1$ and its dual decomposition $(D^i)_{0\leq i \leq \ell}$ of the window $W$. Consider $1\leq i < \ell$ (resp. $1<i \leq \ell$) such that $D^i$ is positive (resp. negative) and $A^{i+1}$ is negative (resp. $A^{i}$ is positive). Let $\gint$ be the interior of $D^i$ and let $\ind{1}$ such that $P_{\ind{1}} =A^{i+1}_{|A^{i+1}|}$ (resp. $P_{\ind{1}} =A^{i}_{|A^{i}|}$) then $P_{[\ind{1},\lastp]}$ is a subgraph of $\gint$ and $V(P)\cap V(D^i)=\{P_{\ind{1}}\}$. 
\end{lemma}

\begin{proof}
We consider the case where $D^i$ is positive and $A^{i+1}$ is negative, see Figure \ref{fig:deal:dual:oppo}. In this case, let $1 \leq \ind{2} \leq \lastp$ such that $P_{\ind{2}}=A^i_{|A^i|}$ if $A^i$ is positive or  $P_{\ind{2}}=A^i_{1}$ otherwise. Since $D^i$ is negative, we have $\ind{2}< \ind{1}$. Moreover, let $C$ be the window of $\gint$, then $C=P_{[\ind{2},\ind{1}]} \cdot D^i$. Now consider the decomposition $(B^i)_{1\leq i \leq \ell'}$ of $P_{[1,\ind{1}]}$ in extremum arcs in $\free$ and let $E^{i}_{0\leq i \leq \ell'}$ be its dual decomposition of $W'$. Since $\ind{2}<\ind{1}$ there exists $j$ such that $B^j=A^i$ and $B^{j+1}=A^{i+1}$. Thus $D^i=E^j$ and the interior of $E^j$ is $\gint$. Moreover, by lemma \ref{lem:exit:intoint}, $P_{\ind{1}+1} \in V(\gint)$. Since $P$ is simple and by lemma \ref{lem:decompo:path} then $P_{[\ind{1},\lastp]}$ is a subgraph of $\gint$.

\end{proof}

Since $P$ is simple then a corollary of this result and of lemma \ref{lem:extact:ext:path} is that the end of the path assembly contains not  relevant information.

\begin{corollary}
\label{lem:deal:dual:oppo2}
Consider a cut $(G_1,G_2)$ of the grid, its bi-infinite window $W$, a finite path $P$ such that $P_1 \in V(W)$ and $P_{\lastp} \in V(W)$, the decomposition $(A^i)_{1\leq i \leq \ell}$ of $P$ into extremum arcs in $G_1$ and its dual decomposition $(D^i)_{0\leq i \leq \ell}$ of the window $W$. Consider $1\leq i < \ell$ (resp. $1<i \leq \ell$) such that $D^i$ is positive (resp. negative) and $A^{i+1}$ is negative (resp. $A^{i}$ is positive). Let $\gint$ be the interior of $D^i$ and let $\ind{1}$ such that $P_{\ind{1}} =A^{i+1}_{|A^{i+1}|}$ (resp. $P_{\ind{1}} =A^{i}_{|A^{i}|}$) then the decomposition of $P_{[1,\ind{1}]}$ is $(A^i)_{1\leq i \leq \ell}$. 
\end{corollary}

Same reasoning could be done after switching the indexing of path $P$ and we obtain the following result, see Figure \ref{fig:deal:dual:oppo2}a.

\begin{corollary}
\label{lem:deal:dual:oppo3}
Consider a cut $(G_1,G_2)$ of the grid, its bi-infinite window $W$, a finite path $P$ such that $P_1 \in V(W)$ and $P_{\lastp} \in V(W)$, the decomposition $(A^i)_{1\leq i \leq \ell}$ of $P$ into extremum arcs in $G_1$ and its dual decomposition $(D^i)_{0\leq i \leq \ell}$ of the window $W$. Consider $1\leq i < \ell$ (resp. $1<i \leq \ell$) such that $D^i$ is positive (resp. negative) and $A^{i}$ is negative (resp. $A^{i+1}$ is positive). Let $\gint$ be the interior of $D^i$ and let $\ind{1}$ such that $P_{\ind{1}} =A^{i}_{1}$ (resp.  $P_{\ind{1}} =A^{i+1}_{1}$) then the decomposition of $P_{[\ind{1},\lastp]}$ is $(A^i)_{1\leq i \leq \ell}$. 
\end{corollary}

%\begin{corollary}
%\label{lem:deal:dual:coro1}
%Consider a cut $(G_1,G_2)$ of the grid, its bi-infinite window $W$, a finite path $P$ such that $P_1 \in V(W)$ and $P_{\lastp} \in V(W)$, the decomposition $(A^i)_{1\leq i \leq \ell}$ of $P$ into extremum arcs in $G_1$ and its dual decomposition $(D^i)_{0\leq i \leq \ell}$ of the window $W$. Consider $1\leq i < \ell$ such that $D^i$ is positive and $A^{i+1}$ is negative. Let $\gint$ be the interior of $D^i$ and let $\ind{1}$ such that $P_{\ind{1}} =A^{i+1}_{A^{i+1}}$ then $P_{[\ind{1},\lastp]}$ is a subgraph of $\gint$. 
%\end{corollary}
%
The two previous corollaries lead to a special case occurring when a positive dual segment is surrounded between two positive arcs, see Figure \ref{fig:deal:dual:oppo2}b.

\begin{corollary}
\label{lem:deal:dual:coro2}
Consider a cut $(G_1,G_2)$ of the grid, its bi-infinite window $W$, a finite path $P$ such that $P_1 \in V(W)$ and $P_{\lastp} \in V(W)$, the decomposition $(A^i)_{1\leq i \leq \ell}$ of $P$ into extremum arcs in $G_1$ and its dual decomposition $(D^i)_{0\leq i \leq \ell}$ of the window $W$. Consider $1\leq i < \ell$ such that $A^i$ is positive, $A^{i+1}$ is positive and $D^{i}$ is negative. Let $\gint$ be the interior of $D^i$ and let $\ind{1}$ such that $P_{\ind{1}} =A^{i}_{A^{i}}$ and $\ind{2}$ such that $P_{\ind{2}}=A^{i+1}_1$ then the decomposition of $P_{[\ind{1},\ind{2}]}$ is $(A^i)_{1\leq i \leq \ell}$. 
\end{corollary}

Now we deal with the orientation of the arcs.

\begin{lemma}
\label{lem:deal:ori:arc}
Consider a cut $(G_1,G_2)$ of the grid, its bi-infinite window $W$, a finite path $P$ such that $P_1 \in V(W)$ and $P_{\lastp} \in V(W)$ and the decomposition $(A^i)_{1\leq i \leq \ell}$ of $P$ into extremum arcs in $G_1$. Consider $1\leq i < j < k \leq \ell$ such that $A^i$ and $A^k$ are both positive (resp. negative) arcs. If $P$ is positive (resp. negative) then $A^j$ is positive (resp. negative). 
\end{lemma}

\begin{proof}
We suppose that $P$ is positive, the other case is symmetric. Consider the six indices $\ind{1},\ind{2},\ind{3},\ind{4},\ind{5}$ and $\ind{6}$ such that $P_{\ind{1}}=A^i_1$, $P_{\ind{2}}=A^i_{|A^i|}$,  $P_{\ind{3}}=A^j_1$, $P_{\ind{4}}=A^j_{|A^j|}$,  $P_{\ind{5}}=A^k_1$ and $P_{\ind{6}}=A^k_{|A^k|}$. By definition $\ind{1} \leq \ind{2}$, $\ind{3} \leq \ind{4}$ and $\ind{5} \leq \ind{6}$. Moreover, by lemma \ref{lem:deal:dual}, we have $\ind{2} < \ind{5}$. For the sake of contradiction, we suppose that $A^j$ is negative. We now distinguish three cases which are $\ind{4} < \ind{1}$, $\ind{2} < \ind{3} \leq \ind{4} < \ind{5}$ and $\ind{6} < \ind{3}$, see Figure \ref{fig:deal:ori:arc}a.

\vspace{+0.5em}

\noindent \textbf{Case $\ind{4} < \ind{1}$}: see Figure \ref{fig:deal:ori:arc}b for an illustration of this case. Consider the decomposition $(B^i)_{1\leq i \leq \ell'}$ of $P_{[\ind{3},\ind{2}]}$ and let $(D^i)_{0\leq i \leq \ell'}$ be the dual decomposition of $W$ according to $(B^i)_{1\leq i \leq \ell'}$. Note that there exists $1\leq i' < j' \leq \ell'$ such that $B^{i'}=A^{i}$ and $B^{j'}=A^{j}$. Let $\gint$ be the interior of $D^{i'}$. Since $\ind{2}<\ind{5}$, we have $\ind{2}<\lastp$ and since $B^{i'}$ is positive, then by lemma \ref{lem:exit:intoint} the edge $(P_{\ind{2}},P_{\ind{2}+1})$ belongs to $E(\gint)$. Moreover, since $i<j<k$ there exists $k'\geq j'$ such that $P_{\ind{5}} \in V(D^{k'})$. Then there exists $\ind{7}=\min\{i>\ind{2}: \text{there exists $a$ such that } P_i \in D^{a} \text { with } a<i' \text{ or } a \geq j' \}$.
By definition of $\ind{7}$, there exists no arc which starts in $D^0$ and ends in $D^{\ell'}$ and which is a subgraph of $P_{[\ind{2}+1,\ind{7}]}$. Moreover, since $P$ is simple then $P_{[\ind{2}+1,\ind{7}]}$ does not intersect with $P_{[\ind{3},\ind{2}]}$. Then by applying lemma \ref{lem:conc:toolbox} on $(B^i)_{1\leq i \leq \ell'}$ and $P_{[\ind{2}+1,\ind{7}]}$, there exists an arc $A$ which dominates either $B^{i'}=A^i$ or $B^{j'}=A^j$ and which is a subgraph of $P_{[\ind{2}+1,\ind{7}]}$ which is a contradiction.

\vspace{+0.5em}

\noindent \textbf{Case $\ind{2} < \ind{3} \leq \ind{4} < \ind{5}$}: see Figure \ref{fig:deal:ori:arc}c for an illustration of this case. Consider the decomposition $(B^i)_{1\leq i \leq \ell'}$ of $P_{[\ind{1},\ind{4}]}$ and let $(D^i)_{0\leq i \leq \ell'}$ be the dual decomposition of $W$ according to $(B^i)_{1\leq i \leq \ell'}$. Note that there exists $1\leq i' < j' \leq \ell'$ such that $B^{i'}=A^{i}$ and $B^{j'}=A^{j}$. Let $\gint$ be the interior of $D^{j'-1}$. Since $\ind{4}<\ind{5}$, we have $\ind{4}<\lastp$  and since $B^{i'}$ is negative, then by lemma \ref{lem:exit:intoint} the edge $(P_{\ind{4}},P_{\ind{4}+1})$ belongs to $E(\gint)$. Moreover, since $i<j<k$ there exists $k'\geq j'$ such that $P_{\ind{5}} \in D^{k'}$. Then there exists $\ind{7}=\min\{i>\ind{4}: \text{there exists $a$ such that } P_i \in D^{a} \text { with } a<i' \text{ or } a \geq j' \}$.
By definition of $\ind{7}$, there exists no arc which starts in $D^0$ and ends in $D^{\ell'}$ and which is a subgraph of $P_{[\ind{4}+1,\ind{7}]}$. Moreover, since $P$ is simple then $P_{[\ind{4}+1,\ind{7}]}$ does not intersect with $P_{[\ind{1},\ind{4}]}$. Then by applying lemma \ref{lem:conc:toolbox} on $(B^i)_{1\leq i \leq \ell'}$ and $P_{[\ind{4}+1,\ind{7}]}$, there exists an arc $A$ which dominates either $B^{i'}=A^i$ or $B^{j'}=A^j$ and which is a subgraph of $P_{[\ind{2}+1,\ind{7}]}$ which is a contradiction.

\vspace{+0.5em}

\noindent \textbf{Case $\ind{6} < \ind{3}$}: see Figure \ref{fig:deal:ori:arc}d for an illustration of this case. Consider the decomposition $(B^i)_{1\leq i \leq \ell'}$ of $P_{[\ind{1},\ind{6}]}$ and let $(D^i)_{0\leq i \leq \ell'}$ be the dual decomposition of $W$ according to $(B^i)_{1\leq i \leq \ell'}$. Note that there exists $1\leq i' \leq k' \leq \ell'$ such that $B^{i'}=A^{i}$ and $B^{k'}=A^{k}$. Let $\gint$ be the interior of $D^{k'}$. Since $\ind{6}<\ind{3}$, we have $\ind{6}<\lastp$  and since $B^{k'}$ is positive, then by lemma \ref{lem:exit:intoint} the edge $(P_{\ind{6}},P_{\ind{6}+1})$ belongs to $E(\gint)$. Moreover, since $i<j<k$ there exists $i' \leq j'<k'$ such that $P_{\ind{5}} \in D^{j'}$. Then there exists $\ind{7}=\max\{i<\ind{5}: \text{there exists $a$ such that } P_i \in D^{a} \text { with } a<i' \text{ or } a \geq j' \}$.
By definition of $\ind{7}$, we have $\ind{6}+1 \leq \ind{7}$ and there exists no arc which starts in $D^0$ and ends in $D^{\ell'}$ and which is a subgraph of $P_{[\ind{7},\ind{5}]}$. Moreover, since $P$ is a simple path then $P_{[\ind{7},\ind{5}]}$ does not intersect with $P_{[\ind{1},\ind{6}]}$. Then by applying lemma \ref{lem:conc:toolbox} on $(B^i)_{1\leq i \leq \ell'}$ and $P_{[\ind{7},\ind{5}]}$, there exists an arc $A$ which dominates either $B^{i'}=A^i$ or $B^{k'}=A^k$ and which is a subgraph of $P_{[\ind{7},\ind{5}]}$ which is a contradiction.

\end{proof}

Now, we can prove the main lemma of this section.

\begin{lemma}
\label{lem:strong}
Consider a cut $(G_1,G_2)$ of the grid, its bi-infinite window $W$, a finite path $P$ such that $P_1 \in V(W)$ and $P_{\lastp} \in V(W)$ and the decomposition $(A^i)_{1\leq i \leq \ell}$ of $P$ into extremum arcs in $G_1$. Let $(D^i)_{0\leq i \leq \ell}$ be the dual decomposition of $W$ according to $(A^i)_{1\leq i \leq \ell}$. If $P$ is positive (resp. negative), then there exist $1\leq j \leq \ell$ and $1\leq k \leq \ell$ such that for all $i < j$ or $i> k$ the arc $A^i$ is negative (resp. positive) and for all $j \leq i \leq k$ the arc $A^i$ is positive (resp. negative). Moreover, for all $j \leq i \leq k-1$ the dual segment $D^{i}$ is positive (resp. negative), for all $i <j-1$ or $i> k$ the dual segment $D^i$ is negative (resp. positive).
\end{lemma}

\begin{proof}
We consider that $P$ is positive, the other case is symmetric. If for all $1\leq i \leq \ell$, the arc $A^i$ is negative then let $\ind{1}$ such that $P_{\ind{1}}=A^1_{|A^1|}$ and let $\ind{2}$ such that $P_{\ind{2}}=A^\ell_1$ (see Figure \ref{fig:strong}). Since $P$ is positive and $\ind{1}< \ind{2}$ then there exists $1\leq i \leq \ell-1$ such that the dual segment $D^i$ is positive. Now, suppose for the sake of contradiction that there exists $1\leq j \leq \ell-1$ such that the dual segment $D^j$ is positive and $j\neq i$. Then let $\ind{3}$ (resp. $\ind{4}$) such that $P_{\ind{3}}=A^i_1$ (resp. $P_{\ind{4}}=A^j_1$). Without loss of generality, we suppose that $\ind{3}<\ind{4}$ then $A^i$ is not a subgraph of $P_{[\ind{4},\lastp]}$ but by corollary \ref{lem:deal:dual:coro2}, the decomposition of $P_{[\ind{4},\lastp]}$ in extremum arc in $G_1$ is $(A^i)_{1\leq i \leq \ell}$ which is a contradiction. Then $D^i$ is the only positive dual segment. Let $j=i+1$ and $k=i$ and the lemma is true. 

Now suppose that there exists at least one positive arc, then we can define $j=\min\{i:A^i \text{ is positive}\}$ and $k=\max\{i:A^i \text{ is positive}\}$. Then by lemma \ref{lem:deal:ori:arc}, for all $j \leq i \leq k$, the arc $A^i$ is positive. By definition of $j$ and $k$, for all $i < j$ or $i> k$ the arc $A^i$ is negative. Now, by lemma \ref{lem:deal:dual}, for all $j \leq i \leq k-1$ the dual segment $D^{i}$ is positive. Now for the sake the contradiction suppose that there exists $1\leq i \leq \ell-1$ such that $i <j-1$ or $i> k$ and $D^i$ is positive. We consider that $i>k$, the other case is symmetric. We also consider that $D^k$ is positive, the other case is symmetric. Then let $\ind{5}$ such that $P_{\ind{5}}=A^{k+1}_{|A^{k+1}|}$ and let $\ind{6}$ such that $P_{\ind{6}}=A^{i+1}_{|A^{i+1}|}$. If $\ind{5}<\ind{6}$ (resp.  $\ind{5}>\ind{6}$) then $A^{i+1}$ (resp. $A^{k+1}$) is not a subgraph of $P_{[1,\ind{5}]}$ (resp. $P_{[1,\ind{6}]}$) but by corollary \ref{lem:deal:dual:oppo} (resp. corollary \ref{lem:deal:dual:coro2}), the decomposition of $P_{[1,\ind{5}]}$ (resp. $P_{[1,\ind{6}]}$) in extremum arc in $G_1$ is $(A^i)_{1\leq i \leq \ell}$ which is a contradiction. Then for all $i <j-1$ or $i> k$ the dual segment $D^i$ is negative.

%\begin{figure}[th]
%\centering
%\begin{tikzpicture}[x=0.25cm,y=0.25cm]
%
%
%\path [draw,fill=gray!60] (0,0) -| (35,1) -| (0,0);  
%
%\draw (5,5) node {$A_{1}$};
%\draw (3.5,-1) node {$P_{\ind{1}}$};
%\path [draw,pattern=north west lines] (3,0) |- (7,4) |- (6,0) |- (4,3) |- (3,0);  
%
%\draw (13,5) node {$A_{2}$};
%\draw (14.5,-1) node {$P_1$};
%\path [draw,pattern=north west lines] (11,0) |- (15,4) |- (14,0) |- (12,3) |- (11,0);  
%
%\draw (17,2) node {$D_{2}$};
%
%\draw (21,5) node {$A_{3}$};
%\draw (19.5,-1) node {$P_{\lastp}$};
%\path [draw,pattern=north west lines] (19,0) |- (23,4) |- (22,0) |- (20,3) |- (19,0);  
%
%\draw (29,5) node {$A_{4}$};
%\draw (30.5,-1) node {$P_{\ind{2}}$};
%\path [draw,pattern=north west lines] (27,0) |- (31,4) |- (30,0) |- (28,3) |- (27,0);  
%
%\path [draw] (11,1) |- (7,-3) |- (6,1) |- (12,-4) |- (11,1);  
%\path [draw] (27,1) |- (23,-3) |- (22,1) |- (28,-4) |- (27,1);  
%\path [draw] (30,1) |- (4,-5) |- (3,1) |- (31,-6) |- (30,1);  
%
%\draw[fill] (30.5,0.5) circle (0.25) ;
%\draw[fill] (19 .5,0.5) circle (0.25) ;
%\draw[fill] (3.5,0.5) circle (0.25) ;
%\draw[fill] (14.5,0.5) circle (0.25) ;
%
%\path [dotted, draw, thin] (0,-7) grid[step=0.25cm] (35,8);
%
%\end{tikzpicture}
%
%
%\caption{Illustration of the first case of lemma \ref{lem:strong}. A path $P$ such that $A^1$, $A^2$, $A^3$ and $A^4$ are all negative. The dual segment $D^2$ is the only positive one. With $j=3$ and $k=2$, the lemma is true.}
%\label{fig:strong}
%\end{figure}

\end{proof}

Applying the previous lemma on an extremum path $P$ leads to the following result.

\begin{lemma}
\label{lem:conc:prel}
Consider a cut $(G_1,G_2)$ of the grid, its window $W$ and a simple finite extremum path $P$ then the decomposition of $P$ in extremum arcs is positive (resp. negative) if $P$ is positive (resp. negative). 
\end{lemma}

\begin{proof}
We consider that $P$ is positive, the other case is symmetric. Let $(A^i)_{1\leq i \leq \ell}$ be the decomposition of $P$ into extremum arcs and let $(D^i)_{0\leq i \leq \ell}$ be the dual decomposition of $W$ according to $(A^i)_{1\leq i \leq \ell}$. By fact \ref{lem:firstlastarc}, the arcs $A^1$ and $A^\ell$ are both positive. Then by lemma \ref{lem:strong}, for all $1\leq i \leq \ell$ the arc $A^i$ is positive and for all $1\leq i \leq \ell-1$ the dual segment $D^{i}$ is positive. Then by fact \ref{fact:sufficient:positive}, the decomposition of $P$ in extremum arcs is positive. 
\end{proof}

Note that we have defined the decomposition into extremum arcs for a path $P$ such that $P_{1} \in V(W)$ and $P_{\lastp} \in V(W)$. Now we generalize our result to a path $P$ such that $V(P) \cap V(W)$ is finite. In this case, there exists $\ind{1}$ (resp. $\ind{2}$) such that $\ind{1}=\min\{i:P_i \in V(W)\}$ (resp. $\ind{2}=\max\{i:P_i \in V(W)\}$), we define the decomposition of $P$ into extremum arcs as the decomposition of $P_{[\ind{1},\ind{2}]}$ into extremum arcs. 

We end this section with a sequence of three technical lemmas/remarks which will be useful later. These lemma deals with the position of the first edge of a dual segment. Our final aim is illustrated in Figure \ref{fig:cor:dual:firstedge}. %We give now a sufficient condition for a path to have the same decomposition into extremum arcs than the extremum path extracted from it. %Before this result we need a technical one.

\begin{lemma}
\label{cor:dual:firstedge:prel}
Consider a cut $(G_1,G_2)$ of the grid, its bi-infinite window $W$, a simple path $P$. We suppose that $G_1$ is the left side of the cut and that there exists $1\leq \ind{1} \leq \ind{2} \leq \lastp$ such $P_{[\ind{1},\ind{2}]}=A$, $P_{[\ind{1},\ind{2}]}$ is a positive arc of $P$, that $1<\ind{2}<\lastp$. Now, consider the window $W'$ associated to arc $P_{[\ind{1},\ind{2}]}$ and its cut $(G_3,G_4)$ of the grid. We suppose that $G_3$ is a subgraph of $G_1$ and that $G_2$ is a subgraph of $G_3$. Now assume that $P_{[\ind{2},\ind{2}+1]}$ is strictly on the right side of $W'$. Then there exists $\ind{3}$ such that $W_{\ind{3}}=P_{\ind{2}}$ and moreover $(W_{\ind{3}},W_{\ind{3}+1})$ is strictly on the right side of~$P$.
\end{lemma}

%\begin{lemma}
%\label{cor:dual:firstedge:prel}
%Consider a cut $(G_1,G_2)$ of the grid, its bi-infinite window $W$ and a simple path $P$. Let $1\leq \ind{1} \leq \ind{2} \leq \lastp$ such that $P_{[\ind{1},\ind{2}]}$ is a positive extremum arc of $G_1$ and let $\ind{3}$ such that $W_{\ind{3}}=P_{\ind{2}}$. Let $(G_3,G_4)$ be the cut of the grid associated to $P_{[\ind{1},\ind{2}]}$ and let $W'$ be its window. Suppose that $G_1$ is the left side $W$, that $G_3$ is the left side of $W'$ and that $1<\ind{2}<\lastp$. If the edge $(P_{\ind{2}},P_{\ind{2}+1})$ is strictly on the right side of $W$ then either the edge $(W_{\ind{3}},W_{\ind{3}+1})$ is strictly on the left side of $P_{[\ind{2}-1,\ind{2}+1]}$ or the edge $(P_{\ind{2}},P_{\ind{2}+1})$ belongs to $E(G_3)$. 
%%Consider a cut $(G_1,G_2)$ of the grid, its bi-infinite window $W$ and a simple positive path $P$ such that $V(P) \cap V(W)$ is finite. Let $1\leq \ind{1} <\ind{2} \leq \lastp$ such that $P_{[\ind{1},\ind{2}]}$ is the extremum path extracted from $P$. Let $(G_3,G_4)$ be the cut of the grid associated to $P$. Suppose that $G_1$ is the left side $W$, that $G_3$ is the left side of $W'$, that $\ind{2}<\lastp$ and that $|A^1|>1$. If $P_{[1,\ind{1}]}$ and $P_{[\ind{2},\lastp]}$ are both subgraph of $G_4$ then for all $1\leq i \leq \ell$, there exists $1< \ind{3} < \lastp$ such that the edge $(D^i_1,D^i_2)$ is strictly on the left side of $P_{[\ind{3}-1,\ind{3}+1]}$. 
%\end{lemma}

\begin{proof}
This proof is made by a local reasoning around tile $P_{\ind{2}}$. By definition there exists $\ind{4}$ such that $W'_{\ind{4}}=P_{\ind{2}}=W_{\ind{3}}$ and since $1<\ind{2}<\lastp$ the positions $P_{\ind{2}-1}$, $P_{\ind{2}+1}$, $W'_{\ind{4}-1}$ and $W'_{\ind{4}+1}$ are neighbors of $P_{\ind{2}}$ in $G$. Since $P_{[\ind{1},\ind{2}]}$ is positive and extremum then $P_{\ind{2}-1}\neq W'_{\ind{4}+1}$. Since the edge $(P_{\ind{2}},P_{\ind{2}+1})$ is strictly on the right side of $W'$ then $P_{\ind{2}+1}\neq W'_{\ind{4}-1}$ and $P_{\ind{2}+1}\neq W'_{\ind{4}+1}$. Then $W'_{\ind{4}+1}=W_{\ind{3}+1}$ and without loss of generality, we suppose that $W'_{\ind{4}+1}=W'_{\ind{4}}+ (0,1)$. Finally remark that the edge $(P_{\ind{2}},P_{\ind{2}+1})$ cannot belongs to $E(G_3)\setminus E(W')$ otherwise it would contradict the definition of $W'$. Figure \ref{fig:cor:dual:firstedge:prel} represents the different possible cases and in all possible cases the edge $(W_{\ind{3}},W_{\ind{3}+1})=(W'_{\ind{4}},W'_{\ind{4}+1})$ is strictly on the left side of $P_{[\ind{2}-1,\ind{2}+1]}$.

\end{proof}

Consider a cut $(G_1,G_2)$ of the grid, a finite path $P$ and its decomposition $(A^i)_{1\leq i \leq \ell}$ into extremum arcs in $G_1$. Consider $1\leq i \leq \ell$ then around tile $A^i_{|A^i|}$ the window associated to $(A^i)_{1\leq i \leq \ell}$ and the window associated to $A^i$ are identical. Then, a corollary of lemma \ref{cor:dual:firstedge:prel} and lemma \ref{lem:exit:notfree} leads to the following result.
 
\begin{corollary}
\label{cor:dual:firstedge:prel2}
Consider a cut $(G_1,G_2)$ of the grid, its bi-infinite window $W$, a finite path $P$ such that $P_1 \in V(W)$ and $P_{\lastp} \in V(W)$ and the decomposition $(A^i)_{1\leq i \leq \ell}$ of $P$ into extremum arcs in $G_1$ and its dual decomposition $(D^i)_{0\leq i \leq \ell}$. Consider $1\leq i \leq \ell$ and $1\leq \ind{1} \leq \lastp$ such that $P_{\ind{1}}=A^i_{|A^i|}$. Suppose that $1<\ind{1} < \lastp$, that $A^i$ is positive and that $G_1$ is the left side of $W$. Let $\ind{3}$ such that $W_{\ind{3}}=P_{\ind{1}}$, then the edge $(W_{\ind{3}},W_{\ind{3}+1})$ is strictly on the left side of $P_{[\ind{1}-1,\ind{1}+1]}$.
\end{corollary}

\begin{lemma}
\label{cor:dual:firstedge}
Consider a cut $(G_1,G_2)$ of the grid, its bi-infinite window $W$ and a simple positive path $P$ such that $V(P) \cap V(W)$ is finite. Suppose that there exists $1 <\ind{1} < \lastp$ such that $P_{[1,\ind{1}]}$ is the extremum path extracted from $P$. Let $(G_3,G_4)$ be the cut of the grid associated to $P$. Suppose that $G_1$ is the left side $W$, that $G_3$ is the left side of $W'$ and that $|A^1|>1$. If $P_{[\ind{1},\lastp]}$ is a subgraph of $G_4$ then for all $1\leq i \leq \ell$, there exists $1< \ind{2} < \lastp$ such that the edge $(D^i_1,D^i_2)$ is strictly on the left side of $P_{[\ind{2}-1,\ind{2}+1]}$. 
\end{lemma}

\begin{proof}
By lemma \ref{lem:extact:ext:path}, the decomposition of $P_{[1,\ind{1}]}$ in extremum arcs in $G_1$ is the same as the one of $P$. Since by hypothesis $P$ is positive then by lemma \ref{lem:conc:prel}, the decomposition $(A^i)_{1\leq i \leq \ell}$ is positive. Consider $1\leq i \leq \ell-1$ and since $A^i$ is positive then $A^i_{|A^i|}=D^i_1$. Let $\ind{3}$ such that $P_{\ind{3}}=A^i_{|A^i|}$. By lemma \ref{lem:firstlastarc}, $P_{1}=A^1_1$ and $P_{\ind{1}}=A^\ell_{|A^\ell|}$ then $\ind{3}<\ind{1}$. Since by hypothesis $|A^1|>1$ then $\ind{3}>1$. Then by corollary \ref{cor:dual:firstedge:prel2}, the edge $(D^i_1,D^i_2)$ is strictly on the left side of $P_{[\ind{3}-1,\ind{3}+1]}$. Now if $|A^\ell|>2$ then $A^\ell$ is a positive arc. Moreover since $P_{[\ind{1},\lastp]}$ is a subgraph of $G_4$ then the edge $(P_{\ind{1}},P_{\ind{1}+1})$ is strictly on the left side of $W'$. Then by lemma \ref{cor:dual:firstedge:prel}, the edge $(D^\ell_1,D^\ell_2)$ is strictly on the left side of $P_{[\ind{1}-1,\ind{1}+1]}$ (remark that locally $W'$ and the window associated to $A^\ell$ are identical). Finally, if $|A^\ell|=1$ then if the arc $A^\ell$ is negative in $P$, let $(G_5,G_6)$ be the cut of the grid associated to $(A^i)_{1\leq i \leq \ell}$. Suppose that $G_5$ is the left side of this cut, then $G_5$ is a subgraph of both $G_1$ and $G_3$. Then by lemma \ref{lem:exit:intoint} the edge $(P_{\ind{1}},P_{\ind{1}+1})$ is either in the interior of $D^\ell-1$ which is subgraph of $G_3$ or in $G_5$ which is a subgraph of $G_3$, both cases contradict the hypothesis of the lemma. Then $A^\ell$ is also a positive arc of $P$. Then by lemma \ref{cor:dual:firstedge:prel}, the edge $(D^\ell_1,D^\ell_2)$ is strictly on the left side of $P_{[\ind{1}-1,\ind{1}+1]}$.
%By hypothesis, we have $\ind{2}<\lastp$ and then $1<\ind{3} \leq \ind{2}<\lastp$. If $i<\ell$, then by lemma \ref{lem:decompo:path} the edge $(D^i_1,D^i_2)$ is either strictly on the right side of $P_{[\ind{3}-1,\ind{3}+1]}$ or strictly on the left side of $P_{[\ind{3}-1,\ind{3}+1]}$. By lemma \ref{lem:dual:bon:cote}, the edge $(D^i_1,D^i_2)$ is strictly on the right side of $P_{[\ind{3}-1,\ind{3}+1]}$. Finally, if $i=\ell$ then by lemma \ref{lem:firstlastarc} we have $\ind{3}=\ind{2}$. If the edge $(D^\ell_1,D^\ell_2)$ is on the right side of $P_{[\ind{2}-1,\ind{2}+1]}$ then $P_{\ind{2}+1} \in V(G_3)$ (by a reasoning similar to the one done in lemmas \ref{lem:exit:notfree} and \ref{lem:exit:intoint}) which contradicts the hypothesis of the lemma. Thus the edge $(D^\ell_1,D^\ell_2)$ is strictly on the left side of $P_{[\ind{2}-1,\ind{2}+1]}$.
\end{proof}

\subsection{Application of the toolbox}

In this subsection, we prove a technical result which will a key argument of theorem \ref{lemma:exit:straddle}. This result is long to prove and several technical difficulties need to be overcome. We prove it here to illustrate an application of our toolbox and to avoid making a lengthy technical interlude in section \ref{sec:Uturn}. 

This result requires a very specific setting, see Figure \ref{fig:def:shrink1} of appendix \ref{app:shrink}. We consider a cut of the grid $(G_1,G_2)$ and its bi-infinite window $W$ (in fact, this cut of this grid will be a cut of the grid done by a path with visible extremities). Next, we consider a good path $C$ such that the extension $\omeplus{C}$ splits $G_1$ into $(G^-_1,G^+_1)$. Finally, we consider a good arc $A$ of $G_1^+$ such that $A_0=C_0$, $A_{\lastp}=C_{|C|}$. We aim to show that either the extension $\omeplus{A}$ is a subgraph of $G^+_1$ or we can find another path $R$ which splits $G_1^+$. Later, this lemma will be used as follow: the path $\omeplus{C}$ corresponds to a failed attempt of pumping a candidate segment whereas the arc $A$ correspond to a new attempt to pump another candidate segment. If $\omeplus{A}$  is a subgraph of $G_1^+$ then the candidate segment $A$ will be pumpable otherwise we can find the path $R$ which corresponds to the failed attempt of pumping $A$. In this case, $\omeplus{R}$ will replace $\omeplus{C}$ and we will find another candidate segment. This reasoning will be repeated until a pumpable candidate segment is found. This lemma is proven in three steps. In the first step, we show that the good arc $A$ can be decomposed in a sequence of elementary arcs of $G_1^+$. In the second step, we show that for any basic path, we can find a basic arc of $G_1^+$ such that the restrained interior of this arc is a subgraph of the restrained interior of the path. Finally, in the last step we conclude this subsection using the two previous results.

%Now, we show that a simple arc can be decomposed into a sequence of simple arcs (see Figure \ref{fig:cut:deco:arc}).

In the following definition, we introduce the decomposition of an extremum arc in a sequence of elementary arcs. Here we consider a sequence of arcs $(A^i)_{1\leq i \leq \ell}$ such that for all $1\leq i \leq \ell-1$ the arc $A^{i+1}$ is consecutive to $A^i$. The decomposition and the dual decomposition of the window associated to $(A^i)_{1\leq i \leq \ell}$ is defined as in a decomposition in extremum arcs. Nevertheless, in this case a dual segment could be of length one since the arcs $(A^i)_{1\leq i \leq \ell}$ may not be strictly consecutive, see Figure \ref{fig:cut:deco:arc} for an illustration of these notions and of the following definition.

\begin{definition}
\label{def:decompo:ext}
Consider a cut $(\cutp,\cutm)$ of the grid, its bi-infinite window $W$, two indexes $\ind{1} \leq \ind{4}$ and a positive extremum arc $A$ of $\cutp$ which starts in $W_{\ind{1}}$ and ends in $W_{\ind{4}}$. Consider a sequence of arcs $(A^i)_{1\leq i \leq \ell}$ such that for all $1\leq i \leq \ell-1$ the arc $A^{i+1}$ is consecutive to $A^i$. Consider the dual decomposition $(D^i)_{0\leq i \leq \ell}$ of the window $W$ associated to $(A^i)_{1\leq i \leq \ell}$. This sequence is a decomposition of $A$ into elementary arcs if and only if for all $1\leq i \leq \ell$, $A^i$ is a subpath of $A$, $A^i$ is a positive elementary arc and there exists two indices $\ind{1} \leq \ind{2} \leq \ind{3} \leq \ind{4}$ such that $A=W_{[\ind{1},\ind{2}]} \cdot_{1\leq i \leq \ell-1} (A^i \cdot D^i) \cdot A^{\ell} \cdot W_{[\ind{3},\ind{4}]}$. %Moreover this decomposition is positive if $A$ is positive and negative if $A$ is negative.
\end{definition}

We now prove that for any extremum arc, there always exists a decomposition of this arc into elementary arcs.

\begin{lemma}
\label{lem:decompo:ext}
Consider a cut $(\cutp,\cutm)$ of the grid, its bi-infinite window $W$ and a positive extremum arc $A$ of $\cutp$ then there exists a decomposition of $A$ into elementary arcs. %Moreover this decomposition is positive if $A$ is positive and negative if $A$ is negative.
\end{lemma}

\begin{proof}
Let $m=w(A)$ be the width of $A$. Without loss of generality, we suppose that the arc $A$ starts in $W_0$ and ends in $W_{m}$. Since $A$ is positive and extremum then $A_1=W_0$ and $A_{|A|}=W_m$. Let $\ell=|\{i:1\leq i \leq |A|-1 \text{ and } A_i \in V(W) \text{ and } (A_{i},A_{i+1}) \notin E(W)\}|$ and we define $\ind{1}=1$ and $\ind{2\ell+2}=|A|$, for all $1\leq i \leq \ell$ we define $1\leq \ind{2i} \leq {|A|}$ as $\ind{2i}=\min\{j:j \geq \ind{2i-1} \text{ and } (A_{j},A_{j+1}) \notin E(W)\}$ and for all $1 \leq i \leq \ell$ we define $1\leq \ind{2i+1} \leq {|A|}$ as $\ind{2i+1}=\min\{j:j>\ind{2i} \text{ and } A_{j} \in V(W)\}$. By definition of $\ell$ this sequence of indexes is correctly defined. Now, for all $1 \leq i \leq \ell$ we define the arc $A^i$ as $A_{[\ind{2i},\ind{2i+1}]}$. Note that for all $1\leq i \leq \ell$, the arc $A^{i}$ is elementary and $|A^i|>1$ and for all $0\leq i \leq 2\ell$, $A_{[\ind{2i+1},\ind{2(i+1)}]}$ is a subgraph of the window $W$. Also since $A$ is a simple path then for all $1\leq j < k \leq \ell$, then either $V(A^j) \cap V(A^k)=\emptyset$ or $V(A^j) \cap V(A^k)=A^j_{|A^j|}=A^k_1$ and $k=j+1$. Moreover, by definition of $\ind{1}$ and $\ind{2\ell+1}$, we have $A=\cdot_{1\leq i \leq 2\ell} A_{[\ind{i},\ind{i+1}]}$. % Now consider the following hypothesis of recurrence for $1\leq i \leq \ell$ $H(i)=$"$(A^j)_{1\leq i \leq i}$ is a coherent decomposition of $\cdot_{1\leq j \leq 2i} P_{[\ind{j},\ind{j+1}]}$".
Now if for all $1\leq i\leq \ell$, $A^i$ is positive and if for all $1\leq i\leq \ell-1$, $A^{i+1}$ is consecutive to $A^i$ then for all $1\leq i \leq \ell-1$, we have $A_{[\ind{2i+1},\ind{2(i+1)}]}=D^i$ (where $(D^i)_{0\leq i \leq \ell}$ is the dual decomposition of $W$ according to $(A^i)_{1\leq i \leq \ell}$). In this case, $(A^i)_{1\leq i \leq \ell}$ is a decomposition of $A$ into elementary arcs.

Otherwise for the sake of contradiction, suppose that there exists $1\leq k\leq \ell$ such that $k=\min\{1\leq i \leq \ell: A^i\text{ is negative or $i>1$ and $A^i$ is not consecutive to $A^{i-1}$} \}$. 
Then consider the arc $B=\cdot_{1\leq i \leq 2(k-1)} A_{[\ind{i},\ind{i+1}]}$ (see Figure \ref{fig:deco:arc:proof}). Note that $B$ is an extremum arc and that $(A^i)_{1\leq i \leq k-1}$ is a decomposition of $B$ in elementary arcs. Moreover, $B$ starts in $W_0$ and ends in $A_{\ind{2k-1}}=A^{k-1}_{|A^{k-1}|}$. Now let $w,x,y$ and $z$ such that  $W_w=A_{\ind{2k-2}}=A^{k-1}_{1}$, $W_x=A_{\ind{2k-1}}=A^{k-1}_{|A^{k-1}|}$, $W_y=A_{\ind{2k}}=A^{k}_{1}$ and $W_z=A_{\ind{2k+1}}=A^k_{|A^k|}$. Since $A^{k-1}$ is positive then $w<x$. To conclude this proof, we consider two cases: $y>x$ or $x \leq y$.

\vspace{+0.5em}

\noindent \textbf{Case $y>x$:} see Figure \ref{fig:deco:arc:proof}b for an illustration of this case. Since $y>x$ then by definition of $k$, the arc $A^k$ is negative and we have $z<y$ and since $A$ is extremum then $z>0$. Thus, the arc $A^{k}=A_{[\ind{2k},\ind{2k+1}]}$ intersects with $B \cdot A_{[\ind{2k-1},\ind{2k}]}$ which contradicts the fact that $A$ is a simple path.  

\vspace{+0.5em}

\noindent \textbf{Case $x< y$:} see Figure \ref{fig:deco:arc:proof}c for an illustration of this case. If $y\leq w$ then since $A_{[\ind{2k-1},\ind{2k}]}=W_{[y,x]}$ then $A_{\ind{2k-2}} \in A_{[\ind{2k-1},\ind{2k}]}$ and thus $A$ is not simple which is a contradiction. Otherwise, $w <y$. Since $A$ is an arc of $G_1$, then $A_{[\ind{2k+1},|A|]}$ is an arc of $G_1$. Moreover, since $w<y<x$ and since $A_{|A|} \notin W_{[x,y]}$ then $A_{[\ind{2k},|A|]}$ intersects with $A^{k-1}$ which is a contradiction.

\vspace{+0.5em}

\noindent \textbf{Case $x= y$:} in this case, either $z\leq w$ and this case is similar to the case $y>x$ or $z>w$ and this case is similar to the case $x< y$.

\end{proof}

Now we prove the second step of our result.

\begin{lemma}
\label{lem:shrink:path:tool}
Consider a window $W$ and its cut of the grid $(G_1,G_2)$. Consider a basic path $P$, let $\ind{1}$ (resp. $\ind{2}$) such that $P_{1}=W_{\ind{1}}$ (resp. $P_{\lastp}=W_{\ind{2}}$). Suppose that $P$ starts in $W_{\ind{2}}$ and that the edge $(P_{\lastp-1},P_{\lastp})$ belongs to $E(G_1)$. Then there exists a basic arc $A$ of $G_1$ such that the restrained interior of $A$ is a subgraph of the restrained interior of $P$ and if $P$ is not an arc of $G_1$ then $V(A) \cap V(W_{[\ind{1}+1,+\infty]}) \neq \emptyset$. 
\end{lemma}

\begin{proof}
See Figure \ref{fig:cut:deco:arc:proof}a, for an illustration of a path $P$ which satisfies the hypothesis of this lemma. Since $P$ starts in $W_{\ind{2}}=P_{\lastp}$ and since $P$ is basic then $P$ is negative and $V(P) \cap V(W_{[-\infty,\ind{1}]})=\{P_1,P_{\lastp}\}$. Since $P_1\in V(W)$, there exists $\ind{4}=\max\{i <\lastp: P_i \in V(W )\}$. Moreover, since $(P_{\lastp-1},P_{\lastp}) \in E(G_1)$ then $P_{[\ind{4},\lastp]}$ is an elementary arc of $G_1$. Remark that if $\ind{4}=1$ then $P$ is an elementary arc of $G_1$ and the lemma is true with $A=P$. Otherwise $\ind{4}>1$ and let $\ind{3}$ be such that $W_{\ind{3}}=P_{\ind{4}}$ and then $\ind{3} > \ind{1}$. Consider a decomposition $(A^i)_{1\leq i \leq \ell}$ of $P_{[1,\ind{4}]}$ into extremum arcs in $G_1$ (see Figure \ref{fig:cut:deco:arc:proof}b) and its dual decomposition $(D^i)_{0\leq i \leq \ell}$ of the window $W$. Also, let $\gint$ be the restrained interior of $P$. To conclude the proof we proceed in three steps, first we analyze some properties of the decomposition $(A^i)_{1\leq i \leq \ell}$. Then we explain how we build the arc $A$ and we conclude by showing that $A$ is a subgraph of $\gint$.

\vspace{+0.5em}

\noindent \textbf{Analysis of the decomposition $(A^i)_{1\leq i \leq \ell}$:} firstly, since $\ind{4}<\lastp$ then $P_{[1,\ind{4}]} \cap W_{[-\infty,\ind{1}]} = \{P_{1}\}$ and by lemma \ref{lem:firstlastarc:prel}, we have $A^1_1=P_{1}$  and $A^1$ is a positive arc of $P_{[1,\ind{4}]}$. Secondly, by lemma \ref{lem:decompo:path}, there exists $1\leq j \leq \ell$ such that $P_{\ind{4}}$ belongs to the interior of $A^j$. Now if $P_{\ind{4}} \notin V(A^j)$ then since $P_{\lastp}=W_{\ind{2}}$, since $\ind{2}<\ind{1}$ and since $P_{[1,\ind{4}]}$ starts in $W_{\ind{1}}$ then the arc $A^j$ intersects with the arc $P_{[\ind{4}, \lastp]}$; this fact would contradict that $P$ is simple. Then, $P_{\ind{4}} \in V(A^j)$ and since  $(A^i)_{1\leq i \leq \ell}$ is the decomposition of $P_{[1,\ind{4}]}$ then $A^j_{|A^j|}=P_{\ind{4}}$. Now if $j=1$ then $\ell=1$ and $P=A^1\cdot P_{[\ind{4},\lastp]}$ then $P$ is an arc of $G_1$ and the lemma is true with $A=P$. Otherwise $j>1$ and since $A^j_{|A^j|}=P_{\ind{4}}$ then the dual segment $D^{j-1}$ is positive. Moreover, since $A^1$ is positive by lemma \ref{lem:strong} for all $1\leq i \leq j-1$, the arc $A^i$ is positive. 

\vspace{+0.5em}

\noindent \textbf{Definition of the arc $A$:} if $A^j$ is negative then let $A=\cdot_{1\leq i \leq j-1} (A^i \cdot {D^i}) \cdot P_{[\ind{4},\lastp]}$, otherwise let $A=\cdot_{1\leq i \leq j-1} (A^i \cdot {D^i}) \cdot A^j \cdot P_{[\ind{4},\lastp]}$ (see Figure \ref{fig:cut:deco:arc:proof2}a). Since $P$ is a basic path and since $P_{[\ind{4},\lastp]}$ is an elementary arc of $G_1$ then $A$ is a basic arc of $G_1$ such that $A_1=P_1$, $A_{\lastp}=P_{\lastp}$ and $W_{\ind{3}} \in (V(A) \cap V(W_{[\ind{1}+1,+\infty]}))$. 

\vspace{+0.5em}

\noindent \textbf{The arc $Q$ is a subgraph of $\gint$:}  let $C$ be the window of $\gint$ (see Figure \ref{fig:cut:deco:arc:proof2}b). Let $Q=W_{[\ind{2},\ind{1}]} \cdot A$, then $Q$ is a simple cycle and is the window of the restrained interior of $A$ (see Figure \ref{fig:cut:deco:arc:proof2}c). Let $m=\ind{1}-\ind{2}$ then $m>0$. Without loss of generality we suppose that $G_1$ is the left side of the cut of the grid and we consider an indexing of $C$ such that $C_0=W_{\ind{2}}$ and $C_{m}=W_{\ind{1}}$ then $\gint$ is the left side of $C$. Now remark that $Q_{[0,m]}=W_{[\ind{2},\ind{1}]}=C_{[0,m]}$, that $P_{[\ind{4},\lastp]}$ is a subgraph of $P$ and thus a subgraph of $C$ and that for all $1\leq i \leq j$, the arc $A^i$ is a subpath of $P$ and thus a subpath of $C$. Then, all these paths are subgraphs of $\gint$. Now, let $(B^i)_{1\leq i \leq \ell}$ be the sequence of arcs such that $B^1=C_{[0,m]} \cdot A^1$ and for all $2\leq i \leq j$, we have $B^i=A^i$. Then this sequence is the decomposition of $C_{[0,m]}\cdot P_{[1,\ind{4}]}$ in extremum arcs in $G_1$. Since $m>0$ then $|B^1|>1$ and remark that $C_{[0,m]}\cdot P_{[1,\ind{4}]}$ is a subgraph of $C$. Now, consider $1\leq i \leq j-1$, since $|B^1|>1$ and $A^i$ is positive, then by lemma \ref{cor:dual:firstedge:prel2} the edge $(D^i_1,D^i_2)$ is strictly on the left side of $C$ and thus in $E(\gint)$. Moreover, by lemma \ref{lem:decompo:path}, $V(D^i)\cap V(P)=\{D^i_1,D^i_{|D^i|}\}$ then $V(D^i)\cap V(C)=\{D^i_1,D^i_{|D^i|}\}$ and thus $D^i$ is a subgraph of $\gint$. Then the arc $A$ is a subgraph of $\gint$ and the restrained interior of $A$ is a subgraph of $\gint$.

\end{proof}

%This decomposition of an extremum arc is called \emph{the decomposition of $A$ into simple arcs}. It is easy to prove that this decomposition is unique but we do not require this property for the rest of the proof. 

Finally we can state the main lemma of this subsection.

\begin{lemma}
\label{lem:shrink:path}
Consider a cut of the grid $(G_1,G_2)$ by a bi-infinite window $W$, an index $\ind{1}$ and a simple good path $C$ of direction $\vu $ such that $y_{\vu}\geq 1$ and $\omeplus{C}$ splits $G_1$ into $(G_1^-,G_1^+)$ at position $W_{\ind{1}}$. Consider a good arc $A$ of $G_1^+$ of direction $\vu$ such that $A_1=C_1$ and $A_{|A|}=C_{|C|}$ then either $\omeplus{A}$ is a subgraph of $G_1^+$ or there exists a good path $R$ of direction $\vu$ such that $R_1=C_1$, $\omeplus{R} $ is a subgraph of $G_1^+$ and there exists $\ind{2}>\ind{1}$ and $\ind{3}\geq |R|$ such that $\omeplus{R}_{\ind{3}} =W_{\ind{2}}$ and such that $y_{W_{\ind{2}}} \geq \min\{y_C,y_Q\}+y_{\vu}$. % and $\omeplus{R}_{[\ind{3}+1,+\infty]} \cap W= \emptyset$.
\end{lemma}

\begin{proof}
This proof is done in three steps and relies on an algorithm to obtain the desired result. An example of this algorithm is illustrated in the Figures of appendix \ref{app:shrink}, the input of the algorithm is Figure \ref{fig:def:shrink1}. During the first step of the proof, we decompose the arc $A$ into several elementary arcs to obtain a correct input for our algorithm (see Figure \ref{fig:def:shrinkOublie}). Now for each arc of this decomposition, the second step introduces a loop of the algorithm which modifies this arc (see Figures \ref{fig:def:shrink2} and \ref{fig:def:shrink3} for arc $A^1$ and Figure \ref{fig:def:shrink4} for arc $A^2$). The third step puts together all the different modified arcs of $A$ to build the path $R$ and proves that the path $R$ matches the hypothesis of the lemma (see Figure \ref{fig:def:shrink5}). Without loss of generality, we consider that $G_1$ is the left side of $W$ and that $G_1^+$ is the right side of $\omeplus{C}$. 

\vspace{+0.5em}
\noindent \textbf{Input of the algorithm:}
consider the cut of the grid $(\cutp,\cutm)$ associated to the periodic bi-infinite window $W'=\ome{C}$. We suppose that $\cutp$ is the right side of the cut. Let $p=|C|-1$ and consider an indexing of $W'$ such that $W'_0=C_1=A_1$ and  $W'_p=C_{|C|}=A_{|A|}$. Now let $\ind{4}=\min\{i: A_{[\ind{4},|A|]}$ is a subgraph of $\cutp\}$. Since $A$ is a subgraph of $G_1^+$ then the edge $(A_{|A|-1},A_{|A|}) \in E(\cutp)$ and thus $\ind{4}<|A|$. If $\ind{4}=1$ then $A$ is an arc of $\cutp$ otherwise by contradiction suppose that $1< \ind{4} < |A|$. Then, there exists $\ind{5}$ such that  $A_{\ind{4}} = W'_{\ind{5}}$. 
Since  $A$ is a subgraph of $G_1^+$  then $\ind{5}<0$ and then the width of $A_{[\ind{4},|A|]}$ is strictly greater than $p$. Then, by lemma \ref{lem:per:arc}, $A_{[\ind{4},|A|]}$ intersect with $A_{[\ind{4},|A|]}+\vu$ which contradicts the fact that $A$ is a good path. Then $A$ is an arc of $\cutp$. Now if $A$ is not extremum then let $A'$ be the extremum arc extracted from $A$. Similarly to the previous reasoning, the width of $A'$ is strictly greater than $p$. Thus, arcs $A'$ and $A'+\vu$ intersect which is a  contradiction. Thus $A$ is an extremum positive arc of $\cutp$. Now, consider a decomposition $(A^i)_{1\leq i \leq \ell}$ of $A$ in elementary arcs in $\cutp$ and its dual decomposition $(D^i)_{1\leq i \leq \ell}$ of $W'$. Then by lemma \ref{lem:decompo:ext}, there exists $0\leq \ind{5} \leq \ind{6}\leq p$ such that $A=W'_{[0,\ind{5}]} \cdot_{1\leq i \leq \ell-1} (A^i \cdot D^i)\cdot  A^\ell \cdot W'_{[\ind{6},p]}$. Also remark that $W'_{[0,\ind{5}]}$ and $W'_{[\ind{6},p]}$ are subpath of $C$ and for all $1\leq i \leq \ell$, the dual segment $D^i$ is a subpath of $C$. Remark that for all $1\leq i \leq \ell$, the path $A^i$ is an arc of $G_1^+$.

\vspace{+0.5em}
\noindent \textbf{Shrinking an arc:}
Consider $1\leq i \leq \ell$, if for all $k\geq 0$ the path $A^i+k\vu$ is a subgraph of $G_1^+$ then let $B^i=A^i$. Otherwise, there exists $k>0$ such that $A^i+k\vu$ is not an elementary arc of $G_1^+$ then remark that $A^i+k\vu \cap V(\omeplus{C})=\{A^i_1+k\vu, A^i_{|A^i|}+k\vu\}$ then $A^i+k\vu$ is a basic path of $G_1^+$ which ends in $A^i_{|A^i|}+k\vu$. Then by applying lemma \ref{lem:shrink:path:tool} on the basic path $A^i+k\vu$ of $G_1^+$, there exists a basic arc $B^i$ of $G_1^+$ such that the restrained interior $B^i+k\vu$ is a subgraph of the restrained interior of $A^i+k\vu$, such that $B^i_1=A^i_1$, such that $B^i_{|B^i|}=A^i_{|A^i|}$ and such that $V(B^i+k\vu) \cap V(W_{[\ind{1}+1,+\infty]}) \neq \emptyset$. Then, there exists $\ind{2}>\ind{1}$ such that $W_{\ind{2}} \in B^i+k\vu$ and since $y_{\vu}\geq 1$ and since $B^i+k\vu$ is a subgraph of the restrained interior of $A^i+k\vu$ then $y_{W_{\ind{2}}}\geq \min\{y_C,y_Q\}+y_{\vu}$. Also, $B^i$ is a subgraph of the restrained interior of $A^i$ which is a subgraph of $G_1^+$, then $B$ is a subgraph of $G_1^+$. Remark that the number of edges of the restrained graph of $B^i$ is strictly less than the number of edges in the restrained graph of $A^i$. By iterating this reasoning we can suppose for all $k\geq 0$, the arc $B^i+k\vu$ is a subgraph of $G_1^+$. %The path $W'_{[0,\ind{1}]} \cdot_{1\leq i \leq \ell-1} (A^i \cdot D^i) \cdot A^{\ell} \cdot W'_{[\ind{2},p]}$ is a good path which belong to the interior of $Q$. 

\vspace{+0.5em}
\noindent \textbf{Iterating for all arcs:}
by doing the same reasoning for all elementary arcs, we obtain that a path $R=W'_{[0,\ind{5}]} \cdot_{1\leq i \leq \ell-1} (B^i \cdot D^i) \cdot B^{\ell} \cdot W'_{[\ind{6},p]}$. Now since for all $1\leq i \leq \ell$ and for all $k \geq 0$, the arc $B^i+k\vu$ is a subgraph of the restrained interior of $A^i+k\vu$ and since $A$ is a good path, then $R$ is a good path. Moreover $\omeplus{R}$ is a subgraph of $G_1^+$. Now if $R=Q$ then $\omeplus{Q}$ is a subgraph of $G_1^+$. Otherwise, there exists $\ind{2}>\ind{1}$ and $\ind{3}\geq |R|$ such that $\omeplus{R}_{\ind{3}} =W_{\ind{2}}$ and such that $y_{W_{\ind{2}}} \geq \min\{y_C,y_Q\}+y_{\vu}$.  \end{proof}

\section{Modified window lemma and jailed path assembly}
\label{sec:uwl}
In this part, we modify a previous result known as the window movie lemma \cite{Meunier-2014}.  The previous result is general but not powerful enough for our study. We modify it in order to be more powerful but less general and we give two applications of this modified window movie lemma. One is about thin path assembly which will be later use to prove the stakes lemma. The second one is the proof of the jail lemma, one of the fourth macroscopic lemmas of our toolbox. In this section, we will cut the grid with discrete line and only this kind of cut will be used. The figures of this section are in Appendix \ref{app:uwl}.

\subsection{Output glues, movies, stripes and window movie lemma}

First, we adapt the window movie lemma \cite{Meunier-2014} to our notations. Note that, all of the notations introduced in this section are specific to this section and are not used in other parts of the article. Here, we will cut the grid with several lines to partition the set of vertices of the grid (see figure \ref{fig:uwl:partition}). Consider a non null vector $\vu \in \mathbb{Z}$ and the height function associated to $\vu$, let $p=x_{\vu}+y_{\vu}$. For all $h \in \mathbb{Z}$, let $(\lmink{h},\lplusk{h})$ be the cut by the line of direction $\vu$ and height $hp$ and let $L^h$ be its window. Remark that the set of vertices of all these windows is a partition of $V(\grid)$, \emph{i.e.} $V(G)=\bigcup_{h \in \mathbb{Z}}(V(L^h))$ and for all $h,h' \in \mathbb{Z}$ such that $h\neq h'$, we have $V(L^h)\cap V(L^{h'})=\emptyset$. Note that this property is not true for the edges of $E(\grid)$, $E(G)\neq \bigcup_{k \in \mathbb{Z}}(E(L^h))$. Indeed, there exists two kinds of edges of $E(\grid)$: the ones where there exists $k \in \mathbb{Z}$ such that they belong to $E(L^h)$ and the ones where there exists $h \in \mathbb{Z}$ such that one extremity in $E(L^h)$ and the other extremity is in $E(L^{h+1})$ (see figure \ref{fig:uwl:partition}). The window movie lemma focusses on the edges of the second kind. For a given $h \in \mathbb{Z}$, we will consider all the edges of a path $P$ with an extremity in $L^h$ and the other one in $L^{h+1}$. Before pursuing, remark that since the last section, we have worked on path. Now, we need to work again on path assembly. To achieve this goal, we introduce output glues. Intuitively, an output glue of a path assembly is an edge of its underlying path labeled by a type of glue.

Consider a tiling system $\tiling=(T,\seed,1)$, an \emph{output glue} is a couple $(g,e)$ where $g \in \glue$ is a glue and $e \in E(G)$ is an oriented edge of the grid. Intuitively, an output glue is the glue linking two adjacent tiles positioned on the grid. We define by $\type((g,e))=g$ and $edge((g,e))=e$. Consider a path assembly $P$ producible by a tiling system $\tiling$. For all $1\leq i \leq \lastp-1$, the \emph{output glue} $g_i$ of $P$ is the output glue where $\type(g_i)$ is the type of the glue on the abuttal side of $P_i$ and $P_{i+1}$ and $edge(g_i)=(\pos(P_i),\pos(P_{i+1}))$. %An output glue is \emph{oriented to the north} if $y_{P_{i+1}}=y_{P_i}+1$, to the south if $y_{P_{i+1}}=y_{P_i}-1$, to the east if $x_{P_{i+1}}=x_{P_i}+1$ and to the west if $x_{P_{i+1}}=x_{P_i}-1$. Moreover an output glue is \emph{vertical} if it is oriented to the north or to the south, and it is \emph{horizontal} if it is oriented to the west or to the east. 
Remark that the edge of an output glue is an edge of the binding graph of $P$. Similarly to previous definitions, the translation of an output glue $(g,e)$ by a vector $\wu$ is $(g,e+\wu)$.

%Consider a path assembly $P$ producible by a tiling system $\tiling=(T,\seed,1)$. An \emph{output glue} is couple $(g,A)$ where $g$ is a glue ($g \in \glue$) and $A$ is a position ($A \in \mathbb{Z}^2$). Intuitively, an output is the glue linking two tiles positioned on the grid. More formally, for $1\leq i \leq \lastp-1$, the output glue $g_i$ of $P$ is the output glue where $glue(g)$ is the type of the glue in the abuttal side of $P_i$ and $P_{i+1}$ and $\pos(g_i)=(\min\{x_{P_i},x_{P_{i+1}}\}, \min\{y_{P_i},y_{P_{i+1}}\})$. This definition using $\min$ may seems surprising but is the more effective (see figure \ref{}). Since output glue are between two tiles, a more accurate definition would be $\frac{x_{P_i}+x_{P_{i+1}}}{2}$ but with this definition we would need to work with non integer position which is a problem. Also, the definition $\pos(g_i)=\pos(P_i)$ allow output glues with different ordinates to be on the same vertical line, see figure \ref{}.

%Consider $h \in \mathbb{Z}$, $\vu \in \mathbb{Z}^2$ and the cut $(\cutLinep,\cutLinem)$ of $\grid$ by the line of direction $\vu$ of height $h$, 
Now, we define the movie of a path assembly $P$. Intuitively a movie of heigh $h$ contains all the output glues of $P$ which have on extremity in $L^h$ and the other one in $L^{h+1}$. The \emph{movie} $\movie$ of path assembly $P$ of height $h$ and direction $\vu$ is the sequence $\movie=(\movie_i)_{1\leq i \leq |\movie|}$ of output glues such that (see Figure \ref{fig:uwl:movstri}):
\begin{itemize}
\item for all $1 \leq i \leq |\movie|$, there exists $1\leq j \leq \lastp-1$ such that $g_j=\movie_i$ and either $P_{j}\in L^h$ and $P_{j+1}\in L^{h+1}$ or $P_{j}\in L^{h+1}$ and $P_{j+1}\in L^{h}$;
\item for all $1\leq j \leq \lastp-1$ such that either $P_{j}\in L^h$ and $P_{j+1}\in L^{h+1}$ or $P_{j}\in L^{h+1}$ and $P_{j+1}\in L^{h}$, there exists $1 \leq i \leq |\movie|$ such $\movie_i=g_j$ ;
\item if $1\leq i<j \leq |\movie|$ and $\movie_i=g_{i'}$ and $\movie_j=g_{j'}$ then $i'<j'$.
\end{itemize}
The translation of a movie by a vector $\wu$ is obtained by translating all output glues of $\movie$ by $\wu$ and keeping the same order. The movie $\movie$ is identical to the movie $\movie'$ up to translation if only if there exists a vector $\wu$ such that $\movie=\movie'+\wu$.  We denote by $\movie(h,\vu)$ the movie of $P$ according to the cut $(\lmink{h},\lplusk{h})$ of direction $\vu$. Remark that for an output glue $g$ of $P$ and for a given non-null vector $\vu$, there is at most one $h$ such that $g$ is an output glue of $\movie(h,\vu)$.

For a path assembly $P$ and a subgraph $Q$ of $G$, we define $\alpha_{(P,Q)}$ as the function whose domain is $V(Q)\cap \dom{P}$ for all $(x,y) \in V(Q) \cap \dom{P}$, $\alpha_{(P,Q)}(x,y)=\type(P_i)$ where $\pos(P_i)=(x,y)$. With this definition, for all  $h \in \mathbb{Z}$, we have $\alpha_P=\alpha_{(P,U^{h+1})}\cup \alpha_{(P,D^{h})}$ (see figure \ref{fig:uwl:movstri}). Note that the function $\alpha_{(P,Q)}$ is not necessary an assembly because its domain may not be connected. Now, we can state the window movie lemma \cite{Meunier-2014} applied to our setting. Note that, this result is a special case of the window movie lemma which is more general. For an illustration of the following lemma, see Figures \ref{fig:movie:lemma} and \ref{fig:movie:lemma2} of appendix \ref{app:modi}.

\begin{lemma}
\label{window:movie:lemma}
Consider a path assembly $P$ producible by a tiling system $\tiling=(T,\seed,1)$, a non null vector $\vu \in \mathbb{Z}^2$, two integers $h,h' \in \mathbb{N}$ such that $h'>h$ and two non empty movies $\movie=\movie(\vu,h)$ and $\movie'=\movie(\vu,h')$ of $P$. If there exists a vector $\wu$ such that $\movie'$ is equal to $\movie+\wu$ and if $\dom{\seed} \subset V(\lmink{h})$ (resp. $\dom{\seed} \subset V(\lplusk{h'+1})$) 
%Consider a path assembly $P$ producible by a tiling system $\tiling=(T,\seed,1)$, a vector $\vu$, two non empty movies $\movie$ of height $h$ and $\movie'$ of height $h'$ with $h'>h$ and the stripe $\stripe$ of direction $\vu$ between height $h$ and $h'$. If the following hypothesis are true:
%\begin{itemize}
%\item there exists a non-null vector $\wu \in \mathbb{Z}$ such that $\movie'=\movie+\wu$;
%\item $\dom{\seed}\cap V(\stripe)=\emptyset$;
%\end{itemize}
then $\alpha_{(P,\lmink{h'})}\cup (\alpha_{(P,\lplusk{h+1})}+\wu)$  (resp. $(\alpha_{(P,\lmink{h'})}-\wu) \cup \alpha_{(P,\lplusk{h+1})}$) is an assembly producible by~$\tiling$. 
\end{lemma}

Note that the definition of $\wu$ implies that $\wu$ is non-null and that $\lplusk{h+1}+\wu=\lplusk{h'+1}$. Thus the movies $\movie(\vu,h)$ and $\movie(\vu,h')$ of $P$ are identical to the movies $\movie(\vu,h)$ and $\movie(\vu,h')$ of the assembly built by the window movie lemma. Thus it is possible to apply the window movie lemma again to obtain this more general result (see Figure~\ref{fig:movie:lemma}, Figure~\ref{fig:movie:lemma3} and Figure~\ref{fig:movie:lemma4} of appendix \ref{app:modi}).

 \begin{corollary}
 \label{window:movie:lemma:iterated}
Consider a path assembly $P$ producible by a tiling system $\tiling=(T,\seed,1)$, a non null vector $\vu \in \mathbb{Z}^2$, two integers $h,h' \in \mathbb{N}$ such that $h'>h$ and two non empty movies $\movie=\movie(\vu,h)$ and $\movie'=\movie(\vu,h')$ of $P$. Suppose that there exists a vector $\wu$ such that $\movie'$ is equal to $\movie+\wu$ and that $\dom{\seed} \subset V(\lmink{h})$ (resp. $\dom{\seed} \subset V(\lplusk{h'+1})$). We define the infinite sequence $(P^{*i})_{0\leq i}$ as $P^{*0}=P$ and for all $i\geq 0$, $P^{*(i+1)}=\alpha_{(P^{*i},\lmink{h'})}\cup (P^{*(i+1)}=\alpha_{(P^{*i},\lplusk{h+1})}+\wu)$  (resp. $(\alpha_{(P^{*i},\lmink{h'})}-\wu) \cup \alpha_{(P^{*i},\lplusk{h+1})}$). Then for all $i>0$, $P^{*i}$ is an assembly producible by~$\tiling$. 
 \end{corollary}

As previously mentioned, we have $\lplusk{h+1}+\wu=\lplusk{h'+1}$. Now, since $h'>h$ then $\lplusk{h'+1}$ is a subgraph of $\lplusk{h+1}$ and then a simple recurrence leads to the fact that any segment of $P$ which is a subgraph of $\lplusk{h+1}$ appears translated by $i\wu$ in $P^{*i}$ for all $i\geq 0$ (see Figures \ref{fig:movie:lemma5}, \ref{fig:movie:lemma6} and \ref{fig:movie:lemma7} of appendix \ref{app:modi}).

 \begin{fact}
 \label{window:movie:lemma:iterated:fact}
Consider a path assembly $P$ producible by a tiling system $\tiling=(T,\seed,1)$, a non null vector $\vu \in \mathbb{Z}^2$, two integers $h,h' \in \mathbb{N}$ such that $h'>h$ and two non empty movies $\movie=\movie(\vu,h)$ and $\movie'=\movie(\vu,h')$ of $P$. Suppose that there exists a vector $\wu$ such that $\movie'$ is equal to $\movie+\wu$ and that $\dom{\seed} \subset V(\lmink{h})$ (resp. $\dom{\seed} \subset V(\lplusk{h'+1})$). Consider the infinite sequence  of assemblies $(P^{*i})_{0\leq i}$ obtained by lemma \ref{window:movie:lemma:iterated}. Consider $1\leq \ind{1} \leq \ind{2} \leq \lastp$ such that the underlying path of $\dom{P_{[\ind{1},\ind{2}]}}$ is a subgraph of $U^{h+1}$ (resp. $D^{h'}$) then for all $i\geq 0$, $P_{[\ind{1},\ind{2}]}+i\wu$ (resp. $P_{[\ind{1},\ind{2}]}-i\wu$) is a subassembly of $P^{*i}$. 
 \end{fact}

\subsection{The modified window movie lemma}

Now we present an original result, We show that in our setting, a path assembly $P$ is in fact pumpable or fragile if two identical movies of $P$ can be found. %To prove this result, we show that if two identical movies exists then there exists a candidate segment in $P$ of direction $\wu$ which is a subgraph of $U^{h+1}$ and fact \ref{window:movie:lemma:iterated:fact} leads to the fragility of $P$ or the pumpability of this candidate segment (see Figures \ref{fig:movie:lemma5} and \ref{fig:movie:lemma8} of appendix \ref{app:modi}).

 \begin{lemma}
 \label{lem:samemovie}
 Consider a path assembly $P$ producible by a tiling system $\tiling=(T,\seed,1)$, a non null vector $\vu \in \mathbb{Z}^2$, two integers $h,h' \in \mathbb{N}$ such that $h'>h$ and two non empty movies $\movie=\movie(\vu,h)$ $\movie'=\movie(\vu,h')$ of $P$. If there exists a vector $\wu$ such that $\movie'$ is equal to $\movie+\wu$ and if $\dom{\seed} \subset D^{h}$ or $\dom{\seed} \subset U^{h'+1}$ then $P$ is pumpable or fragile. 
 \end{lemma}

 \begin{proof}
See Figures \ref{fig:movie:lemma5} and \ref{fig:movie:lemma8} of appendix \ref{app:modi} for a graphical representation of this proof. %Consider the partition $(\cutLinep,\stripe,\cutLinem)$ of $G$ associated to movies $\movie$ and $\movie'$. 
Since $V(S) \cap \dom{\seed}=\emptyset$ and since the seed is connected then we can assume without loss of generality that $\dom{\seed} \subset V(\lmink{h})$. We start by localizing a candidate segment of the path assembly $P$ of direction $\wu$ which is a subgraph of $\lmink{h+1}$. Then using fact \ref{window:movie:lemma:iterated:fact}, we prove that either this segment is pumpable or that the path assembly $P$ is fragile. Note that this is the only time in this article that we will try to pump a candidate segment which is not necessarily a good candidate segment.

 We start by proving the existence of indices $1\leq k\leq |\movie|$ and $1\leq \ind{1} \leq \ind{2} \leq \lastp$ such that:
\begin{itemize}
\item there is no tile of the segment $P_{[\ind{1},\ind{2}]}$ positioned in $\lmink{h}$, \emph{i.e.} ($\dom{P_{[\ind{1},\ind{2}]}} \cap V(\lmink{h})=\emptyset$);
\item the output glue of $P_{\ind{1}-1}$ is $\movie_k$;
\item the output glue of $P_{\ind{2}-1}$ is $\movie'_k$. 
\end{itemize}
Consider the following invariant $I(k)$="there exists $1\leq \ind{1} \leq \ind{2} \leq \lastp$ such that the output glue $g_{\ind{1}-1}$ of $P$ is $\movie_k$ and the output glue $g_{\ind{2}}$ of $P$ is $\movie'_k"$. First, we prove that $H(1)$ is true. By hypothesis, the domain of the seed is in $\lmink{h}$, \emph{i.e.} $\dom{\seed} \subset V(\lmink{h})$ and thus $h(\pos(P_1)) < (h+1)p < (h'+1)p$. Since the movies $\movie$ and $\movie'$ are not empty, consider the first index $1\leq \ind{1} \leq \lastp$ such that the height of $P_{\ind{1}}$ is greater than $(h+1)p$ then the output glue $g_{\ind{1}-1}$ of $P$ is $\movie_1$. Similarly, consider the first index $\ind{2}$ such that the height of $P_{\ind{2}}$ is greater than $(h'+1)p$ then the output glue $g_{\ind{2}-1}$ of $P$ is $\movie'_1$. Moreover $h(P_{\ind{2}-1})\geq (h'+1)p-\max(x_{\vu},y_{\vu})>(h+1)p$ and thus we have $\ind{1} \leq \ind{2}-1<\ind{2}$ and $I(1)$ is true. Now, suppose that $I(k)$ is true for $1\leq k \leq |\movie|-1$. Then, there exists $1\leq \ind{1} \leq \ind{2} \leq \lastp$ such that the output glue $g_{\ind{1}-1}$ of $P$ is $\movie_k$ and the output glue $g_{\ind{2}}$ of $P$ is $\movie'_k$. Two cases can occur: either $\dom{P_{[\ind{1},\ind{2}]}} \cap V(\lmink{h})=\emptyset$ and the desired indices $\ind{1},\ind{2}$ and $k$ have been found or $P_{[\ind{1},\ind{2}]}$ contains tiles positioned in $\lmink{h}$. In the second case, there exists $\ind{1}\leq \ind{3}<\ind{2}$ such that $h(P_{\ind{3}})< (h+1)p$ and since the output glue of $P_{\ind{2}-1}$ is $\movie'_k$ then either $h(P_{\ind{2}})>(h'+1)p>(h+1)p$ or $h(P_{\ind{2}-1})>(h'+1)p>(h+1)p$, then there $k'>k$ and $\ind{3}<\ind{4} \leq \ind{2}$ such that the output glue $g_{\ind{4}-1}$ of $P$ is $\movie_{k'}$. By definition of a movie there exists $\ind{5}>\ind{4}$ such that the output glue of $P_{\ind{5}-1}$ is $\movie'_{k'}$ and this means that $I(k')$ is true. Thus either we have found the desired index $k$ or $I(k')$ is true with $k'>k$. If $I(|\movie|)$ is true then there exists $1\leq \ind{1} \leq \ind{2} \leq \lastp$ such that the output glue $g_{\ind{1}-1}$ of $P$ is $\movie_{|\movie|}$ and the output glue $g_{\ind{2}-1}$ of $P$ is $\movie'_{|\movie|}$. If $P_{[\ind{1},\ind{2}]}$ contains positions of $\lmink{h}$, then $\movie_{|\movie|}$ could not be the last output glue of the movie which is a contradiction. Thus, the domain of the segment $P_{[\ind{1},\ind{2}]}$ cannot contain a vertex of $\lmink{h}$ and the desired indices $\ind{1},\ind{2}$ and $k=|M|$ have been found. Thus in all cases, there exist $1\leq k \leq |\movie|$ and $1\leq \ind{1}<\ind{2} \leq \lastp$ such that the output glue of $P_{\ind{1}-1}$ is $\movie_k$, the output glue of $P_{\ind{2}-1}$ is $\movie'_k$ and $\dom{P_{[\ind{1},\ind{2}]}} \cap V(\lmink{h-1})=\emptyset$. 

Moreover, since $\movie'_k=\movie_k +\wu$ and since the edges of the output glue are oriented then the direction of $P_{[\ind{1},\ind{2}]}$ is $\wu$. Now, remark that the path assembly $Q$ such that $Q_{[1,|Q|-1]}=P_{[1,\ind{2}-1]}$ and $Q_{|Q|}=P_{\ind{1}}+\wu$ is a path assembly producible by $\tiling$. Thus if $\type(P_\ind{2}) \neq \type (P_\ind{1})$ then $P$ is fragile. Then either $P$ is fragile or $P_{[\ind{1},\ind{2}]}$ is a candidate segment of $P$ of direction $\wu$. If this candidate segment is pumpable then the lemma is true. Otherwise by lemma \ref{lem:good:week}, there exists $j>0$ such that there exists a conflict between either $P_{[\ind{1},\ind{2}]}+j\wu$ and $\seed$ or between $P_{[\ind{1},\ind{2}]}+j\wu$ and $P_{[1,\ind{2}]}$. Now consider the sequence of assemblies $(P^{*i})_{i\geq 0}$ created by the window movie lemma (Lemma \ref{window:movie:lemma:iterated}). Then these assemblies are producible by $\tiling$ and by fact \ref{window:movie:lemma:iterated:fact}, $P_{[\ind{1},\ind{2}]}+j\wu$ is a subassembly of $P^{*j}$. Then $P$ and $P^{*j}$ are in conflict. Thus the path assembly $P$ is fragile in this final case.

 \end{proof}
 
To use this modified windows movie lemma, we need two identical movies up to translation. To achieve this goal, we will just consider a lot of different movies until we find two identical ones.  The following result is a more practical setting of the modified windows movie lemma. 

\begin{lemma} 
\label{lem:uwl:final}
Consider a finite path assembly $P$ producible by a tiling system $\tiling=(T,\seed,1)$, a non null vector $\vu \in \mathbb{Z}^2$. Let $p=x_{\vu}+y_{\vu}$ and suppose that:
\begin{itemize}
\item for all $i \in \mathbb{Z}$, the width of the binding graph $\alpha_P$ for window $L^i$ is bounded by $c$;
\item $\lastp \geq 4cp(2c!)(2|T|+1)^{2c}+2c|\seed|$;
\end{itemize}
then $P$ is pumpable or fragile.
\end{lemma}

\begin{proof}
Consider the window $L^0$ and an indexing of $L^0$ such that for all $i \in \mathbb{Z}$, we have $L^0_{i+p}=L^0_i+\vu$, we will use this window and this indexing as a reference to compare different movies. Now, consider $h,h' \in \mathbb{Z}$, the two windows $L^{h}$ and $L^{h'}$ and the two non-empty movies $\movie=\movie(\vu,h)$ and $\movie'=\movie(\vu,h')$ of~$P$. There exist a vector $\wu$ (resp. $\wu'$) and an indexing of $L^{h}$ (resp. $L^{h'}$) such that $L^0+\wu=L^{h}$ (resp. $L^0+\wu'=L^{h'}$), such that $L^0_0+\wu=L^{h}_0$ (resp. $L^0_0+\wu=L^{h'}_0$) and such that $0\leq \min\{i \in \mathbb{Z}: L^{h}_i \in \dom{P}\} <p$ (resp. $0\leq \min\{i \in \mathbb{Z}: L^{h'}_i \in \dom{P}\} <p$); this fact is due to the periodicity of $L^{h}$ and $L^{h'}$, see Figure \ref{fig:uwl:simple}. If $\movie^{h}-\wu=\movie^{h'}-\wu'$ then these two movies are identical up to translation. Now, we enumerate the number of possible different movies. First, $\min\{i \in \mathbb{Z}: L^{h}_i \in \dom{P}\}$ can take $p$ different values. Since for all windows, the width of $\alpha_P$ is bounded by $c$ then there exists at most $c$ tiles of $P$ positioned on the window $L^{h}$. Any of these tiles can belong to at most two output glues of the movie $\movie^{h}$. Then the size of all movies is bounded by $2c$. Now, there exists $(2c!)$ possible ways to order these $2c$ output glues. For each of this ordering there exists $(2|T|+1)^{2c}$ ways to color it (for this enumeration we consider the absence of an output glue has a type of glue and there is at most $|T|$ types of glues multiplied by two for the orientation of the edge). Then, there exists at most $p(2c!)(2|T|+1)^{2c}$ different movies up to translation. Now, there exists three kind of movies: the movies which contains output glues of the seed, the movies of height $h$ such that the domain of the seed is a subgraph of $\lmink{h}$ and the movies of height $h$ such that the domain of the seed is a subgraph of $\lplusk{h+1}$. There exists at most $|\seed|$ movies of the first kind. Since the length of the movies is bounded by $2c$ and since $\lastp \geq 4cp(2c!)(2|T|+1)^{2c}+2c|\seed|$ then there exists at least $2p(2c!)(2|T|+1)^{2c}$ non-empty movies of the second or $2p(2c!)(2|T|+1)^{2c}$ non-empty movies of the third kind. Then, either there exists at least two movies of the second kind which are identical up to translation or there exists two movies of the third kind which are identical up to translation. Then by lemma \ref{lem:samemovie}, the path assembly $P$ is fragile or pumpable.
\end{proof}

\subsection{Application of the modified window movie lemma}

\noindent \textbf{Thin path.} Before proving the jail lemma and concluding this section. We present a direct application of the modified window movie lemma to thin path assembly. Intuitively a thin path assembly, is a path assembly whose length is far greater than its width. Such a path assembly is fragile or pumpable by the modified windows movie lemma. This result is a good illustration on how to use this lemma and will be useful to prove the stakes lemma in section \ref{sec:lastpart}.

% This subsection deals with a special case where the length of the path assembly is far larger than its width, \emph{i.e} $P$ leaves the rectangle area $\rect(\fband(Y_P),Y_P)$ by its east side. It is concluded by the proof that such a path assembly is pumpable or fragile (see lemma \ref{lem:band}). This result is achieved by using the window movie lemma which is proven in \cite{}. The followings definitions are simplified versions of the ones in \cite{} which are sufficient to deal with the cases occurring in this article.

\begin{definition}[Thin path assembly]
\label{def:thin}
Consider a tiling system $\tiling=(T,\seed,1)$, let $\fband:\mathbb{N}\rightarrow \mathbb{N}$ be the function $\fband(x)=4x(2x!)(2|T|+1)^{2x}+2x|\seed|$. A path assembly $P$ is thin if and only if $Y_P-y_P > \fband(X_P-x_P)$ or $X_P-x_P > \fband(Y_P-y_P)$.%, \emph{i.e.} $P$ does not fit into the rectangle area $\rect(X_P,\fband(X_P))$ (resp. $\rect(\fband(Y_P),Y_P)$).
 \end{definition}

\begin{lemma}
\label{lem:thin}
If a path assembly $P$ producible by a tiling system $\tiling=(T,\seed,1)$ is thin then it is fragile or pumpable.
\end{lemma}

 \begin{proof}
Without loss of generality, we assume that $Y_P-y_P > \fband(X_P-x_P)$. Let $\vu=(1,0)$ and then by definition of thin path assembly for all $h\in \mathbb{Z}$, the width of $P$ for window $L^h$ of direction $\vu$ is bounded by $X_P-x_P$. Since $\lastp \geq Y_P-y_P\geq \fband(X_P-x_P)$ and then by lemma \ref{lem:uwl:final}, $P$ is pumpable or fragile.
%We consider all the movies of height greater than $2|\glue|+|\seed|$, there are $\fband(X^P) -  2|\glue|+|\seed|-1$ of them. Each of these movies contains less than $X^P$ output glues. Then there exists less than $|\glue|^{4(X^P+2)}(4(X^P+2))!$ different possible non-empty movies (see \cite{} for more details on this bound) and thus by definition of $\fband(X^P)$ there exists two movies $M$ of height $h$ and $M'$ of height $h'$ with $Y_{\seed}<h<h'<Y_P$ such that $M'=M+\wu$. Finally, by lemma \ref{lem:samemovie}, the path is pumpable or fragile.
 \end{proof}

% \begin{figure}[h]
% \begin{center}
% \includegraphics[scale=0.4]{./figure/fitband}
% \end{center}
%  \caption{A thin path assembly, and its movie of height $h$. The south and north output glues alternates along the movie.}
%  \label{fig:def:band}
% \end{figure}

\noindent \textbf{Jail lemma.} Consider a path $P$, a vector $\vu$ of $\mathbb{Z}^2$ such that $P$ and $P-\vu$ does not intersect and let $p=x_{\vu}+y_{\vu}$. Then consider a cut $(U,D)$ of the grid by a discrete line of direction $\vu$ and its window $L$. First we prove that the width of $P$ according to $L$ is bounded by $2p^2+2p$. This result and the lemma \ref{lem:uwl:final} lead to the jail lemma. Bounding the width of $P$ is not an easy task. Here is a quick  sketch of the proof. In the previous section, we have shown that if there exists an arc of width greater than $p$ which is a subgraph of $P$ then $P$ and $P-\vu$ intersect (lemma \ref{lem:jail:arc}). Then, Figure \ref{fig:uwl:worst} shows what we conjecture to be the worst case scenario. In this case, consider a decomposition $(A^i)_{1\leq i \leq \ell}$ of $P$ into extremum arcs. From this decomposition, we can extract a sequence of arc $(B^i)_{1\leq i \leq \ell'}$ such that for all $1\leq i \leq \ell'-1$ the arc $B^i$ is dominated by $B^{i+1}$ (lemma \ref{lem:jail:conc}). This result implies that the widths of the arcs of this sequence increase by at least two each time, \emph{i.e.} for all  $1\leq i \leq \ell'-1$  the width of $B^{i+1}$ is at least the width of $B^{i}$ plus $2$. Then if this sequence contains more than $\frac{p}{2}$ arcs, it also contains an arc of width greater than $p$. By the previous lemma, $P$ and $P-\vu$ intersect in this case. These results provides a bound of $2p^2+2p$ for the width of $P$ according to window $L$.

\begin{lemma}
\label{lem:jail:conc}
Consider a path $P$, a non null vector $\vu$ of $\mathbb{Z}^2$ and a cut $(U,D)$ of the grid by a line of direction $\vu$ and its window $L$. Let $p=x_{\vu}+y_{\vu}$. Suppose that $P$ and $P-\vu$ does not intersect, then the width of $P$ according to the window $L$ is bounded by $2p^2+2p$.
\end{lemma}

\begin{proof}
By contradiction, suppose that the width of $P$ is strictly greater than $2p^2+2p$. First, we show that without loss of generality we can consider that $P$ is an extremum finite path. Remark that if the width of $P$ is infinite, then it is possible to find a finite subpath of $P$ such that its width is strictly greater than $2p^2+2p$. Then, without loss of generality, we can suppose that $P$ is finite and such is its width. Now, consider an index of $L$ such that $0=\min\{i\in \mathbb{Z}: L_i \in V(P)\}$ and let $m=\max\{i\in \mathbb{Z}: L_i \in V(P)\}$ then the width of $P$ is $m>2p^2+2p$. Now, consider the extremum path extracted from $P$, %let $1\leq z \leq \lastp$ and $1\leq z' \leq \lastp$ such that $P_{z}=L_0$ and $P_{z'}=L_m$. Without loss of generality, we can assume that $z<z'$ and then the subpath $P_{[z,z']}$ of $P$ is an extremum subpath of $P$ 
this path is extremum and its width is $m$. Then, we can assume without loss of generality that $P$ is an extremum finite path of width strictly greater than $2p^2+2p$.

Let $Q=P-\vu$ be the path $P$ translated by $-\vu$. Our aim is to prove that $P$ and $Q$ intersect. The path $Q$ starts in $L_{-p}=Q_1$ and ends in $L_{m-p}=Q_{|Q|}$. This proof will require to define eight indices. %which will defined such that $1\leq \ind{1} \leq \ind{2} \leq \lastp$, $1\leq \ind{3} \leq \ind{4} \leq \ind{5} \leq \ind{6} \leq |Q|$ and $\ind{7}, \ind{8}, \ind{9}, \ind{10}, \ind{11}, \ind{12}  \in \mathbb{Z}$ such that $P_{\ind{1}}=L_{\ind{7}}$, $P_{\ind{2}}=L_{\ind{8}}$, $Q_{\ind{3}}=L_{\ind{9}}$, $Q_{\ind{4}}=L_{\ind{10}}$, $Q_{\ind{5}}=L_{\ind{11}}$ and $Q_{\ind{6}}=L_{\ind{12}}$ and $\ind{9} <0 \leq \ind{7} \leq \ind{11} \leq \ind{10} \leq \ind{12} \leq \ind{8}$.
%First let $\ind{3}=\max\{1\leq i \leq |Q|: \text{ there exists } -p\leq j \leq -1 \text{ such that } Q_i=L_j\}$ and $Q'=Q_{[\ind{3},|Q|]}$. 
%Consider the cut $(G_1,G_2)$ associated to the extremum path $P$ and $W'$ be its window. Then $W'=W_{[-\infty,0]} \cdot P \cdot W_{[m,+\infty]}$, since $Q_{|Q|}=W_{m-p}$ and $m-p>0$ then if $W_{m-p} \in V(W')$ then $W_{m-p} \in V(P)$ and $P$ and $Q$ intersect. Otherwise, without loss of generality, we suppose that $Q_{|Q|} \in V(G_1) \setminus V(W)$ and that $G_1$ is the left side of $W'$ and that $U$ is the left side of $W$. 
Consider the decomposition $(A^i)_{1\leq i \leq \ell}$ of $P$ in extremum arcs in $U$ and let $(S^i)_{0\leq i \leq \ell}$ be the decomposition of the window $L$ according to $(A^i)_{1\leq i \leq \ell}$ and its dual decomposition $(D^i)_{0\leq i \leq \ell}$. %Consider the decomposition $(B^i)_{1\leq i \leq \ell}$ of $Q$ in extremum arcs in $U$. 
%Then either $P$ and $Q'$ intersect or $V(Q') \cap V(W')=Q'_1$. In the second case, without loss of generality, we can suppose that $Q'$ is a subgraph of $G_1$ and that $G_1$ is the left side of $W'$ and that $G_2$ is the left side of $W$. Then, consider a decomposition $(A^i)_{1\leq i \leq \ell}$ of $P$ in extremum arcs in $U$ and let $(S^i)_{0\leq i \leq \ell}$ be the decomposition of the window $L$ according to $(A^i)_{1\leq i \leq \ell}$ and its dual decomposition $(D^i)_{0\leq i \leq \ell}$. 
Let $e=\max\{0\leq i \leq \ell: \text{there exists $1\leq j \leq |Q|$ such that $Q_j \in V(D^i)$}\}$ ($e$ is correctly defined since $Q_0 \in V(D^0)$). %Now we deal with the special case where $e=0$
%\vspace{+0.5em} 
%
%\noindent \textbf{Special case (e=0):} 
%Remark that if $e=0$ then since $Q'_{|Q'|} -\vu =W_{m-p}$ then there exists  $\ind{3} < \ind{4} \leq |Q|$ such that $Q_{\ind{4}}=\min\{i>\ind{3}: Q_{i} \in \{V(W_{[1,m-p]})\}$. Then, there exists $1\leq i \leq \ell$ such that $Q_{\ind{4}} \in V(S^i)$. In this case by definition of $\ind{3}$ and $\ind{4}$, the path $Q_{[\ind{3},\ind{4}]}$ is an arc of $U$ which starts in $W_{\ind{9}}$ with $\ind{9}<0$ and ends in $W_{\ind{10}}$. Moreover, the arc $A^i$ associated to segment $S^i$ starts in $W_{x}$ and ends in $W_y$ such that $0\leq x < \ind{10}< y$ and then by fact \ref{fact:arc}, the arcs $Q_{[\ind{3},\ind{4}]}$ and $A^i$ intersect. 
%
%\vspace{+0.5em}
%Then $e>0$ and 
Let $1 \leq \ind{1} \leq |Q|$ be such that $\ind{1}=\min\{1 \leq i \leq |Q|: Q_{\ind{1}} \in V(D^e)\}$ and let $\ind{2}$ such that $W_{\ind{2}}=Q_{\ind{1}}$. Now two cases may occur either $\ind{2}>p^2+p$ or $\ind{2}\leq p^2+p$.

\vspace{+0.5em}

\noindent \textbf{First case} ($\ind{2}>p^2+p$): %let $Q''=Q_{[\ind{3},\ind{4}]}$ and let $(B^i)_{1\leq i \leq \ell'}$ be the decomposition of $Q''$ in extremum arcs in $U$. 
see Figure \ref{fig:uwl:jail:part1} for a graphical representation of this case. Let $Q'=Q_{[1,\ind{1}]}$ and let $(B^i)_{1\leq i \leq \ell'}$ be the decomposition of $Q'$ in extremum arcs in $U$. Since $Q \cap W_{[m,+\infty]}=\emptyset$ then by lemma \ref{lem:conc:toolbox} either $P$ and $Q'$ intersect or for all $1\leq i \leq e$, there exists $1\leq j \leq \ell'$ such that arc $B^j$ dominates arc $A^i$. Now, note that for all $0\leq i \leq \ell'$, the arc $B^i+\vu$ is an arc of $U$ and a subgraph of $P$ and then there exists $1 \leq j \leq \ell$ such that either $A^j=B^i+\vu$ or $A^j$ dominates $B^i+\vu$. Now, we define a sequence of arcs $(C^i)_{1 \leq i \leq \ell''}$ of $U$ such that $C^1=A^1$ and for all $1 \leq i \leq \ell''$, if there exists $0\leq j \leq \ell'$ such that $B^j$ dominates $C^i$ then the arc $C^{i+1}$ is defined as the arc $A^{j'}$ such that either $A^{j'}=B^j+\vu$ or $A^{j'}$ dominates $B^j+\vu$; if there exists no $0\leq j \leq \ell'$ such that $B^j$ dominates $C^i$ then $i=\ell''$ and $C^i$ is the last arc of the sequence. Let $\ind{3} \in \mathbb{Z}$ such that $C^{\ell''}$ ends in $L_\ind{3}$ then $\ind{3}>\ind{2}$ otherwise the arc $C^{\ell''}$ would be dominated by an arc of $(B^j)_{1\leq j\leq \ell'}$. Then, for all $1 \leq i \leq \ell''$, the arc $C^i$ is a subgraph of $P$ and for all $1\leq i \leq \ell''-1$ the arc $C^{i+1}-\vu$ dominates $C^i$. For all $1\leq i \leq \ell''$, let $w_i$ be the width of arc $C^i$. %If there $1\leq i \leq \ell''$ such that $w_i \geq p$ then by lemma \ref{}, $A^i$ and $A^i-\vu$ intersect which is a contradiction. Then we suppose that for all $1\leq i \leq \ell''$, $w_i $ be

Now, we prove by recurrence that for all $1 \leq i \leq \ell''$, the hypothesis $H(i):"$ $w_i \geq 2(i-1)$. Moreover for all $0\leq j <i $, we have $w_i\geq w_j$ and there exists $k \in \mathbb{Z}$ such that $C^i$ ends in $L_k$ with $k\leq i(w_i)+(i-1)p"$. The arc $C^1$ is the arc $A^1$ which starts in $L_0$ and $w_1\geq 0$ and $C^1$ ends in $L_{w_1}$. Thus the initialization of the recurrence is done. Now, suppose that $H(i)$ is true for $1 \leq i \leq \ell''-1$, then $C^i$ is dominated by $C^{i+1}-\vu$. Then there exists $\ind{4},\ind{5},\ind{6},\ind{7} \in \mathbb{Z}$ such that $C^i$ starts in $L_{\ind{4}}$ and ends in $L_{\ind{5}}$ and  $C^{i+1}$ starts in $L_{\ind{6}}$ and ends in $L_{\ind{7}}$ with $\ind{6}-p<\ind{4}\leq \ind{5}<\ind{7}-p$. This means that $\ind{7}-\ind{6} \geq \ind{5}-\ind{4}+2\geq w_i+2\geq 2i$. Moreover, by recurrence we have that for all $1\leq j \leq i $, $w_{i+1} > w_i \geq w_j$. By recurrence, we have $\ind{5} \leq i(w_i)+(i-1)p$. Moreover $\ind{7}=\ind{6}+w_{i+1}$ and $\ind{6}-p<\ind{5}$ then $\ind{7} \leq i(w_i)+(i-1)p+w_{i+1}+p \leq (i+1)(w_{i+1})+ip$ and the recurrence is true. Now, if $\ell'' \geq \frac{p}{2}+1$ then $w_{\ell''}\geq p$ and by lemma \ref{lem:jail:arc}, $P$ and $Q$ intersect. Otherwise, suppose that $\ell'' \leq \frac{p}{2}$ and $w_{\ell''}<p$ then $C^{\ell'}$ ends in $L_{\ind{3}}$ with $\ind{3} < p^2+p$ which is a contradiction.

\vspace{+0.5em}

\noindent \textbf{Second case ($\ind{2}\leq p^2+p$)}: see Figure \ref{fig:uwl:jail:part2} for a graphical representation of this case.
Since $Q=P-\vu$ then the decomposition of $Q$ in extremum arcs in $U$ is $(B^i)_{1\leq i \leq \ell}$ with for all $1\leq i \leq \ell$, $B^i=A^i-\vu$. Consider $1\leq i\leq \ell$ such that $B^i$ starts in $W_{\ind{8}}$ with $\ind{8} > \ind{2}$ then there exists $1\leq j \leq \ell$ such that $W_\ind{8} \in S^j$ (otherwise this would contradicts the definition of $e$) then either $A^j$ and $B^i$ intersect or $A^j$ dominates $B^i$. Now, we define a sequence of arcs $(C^i)_{1 \leq i \leq \ell'}$ of $U$ such that $C^{1}=A^\ell$ and for all $1 \leq i \leq \ell'$, if there exists $1\leq j \leq \ell$ such that $A^j$ dominates $C^i-\vu$ then the arc $C^{i+1}$ is defined as the arc $A^{j}$; if there exists no $1\leq j \leq \ell$ such that $C^j$ dominates $C^i-\vu$ then $i=\ell'$ and $C^i$ is the last arc of the sequence. Let $\ind{3} \in \mathbb{Z}$ such that $C^{\ell'}$ starts in $L_\ind{3}$ then $\ind{3}\leq \ind{2}+p$ otherwise the arc $C^{1}-\vu$ would be dominated by an arc of $(A^j)_{1\leq j\leq \ell'}$. Then, for all $1 \leq i \leq \ell'$, the arc $C^i$ is a subgraph of $P$ and for all $1\leq i \leq \ell'-1$ the arc $C^{i+1}$ dominates $C^i-\vu$. For all $1\leq i \leq \ell'$, let $w_i$ be the width of arc $C^i$.

Now, we prove by recurrence that for all $1 \leq i \leq \ell'$, the hypothesis $H(i):"$ $w_i \geq 2(i-1)$. Moreover for all $1\leq j \leq i$, we have $w_i \geq w_j$ and there exists $k \in \mathbb{Z}$ such that $C^i$ starts in $L_k$ with $k\geq m - (i(w_i)+(i-1)p)"$. The arc $C^{1}$ is the arc $A^{\ell}$ which ends in $W_{m}$ and thus $C^{1}$ starts in $W_{m-w_{\ell'}}$. Thus the initialization of the recurrence is done. Now, suppose that $H(i)$ is true for $1 \leq i \leq \ell'-1$, then $C^{i}-\vu$ is dominated by $C^{i+1}$. Then there exists $\ind{4},\ind{5},\ind{6},\ind{7} \in \mathbb{Z}$ such that $C^i$ starts in $L_{\ind{4}}$ and ends in $L_{\ind{5}}$ and  $C^{i+1}$ starts in $L_{\ind{6}}$ and ends in $L_{\ind{7}}$ with $\ind{6}<\ind{4}-p \leq \ind{5}-p < \ind{7}$. This means that $\ind{7}-\ind{6}\geq \ind{5}-\ind{4}+2\geq w_i+2\geq 2i$. Moreover, by recurrence we have that for all $1\leq i \leq j$, $w_{i+1} > w_i \geq w_j$. By recurrence, we have $\ind{4} \geq m-(i(w_i)+(i-1)p)$. Moreover $\ind{6}=\ind{7}-w_{i+1}$ and $\ind{7}>\ind{4}-p$ then $\ind{6} \geq m-(i(w_i)+(i-1)p)-w_{i+1}-p \geq m-((i+1)(w_{i+1})+ip)$ and the recurrence is true. Now, if $\ell' \geq \frac{p}{2}+1$ then $w_{1}\geq p$ and by lemma \ref{lem:jail:arc}, $P$ and $Q$ intersect. Otherwise, suppose that $\ell' \leq \frac{p}{2}$ and $w_{\ell'}<p$ then $C^{1}$ starts in $L_{\ind{3}}$ with $\ind{3} > m-p^2 \geq \ind{2}+p$ which is a contradiction.

\end{proof}

%\begin{proof}
%Consider two consecutives movies, the stripes defined by their cut is in fact a border which is connected path. This border is made by a periodic path of period $x_{\vu}+y_{\vu} \leq 2||\vu||$ thus if the path assembly does not cross these movies after $2\vu$ iteration then there exists $i,j$ such that $P_i=P_j-\vu$ which is a contradiction.
%\end{proof}
%
%Consider of cut $(\cutLinep,\cutLinem)$ by the line of direction $\vu$, the key point is to show that if $P$ and $P-\vu$ does not intersect then the window of $P$ through $(\cutLinep,\cutLinem)$ contains no site at distance more than $||\vu||^2$.
%
%\begin{lemma}
%There exists no movies with output glues at distance more than $||\vu||^2$.
%\begin{lemma} 

We can now prove the jail lemma.

\begin{lemma}[Jail lemma]
Let $P$ be a path assembly producible by $\mathcal{T}=(T,\sigma,1)$ satisfying macroscopic initial conditions and $1\leq \ind{1} \leq \lastp$. If the path assembly $P$ is jailed at index $\ind{1}$ then it is fragile or pumpable.
\end{lemma} 

\begin{proof}
By the definition of the jail constraints, there exists $\ell$ such that $\dom{P_{[1,\ind{1}]}} \subset \rect(\ell)$ and $1\leq \ind{1} \leq \ind{2} \leq \lastp$ such that $\pos(P_{\ind{2}})\notin \rect(\fjail(\ell))$. Then the length of $P_{[\ind{1},\ind{2}]}$ is at least $\fjail(\ell)-\ell>4(\border^2(2\border+1))(2(2\border^2+\border))!)(2|T|+1)^{2\border^2+\border}+\border\ell^2$ (see appendix \ref{app:bound} for the definition of $\fjail$). By definition of the jail constraints $\dom{\seed} \subset \rect(\ell)$ and $\dom{P_{[1,\ind{1}]}}\subset \rect(\ell)$ then $|\dom{\seed} \cup \dom{P_{[1,\ind{1}]}}| \leq \ell^2$. Now, let $\seed'=\seed \cup \alpha_{P_{[1,\ind{1}]}}$ then $P_{[\ind{1},\ind{2}]}$ is producible by the tiling system $(T, \seed', 1)$. Moreover by definition of the jail constraints, $||\vu||\leq \border$  and then by lemma \ref{lem:jail:conc} and \ref{lem:uwl:final}, the path $P_{[\ind{1},\ind{2}]}$ is pumpable or fragile. Then $P$ is fragile or pumpable.
\end{proof}

\section{Microscopic reasoning}
\label{sec:micro}
In this section, we give definitions and prove results necessary to obtain the three remaining macroscopic lemmas. Proofs of these macroscopic lemmas is left for the next final section. We start by a roadmap of the last part of the proof. The figures of this section are in Appendix \ref{app:micro}.

\subsection{Roadmap of the microscopic reasoning}

In fact, our final aim is to prove the stakes lemma. The seed lemma and the reset lemma will be proven along the way. The main idea of the stakes lemma is simple, consider a path assembly $P$ producible by a tiling system $\tiling=(T,\seed,1)$. Finding a candidate segment of $P$ is easy since a segment of $P$ of length at least $|T|+1$ always contains two tiles of the same type. Then, consider $1\leq \ind{1} < \ind{2} \leq \lastp$ such that the segment $P_{[\ind{1},\ind{2}]}$ is a candidate segment of $P$. Now, we can try to build two path assemblies, one by removing the segment $P_{[\ind{1},\ind{2}]}$, \emph{i.e.} $P_{[1,\ind{1}]}\cdot P_{[\ind{2},\lastp]}$ and the other one by pumping the segment $P_{[\ind{1},\ind{2}]}$ one time, \emph{i.e.} $P_{[1,\ind{2}]}\cdot P_{[\ind{1},\lastp]}$. If these constructions are possible then the stakes lemma is true. Unfortunately, these two path assemblies does not always exist.

To solve this problem, we introduce \emph{visible} candidate segment. As we will see in the next section, visible candidate segments of $P$ are also easy to find and allow us to find an infinite zone (called the \emph{free} zone) of the grid where we can try to build our stakes without fearing intersection with the seed or the beginning of the path assembly $P$. Then, from a visible candidate segment, we built the stakes in two steps. In the first step, we construct forks which share some characteristic with stakes but not all of them and then in the second step these forks are refined into stakes. Here lies the main particularity of the proof. Building the forks is not particularly difficult but refining them into stakes is hard. In particular, the refining process requires the three other macroscopic lemmas. Note that the jail lemma has already been proven in the previous section. To prove the other two lemmas, we need another result which is that a path assembly making a U-turn is fragile or pumpable and in order to prove this result, we need forks. 

To summarize the previous paragraph, after defining visible candidate segments, we give a method to build forks. This method is the first half of the method to build stakes. Then, we use these forks to prove a lemma about U-turn. In the following section, we use U-turn to prove the seed lemma and the reset lemma. Finally, the proof can now be concluded by giving the second half of the method to build stakes from forks. 

\subsection{Visible tiles}
\label{sec:sub:vis}

In this part, we introduce visible tiles and visible candidate segments and we prove some fundamental properties about them. Consider a simple path assembly $P$ producible by a tiling system $\tiling=(T,\seed,1)$ and $1\leq \ind{1} \leq \lastp$, then the tile $P_{\ind{1}}$ is \emph{visible from the west} (resp. east) if and only if there exists no $1\leq \ind{2} \leq \lastp$ such that $x_{P_{\ind{2}}} < x_{P_{\ind{1}}}$ (resp. $x_{P_{\ind{2}}} > x_{P_{\ind{1}}}$) and $y_{\ind{2}}=y_{P_{\ind{1}}}$ (see Figure \ref{fig:vis:def}). Intuitively, a tile is visible from the west if and only if no tile of $P$ occupies a position west of its. Moreover a visible tile $P_{\ind{1}}$ of $P$ is \emph{hidden} by the seed if and only if there exists $(x,y) \in \dom{\seed} $ such that $x< x_{P_{\ind{1}}}$ (resp. $x> x_{P_{\ind{1}}}$) and $y=y_{P_{\ind{1}}}$.
%Remark that there exists at least $Y_P-y_p-|\seed|$ tiles of $P$ which are visible from the west. 
%In this part, we introduce visible tiles and visible candidate segments and we prove some fundamental properties about them. Consider a simple path assembly $P$ producible by a tiling system $\tiling=(T,\seed,1)$ and $1\leq i \leq \lastp$, then the tile $P_i$ is \emph{visible from the west} (resp. east) if and only if there exists no $(x,y) \in \dom{\seed} \cup \dom{P}$ such that $x < x_{P_i}$ (resp. $x > x_{P_i}$) and $y=y_{P_i}$. Intuitively, a tile is visible from the west if and only if no tile of the seed or $P$ occupies a position west of its. %Remark that there exists at least $Y_P-y_p-|\seed|$ tiles of $P$ which are visible from the west. 
We denote by $\vset{P}$ the set of indices of $P$ such that the corresponding tile is visible from the west, \emph{i.e.} $\vset{P}\subset \{1,\ldots, \lastp\}$ and for all $i \in \vset{P}$, $P_i$ is visible from the west. We denote by $\vsets{P}$ the set of indices of $P$ such that the corresponding tile is visible from the west and not hidden by the seed. Similarly, a position $(x,y) \in \dom{\seed}$ which belongs to the domain of the seed $\seed$ is \emph{visible from the west} if and only if there exists no $(x',y') \in \dom{\seed} \cup \dom{P}$ such that $x' < x$ and $y'=y$. We denote by $\vset{\seed}$ the set of positions which belong to the domain of the seed and which are visible from the west. Note that $|\vset{\seed}| \leq |\seed|$ and remark that if $\ind{1} \in \vset{P}$ and $P_{\ind{1}}$ is hidden by the seed then there exists $(x,y)\in \vset{\seed}$ such that $y_{P_{\ind{i}}}=y$. Also, at most $|\vset{\seed}|$ visible tiles are hidden by the seed. Since the binding graph of $P$ is connected, for all $y_P\leq h \leq Y_{P}$, there exists a unique tile visible from the west of ordinate $h$. This remark is formalized in the following fact.

\begin{fact}
\label{fact:heightvis}
Consider an simple path assembly $P$ producible by a tiling system $\tiling=(T,\seed,1)$ then for all $y_P\leq h \leq Y_{P}$, there exists a unique index $\ind{1}\in \vset{P}$ such that $y_{P_{\ind{1}}}=h$. Then, we have $|\vset{P}|=Y_{P}-y_P+1$ and $|\vsets{P}| \geq Y_{P}-y_P-|\seed|+1$.
\end{fact}

Visibility from the west or from the east impose some global constraints on the path assembly $P$ and some local constraints around a tile $P_i$ with $i \in \vset{P}$. This combination of constraints will lead to powerful results. We start by studying the local constraints. consider $1\leq \ind{1} \leq \lastp$ then $P_{\ind{1}}$ is \emph{directed to the north} (resp. to the south) if and only if $\pos(P_{\ind{1}+1})=(x_{P_{\ind{1}}},y_{P_{\ind{1}}}+1)$ (resp. $\pos(P_{\ind{1}-1})=(x_{P_{\ind{1}}},y_{P_{\ind{1}}}+1)$) or $\pos(P_{\ind{1}-1})=(x_{P_{\ind{1}}},y_{P_\ind{1}}-1)$ (resp. $\pos(P_{\ind{1}+1})=(x_{P_{\ind{1}}},y_{P_{\ind{1}}}-1)$). Two remarks, first a tile cannot be directed to the north and to the south at the same time. Secondly, a tile is not necessary oriented to the north or to the south. Now, consider $1\leq \ind{1} < \lastp$ then a tile $P_\ind{1}$ exits to the north if $\pos(P_{\ind{1}+1})=(x_{P_\ind{1}},y_{P_\ind{1}}+1)$, to the south if $\pos(P_{\ind{1}+1})=(x_{P_{\ind{1}}},y_{P_{\ind{1}}}-1)$, to the east if $\pos(P_{\ind{1}+1})=(x_{P_\ind{1}}+1,y_{P_\ind{1}})$ and to the west if  $\pos(P_{\ind{1}+1})=(x_{P_{\ind{1}}}-1,y_{P_{\ind{1}}})$. Remark that if $1< \ind{1} <\lastp$, there exists eight kinds of tiles (exiting to north implies directed to the north) but if the tile $P_{\ind{1}}$ is visible from the west then there are only four possibilities. Indeed, a position $\pos(P_{\ind{1}})$ has four neighbors in $V(\grid)$ and if $\ind{1} \in \vset{P}$, then the position $(x_{P_\ind{1}}-1,y_{P_{\ind{1}}})$ cannot be occupied by a tile of the path assembly $P$. Then if $1<\ind{1}<\lastp$, two of the three remaining neighbors of $\pos(P_{\ind{1}})$ in $V(\grid)$ have to be occupied by the tiles $P_{\ind{1}-1}$ and $P_{\ind{1}+1}$. One of these two occupied positions is either $(x_{P_{\ind{1}}},y_{P_{\ind{1}}}-1)$ or $(x_{P_{\ind{1}}},y_{P_{\ind{1}}}+1)$. This reasoning is summarized in the following fact:

\begin{fact}
\label{lem:visible:orient}
Consider a simple path assembly $P$ producible by a tiling system $\tiling=(T,\seed,1)$ and $i \in \vset{P}$ such that $1<\ind{1}<\lastp$. Then there exists $\ind{2}\in \{\ind{1}-1,\ind{1}+1\}$ such that either $\pos(P_{\ind{2}})=(x_{P_{\ind{1}}},y_{P_{\ind{2}}}-1)$ or $\pos(P_{\ind{2}})=(x_{P_{\ind{1}}},y_{P_{\ind{2}}}+1)$.
\end{fact}

The fact \ref{lem:visible:orient} implies that if $1< \ind{1} < \lastp$ and $\ind{1} \in \vset{P}$ then it is oriented to north or to the south. If $\ind{1}=1$ or if $\ind{1}=\lastp$ then the tile may not be oriented to the north or the south but this case is limited to at most two tiles in the path assembly. Finally we obtain four kinds of tile visible from the west (see Figure \ref{fig:vis:ori}): directed to the north and exiting to the north, directed to the north and exiting to the east, directed to the south and exiting to the south, directed to the south and exiting to the east.

Now we study global constraints imposed by visible tiles. In this section, our reasonings will often require to cut the grid with a simple path whose extremities are visible (see section \ref{sec:sub:cut}, we recall that $\cutp$ is the left side of the cut and that $\cutm$ is the right side). Tiles visible from the west or from the east are effective to define such cuts of the grid. Indeed, consider $1\leq \ind{1} < \ind{2} \leq \lastp$ such that $\ind{1} \in \vset{P}$ and such that $P_{\ind{2}}$ is either visible from east or from the west then it is possible to define a cut $(\cutp,\cutm)$ such that $P_{[\ind{1},\ind{2}]}$ belongs to the window of this cut (see Figure \ref{fig:vis:cut}). Now, if $P_{\ind{2}} \in \vset{P}$ then remark that a tile $P_{\ind{3}}$ such that $\min\{y_{P_{\ind{1}}},y_{P_{\ind{2}}}\} \leq y_{P_{\ind{3}}} \leq \max\{y_{P_\ind{1}},y_{P_{\ind{2}}}\}$ has to belong to $\cutm$ to be visible from the east otherwise a tile of $P_{[\ind{1},{\ind{2}}]}$ would hide $P_{\ind{3}}$. Similarly, a tile $P_{{\ind{3}}}$ such that $\pos(P_{\ind{3}}) \in V(\cutp) \setminus \dom{P_{[\ind{1},\ind{2}]}}$  cannot be visible from the east, a tile of $P_{[{\ind{1}},{\ind{2}}]}$ will always hide it. These two observations are summarized in the two following facts.

\begin{fact}
\label{fact:cut:visun}
Consider a simple path assembly $P$ producible by a tiling system $\tiling=(T,\seed,1)$ and $\ind{1},\ind{2} \in \vset{P}$ such that $\ind{1}<\ind{2}$. Consider the cut $(\cutp,\cutm)$ defined by the underlying path of $P_{[\ind{1},\ind{2}]}$ whose extremities are both visible from the west. If there exists  $\ind{3} \in \vset{P}$ such that $\min\{y_{P_{\ind{1}}},y_{P_{\ind{2}}}\} \leq y_{P_\ind{3}} \leq \max\{y_{P_{\ind{1}}},y_{P_{\ind{3}}}\}$ then $\pos(P_{\ind{3}}) \in \cutp$. 
\end{fact}

\begin{fact}
\label{fact:cut:visdeux}
Consider a simple path assembly $P$ producible by a tiling system $\tiling=(T,\seed,1)$ and $\ind{1},\ind{2} \in \vset{P}$ such that $\ind{1}<\ind{2}$. Consider the cut $(\cutp,\cutm)$ defined by the underlying path of $P_{[\ind{1},\ind{2}]}$ whose extremities are both visible from the west. If there exists $1\leq \ind{3} \leq \lastp$ such that either $\ind{3}<\ind{1}$ or $\ind{3}>\ind{2}$ and $\pos(P_{\ind{3}}) \in \vset{P}$ then $P_{\ind{3}}$ is not visible from the east.
\end{fact}

Now, we adapt the definition of visibility to the segments of a path assembly $P$. Let $1< \ind{1} \leq \ind{2} < \lastp$ such that $P_{[\ind{1},\ind{2}]}$ is a candidate segment of $P$, then $P_{[\ind{1},\ind{2}]}$ is \emph{visible from the west} if and only if both $P_{\ind{1}}$ and $P_{\ind{2}}$ are visible from the west. Moreover the visible candidate segment is \emph{directed to the north} (resp. south) is both $P_{\ind{1}}$ and $P_{\ind{2}}$ are directed to the north (resp. south), the candidate segment \emph{exits to the north} (resp. south) if both $P_{\ind{1}}$ and $P_{\ind{2}}$ exits to the north (resp. south), the candidate segment \emph{exits to the east} if both $P_{\ind{1}}$ and $P_{\ind{2}}$ exits to the east and the candidate segment is not hidden by the seed if both $P_{\ind{1}}$ and $P_{\ind{2}}$ are not hidden by the seed. Note that, we define all these notations for segments visible from the west. Same definitions could be done for the three other directions. Nevertheless, later we will use symmetry and rotation to modify the path assembly in order to consider only visibility from the west. In this section, we will consider mainly the same kind of candidate segment which is defined as follow: a \emph{visible} candidate segment is a candidate segment which is visible from the west, not hidden by the seed, directed to the north and exits either by the east or the north. If we consider another kind of candidate segment, we will give the full list of hypothesis. 

Now, we introduce \emph{exposed} path assembly, we will show that the set of tiles visible from the west of such path assemblies can be easily described (Corollary \ref{lem:existYnorth} and Lemma \ref{lem:order}) and that an exposed path assembly can be extracted from a path assembly without crucial information (Lemma \ref{lem:existexposed}). An exposed path assembly has to satisfy several hypothesis. Thus before defining them, we need some preliminary definitions. Consider a path assembly such that $Y_P>Y_\seed$, the \emph{first highest tile} of $P$ is defined as $P_t$ with $1\leq  t \leq \lastp$ and $t=\min\{1< i \leq \lastp: y_{P_i}=Y_P\}$. Remark that if $Y_P>Y_\seed$, then $1<t$ and by a reasoning similar to the one done for fact \ref{lem:visible:orient}, we have $\pos(P_{t-1})=(x_{P_t},y_{P_t}-1)$, \emph{i.e.} the tile $P_t$ is always directed to the north. Moreover, if $t<\lastp$ then $y_{P_{t+1}}=y_{P_t}$ and either $x_{P_{t+1}}=x_{P_{t}}-1$ (the tile $P^t$ exits to the west) or $x_{P_{t+1}}=x_{P_{t}}+1$, (the tile $P^t$ exits to the east). If $t=\lastp$, we consider that $P^t$ exits to the east. Now, if the highest tile of $P$ is directed to the east then the \emph{tail} of $P$ is \emph{short} if 
%$y_{P_{[t,\lastp]}} \geq  \min\{y_{P_{[0,t]}},y_\seed\}$ 
$y_{P_{[t,\lastp]}} \geq  \min\{y_{P_{[1,t]}},y_\seed\}$ 
otherwise the tail of $P$ is \emph{long}. If the path assembly $P$ satisfies the following hypothesis then $P$ is \emph{exposed} (see Figure \ref{fig:vis:def} for a example of exposed path and see Figure \ref{fig:vis:expo} for an example of a path assembly which is not exposed):
\begin{itemize} 
\item path assembly $P$ is finite and simple;
\item $\dom{P}\cap \dom{\seed}=\pos(P_1)$;
\item $Y_P>Y_{\seed}$;
\item the highest tile of $P$ exits to the east;
\item  the tail of $P$ is short;
\item $P_{\lastp}$ is visible from the east and this visibility is not hidden by the seed. 
\end{itemize}
If $P$ is an exposed path assembly then the set $\vset{P}$ is rather simple to describe and we now prove a sequence of lemmas about this kind of path assembly. The aim of these lemmas can be summarized as follow (see Figure \ref{fig:vis:ori}): there exists an integer $ws$ called the watershed such that all tiles of $P$ which are visible from the west and over this watershed are directed to the north, \emph{i.e.} if $i\in \vset{P}$ and $y_{P_i}\geq ws$ then $P_i$ is directed to the north. Similarly, all tiles of $P$ which are visible from the west and under this watershed are directed to the south, \emph{i.e.} if $i\in \vset{P}$ and $y_{P_i}< ws-1$ then $P_i$ is directed to the south. Moreover the farther a tile is from this watershed the more its index is great, \emph{i.e.} let $\ind{1},\ind{2} \in \vset{P}$, if $ws \leq y_{P_{\ind{1}}}\leq y_{P_{\ind{2}}}$ (resp. $y_{P_{\ind{1}}}\leq y_{P_\ind{2}}\leq ws$) then $\ind{1}\leq \ind{2}$ (resp. $\ind{2}\leq \ind{1}$).

\begin{lemma}
\label{lem:westglue}
Consider an exposed path assembly $P$ producible by a tiling system $\tiling=(T,\seed,1)$ and $1< t \leq \lastp$ such that $P_t$ is the first highest tile of $P$ then $\vsets{P}=\vsets{P_{[1,t]}}$.
\end{lemma}

\begin{proof}
See Figure \ref{fig:westglue} for graphical representation of this proof. By the definition of an exposed path assembly, there exists a path $Q$ such that $Q$ is a subgraph of the binding graph of $P_{[1,t]} \cup \seed$, such that $Q_{|Q|=P_t}$, $y_{Q_1}=\min\{y_P,y_{\seed}\}$ and $Q_1$ is visible from the west in $Q$. Remark that by the definition of an exposed path assembly $y_{\seed}<y_{P_t}$, then $|Q|>1$. The graph $Q$ is a path whose extremities are both visible from the west and then consider the cut $(\cutp,\cutm)$ of the grid by the path $Q$ whose extremities are both visible from the west. Let $W$ be the window of this cut of the grid, let $m=|Q|-1$ and consider an indexing of the window coherent with $Q$ ($W_0=\pos(Q_1)$ and $W_{m}=\pos(Q_{|Q|})$). By definition of an exposed path assembly, the output glue of $P_t$ exits to the east and by the definition of $t$ then $\pos(P_{t+1}) \notin V(\cutp)$. Now, for the sake of contradiction suppose that $\dom{P_{[t+1,\lastp]}} \cap V(\cutp) \neq \emptyset$ then let $\ind{1}=\min\{i>t:\pos(P_i) \in \cutp\}$ then there exists $\ind{2} \in \mathbb{Z}$ such that $W_{\ind{2}}=\pos(P_{\ind{1}})$. Now, if $\ind{2}<0$ remark that the only neighboring position of $W_{\ind{2}}$ which does not belong to $\cutp$ is $(x_{P_{\ind{1}}},y_{P_\ind{1}}-1)$ and then $\pos(P_{\ind{1}-1})=(x_{P_{\ind{1}}},y_{P_\ind{1}}-1)$ which contradicts the fact that the tail of $P$ is short. If $0\leq \ind{2} \leq m$ then $P_{[1,t]}$ and $P_{[t+1,\lastp]}$ intersect which contradicts the fact that $P$ is simple. If $\ind{2}>m$ then $\pos(P_{\ind{1}-1})=(x_{P_{\ind{1}}},y_{P_\ind{1}}+1)$ which contradicts the definition of $t$. Then, we have $\dom{P_{[t+1,\lastp]}} \cap V(\cutp) = \emptyset$. According to fact \ref{fact:cut:visun}, no tile of $P_{[t+1,\lastp]}$ is visible from the west in $P$, \emph{i.e.} $\vsets{P}\subset \vsets{P_{[1,t]}}$. Now consider $i\in \vsets{P_{[1,t]}}$, then for all $k\in \mathbb{N}$, the position $(x_{P_i}-k,y_{P_i})$ belongs to $V(\cutp)$ (otherwise $P_i$ would not be visible). Since $\dom{P_{[t+1,\lastp]}} \cap V(\cutp)=\emptyset$, then $P_i$ is also visible in $P$ and $\vsets{P_{[1,t]}}\subset \vsets{P}$.

\end{proof}

As a corollary of this result the first highest tile of an exposed path assembly is visible from the west and is not hidden by the seed.

\begin{corollary}
\label{for:high:tile}
Consider an exposed path assembly $P$ producible by a tiling system $\tiling=(T,\seed,1)$ and $1< t \leq \lastp$ such that $P_t$ is the first highest tile of $P$ then $t \in \vsets{P}$ and is directed to the north. Moreover $P_t$ is visible from the east in $P_{[1,t]}$ (but maybe not in $P$) and this visibility is not hidden by the seed.
\end{corollary}

\begin{lemma}
\label{lem:existYnorth}
Consider an exposed path assembly $P$ producible by a tiling system $\tiling=(T,\seed,1)$, consider $\ind{1},\ind{2} \in \vset{P}$ such that $y_{P_{\ind{1}}}\leq y_{P_{\ind{2}}}$ (resp.  $y_{P_{\ind{1}}}\geq y_{P_{\ind{2}}}$) and $P_{\ind{1}}$ is directed to the north (resp. south) then $P_{\ind{2}}$ is also directed to the north (resp. south).
\end{lemma}

\begin{proof}
See Figure \ref{fig:proof:lemma2} for a graphical representation of this proof. We suppose that $\ind{1}\leq \ind{2}$, the other case is symmetric. Since $P_{\ind{1}}$ and $P_{\ind{2}}$ are visible from the west then both extremities of the segment $P_{[\ind{1},\ind{2}]}$ are visible from the west. Consider the cut $(\cutp,\cutm)$ of the grid by the underlying path of  $P_{[\ind{1},\ind{2}]}$ whose extremities are both visible from the west and let $W$ be its window. Since $P$ is simple and since $P_{\ind{1}}$ and $P_{\ind{2}}$ is visible from the west then $\dom{P_{[\ind{2}+1,\lastp]}} \cap V(W)=\emptyset$. Assume for the sake of contradiction that $P_{\ind{1}}$ is directed to the north and that $P_{\ind{2}}$ is directed to the south, then either $P_{\ind{2}}$ exits either to the south or to the east. If $P_{\ind{2}}$ exits to the south then $\pos(P_{\ind{2}+1})=(x_{P_\ind{2}},y_{P_{\ind{2}}}-1)$ and either $\pos(P_{\ind{2}-1})=(x_{P_{\ind{2}}}+1,y_{P_{\ind{2}}})$ or $\pos(P_{\ind{2}-1})=(x_{P_{\ind{2}}},y_{P_{\ind{2}}}+1)$, in both cases $\pos(P_{\ind{2}+1}) \in V(\cutp)\setminus V(W)$. If $P_{\ind{2}}$ exits to the east then $\pos(P_{\ind{2}+1})=(x_{P_{\ind{2}}}+1,y_{P_{\ind{2}}})$ and $\pos(P_{\ind{2}-1})=(x_{P_{\ind{2}}},y_{P_{\ind{2}}}+1)$, in both cases $\pos(P_{\ind{2}+1}) \in V(\cutp)\setminus V(W)$. Then in all cases the tile $P_{\ind{2}+1}$ is positioned in $V(\cutp)\setminus V(W)$, \emph{i.e.} $\pos(P_{\ind{2}+1})\in V(\cutp)\setminus V(W)$. Since $P$ is a simple path assembly then $\dom{P_{[\ind{2}+1,\lastp]}} \subset V(\cutp)\setminus V(W)$. In this case $\pos(P_{\lastp}) \in V(\cutp)\setminus V(W)$ and since $P_{\lastp}$ is visible from the east by the definition of an exposed path assembly, then fact \ref{fact:cut:visdeux} leads to a contradiction.

\end{proof}

\begin{corollary}
\label{cor:watershed}
Consider an exposed path assembly $P$ producible by a tiling system $\tiling=(T,\seed,1)$, there exists a value $ws \in \mathbb{Z}$ called the \emph{watershed} of $P$ such that for all $\ind{1} \in \vset{P}$ if $y_{P_{\ind{1}}}\geq ws$ then $P_{\ind{1}}$ is directed to the north and if $y_{P_{\ind{1}}}< ws-1$ then $P_{\ind{1}}$ is directed to the south.
\end{corollary}

Note that we have the condition $y_{P_{\ind{1}}}< ws-1$ because the the tile $P_1$ may be visible from the west and non-oriented.

\begin{lemma}
\label{lem:order}
Consider an exposed path assembly $P$ producible by a tiling system $\tiling=(T,\seed,1)$ and $\ind{1},\ind{2} \in \vset{P}$, if $y_{P_{\ind{1}}}\leq y_{P_{\ind{2}}}$ and $P_{\ind{1}}$ is directed to the north (resp. $P_{\ind{2}}$ is directed to the south) then $\ind{1} \leq \ind{2}$ (resp. $\ind{2} \leq \ind{1}$). 
\end{lemma}

\begin{proof}
See Figure \ref{fig:proof:lemma3} for a graphical representation of the proof. We consider the case where $P_{\ind{1}}$ is directed to the north, the other case is symmetric. Suppose that $y_{P_\ind{2}}\geq y_{P_\ind{1}}$ then by lemma \ref{lem:existYnorth}, $P_{\ind{2}}$ is also directed to the north. Assume for the sake of contradiction that $\ind{2}<\ind{1}$. Since $P_{\ind{1}}$ and $P_{\ind{2}}$ are visible from the west  then both extremities of the segment $P_{[\ind{2},\ind{1}]}$ are visible from the west. Consider the cut $(\cutp,\cutm)$ of the grid by the underlying path of $P_{[\ind{2},\ind{1}]}$ whose extremities are both visible from the west. Since $P$ is simple and since $P_{\ind{1}}$ and $P_{\ind{2}}$ is visible from the west then $\dom{P_{[\ind{1}+1,\lastp]}} \cap V(W)=\emptyset$. Now, either $P_{\ind{1}}$ exits to the north or to the east. If $P_{\ind{1}}$ exits to the north then $\pos(P_{\ind{1}+1})=(x_{P_\ind{1}},y_{P_{\ind{1}}}+1)$ and either $\pos(P_{\ind{1}-1})=(x_{P_\ind{1}}+1,y_{P_{\ind{1}}})$ or $\pos(P_{\ind{1}-1})=(x_{P_\ind{1}},y_{P_{\ind{1}}}-1)$, in both cases $\pos(P_{\ind{1}+1}) \in \cutp$. If $P_{\ind{1}}$ exits to the east then $\pos(P_{\ind{1}+1})=(x_{P_{\ind{1}}}+1,y_{P_{\ind{1}}})$ and $\pos(P_{\ind{1}-1})=(x_{P_{\ind{1}}},y_{P_{\ind{1}}}-1)$, in this case $\pos(P_{\ind{1}+1}) \in \cutp$. Then in both cases the tile $P_{\ind{1}+1}$ is positioned in $\cutp$, \emph{i.e.} $\pos(P_{\ind{1}+1})\in V(\cutp) \setminus V(W)$. Since $P$ is a simple path assembly and $\ind{1},\ind{2} \in \vset{P}$, then $\dom{P_{[\ind{1}+1,\lastp]}} \subset V(\cutp) \setminus V(W)$. In this case $\pos(P_{\lastp}) \in V(\cutp) \setminus V(W)$ and since $P_{\lastp}$ is visible from the east by the definition of an exposed path assembly, then fact \ref{fact:cut:visdeux} leads to a contradiction. 

\end{proof}

\begin{corollary}
\label{cor:visible:order}
Consider an exposed path assembly $P$ producible by a tiling system $\tiling=(T,\seed,1)$ and $1\leq \ind{1} \leq \ind{2} \leq \lastp$ such that $P_{[\ind{1},\ind{2}]}$ is a visible candidate segment of direction $\vu$, then $y_{\vu}\geq 1$.
\end{corollary}

Let $t$ be the first highest tile of $P$ then the first lowest tile of $P$ is defined as the index $s$ such that $s=\min\{i:y_{P_i}=y_{P_{[1,t]}}\}$. Remark that by definition $s\leq t$, we now prove two properties about $P_s$ if there exists at least one tile visible from the west and directed to the south.

\begin{lemma}
\label{lem:svisible}
Consider an exposed path assembly $P$ producible by $\tiling=(T,\seed,1)$, if there exists a tile of $P$ which is visible from the west, oriented to the south and not hidden by the seed then $y_P<y_{\seed}$.
\end{lemma}

\begin{proof}

See Figure \ref{fig:proof:seed:below} for an illustration of this proof. Let $1 \leq \ind{1} \leq \lastp$ such that $P_{\ind{1}}$ is visible from the west, oriented to the south and not hidden by the seed, let $1\leq  t\leq \lastp$ such $P_t$ is the first highest tile of $P$. % and let $1\leq s \leq \lastp$ such that $P_s$ is the first lowest tile of $P$. %Let $b$ the fist lowest tile of $P$. By lemma \ref{lem:lowest}, $P_b$ exits to the east, is visible from the west and $s<t$. 
By lemma \ref{lem:westglue}, $\vsets{P_{[1,t]}}=\vsets{P}$ and since $P_{\ind{1}}$ is visible from the west and not hidden by the seed then 
 %$s$ is visible from the west and $s<t$. Moreover by lemma \ref{lem:existYnorth}, $\ind{1}\leq s$ and then 
 $\ind{1}< t$. Let $Q$ be the underlying path of $P_{[\ind{1},t]}$ and by hypothesis and fact \ref{for:high:tile}, the path $Q$ has its extremity $Q_1$ visible from the west and not hidden by the seed and its extremity $Q_{|Q|}$ visible from the east and not hidden by the seed. Then let $(\cutp,\cutm)$ be the cut of the grid by the path $Q$ with visible extremities and let $W$ be its window. Consider an indexing of $W$ such that $W_1=Q_1$ and $W_{|Q|}=Q_{|Q|}$. We consider that $\cutp$ is the left side of $W$ and then for any position $(x,y) \in V(\cutp) \setminus V(W)$ we have $y>y_{P_{[\ind{1},t]}}\geq y_P$. Now, since $P_{\ind{1}}$ is not hidden by the seed then $\ind{1}>1$. Moreover since $P_{\ind{1}}$ is visible from the west and oriented to the south then $(P_{\ind{1}-1},P_{\ind{1}})$ is an edge on the left side of $W$ and since $P$ is simple then $\pos(P_{\ind{1}-1}) \in V(\cutp)\setminus V(W)$. Now, since $P_{\ind{1}}$ is visible from the west and not hidden by the seed and since $P_{t}$ is visible from the east in $P_{[1,t]}$ and not hidden by the seed then $\dom{\seed}\cap V(W)=\emptyset$ and $\dom{P_{[1,\ind{1}]}}\cap V(W)=\emptyset$. Since $P_{\ind{1}-1} \in V(\cutp)\setminus V(W)$ then $P_{1} \in V(\cutp)\setminus V(W)$ and since $\pos(P_1)\in \dom{\seed}$ then $\dom{\seed} \subset V(\cutp)\setminus V(W)$. Thus, $y_{\seed}>y_P$.
%Then either $\alpha_{\seed}$ is a subgraph of $\cutm$ or a subgraph of $\cutp$. Note that  $P_{[1,\ind{1}]}$
%Now if $y_{\seed} \leq y_P$ then since $\dom{\seed}\cap V(W)=\emptyset$. 
%Remark that $y_{P_b}=y_{\cutp}<y_{\cutp \setminus W}$. Since $j\geq 2$ then $\dom{\seed}\cap V(W)=\emptyset$ and $\dom{P_{[1,j]}}\cap V(W)=\emptyset$. Moreover $P_{j-1}$ is west of $P_j$ and then $(\pos(P_{j-1}),\pos(P_j))\in E(\cutm)\setminus E(W)$ then $\dom{\seed} \subset V(\cutm) \setminus V(W)$ and $y_{P_b}<Y_{\seed}$.
\end{proof}

\begin{lemma}
\label{lem:lowest}
Consider an exposed path assembly $P$ producible by $\tiling=(T,\seed,1)$, consider $s$ the first lowest tile of $P$ then if there exists $1\leq \ind{1} \leq \lastp$ such that tile $P_{\ind{1}}$ is visible from the west, not hidden by the seed and oriented to the south in $P$ then $y_{P_s}=y_P$ and $P_s$ is visible from the west, oriented to the south and exits to east.
\end{lemma}

\begin{proof}

See Figure \ref{fig:proof:seed:last} for an illustration of this proof. Let $t$ be the fist highest tile of $P$. By definition of $s$, we $s\leq t$. By lemma \ref{lem:svisible}, we have $y_P<y_\seed$, and if $y_{P_s}>y_P$ then $P$ has a long tail which contradicts the definition of an exposed path assembly and then $y_{P_s}=y_P$. For the sake of contradiction, suppose that $P_s$ is not visible from the west. Then there exists $1\leq \ind{2}\leq \lastp$ such that $y_{P_{\ind{2}}}=y_{P_s}$, $x_{P_{\ind{2}}}>y_{P_s}$ and $P_{\ind{2}}$ is visible from the west. By definition of $s$, we have $s<\ind{2}$. Moreover, by lemma \ref{lem:existYnorth} since $P_{\ind{1}}$ is oriented to the south then $P_{\ind{2}}$ is also oriented to the south. Then, $\pos(P_{\ind{2}-1})=\pos(P_{\ind{2}})+(0,1)$ and $\pos(P_{\ind{2}+1})=\pos(P_{\ind{2}})-(1,0)$. Now consider the index $\ind{3}$ such that $s \leq \ind{3} < \ind{2}$ and $y_{P_{\ind{3}}}=y_{P_s}$ and $x_{P_{\ind{3}}}=\max\{x<x_{P_{\ind{2}}}: (x,y_{P_s}) \in \dom{P_{[s,\ind{1}]}}\}$. Now consider the path $Q$ from $P_{\ind{3}}$ to $P_{\ind{2}}$ such that $Q_{i+1}=Q_i+(1,0)$. Then let $R$ be the underlying path of $P_{[\ind{3},\ind{2}]}$, let $C$ be the simple cycle defined by $Q\cup R$ and let $\gint$ be its interior. Then $P_{\ind{2}+1} \in V(\gint)$. Moreover, since $P_{\lastp}$ is visible from the west then $P_{\lastp} \notin V(\gint)$. Thus there exists $\ind{4}>\ind{2}$ such $P_{\ind{4}} \in V(C)$ and $P_{\ind{4}} \notin V(\gint)$. Since $P$ is simple $P_{\ind{4}} \notin \dom{P_{[\ind{3},\ind{2}]}}$ and then $y_{P_{\ind{4}}}=y_{P_{s}}$ and $y_{P_{\ind{4}+1}}=y_{P_{\ind{4}}}-(0,1)$ which contradicts the definition of $s$. Then $s$ is visible from the west and since $y_{P_s}=y_P$ then $s$ has to exit to the east.
%In this case $\pos(P_{b+1})=(x_{P_b}+1,y_{P_b})$. Then let $1\leq j \leq \lastp$ such that $y_{P_j}=y_{P_b}$ and $P_j$ is visible from the west. Then by definition of $b$ we have, $j\geq b$ and since $P_b$ exits to the west, we have $j>b$. Moreover by lemma \ref{}, we have $j<t$. By lemma \ref{}, if the lowest tile of $P$ is oriented to north then there exists no tile visible from the west and oriented to the south in $P$ and thus $P_j$ is oriented to the south by hypothesis. Remark that $y_{P_{j+1}}\neq y_{P_j}-1$ otherwise it would contradict the definition of $b$. And thus since $P_j$ is oriented to the north, we have $\pos(P_{j-1})=(x_{P_j},y_{P_j}+1)$ and since $P$ is simple and $P_j$ is visible by the west we have $\pos(P_{j+1})=(x_{P_j}+1,y_{P_j})$.  Now consider $b\leq k \leq j$ such that $k=\max\{b\leq i \leq j: y_{P_i}=y_{P_b}\}$. Since $P_j$ is visible from the west in $P$, we have $x_{P_j}<x_{P_k}$. Let $Q=\alpha_{P_{[k,j]}}$ and $Q'$ the path such that $Q'_1=P_j$, $Q'_|Q|=P_k$ and $Q_{i+1}=Q_i+(0,1)$.  Then $Q\cup Q'$ is a simple simple and let $\gint$ be its interior. Remark that $Q'_2 =\pos(P_{j+1})$ and thus $\pos(P_{j+1}) \in V(\gint)$. Now by the definition of $t$, $\pos{t}\notin \gint$. Thus, we can define $r=\min\{j+1\leq i \leq t: P_r \notin V(\gint)\}$. Note if $P_{r-1} \in V(Q)$ then $P$ is not a simple path assembly which is a contradiction and if $P_r \in Q'_{[1,|Q'|-1]}$ and $P_{r} \notin V(\gint)$ then $y_{P_r}< y_{P_b}$ which contradicts the definition of $b$. Thus $P_b$ is not directed to the west.
\end{proof}

All the previous results have been shown for an exposed path assembly. We end this subsection by giving a method to extract efficiently an exposed path assembly from a path assembly with a long tail. %Also, this method must not destroy useful information about the path assembly. We introduce now the notion of turn back which will  play a crucial role in the subsection dedicated to U-turn. Consider a path $P$ such that $Y_P>Y_\seed$ and let $P_t$ be the first highest tile of $P$ then the turn back of $P$ is defined has $0$ if $P_t$ exits to the west, as $\min\{y_{P_{[1,t]}},y_\seed\}$ if the tail of $P$ is long and otherwise as $\min\{ y_{P_{[t,\lastp]}} \leq i \leq Y_P:$ there exists $t\leq j \leq \lastp$ such that $P_j$ is visible from the east in $P_{[1,j]}$, this visibility is not hidden by the seed and  $y_{P_j}=i\}$ (see Figure \ref{fig:def:turnback}). Remark that the turn back is correctly defined since $Y_P>Y_{\seed}$ then the tile $P_t$ is always visible from the east in $P_{[1,t]}$ and this visibility is not hidden by the seed. The turn back will not be useful in the next subsection but we have to take it into account for future work.
%any path assembly. Also, this method must not destroy useful information about the path assembly. We introduce now the notion of turn back which will  play a crucial role in the subsection dedicated to U-turn. Consider a path $P$ such that $Y_P>Y_\seed$ and let $P_t$ be the first highest tile of $P$ then the turn back of $P$ is defined has $0$ if $P_t$ exits to the west, as $\min\{y_{P_{[1,t]}},y_\seed\}$ if the tail of $P$ is long and otherwise as $\min\{ y_{P_{[t,\lastp]}} \leq i \leq Y_P:$ there exists $t\leq j \leq \lastp$ such that $P_j$ is visible from the east in $P_{[1,j]}$, this visibility is not hidden by the seed and  $y_{P_j}=i\}$ (see Figure \ref{fig:def:turnback}). Remark that the turn back is correctly defined since $Y_P>Y_{\seed}$ then the tile $P_t$ is always visible from the east in $P_{[1,t]}$ and this visibility is not hidden by the seed. The turn back will not be useful in the next subsection but we have to take it into account for future work.%Finally, the last lemma shows that an exposed path assembly could be extracted easily from any path assembly without altering too much of its properties. 

%\begin{figure}[h]
%\centerline{
%\includegraphics[width=6cm]{./photoFinal/visibility/turnback.JPG}
%}
%\caption{Illustration of the definition of turn back.}
% \label{fig:def:turnback}
%\end{figure}

\begin{lemma}
\label{lem:existexposed}
Consider a path assembly $P$ producible by the tiling system $\tiling=(T,\seed,1)$ such that $Y_P>Y_{\seed}$, $\dom{P}\cap \dom{\seed}=P_1$ and $P$ has a long tail. Let $1\leq t \leq \lastp$ such that $P_t$ is the highest tile of $P$ then there exists $t \leq \ind{1} \leq \lastp$ such that  $P_{[1,\ind{1}]}$ is an exposed path assembly and $y_{P_{\ind{1}}}=\min\{y_{P_{[1,\ind{1}]}},y_{\seed}\}$.
\end{lemma}

\begin{proof}
%If $P_t$ is directed to the west than $P$ has a turn back of $0$ and $P_{[1,t]}$ is an exposed path with a turn back of $0$. Now, if $P_t$ is directed to east then either $P$ has a short or long tail. If the tail of $P$ is short let $h$ be the turn back $P$, by definition of the turn back of $P$ there exists $t\leq \ind{1} \leq \lastp$ such that $y_{P_{\ind{1}}}=h$ and $P_{\ind{1}}$ is visible from the east and not hidden by the seed. Then $P_{[1,\ind{1}]}$ satisfies the conditions of our lemma. 
See figure \ref{fig:proof:lemma4} for a graphical representation of this part of the proof. Since the tail of $P$ is long, then let $\ind{3}=\min\{t<i \leq \lastp: y_{p_i}= \min\{y_{P_{[1,t]}},y_\seed\}\}$. If $P_{\ind{3}}$ is visible from the east in $P_{[1,\ind{3}]}$ and not hidden by the seed, then $P_{[1,\ind{3}]}$ satisfies the conditions of our lemma. By contradiction suppose that $P_{\ind{3}}$ is not visible from the east or hidden from the seed. Then let $t\leq \ind{2} <\ind{3}$ such that $P_{\ind{2}}$ is visible from the east in $P_{[1,\ind{3}]}$ and $y_{P_{\ind{2}}}=y_{P_t}$. Now, remark that $P_{\ind{3}}$ is visible from the east in $P_{[\ind{2},\ind{3}]}$. Then, both extremities of $P_{[\ind{2},\ind{3}]}$ are visible from the east. Consider the cut $(\cutp,\cutm)$ of the grid by the underlying path of $P_{[\ind{2},\ind{3}]}$ whose extremities are both visible from the east. Let $W$ be the window of this cut of the grid, let $m=\ind{2}-\ind{1}$ and consider an indexing of the window coherent with $P_{[\ind{2},\ind{3}]}$ ($W_0=\pos(P_\ind{2})$ and $W_m=\pos(P_{\ind{3}})$). Now, remark that either $\ind{2}=t$ and in this case $P_{t-1}$ does not belong to $\cutp$ or $\ind{2}>t$ and in this case $P_{t}$  does not belong to $\cutp$. If $P_{\ind{3}}$ is not visible from the east in $P_{[1,\ind{3}]}$ or hidden by the seed then there exists a path $R$ such that $R$ is a subgraph of the binding graph of $P_{[1,\ind{2}-1]}\cup \seed$, $R_1 \notin \cutp$ and $R_{|R|} \in V(W)$. Then let $\ind{4}=\min\{1\leq i\leq |R|:\pos(P_i) \in \cutp\}$ then there exists $\ind{5} \in \mathbb{Z}$ such that $W_{\ind{5}}=\pos(R_{\ind{4}})$. Now, if $\ind{5}<0$ remark that the only neighboring position of $W_{\ind{5}}$ which does not belong to $\cutp$ is $(x_{W_{\ind{5}}},y_{W_\ind{5}}+1)$ and then $\pos(R_{\ind{4}-1})=(x_{R_{\ind{4}}},y_{R_\ind{4}}-1)$ which contradicts the definition of $t$. If $0\leq \ind{5} \leq m$ then $R$ and $P_{[\ind{2},\ind{3}]}$ intersect which contradicts the fact that $P$ is simple or the fact that $V(P)\cap V(\seed)=\pos(P_1)$. If $\ind{5}>m$ then $\pos(R_{\ind{4}-1})=(x_{R_{\ind{4}}},y_{R_\ind{4}}-1)$ which contradicts the definition of $\ind{3}$. All cases lead to contradictions.
\end{proof}

%\begin{definition}
%\label{def:micro}
%Consider a tiling system $\tiling=(T,\seed,1)$, the path assembly $P$ satisfies microscopic initial conditions if:
%\begin{itemize}
%\item $P$ is finite and simple;
%\item $\dom{P} \cap \dom{\seed}=pos(P_1)$;
%\item $P$ satisfies the first quadrant convention and the axis convention.
%\item the output glue of $P_t$ is directed to the east;
%\item the west face of $P$ is exposed.
%\item the tail of $P$ is short.
%\item the last tile is visible from the west.
%\end{itemize}
%\end{definition}
%
%\begin{lemma}
%\label{lem:micro}
%Consider a path assembly $P$ producible by $\tiling=(T,\seed,1)$ such that TODO then there exists a path assembly $P'$ producible by $\tiling'=(T',\seed',1)$ such that $P'$ satisfies macroscopic initial conditions with the same feedback (TODO) and $P'$ is pumpable or fragile if and only if $P$ is pumpable or fragile.
%\end{lemma}
%
%\begin{proof}
%on fait la meme chose qu'avant
%on inverse éventuellement les arcs x et y
%si Pt orient ouest alors on tronque
%si oriente est on regarde si on passe en dessous dans ce cas, on coupe 
%sinon on coupe au minimum visible
%\end{proof}
%
%\begin{fact}
%\label{lem:mtoM}
%Consider a path assembly $P$ producible by $\tiling=(T,\seed,1)$ satisfying microscopic initial conditions such that TODO then $P$ also satisfies macroscopic initial conditions.
%\end{fact}

\subsection{Fork}

In this section, we introduce \emph{fork} which are useful to build stakes and to prove a powerful lemma about U-turn which will lead to all the final results. We give here the definition of fork and one method relying on cuts of the grid to build them. %One is an algorithm and the other one is based on cuts of the grid. Both lead to the same forks but since this part is the most technical one of the proof, we prefer to give several interpretations for the reader. 

\begin{definition}
\label{def:prestakes}
Let $P$ be a path assembly producible by $\tiling=(T,\seed,1)$. Consider two indices $1\leq \indm < \indp \leq \lastp$, the vector $\vu=\overrightarrow{P_{\indm}P_{\indp}}$ of $\mathbb{Z}^2$ and a path assembly $\pstakem$. The path assembly $\pstakem$ is a fork of $P$ of direction $\vu$ at indices $\indm$ and $\indp$ if and only if:
\begin{itemize}
\item $P_{[1,\indm]}\cdot \pstakem$ and $P_{[1,\indp]}\cdot (\pstakem+\vu)$ are simple path assemblies producible by $\tiling$;
\item for all edges $e$ of the underlying path of $\pstakem$, either there exists an edge $e'$ of the underlying path of $P_{[\indm,\lastp]}$ such that $e=e'$ or there exists an edge $e'$ of the underlying path of $P_{[\indp,\lastp]}$ such that $e=e'-\vu$.
\item for all $1\leq \ind{1} \leq \ind{2} \leq |\pstakem|$, if $\pstakem_{\ind{1}}+\vu \in \dom{P}$ (resp. $\pstakem_{\ind{1}} \in \dom{P}$) and $\pstakem_{\ind{2}}+\vu \in \dom{P}$  (resp. $\pstakem_{\ind{2}} \in \dom{P}$) then there exists $\indp \leq \ind{3} \leq \ind{4} \leq \lastp$ (resp. $\indm \leq \ind{3} \leq \ind{4} \leq \lastp$) such that $\pstakem_{\ind{1}} =P_{\ind{3}}-\vu$ (resp. $\pstakem_{\ind{1}} =P_{\ind{3}}$) and $\pstakem_{\ind{2}} =P_{\ind{4}}-\vu$ (resp. $\pstakem_{\ind{2}} =P_{\ind{4}}$).
\end{itemize}
\end{definition}

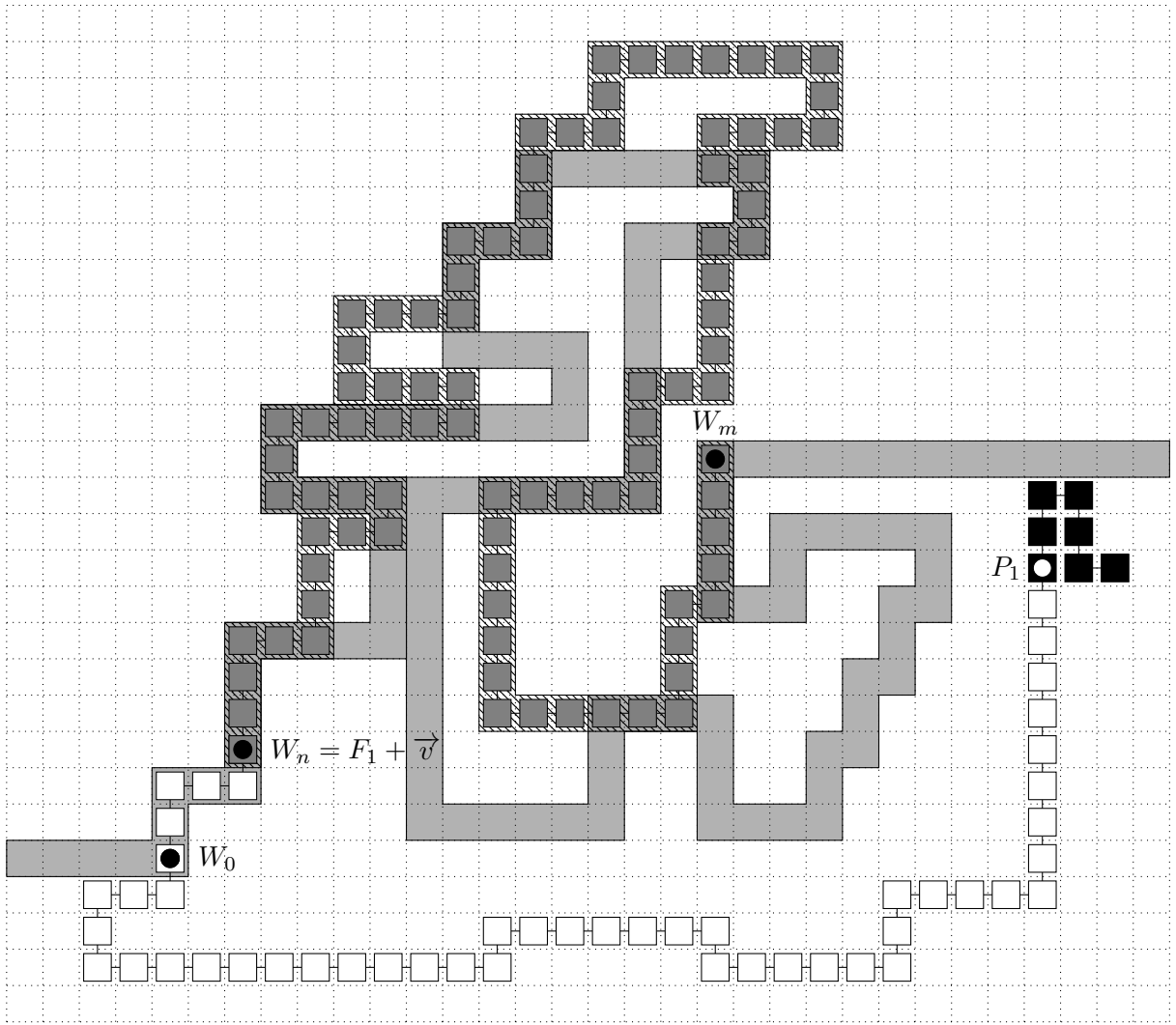
According to the second and third items of this definition, the fork $\pstakem$ is a path assembly made of segments of $P_{[\indm,t]}$ and segments of $P_{[\indp,t]}-\vu$ which also means that $F+\vu$, the translation of the fork by $\vu$, is made of segments of $P_{[\indp,\lastp]}$ and $P_{[\indm,\lastp]}+\vu$. Intuitively, when a fork is found then two path assemblies can be created from $P$. The first one is an hybrid between $P$ and $P-\vu$ and the second one is an hybrid between $P$ and $P+\vu$. The third item of this definition also means that the tiles of $F$ appear in the same order as in $P_{[\indm,t]}$ or as in $P_{[\indp,t]}-\vu$. Also remark that if there exist three indexes such that $1\leq \ind{1} \leq |F|$, $\indm \leq \ind{2} \leq \lastp$, $\indp \leq \ind{3} \leq \lastp$, $\pos(F_\ind{1})=\pos(P_{\ind{2}})$ and $\pos(F_\ind{1})=\pos(P_{\ind{3}})-\vu$ then by the third item of this definition we have $\type(F_\ind{1})=\type(P_{\ind{2}})=\type(P_{\ind{3}})$. If this property is not true then either $P_{[1,\indm]} \cdot F$ is in conflict with $P$ or $P_{[1,\indp]} \cdot (F+\vu)$ is in conflict with $P$. Then if there exists a fork of length at least one, this remark means that $P_{[\indm,\indp]}$ is a candidate segment. See figure \ref{fig:def:fork} for an illustration of the following examples of forks. Consider a candidate segment $P_{[\ind{1},\ind{2}]}$ of direction $\vu$ such that $P'=P_{[1,\ind{1}]}\cdot P_{[\ind{2},\lastp]}$ is a path assembly producible by $\tiling$ then $P_{[\ind{2},\lastp]}-\vu$ is a fork of $P$ of direction $\vu$ at indices $\ind{1}$ and $\ind{2}$. Remark that in this case, the path assembly $P_{[1,\ind{1}]}$ satisfies the conditions for being a stake, half of the work is done. Nevertheless, the method to find the other stake has to wait until the end of the article. Also, remark that $P_{[\ind{1},\ind{1}]}=P_{[\ind{2},\ind{2}]}-\vu$ is always a fork of length $1$. Thus, building a fork is not hard but building long fork is. 

We present now a method to build long fork. This method takes as input an exposed path assembly $P$ and a visible candidate segment $P_{[\indm,\indp]}$ of $P$ (from now now and until the end of the section we will always consider exposed path assembly). The fork is constructed inside a specific area of the plane called the \emph{free zone} $\free$ which is devoid of any obstacle. More precisely, $\free$ is a subgraph of $\grid$ defined as follow (see Figure \ref{fig:def:free}).

\begin{definition} 
\label{def:freezone}
Consider an exposed path assembly $P$ producible by $\tiling=(T,\seed,1)$ and two indices $1\leq \indm \leq \indp \leq \lastp$ such that $P_{[\indm,\indp]}$ is a visible candidate segment. Consider the cut of the grid $(\cutp,\cutm)$ defined by the underlying path of $P_{[\indm,\lastp]}$ whose extremity $\pos(P_{\indm})$ is visible from the west and whose extremity $\pos(P_{\lastp})$ is visible from the east. Let $W$ be the window of this cut and consider an indexing of $W$ coherent with $P_{[\indm,\lastp]}$. Then the \emph{free zone} $\free$ of $P$ associated to $P_{[\indm,\indp]}$ is the graph $\cutp$ (the left side of the window $W$). The window $W$ of the free zone is the window of the cut of the grid $(\cutp,\cutm)$.
\end{definition}

We now fix some notations for the window $W$ of the free area $\free$. Let $m=\lastp-\indm$, $n=\indp-\indm$, we always consider the indexing of $W$ coherent with $P$: the indexing such that $W_0=\pos(P_{\indm})$, $W_n=\pos(P_{\indp})$ and $W_m=\pos(P_{\lastp})$. Remark that the window is made of three parts $W_{[-\infty,0]}$, $W_{[0,m]}$ and $W_{[m,+\infty]}$ where $W_{[0,m]}$ is the underlying path of $P_{[\ind{1},\lastp]}$ and for all $\ind{3}\leq 0$ (resp. $\ind{3} \geq m$), we have $x_{W_{\ind{3}}}<x_{P_{\ind{3}}}$ (resp. $x_{W_{\ind{3}}}>x_{P_{\lastp}}$) and $y_{W_{\ind{3}}}=y_{P_{\ind{1}}}$ (resp. $y_{W_{\ind{3}}}=y_{P_{\lastp}}$). Of course, a path starting in the free zone has to intersect the window in order to leave it.
%\begin{fact}
%\label{fact:freeleft}
%Consider an exposed path assembly $P$ producible by $\tiling=(T,\seed,1)$ and two indices $1\leq \indm \leq \indp \leq \lastp$ such that $P_{[\indm,\indp]}$ is a visible candidate segment. Let $\free$ be the free zone of $P$ associated to $P_{[\indm,\indp]}$ and $W$ its window. Consider a coherent indexing of $W$ then the free zone $\free$ is the left side of $W$.
%\end{fact}
We now show that the free zone of $P$ associated to $P_{[\indm,\indp]}$ is occupied only by tiles of $P_{[\indm,\lastp]}$ (see Figure \ref{fig:def:free}). This remark allow us to build a fork in the free zone without fearing collision with the seed or with $P_{[1,\indm-1]}$ (if a collision occurs between the fork and $P_{[\indm,\lastp]}$ then $P$ is fragile).

\begin{lemma}
\label{lem:freezone}
Consider an exposed path assembly $P$ producible by $\tiling=(T,\seed,1)$ and two indices $1\leq \indm < \indp \leq \lastp$ such that $P_{[\indm,\indp]}$ is a visible candidate segment. Let $\free$ be the free zone of $P$ associated to $P_{[\indm,\indp]}$ then $\dom{\seed} \cap V(\free)=\emptyset$ and $\dom{P_{[1,\indm-1]}} \cap V(\free)=\emptyset$.
\end{lemma}

\begin{proof}
Consider the window $W$ of the free zone $\free$ and a coherent indexing of $W$. By definition $\free$ is the left side of this cut and we also have $W_{-1}=(x_{P_{\indm}}-1,y_{P_{\indm}})$, $W_0=P_{\indm}$ and $W_1=P_{\indm+1}$. Moreover, since $P_{\indm}$ is visible from the west then $W_{-1} \neq \pos(P_{\indm-1})$. Since $P_{\indm}$ is oriented to the north then either $P_{\indm}$ exits to the north or to the east. See Figure \ref{fig:proof:free} for a graphical representation of the following reasoning. If $P_{\indm}$ exits to the north then $\pos(P_{\indm+1})=(x_{P_\indm},y_{P_{\indm}}+1)$ and either $\pos(P_{\indm-1})=(x_{P_\indm}+1,y_{P_\indm})$ or $\pos(P_{\indm-1})=(x_{P_\indm},y_{P_{\indm}}-1)$, in both cases $\pos(P_{\indm-1}) \notin V(\free)$. If $P_{\indm}$ exits to the east then $\pos(P_{\indm+1})=(x_{P_\indm}+1,y_{P_{\indm}})$ and $\pos(P_{\indm-1})=(x_{P_\indm},y_{P_{\indm}}-1)$, in this case $\pos(P_{\indm-1}) \notin V(\free)$. Then in both cases the tile $P_{\indm-1}$ does not belong to the free zone, \emph{i.e.} $\pos(P_{\indm-1}) \notin V(\free)$. Since $P$ is a simple path assembly, $\indm \in \vset{P}$ and $P_{\lastp}$ is visible from the east then $\dom{P_{[1,\indm-1]}} \cap V(W)=\emptyset$. Moreover, since $\pos(P_1) \in \dom{\seed}$, since the seed is connected and since the visibility of $P_{\lastp}$ and $P_{\indm}$ are not hidden by the seed then we also have $\dom{\seed} \cap V(W)=\emptyset$. Since $\pos(P_{\indm-1}) \notin V(\free)$ then $\dom{P_{[1,\indm-1]}} \cap V(\free)=\emptyset$ and $\dom{P_{[1,\indm-1]}} \cap V(\free)=\emptyset$

\end{proof}

Also the free zone possesses two important properties: $y_{P_{\indp}}>y_{P_\indm}$ due to corollary \ref{cor:visible:order} and any position west of $P_{\indp}$ belongs to the free zone.

\begin{lemma}
\label{lem:freezone:adv}
Consider an exposed path assembly $P$ producible by $\tiling=(T,\seed,1)$ and two indices $1\leq \indm < \indp \leq \lastp$ such that $P_{[\indm,\indp]}$ is a visible candidate segment. Let $\free$ be the free zone of $P$ associated to $P_{[\indm,\indp]}$ and $\vu$ be the direction of  $P_{[\indm,\indp]}$ then $y_{\vu} \geq 1$ and for any position $(x,y)\in V(G)$ if $y={y_{P_{\indp}}}$ and $x<x_{P_{\indp}}$ then $(x,y) \in \free$.
\end{lemma}

\begin{proof}
Let $W$ be the window of the free zone and consider a coherent indexing of $W$ and let $m=|P|-\indm$. Since $P_{[\indm,\indp]}$ is a visible candidate segment then by corollary \ref{cor:visible:order}, we have $y_{\indp}>y_{\indm}$. Now consider a position $(x,y) \in V(G)$ such that $y=y_{P_{\indp}}$ and $x<x_{P_{\indp}}$. For the sake of contradiction suppose that $(x,y) \in V(W)$ then there exists $\ind{1}$ such that $W_{\ind{1}}=(x,y)$. If $\ind{1}<0$ then $y_{P_{\indm}}=y_{P_{\indp}}$ and since $P$ is simple and $\indm <\indp$ then tiles $P_{\indm}$ and $P_{\indp}$ are two different tiles and cannot be both visible from the west. If $0 \leq \ind{1} \leq m$ then $\ind{1} \neq \indp$ and by definition of $(x,y)$, $P_{\indp}$ is not visible from the west. Finally if $\ind{1} > m$, then $y_{P_{\indp}}=y=y_{P_{\lastp}}$ and $x_{P_{\lastp}}<x < x_{P_{\indp}}$ which contradicts the fact that $P_{\lastp}$ is visible by the east (by definition of an exposed path assembly). Then either for all $(x,y) \in V(G)$ such that $y=y_{P_{\indp}}$ and $x<x_{P_{\indp}}$, $(x,y) \in V(\free)$ or for all $(x,y) \in V(G)$ such that $y=y_{P_{\indp}}$ and $x<x_{P_{\indp}}$, $(x,y) \notin V(\free)$. Now consider a position $(x,y)$ such that $y=y_{P_{\indp}}>y_{P_{\indm}}$ and $x<X_{P}$ (see Figure \ref{fig:def:free}b), then $(x,y)$ is in the west side of the window $W$ and thus in the free zone.
\end{proof}

As a corollary of this result, the ray starting in $P_{\indp}$ and going west splits the free zone into two zones.  Then, any arc starting in $W_{\ind{1}}$ with $\ind{1} \leq \indp$ and ending in $W_{\ind{2}}$ with $\ind{2} \geq \indp$ has to hide the visibility of $P_{\indp}$ (see Figure \ref{fig:proof:arc}). This remark is summarized in the following corollary.

\begin{corollary}
\label{lem:vispj}
Consider an exposed path assembly $P$ producible by $\tiling=(T,\seed,1)$ and two indices $1\leq \indm < \indp < \lastp$ such that $P_{[\indm,\indp]}$ is a visible candidate segment. Let $\free$ be the free zone of $P$ associated to $P_{[\indm,\indp]}$ and $W$ be its window. Consider a coherent indexing of the window and let $n=\indp-\indm$. Then, for any arc $A$ of $\free$ which starts in $W_{\ind{1}}$ with $\ind{1} \leq n$ and ends in $W_{\ind{2}}$ with $\ind{2} \geq n$, there exists $1\leq \ind{3} \leq |A|$ such that $y_{A_{\ind{3}}}=y_{P_{\indp}}$ and $x_{A_{\ind{3}}}\leq x_{P_{\indp}}$.
\end{corollary}

%\begin{proof}
%Remark that since $\indm<\indp$ and since $P_{\indm}$ and $P_{\indp}$ are both visible from the west then by lemma \ref{lem:order}$y_{P_{\indm}} < y_{P_{\indp}}$. We recall that the free zone is the left side of the window by fact \ref{fact:freeleft}. This proof is done only by local argument around the tile $P_{\indp}$. Since $P_{\indp}$ is visible from the west and directed to the north then three cases may occurs (see figure \ref{}). In all cases, the position $(x_{P_{\indp}}-1,y_{P_{\indp}})$ belongs to the free zone $\free$, then for all $k\in \mathbb{N}$ the position $(x_{P_{\indp}}-k, y_{P_{\indp}})$ belongs to the free zone. Consider the cut of the grid $(\cutp,\cutm)$ by the simple path $\alpha_{P_{[\indm,\indp]}}$ whose extremities are both visible from the west. The previous remark leads to the fact that the graph $\cutp$ is a subgraph of the free zone $\free$ and the left side of this cut of the grid. Moreover, $W_{[-\infty,j]}$ belongs to $\cutp$ while $W_{[j,+\infty]}$ belongs to $V(\cutm)\setminus V(\cutp)$. This remark leads to the lemma.
%\end{proof}

Before stating the method to build our fork, we define some properties that our fork will satisfy. 

\begin{definition}
\label{def:long-fork}
Consider an exposed path assembly $P$ producible by $\tiling=(T,\seed,1)$ and two indices $1\leq \indm \leq \indp \leq \lastp$ such that $P_{[\indm,\indp]}$ is a visible candidate segment. Let $\vu=\overrightarrow{P_{\indm}P_{\indp}}$, $m=\lastp-\indm$, $\free$ be the free zone of $P$ associated to $P_{[\indm,\indp]}$ and $W$ be its window. Consider a coherent indexing of $W$. Then, a fork $\pstakem$ of direction $\vu$ at indices $\indm$ and $\indp$ is \emph{long} if and only if: 
\begin{itemize}
\item the underlying paths of $\pstakem$ and $\pstakem+\vu$ are both subgraphs of the free zone $\free$;
\item there exists $\ind{1} \geq m$ such that either $\pos(\pstakem_{|\pstakem|})=W_{\ind{1}}$ or $\pos(\pstakem_{|\pstakem|})+\vu=W_{\ind{1}}$.
\end{itemize}
\end{definition}

The first item of our definition means that our long fork and its translation by $\vu$ both belong to the free zone. The second item means that our fork or its translation by $\vu$ manages to reach with its last tile a position of $V(W_{[m,+\infty]})$, \emph{i.e} a position which is east of $\pos(P_{\lastp})$. Since a fork starts in $P_{\indm}$ and its translation by $\vu$ starts in $P_{\indp}$, they have to get around $P_{[\indm,\lastp]}$ to reach $W_{[m,+\infty]}$. Note that the fork cannot take a shortcut through $P_{[\indm,\lastp]}$ otherwise its underlying path would not be a subgraph to $\free$. Intuitively, a long fork starts on the west side of $P_{[\indm,\lastp]}$ and reaches its east side. Later, this property will assure us to have a long enough fork to be useful. In fact, if a fork $F$ is long then it is possible to prove that $Y_F\geq Y_P -y_{\vu}$ but since this property is not necessary, we will not prove it.

Now, we explain our method to build a long fork and the aim of the following sequence of lemmas. Let $m=\lastp-\inds$ and $n=\indp-\indm$, we remind that the underling path of $P_{[\indp,\lastp]}$ is $W_{[n,m]}$. Before building a long fork, we will look for the underlying path of this fork. We call it a \emph{trajectory} and will be denoted by $\traj$. Since we are looking for a path made of segments of $W_{[0,m]}$ or $W_{[n,m]}-\vu$ and which is a subgraph a $\free$ then the first intuition is to consider the decomposition of $W_{[n,m]}-\vu$ in extremum arcs in $\free$. Unfortunately this decomposition is not always positive (see Figure \ref{fig:def:build:traj}d). This fact will be problematic to satisfy the third property of a fork (see definition \ref{def:prestakes}). To avoid this problem we consider the decomposition $(A^i)_{1\leq i \leq \ell}$ of $W_{[n,\infty]}-\vu$ in extremum arcs in $\free$ (see Figure \ref{fig:def:build:traj}). We will show that this decomposition is always positive (Lemma \ref{lem:fork:ordered}). To define the trajectory, we proceed in three steps. First we show that there exists an index $\ind{2} \geq n$ such that $W_{[n,\ind{2}]}-\vu$ is the extremum path extracted from $W_{[n,+\infty]}-\vu$ and that the decomposition of $W_{[n,\ind{2}]}-\vu$ in extremum arcs is $(A^i)_{1\leq i \leq \ell}$ (Lemma \ref{lem:firststep} and Figure \ref{fig:def:build:traj2}a). As a first corollary of this result, if we consider the window $W'$ associated to $(A^i)_{1\leq i \leq \ell}$, then there exists an indexing of $W'$ such that for all $i\leq 0$, we have $W'_i=W_i$, moreover $W_{0}=W'_0=W_n-\vu$ (see Figure \ref{fig:def:build:traj2}b). As a second corollary of this result, there exists an index $\ind{1} \geq 0$ such that there exists an index $\ind{3} \geq m$ such that $W'_{\ind{1}}=W_{\ind{3}}$ or $W'_{\ind{1}}=W_{\ind{3}}-\vu$ (see Figure \ref{fig:def:build:traj2}b). Without loss of generality we can suppose that $\ind{1}$ is the first index to satisfy this property. The definition of this index is the second step (see Definition \ref{def:trajectory}) and we can now consider the path $W'_{[0,\ind{1}]}$ (see Figure \ref{fig:def:build:traj2}c). By definition, this path is a subgraph of $\free$ and the third step is to prove that $W'_{[0,\ind{1}]}+\vu$ is a subgraph of $\free$ (see Lemma \ref{lem:secondstep} and  Figure \ref{fig:def:build:traj2}d). Finally, $W'_{[0,\ind{1}]}$ satisfies all the properties to be the underlying path of a long fork and is the trajectory we are looking for. Remark that the trajectory is made 
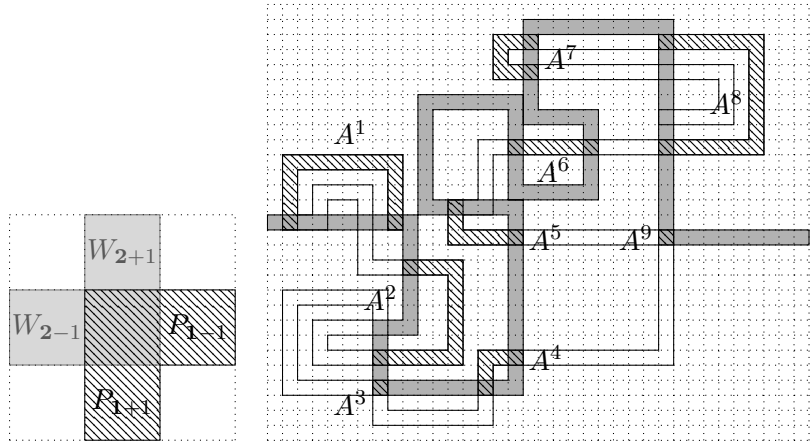
of arcs of the decomposition $(A^i)_{1\leq i \leq \ell}$ and of dual segments of $W$ according to this decomposition. Finally, we try to build a fork using this trajectory. To achieve this aim we use an algorithm illustrated in Figures of appendix \ref{app:buildf}. This algorithm alternates between two phases, one phase tries to tile an arc of the decomposition (see Figures from \ref{fig:def:build:fork5} to \ref{fig:def:build:fork7} and Figures from \ref{fig:def:build:fork11} to \ref{fig:def:build:fork13}) and the other phase tries to tile a dual segment of the decomposition (see Figures \ref{fig:def:build:fork3} and \ref{fig:def:build:fork4} and Figure from \ref{fig:def:build:fork8} to \ref{fig:def:build:fork10}). Eventually, we obtain that either $P$ is fragile or that there exists a long fork. The correctness of this algorithm is proven in Lemma \ref{lem:fork}.

Now, we start by the first step of our method.

\begin{lemma}
\label{lem:firststep}
Consider an exposed path assembly $P$ producible by $\tiling=(T,\seed,1)$ and two indices $1\leq \indm < \indp < \lastp$ such that $P_{[\indm,\indp]}$ is a visible candidate segment. Let $\free$ be the free zone of $P$ associated to $P_{[\indm,\indp]}$, $W$ be its window and consider a coherent indexing of $W$. Let $n=\indp-\indm$ and then the decomposition $(A^i)_{1\leq i \leq \ell}$ of $W_{[n,+\infty]}-\vu$ into extremum arcs in $\free$ is correctly defined. Moreover, let $(\cutp,\cutm)$ be the cut of the grid associated to $W_{[n,+\infty]}-\vu$ and let $\ind{1}\geq n$ and $\ind{2} \geq n$ such that $W_{[\ind{1},\ind{2}]}-\vu$ is the extremum path extracted from $W_{[n,+\infty]}-\vu$. Then $\ind{1}=n$, $W_{[\ind{2},+\infty]}-\vu$ is a subgraph of $\cutm$ and $W_{[n,\ind{2}]}-\vu$ is positive.
%\item $T^e$ is an arc of the weak minimum decomposition of $P+\vu$ according to the free zone $F$.
%\item $T^e_e$ is an arc of the weak minimum decomposition of $P+\vu$ according to the free zone $F$.
\end{lemma}

%\begin{lemma}
%\label{lem:firststep}
%Consider an exposed path assembly $P$ producible by $\tiling=(T,\seed,1)$ and two indices $1\leq \indm < \indp < \lastp$ such that $P_{[\indm,\indp]}$ is a visible candidate segment. Let $\free$ be the free zone of $P$ associated to $P_{[\indm,\indp]}$, $W$ be its window and consider a coherent indexing of $W$. Let $n=\indp-\indm$ and consider a decomposition $(A^i)_{1\leq i \leq \ell}$ of $W_{[n,+\infty]}-\vu$ into extremum arcs in $\free$ and consider the dual decomposition $(D^i)_{0\leq i \leq \ell}$ of $W$ by $(A^i)_{1\leq i \leq \ell}$ then $D^0=W_{[-\infty,0]}$, \emph{i.e.} $A^1_1=W_0$.  
%%\item $T^e$ is an arc of the weak minimum decomposition of $P+\vu$ according to the free zone $F$.
%%\item $T^e_e$ is an arc of the weak minimum decomposition of $P+\vu$ according to the free zone $F$.
%\end{lemma}

\begin{proof}
Firstly, we show that $V(W_{[n,+\infty]}-\vu) \cap V(W)$ is finite. Let $m=\lastp-\indm$ then $V(W_{[n,m]}-\vu) \cap V(W)$ is finite since $W_{[n,m]}$ is finite. Also, $V(W_{[m,+\infty]}-\vu) \cap V(W_{[0,m]})$ is finite since $W_{[0,m]}$ is finite. Moreover since by lemma \ref{lem:freezone:adv}, we have $y_{\vu} \geq 1$ then we have  $V(W_{[m,+\infty]}-\vu) \cap V(W_{[m,+\infty]})=\emptyset$. Finally, $V(W_{[m,+\infty]}-\vu) \cap V(W_{[-\infty,0]})=\emptyset$ otherwise since $W_{m}=P_{\lastp}$, we have $\pos(P_{\lastp}) \in W_{[-\infty,0]}$ and $P_{\indm}$ would not be visible from the west. Then $V(W_{[n,+\infty]}-\vu) \cap V(W)$ is finite and the decomposition of $W_{[n,+\infty]}-\vu$ into extremum arcs in $\free$ is correctly defined. 

Secondly, let $\ind{1}$ and $\ind{2}$ such that $W_{[\ind{1},\ind{2}]}-\vu$ is the extremum path extracted from $W_{[n,+\infty]}-\vu$. Then by definition of $\vu$, we have $P_{\indp}-\vu=P_{\indm}$ and by the definition of a coherent indexing, we have $W_0=P_{\indm}$ then there exists $\ind{3}\leq 0$ such that $W_\ind{3}=W_{\ind{1}}-\vu$. Now, if $\ind{3}<0$ and if $\ind{1} \leq m$ (see Figure \ref{fig:proof:firstsecond}a), then there exists an index $\indp< \ind{4} \leq \lastp$ such that $\pos(P_{\ind{4}}-\vu)=W_{\ind{3}}$ and then $y_{P_{\ind{4}}}-y_{\vu}=y_{P_{\indm}}$ and $x_{P_{\ind{4}}}-x_{\vu}<x_{P_{\indm}}$. Note that, $y_{P_{\ind{4}}}=y_{P_{\indm}}+y_{\vu}=y_{P_{\indp}}$ and $x_{P_{\ind{4}}}<x_{P_{\indm}}+x_{\vu} \leq x_{P_{\indp}}$. Then, this fact contradicts the hypothesis that $P_{\indp}$ is visible from the west. Finally, if $\ind{1}>m$ then there exists $\ind{5}<\ind{1}<0$ such that $\pos(P_{\lastp})=W_{\ind{5}}$ which contradicts the definition of $\ind{1}$. Then $\ind{3}=0$, $\ind{1}=n$ and $W_{[n,\ind{2}]}-\vu$ is positive.

Now, consider the cut $(\cutp,\cutm)$ associated to $W_{[n,+\infty]}-\vu$ and its window $W'$ (see Figure \ref{fig:proof:extremum} for a graphical representation of this part of the proof). We consider that $\cutp$ is the left side of $W'$ and $\cutm$ the right side. Now let $R=(W_{[\ind{2}+1,+\infty]}-\vu)$, since $W_{[n,+\infty]}$ is a simple path then $R$ and $W_{[n,\ind{2}]}-\vu$ does not intersect. Moreover since $W_{[n,\ind{2}]}-\vu$ is the extremum path extracted from $W_{[n,+\infty]}-\vu$ then $V(R) \cap V(W')=\emptyset$. Thus either $V(R) \subset (V(\cutp)\setminus V(W))$ or $V(R) \subset (V(\cutm)\setminus V(W))$. Now, since by lemma \ref{lem:freezone:adv} we have $y_{\vu}\geq 1$, then there exists $\ind{6}$ such $x_{R_{\ind{6}}}>X_{P}$ and $y_{R_{\ind{6}}}<y_{P_{\lastp}}$ and then $R_{\ind{6}} \in V(\cutm)$. Thus $V(R) \subset (V(\cutm)\setminus V(W))$.

\end{proof}

Consider the decomposition $(A^i)_{1\leq i \leq \ell}$ of $W_{[n,+\infty]}-\vu$ into extremum arcs in $\free$ and the dual decomposition $(D^i)_{0\leq i \leq \ell}$ of $W$. Then as a corollary of this result, we have $D^0=W_{[-\infty,0]}$ (see Figure \ref{fig:def:build:traj}c). We now show that this result also implies that the decomposition of $W_{[n,+\infty]}-\vu$ into extremum arcs in $\free$ is positive.  %We start this proof with a quick fact.
%
%\begin{figure}[h]
%\centerline{
%\includegraphics[width=6cm]{./photoFinal/Fork/WrongMethod.JPG}}
%\caption{The decomposition of the translation of $P_{[\indp,\lastp]}$ is not positive: arc $A^7$ is not positive.}
% \label{fig:proof:counter}
%\end{figure}

\begin{lemma}
\label{lem:fork:ordered}
Consider an exposed path assembly $P$ producible by $\tiling=(T,\seed,1)$ and two indices $1\leq \indm < \indp < \lastp$ such that $P_{[\indm,\indp]}$ is a visible candidate segment of direction $\vu$. Let $\free$ be the free zone of $P$ associated to $P_{[\indm,\indp]}$, its window $W$ and consider a coherent indexing of this window. Let $n=\indp-\indm$ and consider the  decomposition $(A^i)_{1\leq i \leq \ell}$ of $W_{[n,+\infty]}-\vu$ into extremum arcs in $\free$, then this decomposition is positive.
\end{lemma}

\begin{proof}
By lemma \ref{fact:pointun}, there exists $\ind{2}\geq n$ such that $Q=W_{[n,\ind{2}]}-\vu$ is the extremum path extracted from $W_{[n,+\infty]}-\vu$. Moreover, the path $Q$ is positive. Consider the cut $(\cutp,\cutm)$ associated to $Q$, its window $W'$ and an indexing of $W'$ coherent with $Q$. We consider that $\cutp$ is the left side of $W'$ and $\cutm$ the right side. By lemma \ref{lem:firststep}, $W_{[\ind{2},+\infty]}-\vu$ is a subgraph of $\cutm$. Since the free zone is the left side of $W$ then by lemma \ref{lem:extact:ext:path}, the decomposition of $Q$ into extremum arcs in $\free$ is the same decomposition as $W_{[n,+\infty]}-\vu$. Finally, since $Q$ is positive then by lemma \ref{lem:conc:prel}, the  decomposition $(A^i)_{1\leq i \leq \ell}$ is positive.
\end{proof}

Lemma \ref{lem:firststep} allows us to define the trajectory $\traj$ which will be the underlying path of the fork we aim to assemble (see Figure \ref{fig:def:build:traj2}c).

\begin{definition}
\label{def:trajectory}
Consider an exposed path assembly $P$ producible by $\tiling=(T,\seed,1)$ and two indices $1\leq \indm < \indp < \lastp$ such that $P_{[\indm,\indp]}$ is a visible candidate segment. Let $\free$ be the free zone of $P$ associated to $P_{[\indm,\indp]}$, $W$ be its window and consider a coherent indexing of $W$. Let $n=\indp-\indm$ and consider a decomposition $(A^i)_{1\leq i \leq \ell}$ of $W_{[n,+\infty]}-\vu$ into extremum arcs in $\free$. Let $W'$ be the window associate to $(A^i)_{1\leq i \leq \ell}$ and consider the indexing of $W'$ such that for all $i\leq 0$, we have $W_i=W'_i$. Let $\ind{1}=\min\{i:\in \mathbb{Z}:W'_i \in V(W_{[m,+\infty]}) \text{ or } W'_i+\vu \in V(W_{[m,+\infty]}) \}$ then the \emph{trajectory} $\traj$ of $P_{[\indp,\lastp]}-\vu$ is $W'_{[0,\ind{1}]}$. 
\end{definition}

The following lemma stipulates that the trajectory $\traj$ is a subgraph to the free zone $\free$ which satisfies most of the conditions requires for being the underlying path of a long fork. %This result is the first step for building our long fork.

\begin{lemma}
\label{fact:pointun}
Consider an exposed path assembly $P$ producible by $\tiling=(T,\seed,1)$ and two indices $1\leq \indm < \indp < \lastp$ such that $P_{[\indm,\indp]}$ is a visible candidate segment of direction $\vu$. Let $\free$ be the free zone of $P$ associated to $P_{[\indm,\indp]}$, $W$ be its window and consider a coherent indexing of this window. Let $\traj$ be the trajectory of $P_{[\indp,\lastp]}-\vu$ then the path $\traj$ is a subgraph of the free zone $\free$ which starts in $W_0$. Moreover, let $m=\lastp-\indm$ and then $\traj_{|\traj|} \in V(W_{[m,+\infty]})$ or $\traj_{|\traj|}+\vu \in V(W_{[m,+\infty]})$ and any edge of $E(\traj)$ belongs to the underlying path of either $P_{[\indm,\lastp]}$ or $P_{[\indp,\lastp]}-\vu$.
\end{lemma}

\begin{proof}
By its definition, the path $\traj$ is a subgraph of the free zone $\free$ such that $\traj_{|\traj|} \in V(W_{[m,+\infty]})$ or $\traj_{|\traj|}+\vu \in V(W_{[m,+\infty]})$ and $E(\traj)\subset (E(W_{[0,+\infty]})\cup E(W_{[n,+\infty]}-\vu))$. Moreover, we also have by definition $V(\traj) \cap V(W_{[m,+\infty]}) \subset \{\traj_{|\traj|}\}$ and $V(\traj)+\vu \cap V(W_{[m,+\infty]}) \subset \{\traj_{|\traj|}+\vu\}$. By lemma \ref{lem:firststep}, $\traj$ starts in $W_0$. Now for the sake of contradiction suppose that there exists $1\leq \ind{1} \leq |\traj|-1$ such that the edge $(\traj_{\ind{1}},\traj_{\ind{1}+1})$ does not belong to the underlying path of $P_{[\indm,\lastp]}$ or to the underlying path of $P_{[\indp,\lastp]}-\vu$ then there exists $\ind{2}\geq m$ such that $(\traj_{\ind{1}},\traj_{\ind{1}+1})=(W_{\ind{2}},W_{\ind{2}+1})$ or $(\traj_{\ind{1}},\traj_{\ind{1}+1})=(W_{\ind{2}},W_{\ind{2}+1})-\vu$ or there exists $\ind{2}> m$ such that $(\traj_{\ind{1}},\traj_{\ind{1}+1})=(W_{\ind{2}},W_{\ind{2}-1})$ or $(\traj_{\ind{1}},\traj_{\ind{1}+1})=(W_{\ind{2}},W_{\ind{2}-1})-\vu$. In both cases, the path $\traj$ should have ended before $\ind{1}$ which is a contradiction.
\end{proof}

Now, we prove that $\traj+\vu$, the trajectory translated by $\vu$, is also a subgraph of the free zone $\free$ (see Figure \ref{fig:def:build:traj2}d). %to achieve this result, we need to explain why during the definition of the trajectory, we consider a decomposition of $W_{[n,+\infty]}-\vu$ into extremum arcs instead one of $P_{[\indp,\lastp]}-\vu$. The reason for this choice is that the extremum decomposition of $W_{[n,+\infty]}-\vu$ into extremum arcs is positive whereas the one of $P_{[\indp,\lastp]}-\vu$ is not, see Figure \ref{fig:proof:counter} for an illustration. %We start this proof with a quick fact.

\begin{lemma}
\label{lem:secondstep}
Consider an exposed path assembly $P$ producible by $\tiling=(T,\seed,1)$ and two indices $1\leq \indm < \indp < \lastp$ such that $P_{[\indm,\indp]}$ is a visible candidate segment. Let $\free$ be the free zone of $P$ associated to $P_{[\indm,\indp]}$, let $W$ be its window and consider a coherent indexing of this window. Let $n=\indp-\indm$ and let $\traj$ be the trajectory of $P_{[\indp,\lastp]}-\vu$ then the path $\traj+\vu$ is a subgraph of the free zone $\free$ which starts in $P_{\indp}$ (\emph{i.e.}, $V(\traj+\vu) \cap V(W_{[-\infty,n-1]})=\emptyset$).
%
%Consider an exposed path assembly $P$ producible by $\tiling=(T,\seed,1)$ and two indices $1\leq i < j < \lastp$ such that $P_{[i,j]}$ is a visible candidate segment. Let $F$ be the free zone of $P$ associated to $P_{[i,j]}$ and $W$ its window. The maximum (resp. minimum) decomposition $(A_i)_{1\leq i \ell}$ of $W-\vu$ (resp. $W+\vu$) according to the free zone $F$ and its window $W$ is correctly defined. Then, let $Q$ be the TODO. 
%For any $1\leq i \ell$ if $T^i_1\in V(Q)$ then the edge $(T^i_1+\vu,T^2_1+\vu)$ belongs to $E(F)\setminus E(W)$.
\end{lemma}

%\begin{proof}
%\end{proof}
%
%\begin{fact}
%the path $Q_{[1,e]}+\vu$ is a subgraph of the free zone $F$.
%\end{fact}

\begin{proof}
Let $m=\lastp-\indm$ and consider the decomposition $(A^i)_{1\leq i \leq \ell}$ of $W_{[n,+\infty]}-\vu$ in extremum arcs in $\free$ and consider its complementary decomposition $(D^i)_{0\leq i \leq \ell}$ of the window $W$. By lemma \ref{lem:fork:ordered}, this decomposition is positive. Consider $u \in E(\traj)$, then by the definition of $\traj$ there exists $1\leq i \leq \ell$ such that either $u\in E(A^i)$ or $u \in E(D^i)$. If $u \in E(A^i)$, note that $A^i+\vu$ is a subgraph of $W_{[n,+\infty]}$, which is a subgraph of the window $W$ and of the free zone $\free$, then $u+\vu \in E(\free)$ and no extremities of $u$ are in $W_{[-\infty,n-1]}$. 

If $u \in E(D^i)$, see Figure \ref{fig:proof:firstsecond}b for a graphical representation of the rest of the proof. Since $P_{\indm}$ and $P_{\indp}$ both exit either to the north or to the east and since by lemma \ref{lem:firststep}, we have $A^1_1=P_{\indm}$ then we have $|A^1|>1$. Moreover by lemma \ref{lem:firststep}, we can applied lemma \ref{cor:dual:firstedge} and then the edge $(D^i_1,D^i_2)$ belongs strictly to the west side of $W_{[n,+\infty]}-\vu$. Then the edge $(D^i_1,D^i_2)+\vu$ belongs strictly to the west side of $W_{[n,+\infty]}$ and thus to $E(\free)$. Now, let $\ind{1}>0$ such that $W_{\ind{1}}=A^{i}_{|A^{i}|}$ and let $\ind{5} \geq n$ such that $W_{\ind{5}}-\vu=W_{\ind{1}}$. Also, let $\ind{4}>\ind{1}$ such that $W_{\ind{4}}=D^{i}_{|D^{i}|}$ and let $\ind{6} \geq n$ such that $W_{\ind{6}}-\vu=W_{\ind{4}}$. Since $(A^i)_{1\leq i \leq \ell}$ is positive then $\ind{6}> \ind{5}$. By lemma \ref{lem:decompo:path}, we have $D^i \cap (W_{[n,+\infty]} -\vu) = \{D^i_1, D^i_{|D^i|}\}= \{W_{\ind{1}},W_{\ind{4}}\}$ and then $D^i+\vu \cap W_{[n,+\infty]} = \{W_{\ind{5}},W_{\ind{6}}\}$. Now either $(D^i+\vu)$ is an arc of $\free$ or $(D^i+\vu) \cap W_{[-\infty,n-1]} \neq \emptyset$. In the first case, then $u \in E(\free)$. In the second case, then there exists $\ind{3}$ such that $\ind{3}= \min\{j \geq \ind{1}: W_j+\vu \in V(W_{[-\infty,n-1]})\}$ and let $\ind{7}< n$ such that $W_{\ind{7}}=W_{\ind{3}}+\vu$. Then $W_{[\ind{1},\ind{3}]}+\vu$ is an arc of $\free$ which starts in $W_{1}+\vu=W_{5}$ with $\ind{5}\geq n$ and ends in $W_{\ind{3}}+\vu=W_{7}$ with $\ind{7}<n$. By corollary \ref{lem:vispj}, there exists an index $\ind{2}$ such that $\ind{1}\leq \ind{2} \leq \ind{3}$, such that $x_{W_{\ind{2}}}+x_{\vu} < x_{P_{\indp}}$  and $y_{W_{\ind{2}}}+y_{\vu} = y_{P_{\indp}}$ then $x_{W_{\ind{2}}} < x_{P_{\indm}}$  and $y_{W_{2}} = y_{P_{\indm}}$ and thus $P_{\indm}$ is not visible from the west which is a contradiction. 

To conclude for all $u\in E(D)$, we have $u+\vu \in E(\free)$, then $D+\vu$ is a subgraph of $D$. Moreover, we have $V(\traj+\vu) \cap V(W_{[-\infty,n-1]})=\emptyset$.

\end{proof}

Now, we can try to create a fork whose binding graph matches the trajectory $\traj$.

\begin{lemma}
\label{lem:fork}
Consider an exposed path assembly $P$ producible by $\tiling=(T,\seed,1)$ and two indices $1\leq \indm < \indp < \lastp$ such that $P_{[\indm,\indp]}$ is a visible candidate segment. Let $\free$ be the free zone of $P$ associated to $P_{[\indm,\indp]}$ and $W$ its window. Let $\traj$ be the trajectory of $P_{[\indp,\lastp]}$. Then either $P$ is fragile or there exists a long fork $F$ of $P$ of direction $\vu$ at indices $\indm$ and $\indp$ such that the underlying path of $F$ is $\traj$.
\end{lemma}

\begin{proof}
We prove this lemma by recurrence. Consider the following hypothesis for all $1\leq i \leq |\traj|$, $H(i):$ "either $P$ is fragile or there exists a fork $F$ such that the underlying path of $F$ is $\traj_{[1,i]}$". First note that $P_{[\indp,\indp]}-\vu$ is always a fork and then $H(1)$ is true. Now, suppose that $H(i)$ is true for $1\leq i \leq |\traj|-1$. Then, by the definition of $\traj$ and by the fact that the decomposition of $W_{[n,+\infty]}$ is positive (Lemma \ref{lem:fork:ordered}), either there exists $\indm \leq \ind{3} \leq \lastp$ such that the edge $(\traj_i,\traj_{i+1})$ is equal to $(P_{\ind{3}},P_{\ind{3}+1})$ or there exists $\indp \leq \ind{3} \leq \lastp$ such that the edge $(\traj_i,\traj_{i+1})$ is equal to $(P_{\ind{3}}-\vu,P_{\ind{3}+1}-\vu)$. See Figures of appendix \ref{app:buildf} for the intuition of the two following cases.

\noindent \textbf{Case  $(\traj_i,\traj_{i+1})=(P_{\ind{3}},P_{\ind{3}+1})$}:
then either $\type(P_{\ind{3}})=\type(F_{|F|})$  or $\type(P_{\ind{3}})\neq \type(F_{|F|})$. In the second case, since $F$ is a fork then $P_{[1,\indm]}\cdot F$ is a path assembly producible by $\tiling$ and $P$ is fragile. Otherwise, let $F'=F \cdot P_{[\ind{3},\ind{3}+1]}$ which is a path assembly whose binding graph is $\traj_{[1,i+1]}$ and by lemmas \ref{lem:freezone} and \ref{fact:pointun}, $P_{[1,\indm]}\cdot F'$ is a path assembly producible by $\tiling$ and by lemmas \ref{lem:freezone} and \ref{lem:secondstep}, $P_{[1,\indp}\cdot (F'+\vu)$ is a path assembly producible by $\tiling$. In this case $H(i+1)$ is true. 
 
\noindent \textbf{Case $(\traj_i,\traj_{i+1})=(P_{\ind{3}}-\vu,P_{\ind{3}+1}-\vu)$}: then either $\type(P_{\ind{3}})=\type(F_{|F|})$ or $\type(P_{\ind{3}})\neq \type(F_{|F|})$. If $\type(P_{\ind{3}})\neq \type(F_{|F|})$, since $F$ is a fork then $P_{[1,\indp]}\cdot (F+\vu)$ is a path assembly producible by $\tiling$ and $P$ is fragile. Otherwise $F'=F \cdot (P_{[\ind{3},\ind{3}+1]}-\vu)$ is a path assembly whose binding graph is $\traj_{[1,i+1]}$ and by lemmas \ref{lem:freezone} and \ref{fact:pointun}, $P_{[1,\indm]}\cdot F'$ is a path assembly producible by $\tiling$ and by lemmas \ref{lem:freezone} and \ref{lem:secondstep}, $P_{[1,\indp]}\cdot (F'+\vu)$ is a path assembly producible by $\tiling$. In this case, $H(i+1)$ is true. 
  
To conclude, Since $H(|\traj|)$ is true, then either $P$ is fragile or there exists a fork $F$ such that the binding graph of $F$ is $\traj$ and in this case $F$ is long. 

%Otherwise, consider $i\min\{1\leq i \leq |T|: T_i =\pos(P_j), T_i+\vu=\pos(P_k-\vu), \type(P_j)\neq P_k\}$. Then, consider the path assembly $F$ whose binding graph is $T_{[1,i]}$ and $\type(F_j)=\type(F_k)$. Notice that by lemma \ref{} and \ref{}, $P_[1,\indm] \cdot F$ is a path assembly producible by $\tiling$ and by lemma \ref{} and \ref{},  $P_[1,\indp] \cdot (F+\vu)$ is a path assembly producible by $\tiling$. Then $F$ is a fork and if $i=|T|$, $F$ is a long fork whose binding graph is $T$. Otherwise,  on casse le chemin. 
\end{proof}

\subsection{U-turn}
\label{sec:Uturn}

We now introduce U-turn, an exposed path assembly makes a U-turn if its last tile is low enough (see figure \ref{fig:def:U-turn}). 

\begin{definition}
\label{def:uturn}
Consider an exposed path assembly $P$ producible by $\tiling=(T,\seed,1)$. We say that the path $P$ makes a U-turn if and only if there exists two indices $1\leq \indm < \indp < \lastp$ such that $P_{[\indm,\indp]}$ is a visible candidate segment of direction $\vu$ and $y_{P_{\lastp}} < y_{P_{[\indm,\lastp]}}+y_{\vu}$. We say that $P$ does a U-turn at indices $\indm$ and $\indp$.
\end{definition}

The goal of this part is to show that an exposed path assembly which makes a U-turn is fragile or pumpable (lemma \ref{lem:uturn}). This result is a powerful tool for proving the three remaining macroscopic lemmas. Nevertheless, the proof of this result is the most technical part of the article and lot of details are crucial here. This result is done is five steps. The first step concern some preliminary results on the trajectory on an exposed path assembly which does a U-turn. The second step is to introduce dominant tiles which satisfy several constraints. The third step is to show that a dominant tile always belongs to a good candidate segment and the fourth step is to show that either this segment is pumpable or there exists another dominant tile with a greater index than the previous one. The fifth and final step is to show that a path which does a U-turn contains at least one dominant tile. Thus the last dominant tile of $P$ always belong to a good pumpable candidate segment and this remark concludes the proof. % First remark, that the fork build by lemma \ref{} always in the window $W$ of the free zone $\free$, case as the on the shown in figure \ref{} cannot occur anymore. This remark is due to the fact that $y_{P_{\lastp}}<Y_P$ and cannot belong to the free zone.

The trajectory and fork introduced in the previous section play a key role for finding dominant tile. When the path assembly $P$ does a U-turn, the trajectory can be described more precisely:  the decomposition of $W_{[n,+\infty]}-\vu$ into extremum arcs in $\free$ is equivalent to the one of $P_{[\indp,\lastp]}-\vu$ and thus the trajectory is a extremum arc of the free zone (see Figure \ref{fig:Uturn:prop} and Lemma \ref{lem:uturn:ordered}). Cases like the one in Figure \ref{fig:def:build:traj2}c cannot occur anymore.

\begin{lemma}
\label{lem:uturn:ordered}
Consider an exposed path assembly $P$ producible by $\tiling=(T,\seed,1)$ and two indices $1\leq \indm < \indp < \lastp$ such that $P$ does a U-turn at indices $\indm$ and $\indp$. Let $\free$ be the free zone of $P$ associated to $P_{[\indm,\indp]}$, $W$ be its window  and consider a coherent indexing of this window. Let $n=\indp-\indm$, then the decomposition of the underlying path of $P_{[\indp,\lastp]}-\vu$ in extremum arcs in $\free$ is identical to the decomposition of $W_{[n,+\infty]}-\vu$ into extremum arcs in $\free$. 
\end{lemma}

\begin{proof}
Let $m=\lastp-\indm$. By lemma \ref{lem:firststep}, there exists $\ind{1}$ such that $W_{[n,\ind{1}]}-\vu$ is the extremum path extracted from $W_{[n,+\infty]}-\vu$. Moreover by lemma \ref{lem:extact:ext:path}, the decomposition of $W_{[n,\ind{1}]}-\vu$ is identical to the one of $W_{[n,+\infty]}-\vu$. By definition of a U-turn, for all $i\geq m$, we have $y_{W_i} -y_{\vu} = y_{P_{\lastp}} -y_{\vu} <y_{P_{[\indm,\lastp]}}$. Then for all $i\geq m$, we have $W_{i} \notin V(\free)$ (see Figure \ref{fig:Uturn:prop}b). Since $W_{\ind{1}}-\vu \in V(W)$ then $\ind{1} \leq m$ and then there exists $\indp \leq \ind{2} \leq \lastp$ such that $W_{\ind{1}}=\pos(P_{\ind{2}})$. Thus the decomposition of the underlying path of $P_{[\indp,\ind{2}]}-\vu$ in extremum arcs in $\free$ is identical to the decomposition of $W_{[n,+\infty]}-\vu$ into extremum arcs in $\free$.

\end{proof}

According to the definition of a trajectory $D$ then either $D_{|D|} \in W_{[m,+\infty]}$ or $D_{|D|}+\vu \in W_{[m,+\infty]}$ but when $P$ does a U-turn then only the first case occurs.

\begin{lemma}
\label{lem:lastInEast}
Consider an exposed path assembly $P$ producible by $\tiling=(T,\seed,1)$ and two indices $1\leq \indm < \indp < \lastp$ such that $P$ does a U-turn at indices $\indm$ and $\indp$. Let $\free$ be the free zone of $P$ associated to $P_{[\indm,\indp]}$, $W$ be its window and consider a coherent indexing of this window. Let $\traj$ be the trajectory of $P_{[\indp,\lastp]}-\vu$ then there exists $\ind{1} \geq m$ such that $\traj_{|\traj|}=W_\ind{1}$. 
\end{lemma}

\begin{proof}
By the definition of a trajectory, there exists $\ind{1} \geq m$ such that either $\traj_{|\traj|}=W_{\ind{1}}$ or $\traj_{|\traj|}+\vu=W_{\ind{1}}$. If $\traj_{|\traj|}=W_{\ind{1}}$ then the lemma is true. For the sake of contradiction suppose that $\traj_{|\traj|}+\vu=W_{\ind{1}}$. By lemma \ref{lem:uturn:ordered}, the decomposition of $W_{[n,+\infty]}-\vu$ in extremum arc in $\free$ is identical to the one of the underlying path of $P_{[\indp,\lastp]}-\vu$. Now, by the definition of a trajectory either there exists $\indp \leq \ind{2} \leq \lastp$ such that $\traj_{|\traj|}=\pos(P_{\ind{2}})-\vu$ or there exists $\indm \leq \ind{3} \leq \lastp$ such that $\traj_{|\traj|}=\pos(P_\ind{3})$. In the first case, then $\pos(P_{\ind{2}})=W_{\ind{1}}$. Moreover, since $\ind{1}\geq m$ then $\ind{1}=m$ and $\ind{2}=\lastp$. Then $\traj_{|\traj|}=P_{\lastp}-\vu$ but by definition of a u-turn, we have $y_{P_{\lastp}}-\vu<y_{P_{[\indm,\lastp]}}$ and then $\pos(P_{\lastp}) \notin V(\free)$ which is a contradiction. In the second case, since $\ind{1} \geq m$ then $y_{P_{\ind{3}}} +y_{\vu} =y_{P_{\lastp}}$ then by definition of a U-turn $y_{P_{\ind{3}}}<y_{P_{[\indm,\lastp]}}$. Since $\indm \leq \ind{3} \leq \lastp$ then it is a contradiction. %If $\pos(P_j)=W_i$, since $\dom{P}\cap V(W_{[m,+\infty]})=\pos(P_{\lastp})=W_m$ then $i=m$ and $j=\lastp$. In this case, $\traj_{|\traj|}=P_{\lastp}-\vu$ and $y_{\traj_{|\traj|}}=y_{\lastp}-y_{\vu}<y_{P_{[\indm,\lastp]}}$. Thus, $y_{\traj_{|\traj|}}< y_W$ and then $\traj_{|\traj|} \notin V(\free)$ which is a contradiction of the definition of a trajectory.
\end{proof}

We remind that by the definition of $\traj$, we have $|V(\traj)\cap W_{[m,+\infty]}|\leq 1$ then lemma \ref{lem:lastInEast} implies that $V(\traj)\cap W_{[m,+\infty]}= \traj_{|\traj|}$ and then $D$ is an extremum arc of the free zone (see Figure \ref{fig:Uturn:prop}b). %Thus we can define $W'$ the window associated to $\traj$. Since $\traj$ is a subgraph of $\free$ then in this special the left side of $W'$ is a subset of $\free$, this zone is called the \emph{narrowed free zone} and is denoted by $\nfree$. We summarize this result in the following corollary.

%we can introduce $W'=W_{[-\infty,0]}\cup \traj \cup W_{[i,+\infty]}$ which is a bi-infinite simple path. Remark that $W'$ is a subgraph of the free zone $\free$. Thus by using $W'$ as a window for a cut of the grid, we can define a zone of the grid $\nfree$ such that $\nfree$ is a subset of $\free$ (see fact \ref{TODO}), this zone is called the narrowed free zone.

\begin{corollary}
\label{def:narrow:free}
Consider an exposed path assembly $P$ producible by $\tiling=(T,\seed,1)$ and two indices $1\leq \indm < \indp < \lastp$ such that $P_{[\indm,\indp]}$ is a candidate segment of direction $\vu$ and such that $P$ does a U-turn at indices $\indm$ and $\indp$. Let $\free$ be the free zone of $P$ associated to $P_{[\indm,\indp]}$ and let $W$ be its window, consider a coherent indexing of $W$. Let $m=\lastp-\indm$ and let $\traj$ be the trajectory of $P_{[\indp,\lastp]}-\vu$ then $\traj$ is an extremum arc of $\free$ which starts in $W_0$ and ends in $W_{\ind{1}}$ with $\ind{1}\geq m$. %Moreover the decomposition is positive.%Let $W'$ be the window associated to $\traj$ then the left side of $W'$ is a subgraph of $\free$ and is called the narrowed free zone (denoted by $\nfree$).
%and let $i\geq m$ such that $\traj_{|\traj|}=W_i$. Then, we say that the bi-infinite simple path $W'=W_{[-\infty,0]}\cup \traj \cup W_{[i,+\infty]}$ can be used as a window for a cut of the grid where its left side $\nfree$, called the \emph{narrowed free zone} is a subgraph of $\free$. The path $W'$ is called the window of the narrowed free zone.
\end{corollary}

This corollary concludes the preliminary results about the trajectory of a path doing a U-turn. We now aim to introduce dominant tiles. Such tiles will help us to locate good pumpable segments. Moreover we would like that the extensions of these pumpable segments stay inside the free zone $\free$ in order to use lemma \ref{lem:freezone} to avoid collision with the seed or with $P_{[1,\indm]}$.
%To achieve our result, we are looking for a candidate segment of $P$ whose pumping stays inside $\free$ in order to use lemma \ref{}.  
The binding graph of such an extension would split the configuration in two zones. Thus we have a particular interest in tiles where the free zone $\free$ could be split by the extension a good path of direction $\vu$ (see Figure \ref{fig:def:split}). More formally we are looking for an index $\indm \leq \domi \leq \lastp$ and a good path $C$ of direction $\vu$ such that $C_1=\pos(P_{\domi})$ and $\omeplus{C}$ splits $\free$ into two zones. The fact that the trajectory is an extremum arc (corollary \ref{def:narrow:free}) has several consequences on $P_{\domi}$ which are stated in the following lemma.% In such a case, we now prove that no arc of $P_{[\indm,\lastp]}-\vu$ could start before $P_{\domi}$ and end after $\domi$. This result will implies that $\pos(P_{\domi}) \in \traj$.

\begin{lemma}
\label{fact:arc:split}
Consider an exposed path assembly $P$ producible by $\tiling=(T,\seed,1)$ and two indices $1\leq \indm < \indp < \lastp$ such that $P_{[\indm,\indp]}$ is a candidate segment of direction $\vu$ and such that $P$ does a U-turn at indices $\indm$ and $\indp$. Let $\free$ be the free zone of $P$ associated to $P_{[\indm,\indp]}$,  $W$ be its window and consider a coherent indexing of this window. Consider $\indm \leq \domi \leq \lastp$ and a good path $C$ of direction $\vu$ such that $C_1=\pos(P_{\domi})$ and $\omeplus{C}$ splits the free zone $\free$ into $(\free^-,\free^+)$. Let $\traj$ be the trajectory of $P_{[\indp,\lastp]}-\vu$, then:
\begin{itemize}
\item there exists $1\leq \ind{2} \leq |\traj|$ such that $\traj_{\ind{2}}=\pos(P_{\domi})$;
\item moreover, $V(\traj) \cap V(\omeplus{C})=\{\traj_{\ind{2}}\}$;
\item $\traj_{[1,\ind{2}]}$ is a subgraph of $\free^-$ and $\traj_{[\ind{2},\lastp]}$ is a subgraph of $\free^+$.
%\item there exists no arc $A$ of $\free$ such that $A$ starts in $\traj_{\ind{1}}$ and ends in $\traj_{\ind{3}}$ with $\ind{1}\leq \ind{2} \leq \ind{3}$; 
\end{itemize}
% there exists no arc $A$ of $\free$ such that $A$ starts in $W_{\ind{1}}$ and ends in $W_{\ind{2}}$ with $\ind{1}\leq \domi-\indm \leq \ind{2}$ and $A$ is a subgraph of $P_{[\indp,+\infty]}-\vu$. Moreover $\pos(P_{\domi}) \in V(\traj)$.
%Consider $(A^i)_{1\leq i \leq \ell}$ be a maximum decomposition of $P_{[\indp,\lastp]}$ according to window $W$ and the free zone $\free$, then there exists no $1\leq j \leq \ell$ and $\indm \leq \da  < \domi \leq \db\leq \lastp$ such that $A^j_1=P_\da$ and $A^j_{|A^j|}=P_\db$.
\end{lemma}

\begin{proof}
See Figure \ref{fig:proof:split} for an illustration of this proof. Let $m=\lastp-\indm$. By corollary \ref{def:narrow:free}, $\traj$ is an extremum arc of the free zone which starts in $W_0$ and ends in $W_{\ind{4}}$ with $\ind{4}\geq m$. Moreover, since $\indm \leq d \leq \lastp$ then $W_0 \in V(\free^-)$ and $W_{\ind{4}} \in V(\free^+)$. Thus, the arc $\traj$ intersects with $\omeplus{C}$. Let $1\leq \ind{2} \leq |\traj|$ such that $\traj_{\ind{2}} \in V(\omeplus{C})$. Then, there exists $\indp \leq \ind{5} \leq \lastp$ such that either $P_{\ind{5}}=\traj_{\ind{2}}+\vu$ or $P_{\ind{5}}=\traj_{\ind{2}}$. We remind that by definition of a split, we have $V(P_{[\indm,\lastp]}) \cap V(\omeplus{C})=\{\pos(P_{\domi})\}$. Then, in the first case, $P_{\ind{5}}\in \omepluse{C}$ which is a contradiction and in the second case, we have $\ind{5}=\domi$ and then $\traj_{\ind{2}}=\pos(P_{\domi})$. Thus, we have $V(\traj) \cap V(\omeplus{C})=\{\pos(P_{\domi})\}$. 
Finally by definition of a trajectory, $\traj_{[1,\ind{2}]}$ (resp. $\traj_{[\ind{2},|\traj|]}$) is a subgraph of $\free$ and since $\indm \leq d$ (resp. $d \leq \lastp$ and $\ind{4}\geq m$) then $\traj_1 \in V(\free^-)$ (resp. $\traj_{|\traj|} \in V(\free^+)$). Since $V(\traj_{[1,\ind{2}]}) \cap V(\omeplus{C})=\{\traj_{\ind{2}}\}$ (resp. $V(\traj_{[\ind{2},\lastp]}) \cap V(\omeplus{C})=\{\traj_{\ind{2}}\}$) then $\traj_{[1,\ind{2}]}$ (resp.  $\traj_{[\ind{2},|\traj|]}$) is a subgraph of $\free^-$ (resp $\free^+$).

\end{proof}

Thus splitting $W$ at position $\pos(P_\domi)$ give us some nice properties. Nevertheless, splitting the free zone in two is not sufficient to define dominant tiles. %Consider the decomposition $(A^i)_{1\leq i \leq \ell}$ of $P_{[\indp,\lastp]}-\vu$ in extremum arcs and the decomposition $(S^i)_{1\leq i \leq \ell}$ (resp. dual decomposition $(D^i)_{0\leq i \leq \ell}$) of $W$ according to this decomposition. If $P_{\domi} \in \traf$ then there exists $1\leq i \leq \ell-1$ such that $\pos(P_{\domi}) \in V(D^i)$. 
Indeed, we have to deal with two other problems. The first one is that the hypothesis $P_{\domi}-\vu \notin V(\free^-)$ is required to prove that $P_{\domi}$ belongs to a pumpable segment. Nevertheless a second problem occurs with only this condition: the pumping of this candidate segment could collide with $P_{[\indp,\domi]}$ which is troublesome for our proof. The good definition of a dominant tile is the following (see Figure \ref{fig:def:dom}).

\begin{definition}
\label{def:dominant}
Consider an exposed path assembly $P$ producible by $\tiling=(T,\seed,1)$ and two indices $1\leq \indm < \indp < \lastp$ such that $P_{[\indm,\indp]}$ is a candidate segment of direction $\vu$ and such that $P$ does a U-turn at indices $\indm$ and $\indp$. Let $\free$ be the free zone of $P$ associated to $P_{[\indm,\indp]}$,  $W$ be its window and consider a coherent indexing of this window. Consider $\indp \leq \domi \leq \lastp$ and a good path $C$ of direction $\vu$, the tile $P_\domi$ is \emph{dominant} due to path $C$ if and only if $C_1=\pos(P_\domi)$, $\omeplus{C}$ splits $\free$ into $(\free^-,\free^+)$ and $C-\vu$ is a subgraph of $\free^-$. 
\end{definition}

Remark that no tile of $P_{[\indm,\indp]}$ could be dominant. We start by a preliminary result: a dominant tile always belongs to a dual segment of $W$ according the decomposition of $P_{[\indp,\lastp]}-\vu$ in extremum arcs whereas the translation of a dominant tile by $-\vu$ belongs to an arc of the decomposition. 

\begin{lemma}
\label{lem:existsStraddle:prel}
Consider an exposed path assembly $P$ producible by $\tiling=(T,\seed,1)$ and two indices $1\leq \indm < \indp \leq \lastp$ such that $P$ does a U-turn at indices $\indm$ and $\indp$. Let $\vu=\overrightarrow{P_{\indm}{P_{\indp}}}$, let $n=\indp-\indm$, let $\free$ be the free zone of $P$ associated to $P_{[\indm,\indp]}$. Consider $\indp\leq \domi \leq \lastp$ such that $P_{\domi}$ is a dominant tile  and consider the decomposition $(A^i)_{1\leq i \leq \ell}$ of of $W_{[n,+\infty]}-\vu$ in $\free$ then there exists $1\leq j \leq k \leq \ell$ such that $\pos(P_{\domi}) \in V(D^k)\setminus \{D^k_1,D^k_{|D^k|}\}$ and $\pos(P_{\domi})-\vu \in V(A^j)$.
\end{lemma}

\begin{proof}
Let $C$ such that $P_{\domi}$ is dominant due to path $C$, let $W$ be the window of the free zone,  and consider a coherent indexing of $W$.  By lemma \ref{lem:uturn:ordered}, the sequence of arcs $(A^i)_{1\leq i \leq \ell}$ is equivalent to the decomposition of $P_{[\indp,\lastp]}-\vu$ into extremum arcs in $\free$. Let $\traj$ be the trajectory of $P_{[\indp,\lastp]}-\vu$ and %The proof is divided in three parts. Firstly, we show how to find the two indices $\ind{1}$ and $\ind{2}$ such that $F_{[\ind{1},\ind{2}]}$ is a candidate segment of $F$. Secondly, we show that the translation of this segment by $\vu$ is a subgraph of $\free^+$ and thirdly, we conclude by showing that this segment is a good candidate segment.
by lemma \ref{fact:arc:split}, there exists $1\leq \ind{2} \leq |\traj|$ such that $\traj_{\ind{2}}=\pos(P_{\domi})$. %By corollary \ref{lem:fork:ordered}, this decomposition is positive. 
Then if there exists $1\leq k \leq \ell$ such that $\pos(P_{\domi}) \in V(A^k)$ then there exists $1\leq \ind{1} \leq \lastp$ such that $P_{\ind{1}}-\vu=P_{\domi}$ and thus $\pos(P_{\ind{1}})=C_{|C|}$ which contradicts the definition of a split. Then there exists $1\leq k \leq \ell$ such that $\pos(P_{\domi}) \in V(D^k)\setminus \{D^k_1,D^k_{|D^k|}\}$. Now, by definition of a dominant tile, $C-\vu$ is a subgraph of $\free^-$ and then $\pos(P_{\domi})-\vu \in V(\free^-)$. By lemma \ref{lem:decompo:path}, there exists $1\leq j \leq \ell$ such that $\gint$ is the interior of arc $A^j$ and $\pos(P_{\domi})-\vu \in V(\gint)$ (see Figure \ref{fig:lem:exists}). Thus, either $\pos(P_{\domi})-\vu \in V(A^j)$ or $\pos(P_{\domi})-\vu \in V(\gint)\setminus V(A^j)$. In the second case (see Figure \ref{fig:lem:exists}a), since $C-\vu$ is a subgraph of $\free$ then there exists $2\leq \ind{3} \leq |C|$ such that $C_{\ind{3}}-\vu \in A^j$ then $C_{\ind{3}} \in \dom{P}$ which contradicts the definition of a split. Then $\pos(P_{\domi})-\vu \in V(A^j)$. Now, remark that if $A^j$ is not a subgraph of $\free^-$ then $A^j$ would intersect with $\omeplus{C}$ and in this case, $P$ intersects with $\omepluse{C}$ which is a contradiction. Thus, $A^j$ is a subgraph of $\free^-$ and then $j\leq k$ (see Figure \ref{fig:lem:exists}b).% Moreover, since by lemma \ref{fact:arc:split}, $D_{[\ind{2},|D|]}$ is a subgraph of $\free^+$ then $\ind{1}\leq \ind{2}$. 

\end{proof}

Now, we show that a dominant tile belongs to a good pumpable segment of the fork build by lemma \ref{lem:fork}. In fact, we are interested only in lemma \ref{lem:existsStraddle} but we decompose the proof in a sequence of three lemmas (see Figure \ref{fig:find:cand:fork} for an illustration of these lemmas).

\begin{lemma}
\label{lem:existsStraddle:part1}
Consider an exposed path assembly $P$ producible by $\tiling=(T,\seed,1)$ and two indices $1\leq \indm < \indp < \lastp$ such that $P$ does a U-turn at indices $\indm$ and $\indp$. Let $\vu=\overrightarrow{P_{\indm}{P_{\indp}}}$, let $\free$ be the free zone of $P$ associated to $P_{[\indm,\indp]}$. Consider $\indp\leq \domi \leq \lastp$ and a good path $C$ such that $P_{\domi}$ is a dominant tile due to path $C$ whose extension $\omeplus{C}$ splits $\free$ into $(\free^-,\free^+)$. Let $\traj$ be the trajectory of $P_{[\indp,\lastp]}-\vu$. Then either $P$ is fragile or there exists a fork $F$ and $1\leq \ind{1} \leq \ind{2} \leq |F|$ such that: 
\begin{itemize}
\item the underlying path of $F$ is $\traj$; 
\item $F_{\ind{2}}=P_{\domi}$;
\item $F_{[\ind{1},\ind{2}]}$ is a candidate segment of $F$ of direction $\vu$.
\end{itemize}
%\begin{itemize}
%\item $\traj_{\ind{1}} =P_{\domi}-\vu$ and $\traj_{\ind{2}} =P_{\domi}$;
%\item $\traj_{[\ind{1},\ind{2}]}$ is a subgraph of the binding graph of $P$ or there exists $d<\ind{3} \leq \lastp$ such that $\pos(P_{\ind{3}}) \in V(\traj_{[\ind{1},\ind{2}]})$;
%\item $\traj_{[\ind{1},\ind{2}]}+\vu$ is a subgraph of $\free^+$.
%\end{itemize}
%$P_{[\ind{1},\ind{2}]}$ is a good candidate segment and $P_{[\ind{1},\domi]}+\vu$ is a subgraph of $\free^+$.%straddles at position $P_{\domi}$.
%
%Let $\traj$ be the trajectory of $P_{[\indm,\indp]}-\vu$. Let $1\leq \db \leq |\traj|$ such that $\traj_{\db}=P_{\domi}$. Then there exists $1\leq \da<\db$ and $1\leq k<k' \leq \ell$ such that $\traj_{[\da,\db]}$ is a straddling path at position $\domi$ and $P_{\domi} \in T^{k'}$ and $\traj_{\da} \in A^k$.
\end{lemma}

\begin{proof}
Let $W$ be the window of the free zone and consider a coherent indexing of $W$. Let $n=\indp-\indm$ and consider the decomposition $(A^i)_{1\leq i \leq \ell}$ of $P_{[\indp,\lastp]}-\vu$ in extremum arcs in $\free$. By lemma \ref{lem:uturn:ordered}, this decomposition is equivalent to the one of $W_{[n,+\infty]}-\vu$. %The proof is divided in three parts. Firstly, we show how to find the two indices $\ind{1}$ and $\ind{2}$ such that $F_{[\ind{1},\ind{2}]}$ is a candidate segment of $F$. Secondly, we show that the translation of this segment by $\vu$ is a subgraph of $\free^+$ and thirdly, we conclude by showing that this segment is a good candidate segment.
By lemma \ref{lem:existsStraddle:prel}, there exist $1\leq j \leq k \leq \ell$ such that $\pos(P_{\domi}) \in V(D^k)$ and $\pos(P_{\domi})-\vu \in V(A^j)$. Then there exists $1\leq \ind{1} \leq \ind{2} \leq |\traj|$ such that $\traj_{\ind{1}}=\pos(P_{\domi})-\vu$ and $\traj_{\ind{2}}=\pos(P_{\domi})$. By lemma \ref{lem:fork}, either $P$ is fragile or pumpable or there exists a fork $F$ such that the binding graph of $F$ is $\traj$. Now, since $P_{[1,\indm]} \cdot F$ is producible by $\tiling$ then either $P$ is fragile or $F_{\ind{2}}=P_{\domi}$. %By corollary \ref{lem:fork:ordered}, this decomposition is positive. 
%Now, by definition of a dominant tile, $C-\vu$ is a subgraph of $\free^-$ and then $\pos(P_{\domi})-\vu \in V(\free^-)$. By lemma \ref{lem:decompo:path}, there exists $1\leq j \leq \ell$ such that $\gint$ is the interior of arc $A^j$ and $\pos(P_{\domi})-\vu \in V(\gint)$ (see Figure \ref{fig:lem:exists}a). Thus, either $\pos(P_{\domi})-\vu \in V(A^j)$ or $\pos(P_{\domi})-\vu \in V(\gint)\setminus V(A^j)$. In the second case, since $C-\vu$ is a subgraph of $\free$ then there exists $2\leq \ind{3} \leq |C|$ such that $C_{\ind{3}}-\vu \in A^j$ then $C_{\ind{3}} \in \dom{P}$ which contradicts the definition of a split. Then $\pos(P_{\domi})-\vu \in V(A^j)$. Now, remark that $A^j$ is a subgraph of $\free^-$ otherwise $A^j$ would intersect with $\omeplus{C}$ and in this case, $P$ intersects with $\omepluse{C}$ which is a contradiction. Thus, $A^j$ is a subgraph of $\traj$ and there exists $1 \leq \ind{1} \leq |\traj|$ such that $\traj_{\ind{1}}=\pos(P_{\domi})-\vu$. Moreover, since by lemma \ref{fact:arc:split}, $D_{[\ind{2},|D|]}$ is a subgraph of $\free^+$ then $\ind{1}\leq \ind{2}$. 
%Consider the index $\ind{4}$ such that $A^i$ ends in $W_{\ind{4}}$ (see Figure \ref{fig:lem:exists}b). By lemma \ref{fact:pointun}, we have $\ind{4}\geq 0$. Since $\pos(P_{\domi})-\vu \in \free^-$ by lemma \ref{fact:arc:split}, there exists $\ind{5} \leq \domi$ such that $W_{\ind{4}}=P_{\ind{5}}$. Since $(A^i)_{1\leq i \leq \ell}$ is a positive decomposition, since $D_\ind{1} \in V(A^i)$ and since $\ind{5} \leq \domi$ then $\ind{1} \leq \ind{2}$. 
Also, since by definition of a fork, the path assembly $P_{[1,\indp]}\cdot (F+\vu)$ is producible by $\tiling$ then either $P$ is fragile or $F_{\ind{1}}=P_{\domi}-\vu$ then $F_{[\ind{1},\ind{2}]}$ is a candidate segment of $F$ of direction $\vu$.

\end{proof}

\begin{lemma}
\label{lem:existsStraddle:part2}
Consider an exposed path assembly $P$ producible by $\tiling=(T,\seed,1)$ and two indices $1\leq \indm < \indp < \lastp$ such that $P$ does a U-turn at indices $\indm$ and $\indp$. Let $\vu=\overrightarrow{P_{\indm}{P_{\indp}}}$, let $\free$ be the free zone of $P$ associated to $P_{[\indm,\indp]}$. Consider $\indp\leq \domi \leq \lastp$ and a good path $C$ such that $P_{\domi}$ is a dominant tile due to path $C$ whose extension $\omeplus{C}$ splits $\free$ into $(\free^-,\free^+)$. Let $\traj$ be the trajectory of $P_{[\indp,\lastp]}-\vu$. Then either $P$ is fragile or there exists a fork $F$ and $1\leq \ind{1} \leq \ind{2} \leq |F|$ such that: 
\begin{itemize}
\item the underlying path of $F$ is $\traj$; 
\item $F_{\ind{2}}=P_{\domi}$;
\item $F_{[\ind{1},\ind{2}]}$ is a candidate segment of $F$ of direction $\vu$;
\item the underlying path of $F_{[\ind{1},\ind{2}]}+\vu$ is a subgraph of $\free^+$.
\end{itemize}
%\begin{itemize}
%\item $\traj_{\ind{1}} =P_{\domi}-\vu$ and $\traj_{\ind{2}} =P_{\domi}$;
%\item $\traj_{[\ind{1},\ind{2}]}$ is a subgraph of the binding graph of $P$ or there exists $d<\ind{3} \leq \lastp$ such that $\pos(P_{\ind{3}}) \in V(\traj_{[\ind{1},\ind{2}]})$;
%\item $\traj_{[\ind{1},\ind{2}]}+\vu$ is a subgraph of $\free^+$.
%\end{itemize}
%$P_{[\ind{1},\ind{2}]}$ is a good candidate segment and $P_{[\ind{1},\domi]}+\vu$ is a subgraph of $\free^+$.%straddles at position $P_{\domi}$.
%
%Let $\traj$ be the trajectory of $P_{[\indm,\indp]}-\vu$. Let $1\leq \db \leq |\traj|$ such that $\traj_{\db}=P_{\domi}$. Then there exists $1\leq \da<\db$ and $1\leq k<k' \leq \ell$ such that $\traj_{[\da,\db]}$ is a straddling path at position $\domi$ and $P_{\domi} \in T^{k'}$ and $\traj_{\da} \in A^k$.
\end{lemma}

\begin{proof}
%Consider the cut $(G^1,G^2)$ associated to the extremum path $\traj$ of $W$ and let $W'$ be its window and consider a coherent indexing of $W'$. Since $\traj$ is a subgraph of $\free$, then $G^1$, the left side of $W'$, is a subgraph of $\free$. If $P_{\domi}-\vu \in V(W')$ then there exists $1 \leq \ind{1} \leq |\traj|$ such that $\traj_{\ind{1}}=P_{\domi}-\vu$. Moreover since $P_{\domi}-\vu \notin \free^+$ then by lemma \ref{fact:arc:split}, $1\leq \ind{1}\leq \ind{2} \leq |T|$. Now, if $P_{\domi}-\vu \in V(G^1)\setminus V(W')$ then there exists an arc $A$ of $W'$ which is a subgraph of the binding graph of $P_{[\indm,\laspt]}-\vu$ and contains $P_{\domi}-\vu$ this fact contradicts the definition of a trajectory. Finally, if $P_{\domi}-\vu \notin V(G^1)$ then since $C_|C|-\vu=P_{\domi}$, then  $C_{|C|}-\vu \in V(W')$ then we can define $\ind{3}=\max\{1\leq i \leq |C|: (C_i,C_{i+1} \notin E(G^1)\}$. Since this edge belongs to $E(\free)$ then there exists $\ind{4}$ such that $P_{\ind{4}}\in V(W')$ and $P_{\ind{4}}-\vu=C_{\ind{3}+1}$ then $\pos(P_{\ind{4}})=C_{\ind{3}+1}$ which contradicts the definition of the split. Now, by a definition of a fork $F_{\ind{1}}=F_{\ind{2}}$ and then $F_{[\ind{1},\ind{2}]}$ is a candidate segment of $F$. 

Let $W$ be the window of the free zone and consider a coherent indexing of $W$. Let $n=\indp-\indm$ and consider the decomposition $(A^i)_{1\leq i \leq \ell}$ of $P_{[\indp,\lastp]}-\vu$ in extremum arcs in $\free$. By lemma \ref{lem:uturn:ordered}, this decomposition is equivalent to the one of $W_{[n,+\infty]}-\vu$ and by lemma \ref{lem:fork:ordered}, this decomposition is positive. By lemma \ref{lem:existsStraddle:part1}, either $P$ is fragile or there exists a fork $F$ and $1\leq \ind{1} \leq \ind{2} \leq |F|$ such that the underlying path of $F$ is $\traj$; $F_{\ind{2}}=P_{\domi}$ and $F_{[\ind{1},\ind{2}]}$ is a candidate segment of $F$ of direction $\vu$.
%The proof is divided in three parts. Firstly, we show how to find the two indices $\ind{1}$ and $\ind{2}$ such that $F_{[\ind{1},\ind{2}]}$ is a candidate segment of $F$. Secondly, we show that the translation of this segment by $\vu$ is a subgraph of $\free^+$ and thirdly, we conclude by showing that this segment is a good candidate segment.
By definition of a fork, $\traj_{[\ind{1},\ind{2}]}+\vu$ is a subgraph of $\free$. For the sake of contradiction suppose that the underlying path of $F_{[\ind{1},\ind{2}]}+\vu$ is not a subgraph of $\free^+$. Then, there exists $\ind{4}=\min\{i\geq \ind{1}: (\traj_i,\traj_{i+1})+\vu \notin E(\free^+)\}$ such that $\ind{4}<\ind{2}$. Let $W'$ be the window of $\free^+$, we consider an indexing of $W'$ such that $W'_{0}=\pos(P_{\domi})$ and $W'_{|C|-1}=C_{|C|}$, then $\free^+$ is the right side of $W'$. Now, we proceed in three steps to conclude this part of the proof, first we show that $\ind{4}>1$, then we show that there exists an edge of $C$ which is strictly on the right side of $\traj_{[\ind{1},\ind{2}]}+\vu$ and then we conclude.

%let $W'$ be the window of $\free^+$ and by lemma \ref{}, either $\traj_{[\ind{1},\ind{2}]}+\vu$ is an arc of $\free^+$ or an arc of $\free^-$ or there exists $\ind{1}<\ind{4}<\ind{2}$ and $\ind{5}$ such that either $(W'_{\ind{5}},W'_{\ind{5}+1})$ or $(W'_{\ind{5}},W'_{\ind{5}-1})$ is on the right side of $F_{[\ind{4}-1,\ind{4}+1]}$. In the first case, this part of the lemma is true and now we show that the two other cases lead to contradictions.
\vspace{+0.5em}

\noindent \textbf{Case $\ind{4}=\ind{1}$:} for the sake of contradiction, suppose that $\ind{4}=\ind{1}$. Then, we consider a local reasoning around tile $P_{\domi}$ which has four neighbors: $C_2$, $P_{\domi+1}$, $P_{\domi-1}$ and $\traj_{\ind{1}+1}+\vu$. Since $P$ is simple then $\pos(P_{\domi+1}) \neq \pos(P_{\domi-1})$. By definition of a split, we have $C_2 \neq \pos(P_{\domi+1})$ and $C_2\neq \pos(P_{\domi-1})$. Now if $\traj_{\ind{1}+1}+\vu=C_2$ or $\traj_{\ind{1}+1}+\vu=\pos(P_{\domi+1})$ then the edge $(\traj_{\ind{1}},\traj_{\ind{1}+1})+\vu$ belongs to $\free^+$. %Also, remark that if $\ind{1}=1$ then $P_{\domi}=P_{\indp}$ and since $P_{\indm}$ and $P_{\indp}$ both exists to the north or the east then $\traj_{\ind{1}+1}=\pos(P_{\domi+1})$ which is a contradiction then $\ind{1}>1$. 
Finally, if $\traj_{\ind{1}+1}+\vu=\pos(P_{\domi-1})$ then there exists $1\leq j \leq \ell$ and $1\leq \ind{5} < |A^j|$ such that  $A^j_{\ind{5}}=\pos(P_{\domi})-\vu$ and  $A^j_{\ind{5}+1}=\pos(P_{\domi-1})-\vu$ then the arc $A^j$ is negative which is a contradiction. %Similarly, we have $\traj_{\ind{1}-1} \neq \pos(P_{\domi+1})$. 
Now, without loss of generality we suppose that $C_2=C_1+(0,1)$. Now, since $(C_1,C_2)$ is on the left side of $P_{[d-1,d+1]}$ and since $(\traj_{\ind{1}},\traj_{\ind{1}+1})$ is in $\free^-$ then there exists only one possibility to satisfies all these constraints (see Figure \ref{fig:lem:exists:freep}a): $\pos(P_{\domi-1})=\pos(P_{\domi})-(0,1)$, $\pos(P_{\domi+1})=\pos(P_{\domi})+(1,0)$ and $\traj_{\ind{1}+1}+\vu=\pos(P_{\domi})-(1,0)$. In this case, the edge $(\traj_{\ind{1}},\traj_{\ind{1}+1})+\vu$ is not an edge of the underlying path of $P$ and then there exists $\ind{6}$ such that $W_{\ind{6}}=\pos(P_{\domi})-\vu=\traj_{\ind{1}}$ and $W_{\ind{6}+1}=\traj_{\ind{1}+1}$. Moreover there exists $1\leq k \leq \ell$ such that the edge $(W_{\ind{6}},W_{\ind{6}+1})$ belong to $E(D^k)$. Then two cases occur (see Figure \ref{fig:lem:exists:freep}a), if $W_{\ind{6}-1} \neq C_2-\vu$ then the edge $(C_1,C_2)-\vu$ is strictly on the right side of $W_{[\ind{6}-1,\ind{6}+1]}$ and is not an edge of $E(\free)$, which is a contradiction. Now if $W_{\ind{6}-1} = C_2-\vu$ then the edge $(\pos(P_{\domi}),\pos(P_{\domi+1}))-\vu$ is strictly on the left side of $W_{[\ind{6}-1,\ind{6}+1]}$ and thus to the free zone $\free$. Moreover, since the arc $A^k$ is positive by lemma \ref{lem:fork:ordered} then it starts in $W_{\ind{7}}$ and ends in $W_{\ind{8}}$ such that $\ind{7} \leq \ind{6}$ and $\ind{8}>\ind{6}$. This remark contradicts the fact that the edge $(W_{\ind{6}},W_{\ind{6}+1})$ belongs to $E(D^k)$. Then $\ind{4}>\ind{1}$.

%if $(\traj_{\ind{1}-1})+\vu=C_2$ then the edge $(P_{\domi},P_{\domi+1})-\vu$ is strictly on the left side of $F_{[\ind{1}-1,\ind{1}+1]}$ which contradicts lemma \ref{fact:decompo:newwind}. Finally if $(\traj_{\ind{1}-1})+\vu=P_{\domi-1}$ then the edge $(C_1,C_2)-\vu$ is strictly on the right side of $F_{[\ind{1},\ind{2}]}$. %In this case, since $\pos(P_{\domi}-\vu) = \traj_{\ind{1}}$ and since $\traj_{\ind{1}+1} \neq \pos(P_{\domi}-\vu)$ then the edge $(P_{\domi}-\vu,P_{\domi+1}-\vu)$ belongs to $E(P_{\indm,\lastp})$. 
%This fact implies either that  $(C_1,C_2)-\vu$ is not an edge of $E(\free)$ which is a contradiction or that there exists $1\leq i \leq \ell$ such that the edge $(C_1,C_2)-\vu$ belongs to the interior of $A^i$. In the second case, since $C_{|C|}-\vu=P_{\domi}$ and since $P_{\domi}$ does not belong to the interior of $A^i$ then $C_{[2,|C|]}-\vu$ intersects with $A^i$ and then $P_{\indp,\lastp}$ intersects with $C_{[2,|C|]}$ which contradicts the definition of a split. 

%either the edge $(C_1,C_2)-\vu$ is on the right side of $W$ (and thus not in $E(\free)$) which is a contradiction of the definition of a dominant tile or $P_{\domi+1} = C_2$, the edge $(P_{\domi-1},P_{\domi})$ is on the left side of $\traj_{[\ind{1}-1,\ind{1}+1]}$ which contradicts lemma \ref{lem:decompo:path}.

\vspace{+0.5em}

\noindent \textbf{An edge of $C$ is strictly on the right side of $\traj_{[\ind{1},\ind{2}]}+\vu$:} see Figure \ref{fig:lem:exists:freep}b for an illustration of this part of the proof. Now, if $\ind{4}>1$ then since $D_{[\ind{1},\ind{4}]}+\vu$ is a subgraph of $\free$ then there exist $\ind{11}>0$ such that $W'_{\ind{11}}=D_{\ind{4}}+\vu$. Now, let $\ind{12}=\max\{i<\ind{4}: \traj_i+\vu \in V(W'_{[-\infty,0]})\}$ and let $\ind{9} \leq 0$ such that $W'_{\ind{9}}=\traj_{\ind{12}}$. Now, let $\ind{13}$ such that $\traj_{[\ind{12},\ind{13}]}$ is the extremum path extracted from $\traj_{[\ind{12},\ind{4}]}$. Then, there exists $\ind{11}<\ind{10}$ such that $W'_{\ind{10}}=\traj_{\ind{13}}+\vu$ and there exists $0<\ind{13}$ such that $\omeplus{C}_{\ind{14}}=W'_{\ind{10}}$. Remark that $\traj_{\ind{4}+1} \notin W'_{[-\infty,\ind{9}]}$ then and thus by lemma \ref{cor:dual:firstedge:prel} then the edge $(\omeplus{C}_{\ind{14}},\omeplus{C}_{\ind{14}+1})$ is strictly on the right side of $\traj_{[\ind{13}-1,\ind{13}+1]}+\vu$.

\vspace{+0.5em}

\noindent \textbf{Conclusion:} since there exists $\ind{14}>0$ such that the edge $(\omeplus{C}_{\ind{14}},\omeplus{C}_{\ind{14}+1})$ is strictly on the right side of $\traj_{[\ind{13}-1,\ind{13}+1]}+\vu$ and by definition of $\traj$ there exists $1\leq i \leq \ell$ such that $\traj_{\ind{13}} \in A^i$ or $\traj_{\ind{13}} \in D^i \setminus \{D^{i}_1, D^{i}_{|D^i|}\}$. In the first case, since $\traj_{\ind{13}} \in \dom{P_{[\indp,\lastp]}-\vu}$ and $\traj_{\ind{13}}+\vu =\omeplus{C}_{\ind{14}}$ then this fact contradicts the definition of a split. In the second case, if $\ind{14}>|C|$ then $\omeplus{C}_{\ind{14}}-\vu \in \dom{P}$ and $\omeplus{C}_{\ind{14}}-\vu=C_{\ind{14}-|C|+1}$ which contradicts the definition of a split. Thus $\ind{14}\leq |C|$ and in this case, the edge $(C_{\ind{14}},C_{\ind{14}+1})$ is strictly on the right side of $\traj_{[\ind{13}-1,\ind{13}+1]}+\vu$. Thus, since $\traj_{\ind{13}} \in D^i \setminus \{D^{i}_1, D^{i}_{|D^i|}\}$ then $(C_{\ind{14}},C_{\ind{14}+1})-\vu$ is on the right side of $W$ (and then not in $E(\free)$) which contradicts the definition of a dominant tile.

\end{proof}

\begin{lemma}
\label{lem:existsStraddle}
Consider an exposed path assembly $P$ producible by $\tiling=(T,\seed,1)$ and two indices $1\leq \indm < \indp < \lastp$ such that $P$ does a U-turn at indices $\indm$ and $\indp$. Let $\vu=\overrightarrow{P_{\indm}{P_{\indp}}}$, let $\free$ be the free zone of $P$ associated to $P_{[\indm,\indp]}$. Consider $\indp\leq \domi \leq \lastp$ and a good path $C$ such that $P_{\domi}$ is a dominant tile due to path $C$ whose extension $\omeplus{C}$ splits $\free$ into $(\free^-,\free^+)$. Let $\traj$ be the trajectory of $P_{[\indp,\lastp]}-\vu$. Then either $P$ is fragile or there exists a fork $F$ and $1\leq \ind{1} \leq \ind{2} \leq |F|$ such that: 
\begin{itemize}
\item the underlying path of $F$ is $\traj$; 
\item $F_{\ind{2}}=P_{\domi}$;
\item $F_{[\ind{1},\ind{2}]}$ is a good candidate segment of $F$ of direction $\vu$;
\item the underlying path of $F_{[\ind{1},\ind{2}]}+\vu$ is a subgraph of $\free^+$.
\end{itemize}
%\begin{itemize}
%\item $\traj_{\ind{1}} =P_{\domi}-\vu$ and $\traj_{\ind{2}} =P_{\domi}$;
%\item $\traj_{[\ind{1},\ind{2}]}$ is a subgraph of the binding graph of $P$ or there exists $d<\ind{3} \leq \lastp$ such that $\pos(P_{\ind{3}}) \in V(\traj_{[\ind{1},\ind{2}]})$;
%\item $\traj_{[\ind{1},\ind{2}]}+\vu$ is a subgraph of $\free^+$.
%\end{itemize}
%$P_{[\ind{1},\ind{2}]}$ is a good candidate segment and $P_{[\ind{1},\domi]}+\vu$ is a subgraph of $\free^+$.%straddles at position $P_{\domi}$.
%
%Let $\traj$ be the trajectory of $P_{[\indm,\indp]}-\vu$. Let $1\leq \db \leq |\traj|$ such that $\traj_{\db}=P_{\domi}$. Then there exists $1\leq \da<\db$ and $1\leq k<k' \leq \ell$ such that $\traj_{[\da,\db]}$ is a straddling path at position $\domi$ and $P_{\domi} \in T^{k'}$ and $\traj_{\da} \in A^k$.
\end{lemma}

\begin{proof}

See Figure \ref{fig:find:cand:fork} for an illustration of this part of the proof. 
%Let $W$ be the window of the free zone and consider a coherent indexing of $W$. Let $n=\indp-\indm$ and consider the decomposition $(A^i)_{1\leq i \leq \ell}$ of $P_{[\indp,\lastp]}-\vu$ in extremum arcs in $\free$. By lemma \ref{lem:uturn:ordered}, this decomposition is equivalent to the one of $W_{[n,+\infty]}-\vu$ and by lemma \ref{lem:fork:ordered}, this decomposition is positive. 
By lemma \ref{lem:existsStraddle:part2}, either $P$ is fragile or there exists a fork $F$ and $1\leq \ind{1} \leq \ind{2} \leq |F|$ such that the underlying path of $F$ is $\traj$; $F_{\ind{2}}=P_{\domi}$; $F_{[\ind{1},\ind{2}]}$ is a candidate segment of $F$ of direction $\vu$ and the underlying path of $F_{[\ind{1},\ind{2}]}+\vu$ is a subgraph of $\free^+$. By lemma \ref{fact:arc:split}, the path $\traj_{[\ind{1},\ind{2}]}$ is a subgraph of $\free^-$ and $V(\traj_{[\ind{1},\ind{2}]}) \cap V(\omeplus{C})=\{\pos(P_\domi)\}$. Thus  $V(\traj_{[\ind{1},\ind{2}]}) \cap V(\traj_{[\ind{1},\ind{2}]}+\vu) =\{\pos(P_{\domi})\}$ and $F_{[\ind{1},\ind{2}]}$ is a good candidate segment of $F$. 
\end{proof}

Now we show that either a candidate segment of $P$ is pumpable or there exists $\domi'>\domi$ such that $P_{\domi'}$ is dominant.

\begin{lemma}
\label{lemma:exit:straddle}
Consider an exposed path assembly $P$ producible by $\tiling=(T,\seed,1)$ and two indices $1\leq \indm < \indp < \lastp$ such that $P$ does a U-turn at indices $\indm$ and $\indp$. Let $\vu=\overrightarrow{P_{\indm}{P_{\indp}}}$. Consider $\indp \leq \domi \leq \lastp$ %and a good path $C$ such that $P_{\domi}$ is split at position $P_{\domi}$ by $C$ into $(\free^-,\free^+)$. %Let $(\cutm,\cutp)$ be the cut of the grid by $\ome{C}$. 
such that $P_{\domi}$ is dominant. Then either there exists $\indm \leq \ind{1} \leq \domi \leq \ind{2} \leq \lastp$ such that $P_{[\ind{1},\ind{2}]}$ is a pumpable segment of $P$ or there exists $\domi<\domi'\leq \lastp$ such that $P_{\domi'}$ is dominant or $P$ is fragile.
%Consider a path $Q$ such that $Q$ straddles at position $P_{\domi}$. Then, either $\omepluse{Q}$ is a subgraph of $\free^+$ or there exists $\domi<\domi'<\lastp$ such that $P_{\domi'}$ is a dominant tile.
%\item let $i=\min\{j \in \mathbb{N}: (\omepluse{Q}_j,\omepluse{Q}_{j+1}) \notin E(\free^+)\}$, then there exists $\domi < j < \lastp$ such that $\pos(P_{j})=Q_i$.
\end{lemma}

\begin{proof}
Let $\free$ be the free zone of $P$ and $W$ be its window. Let $n=\indp-\indm$, $m=\lastp-\indm$ and $\ind{3}$ such that $W_{\ind{3}}=P_{\domi}$. Consider $C$ such that $P_{\domi}$ is dominant due to path $C$. The direction of $C$ is $\vu$ and $\omeplus{C}$ splits $\free$ into $(\free^-,\free^+)$. Let $\traj$ be the trajectory of $P_{[\indp,\lastp]}-\vu$. By lemma \ref{lem:existsStraddle}, either $P$ is fragile or there exists a fork $F$ such that the binding graph of $F$ is $\traj$ and there exists $1 \leq \ind{1} \leq \ind{2} \leq |F|$ such that $F_{[\ind{1},\ind{2}]}$ is a good candidate segment of $F$ of direction $\vu$, such that $F_{\ind{2}}=P_{\domi}$ and such that $\traj_{[\ind{1},\ind{2}]}+\vu$ is a subgraph of $\free^+$. By applying lemma \ref{lem:shrink:path} either $\omepluse{\traj_{[\ind{1},\ind{2}]}}$ is a subgraph of $\free^+$ or there exists $R$ and indices $\ind{4},\ind{5} \in \mathbb{N}$ such that:
\begin{itemize}
\item $R$ is a good path of direction $\vu$ such that $R_1=C_1$ and $R_{|R|}=C_{|C|}$;
\item $\omeplus{R}$ is a subgraph of $\free^+$;
\item  $\ind{3}<\ind{4}$, $\ind{5}\geq |R|$ and $\omeplus{R}_{\ind{5}}=W_{\ind{4}}$ and $y_{W_{\ind{4}}}\geq \min\{y_{C},y_{\traj{[\ind{1},\ind{2}]}}+y_{\vu}\}+y_{\vu}$.%$\omeplus{R}_{[\ind{5}+1,+\infty]} \cap W_{[\ind{3},+\infty]}=\emptyset$.
\end{itemize}
\vspace{+0.5em}

%\noindent \textbf{If the path $R$ exists:} %since $\traj_{[\ind{1},\ind{2}]}$ is a subgraph of $\free$ then $\traj_{[\ind{1},\ind{2}]}+y_{\vu} \geq y_{P_{[\indm,\lastp]}}+\vu >y_{P_{\lastp}}$, then we have $V(\omepluse{\traj_{[\ind{1},\ind{2}]}}) \cap W_{[m,+\infty]}=\emptyset$.
\noindent \textbf{If the path $R$ exists:} since $R$ is a subgraph of $\free^+$ then $y_R+y_{\vu} \geq y_{P_{[\indm,\lastp]}}+\vu >y_{P_{\lastp}}$, thus we have $V(\omepluse{R}) \cap W_{[m,+\infty]}=\emptyset$. Then there exists $\ind{6}=\max\{i:\omeplus{R}_i \in V(W)\}$. Moreover, we have $\ind{6}\geq \ind{5}\geq |R|$ and since $\omeplus{R}_{\ind{6}} \in \free^+$ then there exists $\domi<\domi' \leq \lastp$ such that $\pos(P_{\domi'})=\omeplus{R}_{\ind{6}}$. Let $C'=R_{[\ind{6},\ind{6}+|R|-1]}$. By definition of $\ind{6}$ and $R$, the free zone $\free$ is split at position $W_{\ind{6}}=\pos(P_{\domi'})$ by $\omeplus{C'}$ into $(\free'^-, \free'^+)$. Note that since $R_1=\pos(P_d)$ with $d<d'$ then $R_1 \in V(\free^-)$. Moreover, since $R_{[1,\ind{6}]}$ is a path of $\free$ and since $V(R_{[1,\ind{6}]}) \cap V(\omeplus{C'})=\{R_{\ind{6}}\}$ then $R_{[1,\ind{6}]}$ is a subgraph of $\free'^-$ and since $\ind{6}\geq |R|$ then $C'-\vu=R_{[\ind{6}-|R|+1,\ind{6}]}$ is a subgraph of $\free'^-$. Thus $P_{\domi'}$ is a dominant tile and the lemma is true.

\vspace{+0.5em}

\noindent \textbf{If $\omepluse{\traj_{[\ind{1},\ind{2}]}}$ is a subgraph of $\free^+$:} in this case by lemma \ref{lem:freezone} and since $\traj_{[\ind{1},\ind{2}]}$ is a good candidate segment, then the path assembly $P_{[1,\indm]} \cdot F_{[1,\ind{1}]} \cdot \omeplus{F_{[\ind{1},\ind{2}]}}$ is producible by $\tiling$,\emph{i.e.} the segment $F_{[\ind{1},\ind{2}]}$ is pumpable in $P_{[1,\indm]} \cdot F_{[1,\ind{2}]}$. Now, consider the decomposition $(A^i)_{1\leq i \leq \ell}$ of $P_{[\indp,\lastp]}-\vu$ in extremum arcs in $\free$. By lemma \ref{lem:uturn:ordered}, this decomposition is equivalent to the one of $W_{[n,+\infty]}-\vu$. By lemma \ref{lem:fork:ordered}, this decomposition is positive. By lemma \ref{lem:existsStraddle:prel}, there exists $1\leq k < \ell$ such that $\pos(P_{\domi}) \in V(D^k)$ and there exists $1\leq j \leq k$ such that $\traj_{\ind{1}} \in V(A^j)$. Now we consider two different cases: $j=k$ and $j<k$. 

\vspace{+0.5em}

\textbf{Subcase if $j=k$}: see Figure \ref{fig:lem:exists:nextstraddle}b for a graphical representation of this case. Since $\pos(P_d) \in V(D^k)$ then there exists $\ind{8}\leq \domi$ such that $\pos(P_{\ind{8}})=D^k_1$. Moreover, since $A^k$ is positive and since $\pos(P_d)-\vu \in V(A^k)$ then there exists $\ind{9}\geq \domi$ such that $\pos(P_{\ind{9}})-\vu=A^k_{|A^k|}=D^k_1=\pos(P_{\ind{8}})$. Now, let $\ind{1} \leq \ind{7} \leq \ind{2}$ such that $\traj_{\ind{7}}=A^k_{|A^k|}=D^k_1$ and since $P_{[1,\indp]}\cdot (F+\vu)$ is producible by $\tiling$ then either $P$ is fragile or $P_{[\domi,\ind{9}]}=F_{[\ind{1},\ind{7}]}+\vu$. Thus $\omeplus{P_{[\ind{8},\ind{9}]}}$ is a sub-assembly of $\omeplus{F_{[\ind{1},\ind{2}]}}$. Since $\dom{P_{[\indm,\ind{8}]}}\subset \free^-$ then $P_{[1,\ind{8}]} \cdot \omeplus{P_{[\ind{8},\ind{9}]}}$ is producible by $\tiling$ and then the segment $P_{[\ind{8},\ind{9}]}$ of $P$ is pumpable. 

\vspace{+0.5em}

\textbf{Subcase if $j<k$}: see Figure \ref{fig:lem:exists:nextstraddle}a for a graphical representation of this part of the proof. Now, let $n=\ind{2}-\ind{1}$ and $Q=\omeplus{\traj_{[\ind{1},\ind{2}]}}$ then $Q_{[1,n]}$ is a subgraph of $\free^-$, $Q_{[n,+\infty]}$ is a subgraph of $\free^+$ and $y_{Q_{[n,+\infty]}}=y_{F_{[\ind{1},\ind{2}]}}+y_{\vu} \geq y_{P_{[\indm,\lastp]}}+\vu >y_{P_{\lastp}}$, then we have $V(Q) \cap V(W_{[m,+\infty]})=\emptyset$. If $j<k$ then let $\ind{1} \leq \ind{10} \leq \ind{2}$ such $\traj_{\ind{10}}=A^{k}_{|A^k|}$, then there exists $\indp \leq \ind{11} \leq \lastp$ such that $\traj_{\ind{10}}+\vu=\pos(P_{\ind{11}})$. Moreover, since the decomposition $(A^i)_{1\leq i \leq \ell}$ is positive then $\domi<\ind{11}$ and then $\pos(P_{\ind{11}}) \in \free^+$ . Thus, $\pos(P_{\ind{11}}) \in V(Q_{[n+1,\infty]})$. Now, since $V(Q) \cap V(W_{[m,+\infty]})=\emptyset$ then there exists $\ind{12}\geq n$ such that $Q_{\ind{12}}=\max\{i: Q_i \in V(W)\}$. Moreover since $\ind{11}>d$ and $\pos(P_{\ind{11}}) \in V(Q_{[n+1,\infty]})$ then $\ind{12}>n$. Now, since $V(Q) \cap V(W_{[m,+\infty]})=\emptyset$ and $Q_{\ind{12}} \in \free^+$ then there exists $d<d'<\lastp$ such that $\pos(P_{d'})=Q_{\ind{12}}$. Then, let $C'=Q_{[\ind{12},\ind{12}+n]}$ be a path of direction $\vu$. By definition of $\ind{12}$ and $Q$, the free zone $\free$ is split at position $\pos(P_{\domi'})$ by $\omeplus{C'}$ into $(\free'^-, \free'^+)$. Note that since $Q_n=P_d$ with $d<d'$ then $Q_n \in V(\free'^-)$. Moreover, since $Q_{[1,\ind{12}]}$ is a path of $\free$ and since $V(Q_{[1,\ind{12}]}) \cap V(\omeplus{C'})=\{Q_{\ind{12}}\}$ then $Q_{[1,\ind{12}]}$ is a subgraph of $\free'^-$ and since $\ind{12}> n$ then $C'-\vu=Q_{[\ind{12}-n,\ind{12}]}$ is a subgraph of $\free'^-$. Thus $P_{\domi'}$ is a dominant tile and the lemma is true.

\end{proof}

Finally, we prove that there exists at least one dominant tile in a path assembly which does a U-turn.%we need an efficient way to find way to find them. To achieve this goal, we introduce straddling path. Be careful, in the following definition, we require that the free zone could be split at position $P_{\domi}$ but we do not require that $P_{\domi}$ is a dominant tile.

\begin{lemma}
\label{lem:existDom}
Consider an exposed path assembly $P$ producible by $\tiling=(T,\seed,1)$ and two indices $1\leq \indm < \indp < \lastp$ such that $P$ does a U-turn at indices $\indm$ and $\indp$. Then there exists, $\indp \leq \domi \leq \lastp$ such that $P_{\domi}$ is dominant. 
\end{lemma}

\begin{proof}
%on prend le max en hauteur si au dessus de $P_{\indp}$, c'est ok. On crée un chemin de $P_{\domi}$ à $P_{\domi}+\vu$ avec un morceau de ce qui est à l'est au cas où et qui est good. On l'étend et on le réduit, la dernière tuile selon ce chemin est dominante.

This proof is in three parts. In the first one, we define an infinite path which will be useful to split the free zone. In the second part, we show that this path is indeed a subgraph of the free zone. In the last part, we show that the last intersection between this path and $P_{[\indp,\lastp]}$ is a dominant tile. See Figure \ref{fig:lem:exists:dom} for an illustration of this proof.%Then $W_{[-\infty,0]}$ intersects with $L^H$ otherwise the height of $P_{\indm}$ would be strictly greater than $H$ which is a contradiction. Consider an indexing of $L^H$ such that $L^H_0 \in W_{[-m,0]}$ and such that $L^H_{x_{\vu}+y_{\vu}}=L^H_0+\vu$. 

\vspace{+0.5em}

\noindent{\textbf{Definition}}: let $\vu=\overrightarrow{P_{\indm}{P_{\indp}}}$, let $\free$ be the free zone of $P$ associated to $P_{[\indm,\indp]}$ and $W$ be its window, let $p=x_{\vu}+y_{\vu}$. Consider a coherent indexing of this window. Consider the height function $h$ of direction $\vu$. Let $H=\max\{h(\pos(P_i)): \indm \leq i \leq \lastp \text{ and $y_{P_i}\geq y_{P_\indm}$}\}$. Let $(D^H,U^H)$ be the the cut of the grid by the line of direction $\vu$ and height $H$ and let $L^H$ be its window. %Let $p=x_{\vu}+y_{\vu}$. Remark that by definition of $H$, we have $H\geq h(\pos(P_\indm))$ and thus $\pos(P_\indm) \in V(D^h)$. Moreover for all $i < 0$ we have $h(P_i)>h(P_{i+1})$ and thus there exists $j \leq 0$ such $W_j \in V(L^H)$. We consider an indexing of $L^H$ such that $L^H_0=W_j$ and $L^H_{p-1}=W_j+\vu$. Now by a local argument the edge $(L^H_0,L^H_1)\in E(\free)$. 
Now, since the height of $P_{\indm}$ is less than $H$ then the infinite line $W_{[-\infty,0]}$ intersects with $L^H$. Consider an indexing of $L^H$ such that $L^H_0 \in W_{[-m,0]}$, $L^H_1 \notin W_{[-m,0]}$ and $L^H_{p}=L^H_0+\vu$. Now, let $$H'=\max\{h(\pos(P_i)): \indm \leq i \leq \lastp \text{ and there exists $k\geq 0$ such that $y_{P_i}=y_{P_{\indm}}+ky_{\vu}$}\}.$$ By definition $H'\leq H$. Now, we define the path $R$ as follow: we definite the positions $A$ and $B$ such that $y_{A}=y_{P_{\indm}}$ and $h(A)=H'$ and $B=A+(0,1)$ then $R$ is the concatenation of the four paths $R^1\cdot R^2 \cdot R^3 \cdot R^4$ defined as:
\begin{itemize}
\item $R^1$ is the path such that $R^1_1=L^H_{0}$, $R^1_{|R^1|}=A$ and for all $1\leq i \leq |R^1|-1$, we have $R^1_{i+1}=R^1_{i}+(1,0)$;
\item $R^2$ is the path from $A$ to $B$ ($R^2$ is made of two vertices and one edges);
\item $R^3$ is the path such that $R^3_1=B$, $R^3_{|R^3|}=L^H_1$ and for all $1\leq i \leq |R^1|-1$, we have $R^3_{i+1}=R^3_{i}-(1,0)$;
\item $R^4$ is the path $L^H_{[1,p]}$.
\end{itemize}

\vspace{+0.5em}

\noindent\textbf{The path $\omeplus{R}$ is a subgraph of $\free$}:
If $L^H_{[0,+\infty]}$ is a not subgraph of $\free$ then there exists $\indm \leq \ind{1} \leq \lastp$ such that  $h(\pos(P_{\ind{1}}))>H$ and $y_{P_{\ind{1}}}>P_{\indm}$ which contradicts the definition of $H$. Now if $\omeplus{R}$ is not a subgraph of $\free$ then there exists $\indm \leq \ind{1} \leq \lastp$ and $k\geq 0$ such that $y_{P_{\ind{1}}}=y_{P_{\indm}}+ky_{\vu}$ and  $x_{P_{\ind{1}}}>x_{A}+kx_{\vu}$ which contradicts the definition of $H'$.%In the first case, $h(\pos(P_{\ind{1}}))>H$ and  $y_{P_{\ind{1}}}\geq y_{P_{\indm}}$ which contradicts the definition of $H$ and 
% In the second the case $h(\pos(P_{\ind{1}}))>H'$ and $y_{P_{\ind{1}}}=y_{A}+ky_{\vu}$ 

\vspace{+0.5em}

\noindent\textbf{There exists a dominant tile}: let $Q=\omeplus{R}$, by definition of $R$ and $H'$ there exists an index $\ind{3} \in \mathbb{N}$ and $k\geq0$ such that $Q_{\ind{3}} \in \dom{P_{[\indm,\lastp]}}$ and $y_{Q_{\ind{3}}} = y_{P_{\indm}}+k\vu$. Note that if $k=0$ then since $P_{\indm}$ is visible from the west, we have $Q_{\ind{3}}=\pos(P_{\indm})$. In this case, $Q_{\ind{3}+|R|-1}=P_{\indp}$ and then we can suppose that $k>0$ and that $\ind{3}\geq |R|$. Remark that $V(Q_{[|R|,+\infty]}) \cap V(W_{[m,+\infty]})=\emptyset$ otherwise we would have $y_{P_{\lastp}}\geq y_{P_{[\indm,\lastp]}}+y_{\vu}$ which contradicts the definition of a U-turn. Now, let $\ind{4}=\max\{i: Q_i \in V(W_{[0,+\infty]})\}$. By definition $\ind{4} \geq \ind{3} \geq |R|$ and then $y_{Q_{\ind{4}}} \geq y_{P_{\indp}}$. Also, since $V(Q_{[|R|,+\infty]}) \cap V(W_{[m,+\infty]})=\emptyset$, there exists $\indm \leq \domi\leq \lastp$ such that $P_{\domi}=Q_{\ind{4}}$. Now if $\domi < \indp$ then let $(G_1,G_2)$ be the cut of the grid by the underlying path of $P_{[\indm,\indp]}$ whose extremities are both visible from the west and let $W'$ be its window. Consider an indexing of $W'$ coherent with $P_{[\indm,\indp]}$ and suppose that $G_1$ is the left side of $W'$, thus $G_1$ is a subgraph of $\free$. Then $Q_{\ind{4}}$ belongs to $G_1$. Moreover, since $y_{Q_{\ind{4}}} \geq y_{\indp}$, since $Q_{\ind{4}}\neq \pos(P_{\indp})$ and since $Q$ is a subgraph of $\free$ then $Q_{[\ind{4},+\infty]}$ is a subgraph of $G_1$. But $Y_{G_1}=Y_{P_{[\indm,\indp]}}$
 is finite which contradicts the definition of $Q$. Then, $\domi\geq \indp$ and let $C=Q_{[\ind{4},\ind{4}+|R|-1]}$. Since $Q$ is a subgraph of $\free$ and by definition of $\ind{4}$, then $\omeplus{C}$ splits the free zone into $(\free^-,\free^+)$. Moreover $Q_{[0,\ind{4}]}$ is a subgraph of $\free$ such that $Q_0 \in W_{[-\infty,0]}$ and then  $Q_{[0,\ind{4}]}$ is a subgraph of $\free^-$. Since $\ind{4} \geq |R|$ then $C-\vu=Q_{[\ind{4}-|R|+1,\ind{4}]}$ is a subgraph of $\free^-$ and thus $P_{\domi}$ is a dominant tile.

\end{proof}

Finally, we can prove the final result of this section.

\begin{lemma}
\label{lem:uturn}
Consider an exposed path assembly $P$ producible by $\tiling=(T,\seed,1)$. If $P$ makes a U-turn then $P$ is fragile or pumpable.
\end{lemma}

\begin{proof}
Consider two indices $1\leq \indm < \indp < \lastp$ such that $P$ does a U-turn at indices $\indm$ and $\indp$. Consider the set $R=\{\indp \leq i \leq \lastp: P_i\text{ is dominant }\}$ which contains the indices of the dominant tiles of $P$, by lemma \ref{lem:existDom} this set is not empty. Consider $\domi=\max \{i:i \in R\}$ the last dominant tile of $P$, by lemma \ref{lemma:exit:straddle} either $P$ is fragile or there exists $\ind{1} \leq \domi \leq \ind{2}$ such that $P_{[\ind{1},\ind{2}]}$ is a pumpable segment of $P$ or there exists $\domi' \in R$ such that $\domi'>\domi$. Since the last case is not possible, then $P$ is fragile or pumpable. 
\end{proof}

\section{From microscopic reasoning to macroscopic one: proof of the remaining macroscopic lemmas}
\label{sec:lastpart}
\vspace{+0.5em}

The aim of the last section of the proof is to prove the last three remaining macroscopic lemma. We proceed in this order: the seed lemma, the reset lemma, the stakes lemma. This section used several bounds and functions which are all defined in the appendix \ref{app:bound}. The figures of this section are in Appendix \ref{app:last}.

\subsection{The seed lemma}
\label{sec:truelast:seed}

To prove the existence of a U-turn in an exposed path assembly $P$, we need to verify two hypothesis: there exists a visible candidate segment and the last tile of $P$ is low enough. For the second hypothesis if there exists a visible candidate segment and if $y_P=y_{P_{\lastp}}$ then $P$ does a U-turn.

\begin{lemma}
\label{lem:complete}
Consider an exposed path assembly $P$ producible by $\tiling=(T,\seed,1)$, if there exists a visible candidate segment in $P$ and if  $y_{P_{\lastp}}=y_P$ then $P$ is fragile or pumpable.
\end{lemma}

\begin{proof}
Consider $1\leq \indm< \indp \leq \lastp$ such that $P_{[\indm,\indp]}$ is a visible candidate segment of direction $\vu$. By lemma \ref{cor:visible:order}, $y_{\vu} > 0$. Now, remark that $y_{P_{\lastp}}=y_P \leq y_{P_{[\indm,\lastp]}}<y_{P_{[\indm,\lastp]}}+y_{\vu}$. Thus the path assembly $P$ makes a U-turn and by lemma \ref{lem:uturn}, $P$ is fragile or pumpable.
\end{proof}

Now, suppose that we have found a candidate segment in an exposed path assembly. Then this candidate segment may be oriented to the south, we deal with this problem now.%Remark that if there exists more $4|\glue|$ tile visible from the west, oriented to the south and not hidden by the seed. Then at least two of them are of the same type, directed to the south and exit in the same direction. Moreover, these two tiles are before the first highest tile of $P$. Then by making a symmetry on the $y$ axis, we obtain a path with a visible segment and a turn back of $y_P$ which by the previous result is fragile or pumpable.

\begin{lemma}
\label{lem:low:watershed}
Consider an exposed path assembly $P$ producible by $\tiling=(T,\seed,1)$, if there exists $1\leq \indm \leq \indp \leq \lastp$ such that $P_{[\indm,\indp]}$ is a candidate segment, visible from the west, not hidden by the seed, directed to the south and which exits to the south or the east, then $P$ is fragile or pumpable.
\end{lemma}

\begin{proof}
%Let $t$ be the first highest tile of $P$ then $P_t$ is oriented to the north. Let $s$ be the first lowest tile of $P$. We have $s<t$ and $P_s$ exits to the west by lemma \ref{}. 
%If the watershed of $P$ is greater than $y_P+4|\glue|+|\seed|$ then there exists $4|\glue|$ tiles of $P$ which are visible from the west, oriented to the south and not hidden by the seed. Then there exists $1\leq \indm\leq \indp \leq \lastp$ such that $P_{[\indm,\indp]}$ is a candidate segment, visible from the west, directed to the south, and which exits to the south or the east.
See Figure \ref{fig:proof:seed1} for an illustration of this proof. Let $s$ be the first lowest tile of $P$ and $t$ be the first highest of $P$, by definition we have $s\leq t$. Since $P_{\indm}$ is visible from the west, not hidden by the seed and oriented to the south, then by lemma \ref{lem:lowest}, $P_s$ is visible from the west and exits to the east. Also by lemma \ref{lem:order}, we have $y_{P_{\indm}} \geq y_{P_{\indp}} \geq y_{P_{s}}$ and $\indm \leq \indp \leq s$ and by lemmas \ref{lem:svisible} and \ref{lem:lowest}, we have $y_{P_s} =y_P < y_{\seed}$. By definition of $t$, we have $y_{P_t}>P_{[1,s]}$ Now consider $\tiling'=(T',\seed',1)$ where $T'$ is the tile set obtained by switching the north and south glues of the tiles of $T$ and $\seed'$ is obtained by a symmetry of $\seed$ according to the $x$ axis. Let $Q$ be the path assembly obtained by a symmetry of $P_{[1,t]}$ by the the $x$ axis. Then $Q$ is fragile or pumpable if and only if $P$ is fragile or pumpable. Remark that $Q_{[\indm,\indp]}$ is a visible candidate segment of $Q$, that $Q_s$ is the first highest tile of $Q$, that $Q_s$ exits to the east and that $y_{Q_s}>Y_{\seed'}$. Moreover $y_{Q_{|Q|}}=y_{Q_t}<\min\{y_{\seed}, y_{Q_{[1,s]}}\}$  and then the tail of $Q$ is long and by lemma \ref{lem:existexposed}, there exists $\ind{1}>s$ such that $Q_{[1,\ind{1}]}$ is an exposed path assembly with $y_{Q_{\ind{1}}}=y_{Q_{[1,\ind{1}]}}$. Since $\indm\leq \indp\leq s$ then $Q_{[\indm,,\indp]}$ is a candidate segment of $Q_{[1,\ind{1}]}$ and then by lemma \ref{lem:complete}, $Q$ is fragile or pumpable and then $P$ is fragile or pumpable.
\end{proof}

\begin{lemma}
\label{lem:complete:seed}
Consider a path assembly $P$ producible by $\tiling=(T,\seed,1)$, if $Y_P-Y_{|\seed|}>4|T|$ (resp. $X_P-X_{|\seed|}>4|T|$) and $y_{|\seed|}-y_{P}>4|T|$ and (resp. $x_{|\seed|}-x_{P}>4|T|$) then $P$ is fragile or pumpable.
\end{lemma}

\begin{proof}
This proof relies on using symmetries in order to find a U-turn in $P$, Figures \ref{fig:proof:seed2} and \ref{fig:proof:seed3} represent an example where all symmetries are required. First, if $X_P-X_{|\seed|}>4|T|$ and $x_{|\seed|}-x_{P}>4|T|$ then by switching the abscissa and the ordinate axis, we can consider, without loss of generality, that $Y_P-Y_{|\seed|}>4|T|$ and $y_{|\seed|}-y_{P}>4|T|$. Now, let $\ind{1}=\min\{1\leq i\leq \lastp: y_{P_i}>Y_{|\seed|}+4|T| \text{ or } y_{P_i}<y_{|\seed|}-4|T|\}$. Secondly, if $y_{P_{\ind{1}}}<y_{|\seed|}-4|T|$ then by switching north and south (\emph{i.e} by a symmetry of the $x$ axis), we can consider, without loss of generality, that $y_{P_{\ind{1}}}-Y_{|\seed|}>4|T|$. Now, by hypothesis there exists $\ind{1} \leq \ind{3} \leq \lastp$ such that $\ind{3}=\min\{\ind{1} \leq i\leq \lastp: y_{P_i}<y_{|\seed|}-4|T|\}$. Now, let $t$ be the first highest tile of $P_{[1,\ind{3}]}$, note that we have $\ind{1} \leq t < \ind{3}$. Thirdly, if $P_t$ is oriented to the west then by switching west and east (\emph{i.e} by a symmetry of the $y$ axis), we can consider, without loss of generality, that $P_t$ is oriented to the east. Remark that $P$ has a long tail and then by lemma \ref{lem:existexposed}, there exists $t\leq \ind{2} \leq \ind{3}$ such $P_{[1,\ind{2}]}$ is an exposed path assembly with $y_{P_{\ind{2}}}=y_{P_{[1,\ind{2}]}}$. Since $y_{P_t}-Y_{|\seed|}>4|T|$, there exists at least $4|T|$ tiles visible from the west and not hidden by the seed in $P_{[1,\ind{2}]}$. Then there $1\leq \indm \leq \indp\leq t$ such that the segment $P_{[\indm,\indp]}$ of $P_{[1,\ind{2}]}$ is visible from the west, not hidden by the seed, directed to the north or to the south and exits to north, south or west. If  $P_{[\indm,\indp]}$ is oriented to the south then by lemma \ref{lem:low:watershed}, $P$ is fragile or pumpable (Figure \ref{fig:proof:seed3}). Now, if $P_{[\indm,\indp]}$ is oriented to the north then by lemma \ref{lem:complete}, $P$ is fragile or pumpable (Figure \ref{fig:proof:seed2}).

% \begin{figure}[h]
% \begin{center}
% \includegraphics[scale=0.5]{./photoFinal/Divers1/ProofSeed2.JPG}
% \end{center}
%  \caption{A path assembly cannot move at distance $4|T|$ from the seed in one direction and $4|T|$ in the opposite direction without doing a U-turn. Several symmetries may be required to find the desire U-turn.}
%  \label{fig:proof:seed2}
% \end{figure}
\end{proof}

% $P_{[\indm,\indp]}$ is a visible candidate segment of $P_{[1,t]}$ and $P_{[1,\ind{2}]}$ has a long tail. By lemma \ref{}, there exists $t\leq k \leq g$ such that $P_{[0,k]}$ is an exposed path assembly of $y_P$. By lemma \ref{}, $P_{[\indm,\indp]}$ is a visible candidate segment of $P_{[1,g]}$ and  $y_{P_g}=y_{P_{[1,g]}}<y_{P_{[\indm,g]}}+\vu$, and $P_{[0,g]}$ does a U-turn. Then $P$ is fragile or pumpable. Now, if the candidate segment  $P_{[\indm,\indp]}$ is oriented to the south the by lemma \ref{} the first lowest tile $P_{b}$ exit to the west and by lemma \ref{}, we have $1\leq \indm\leq \indp \leq b\leq t$. Then by switching north and south, we obtain a path $P'$ such that $P'_{[\indm,\indp]}$ is visible candidate segment of $P'_{[0,b]}$, $P'_b$ is the first height tile and $P'_{[1,t]}$ has a long tail. As the previous case, we can show that $P'_{[1,t]}$ is fragile or pumpable. Then $P$ is fragile or pumpable in all cases.
%\end{proof}

To conclude this section, consider a path assembly with macroscopic initial condition then if $\dom{\seed}$ is not a subset of $\rseed$. Then either $y_{|\seed|}-y_P>4|T|$ or $x_{|\seed|}-x_P>4|T|$. W.l.o.g. we consider that $y_{|\seed|}-y_P>4|T|$ and by definition of a path assembly with good macroscopic condition, we have $Y_P-Y_{\seed} \geq y_{|\seed|}-y_P>4|T|$ and by lemma \ref{lem:complete:seed}, $P$ is fragile or pumpable. The seed lemma is thus a corollary of this result.% leads to the seed lemma.

%%%%% RESET Lemma

\subsection{Reset lemma}
\label{sec:truelast:reset}

To prove the reset lemma, we assume that the path assembly will grow a lot to the north and then will come back near the seed. Then, we are able to find a lot of tiles visible from the west. The first step is to show that almost all of them are oriented to the north. Then, we can find a type of tile which appears frequently among the tiles visible from the west which allow us to find a candidate segment of direction $\vu$ such that $y_{\vu}$ is big enough to find a U-turn.

\begin{lemma}
\label{lem:prelem:reset}
Consider an exposed path assembly $P$ producible by $\tiling=(T,\seed,1)$. If $Y_{P}-y_P>|\seed|+2(y_{P_{\lastp}}-y_P+3)|T|$ then $P$ is fragile or pumpable. 
\end{lemma}

\begin{proof}
See Figure \ref{fig:proof:reset1} for an illustration of this proof. By fact \ref{fact:heightvis}, we have $\vset{P}>|\seed|+2(y_{P_{\lastp}}-y_P+3)|T|$ and since at most $|\seed|$ tiles of $P$ are hidden by the seed then $\vsets{P}>2(y_{P_{\lastp}}-y_P+3)|T|$. Now if more than $2|T|$ of these tiles are directed to the south then there exists $1\leq \indm< \indp\leq \lastp$ such that $P_{[\indm,\indp]}$ is a candidate segment, visible from the west, not hidden by the seed, directed to the south and exits to the south or east. In this case, by lemma \ref{lem:low:watershed}, $P$ is fragile or pumpable. Otherwise there exists at least $2(y_{P_{\lastp}}-y_P+2)|T|$ tiles visible from the west and directed to the north. Then there exists $o \in T$ such that there exists at least $y_{P_{\lastp}}-y_P+2$ tiles of $P$ which are of type $o$, oriented to the north and either all of them exit to the east or all of them exit to the north. Then, there exists $1\leq \indm<\indp \leq \lastp$ such that $P_{[\indm,\indp]}$ is a visible candidate segment of direction $\vu$ and such that $|y_{\vu}|\geq y_{P_{\lastp}}-y_P+1$. Moreover by lemma \ref{lem:order}, $y_{\vu}>0$ and then $y_{\vu} \geq y_{P_{\lastp}}-y_P+1$. Thus $y_{P_{[\indm,\lastp]}}+y_{\vu} \geq y_P+y_{\vu} \geq y_P +y_{P_{\lastp}}-y_P+1 > y_{P_{\lastp}}$. Then $P$ makes a U-turn and by lemma \ref{lem:uturn}, $P$ is fragile or pumpable.

\end{proof}

%Consider a path assembly $P$, we define the number of times that $P$ crosses over $\rect(\ell)$ as $V(W)\cap \dom{P}$. This value is bounded by $4\ell$.

Consider $\ell \in \mathbb{N}$, the window associated to the rectangle $\rect(\ell)$ is defined as the finite cycle $C=\bigcup_{1\leq i \leq 4} Q^i$ where: 
\begin{itemize}
\item $Q^1$ is the path starting in $(0,0)$ and ending in $(0,\ell)$ with for all $1\leq i \leq |Q^1|-1$, we have $Q^1_{i+1}=Q^1_i+(0,1)$; 
\item $Q^2$ is the path starting in $(0,\ell)$ and ending in $(\ell,\ell)$ with for all $1\leq i \leq |Q^2|-1$, we have $Q^2_{i+1}=Q^2_i+(1,0)$; 
\item $Q^3$ is the path starting in $(\ell,\ell)$ and ending in $(\ell,0)$ with for all $1\leq i \leq |Q^3|-1$, we have $Q^3_{i+1}=Q^3_i+(0,-1)$; 
\item $Q^4$ is the path starting in $(\ell,0)$ and ending in $(0,0)$ with for all $1\leq i \leq |Q^4|-1$, we have $Q^4_{i+1}=Q^4_i+(-1,0)$. 
\end{itemize}
Let $\gint$ be the interior of $C$ then $V(\gint)=\rect(\ell)$. Remark that $V(C)\leq 4\ell$. For the definition of the function $\fresetstep$ and $\freset$ since the appendix \ref{app:bound}. To demonstrate the reset lemma, we show that a path which goes out of the square $\rect(\freset(\ell))$ and then enter $\rect(\ell)$ afterwards is either pumpable or fragile by the previous lemma or admits a subpath with almost the same properties but which intersects the window of rectangle $\rect(\ell)$ one less time. Then a recurrence leads to the desired result. The following lemma is the proof of one step of the recurrence.

%\begin{definition}
%\label{def:fun:reset}
%Consider a tiling system $\tiling=(T,\seed,1)$, then let $\fresetstep$ be the function defined as $\fresetstep(\ell)=2|\seed|+4(\ell+3)|T|+\ell$. Now, consider $\ell \in \mathbb{N}$, then we define $\ell_0=\fresetstep(\ell)$ and for $0\leq i \leq 4\ell-1$, let $\ell_{i+1}=\fresetstep(\ell_{i})$. The function $\freset$ is the function $\freset(\ell)=\ell_{4\ell}$.
%
%\end{definition}

\begin{lemma}
\label{lem:banana:rec}
Consider a simple path assembly $P$ producible by $\tiling=(T,\seed,1)$ such that $\dom{P} \cap \dom{\seed}=\pos(P_1)$ and $\dom{P}$ and $\dom{\seed}$ belong to the first quadrant of the plane. Consider $\ell_1,\ell_2,\ell_3 \in \mathbb{Z}$ such $\ell_2\geq \ell_1$ and $\ell_3 \geq \fresetstep(\ell_2+1)$. Suppose that $\dom{\seed} \subset \rect(\ell_1)$ then if there exists $1\leq \ind{3} \leq \ind{4} \leq \lastp$ such that $\pos(P_\ind{3}) \notin \rect(\ell_3)$ and $\pos(P_{\ind{4}}) \in \rect(\ell_1)$ then either $P$ is pumpable or fragile or there exists $1\leq \ind{1} \leq \ind{2} \leq \ind{3}$ such that $\pos(P_{\ind{1}}) \notin \rect(\ell^2)$ and $\pos(P_{\ind{2}}) \in \rect(\ell^1)$. Moreover, let $C$ be the window of $\rect(\ell_1)$, then $|\dom{P_{[1,\ind{2}]}}\cap V(C)|<|\dom{P}\cap V(C)|$.
\end{lemma}

\begin{proof}
See Figure \ref{fig:proof:reset2} for an illustration of this proof. Since $\pos(P_{\ind{3}}) \notin \rect(\ell_3)$ then either $x_{P_{\ind{3}}}>\ell_3$ or $y_{P_{\ind{3}}}>\ell_3$. If $x_{P_{\ind{3}}}>\ell_3$ then by switching the $x$ and $y$ axis we are in the case $y_{P_{\ind{3}}}>\ell_3$. Then without loss of generality, we can consider that $y_{P_{\ind{3}}}>\ell_3$. Since $\pos(P_{\ind{4}}) \in \rect(\ell_1)$ then there exists an index $\ind{3} \leq \ind{6} \leq \ind{4}$ such that  $\ind{6}=\min\{\ind{3}\leq i \leq \lastp: P_i \in \rect(\ell)\}$. Moreover, $P_{\ind{6}}\in V(C)$. Let $t$ be the first highest tile of $P_{[1,\ind{6}]}$ then $y_{P_t}\geq y_{P_\ind{3}}>\ell^3>\ell^1\geq Y_\seed$. If $P_t$ exits to the west then by a symmetry according to the $y$ axis, we are in a case where $P_t$ exits to east and $P$ and $\seed$ are in the second quadrant of the proof (the important hypothesis which is "$y_P \geq 0$ and $y_{\seed} \geq 0$" is preserved). Then, without loss of generality we can consider that $P_t$ exits to the east. Now, since $P_{\ind{6}} \in \rect(\ell_1)$ then there exists $t \leq \ind{5} \leq \ind{6}$ such  that $y_{\ind{5}}=\ell_2+1$ and $P_{\ind{5}}$ is visible from the east in $P_{[t,\ind{6}]}$. Moreover since $\ell_2>\ell_1 \geq Y_{\seed}$ then $P_{\ind{5}}$ is not hidden by the seed. Then either $P_{\ind{5}}$ is visible from the east in $P_{[1,\ind{5}]}$ or not. If $P_{\ind{5}}$ is visible from the east in $P_{[1,\ind{5}]}$ then $P_{[1,\ind{5}]}$ is an exposed path assembly then since $\ell_3 \geq \fresetstep(\ell_2+1)$ then $\ell_3 \geq |\seed|+2((\ell_2+1)+3)|T|$ and since $y_P>0$ we have $\ell_3 -y_P\geq |\seed|+2((\ell_2+1)-y_P+3)|T|$ and by definition of $t$ and $\ind{5}$, we have $Y_{P_{[1,\ind{5}]}} -y_P\geq |\seed|+2(y_{P_{\ind{5}}}-y_P+3)|T|$. Then, by lemma \ref{lem:prelem:reset}, $P_{[1,\ind{5}]}$ is fragile or pumpable. Otherwise $P_{\ind{5}}$ is not visible from the west in $P_{[1,\ind{6}]}$ and there exists $1\leq \ind{1} \leq t$ such that $y_{P_{\ind{1}}}=y_{P_{\ind{5}}}$ and $x_{P_{\ind{1}}}>x_{P_{\ind{5}}}$. In this case, $P_{\ind{1}} \notin \rect(\ell_2)$. Now let $Q$ be the underlying path of $P_{[t,\ind{6}]}$ remark that by the definition of $t$, $Q_1$ is visible from the north and by the definition of $\ind{6}$, $Q_{|Q|}$ is visible from the south in $Q$. Then let $(\cutp,\cutm)$ be the cut defined by $Q$ with visible extremities and let $W$ be its window. Consider a coherent indexing of this window. Now, remark that $P_{t-1}=P_t-(0,1)$ and $P_{t+1}=P_t+(1,0)$ and then $\pos(P_{t-1})\in V(\cutm)\setminus V(W)$ (where $\cutm$ is the right side of the cut of the grid). Now for all $k>1$, the position $(x_{P_{\ind{5}}}+k,y_{P_{\ind{5}}})$ is not a vertex of $V(W)$, otherwise it would either contradict the definition of $P_t$ or the fact that $P_{\ind{5}}$ is visible from the east in $P_{[t,\ind{6}]}$. Then all these tiles belong to $\cutp$ and then $\pos(P_{\ind{1}}) \in V(\cutp) \setminus V(W)$. Then there exists $\ind{1} \leq \ind{2} \leq t-1$ such that $\ind{2} \in V(W)$. Since $P$ is simple and by definition of $t$, we have $x_{P_{\ind{2}}}=x_{P_{\ind{6}}}$ and $y_{P_{\ind{2}}}<y_{P_{\ind{6}}}$. Since $\pos(P_{\ind{6}}) \in \rect(\ell_1)$ then $x_{P_{\ind{6}}}\leq \ell_1$ and $y_{P_{\ind{6}}} \leq \ell_1$. Then $x_{P_{\ind{2}}}\leq \ell_1$ and $y_{P_{\ind{2}}} \leq \ell_1$ and by hypothesis $x_{P_{\ind{2}}}\geq 0$ and $y_{P_{\ind{2}}} \geq 0$. Then $\pos(P_{\ind{1}}) \in \rect(\ell_1)$. Thus $P_{[1,\ind{2}]}$ and $\ind{1}$ satisfies the desired conditions of the lemma. Moreover since $\ind{2}<\ind{6}$ and $\pos(P_{\ind{6}}) \in V(C)$ then $|\dom{P_{[1,\ind{2}]}}\cap V(C)|<|\dom{P}\cap V(C)|$.

\end{proof}

Using lemma \ref{lem:banana:rec} to do a recurrence on a path assembly with macroscopic initial conditions leads to the reset lemma. We present here the reset lemma in a more general setting.

\begin{lemma}
\label{macro:reset:bis}
Consider a simple path assembly $P$ producible by $\tiling=(T,\seed,1)$ such that $\dom{P} \cap \dom{\seed}=\pos(P_1)$ and both $\dom{P}$ and $\dom{\seed}$ belong to the first quadrant of the plane. Consider $\ell \in \mathbb{N}$ such that $\dom{\seed} \subset \rect(\ell)$. If $P$ does not satisfy the reset constraint for $\rect(\ell)$ then $P$ is fragile or pumpable.
\end{lemma}

\begin{proof}
Let $\ell_0=\fresetstep(\ell+1)$ and for $0\leq i \leq 4\ell-1$, let $\ell_{i+1}=\fresetstep(\ell_{i}+1)$. Consider the following hypothesis $H(i)$:"either $P$ is fragile or pumpable or there exists $1\leq \ind{1} \leq \ind{2} \leq \lastp$ such that $\pos(P_{\ind{1}}) \notin \rect(\ell_{4\ell-i})$ and $\pos(P_{\ind{2}}) \in \rect(\ell)$ and $|\dom{P_{[1,\ind{2}]}}\cap V(C)| \leq |\dom{P}\cap V(C)|-i$". By hypothesis, since $P$ does not satisfy the reset constraint for $\rect(\ell)$ then there exists $1\leq \ind{1} \leq \ind{2} \leq \lastp$ such that $\pos(P_{\ind{1}})\notin \rect(\freset(\ell))$ and $\pos(P_{\ind{2}})\in \rect(\ell)$. Since $\freset(\ell)=\ell_{4\ell}$ then $H(0)$ is true. Now suppose that $H(i)$ is true for $i\leq 4\ell-1$ then either $P$ is fragile or pumpable (then $H(i+1)$ is true) or there exists $1\leq \ind{1}\leq \ind{2}\leq \lastp$ which satisfies the properties of the recurrence and in this case, by lemma \ref{lem:banana:rec}, $H(i+1)$ is true. Then, $H(4\ell)$ is true and either $P$ is fragile or pumpable or there exists $1\leq \ind{1} \leq \lastp$ such $\pos(P_{\ind{1}}) \notin \rect(\ell)$ and $|\dom{P_{[1,\ind{1}]}}\cap V(C)| \leq |\dom{P}\cap V(C)|-4\ell\leq 0$. Since $\dom{\seed} \in \rect(\ell)$ and $\pos(P_1) \in \dom{\seed}$ then this second case is not possible and thus $P$ is fragile or pumpable.

\end{proof}

Consider a path assembly $P$ which satisfies macroscopic initial conditions and $\ell > \seedbound$ such that $P$ does not satisfies the reset constraint for $\rect(\ell)$. Then if the hypothesis $\dom{\seed} \subset \rect(\ell)$ is not satisfied then by the seed lemma (lemma \ref{macro:seed}), the path assembly $P$ is fragile or pumpable. Otherwise by lemma \ref{macro:reset:bis}, the path assembly $P$ is fragile or pumpable. Then, the variant of the reset lemma formulated in lemma \ref{macro:reset} is a corollary of this result and of the seed lemma.

\subsection{Stakes Lemma}

The first step for proving the stakes lemma is to find a candidate segment which is \emph{near} the seed.

\begin{lemma}
\label{lem:true:init}
Consider a path assembly $P$ satisfying macroscopic initial conditions and producible by $\tiling=(T,\seed,1)$. Then either $P$ is fragile or pumpable or there exists $1\leq \indm < \indp \leq \lastp$ such that $P_{[\indm,\indp]}$ is a visible candidate segment of direction $\vu$ with $||\vu|| \leq \border$ and $\dom{P_{[1,\indp]}} \subset \rect(\boundv)$. 
\end{lemma}

\begin{proof}
See figure \ref{fig:proof:stakes1} for an illustration of this proof. By the seed lemma \ref{macro:seed}, either $P$ is fragile or pumpable or $\dom{\seed} \subset \rect(\seedbound)$. Now consider $P'$ and $\seed'$ which are the translation of $P$ and $\seed$ by $(-\border,-\border)$. Then by definition of macroscopic initial conditions $P'$ and $\seed'$ are in the first quadrant of the plane and $\dom{\seed'} \subset \rect(8|T|+|\seed|)$. %Now, either $x_{P_{\lastp}}>\fbound-\border$ or $y_{P_{\lastp}}>\fbound-\border$. %Without loss of generality we suppose that $y_{P_{\lastp}}>\fbound-\border$; otherwise we switch the east and north axis and we are in this case (the hypothesis $"\dom{\seed'} \subset \rect(8|T|+\seed)"$ is preserved by this transformation). 
By definition of macroscopic initial conditions, we have $y_{P_{\lastp}}>\fbound$ and then there exists $\ind{1}=\min\{1\leq i \leq \lastp: y_{P'_i}>8|T|+|\seed|\}$ and $\ind{2}=\min\{\ind{1}\leq i \leq \lastp: \pos(P'_i) \notin \rect(\border)\}$. By definition $1\leq \ind{1} \leq \ind{2} \leq \lastp$. Now either $X_{P'_{[1,\ind{1}]}} \leq \fband(8|T|+\seed)$ or $X_{P'_{[1,\ind{1}]}} > \fband(8|T|+\seed)$. In the second case, the path assembly $P'$ fits in a band and by lemma \ref{lem:thin}, $P'$ is fragile or pumpable (and then $P$ is also fragile or pumpable). Now, if $X_{P'_{[1,\ind{1}]}} \leq \fband(8|T|+\seed)$ then $\dom{P'_{[1,\ind{1}]}} \subset \rect(\fband(8|T|+\seed))$. 
Now, there exists more than $4|T|$ tiles visible from the west in $P'_{[1,\ind{1}]}$, then there exists at least $4|T|$ tiles visible from the west in $P'_{[1,\ind{2}]}$ and which belong to $\rect(\fband(8|T|+\seed))$. Then there $1\leq \indm \leq \indp \leq \ind{2}$ such that $P_{\indm}$ and $P_{\indp}$ both belong to $\rect(\fband(8|T|+\seed))$ and either $P'_{[\indm,\indp]}$ is a visible candidate segment of $P'_{[1,\ind{2}]}$ or $P'_{[\indm,\indp]}$ is a candidate segment of $P'_{[1,\ind{2}]}$, visible from the west, oriented to the south, not hidden in the seed and which exits by the south or the east. In the second case, by lemma \ref{lem:low:watershed}, $P'$ is pumpable or fragile. In the first case, either $P'_{[\indm,\indp]}$ is also a visible candidate segment of $P'$ or not. If $P'_{[\indm,\indp]}$ is not visible in $P'$ then there $\ind{3}>\ind{2}$ such that $P'_{\ind{3}}\in \rect(\fband(8|T|+\seed))$. By definition $\pos(P'_{\ind{2}}) \notin \rect(\border)$ and since $\border=\freset(\fband(8|T|+\seed))$, then $P'$ does not satisfy the reset constraint for $\rect(\fband(8|T|+\seed))$ and by lemma \ref{macro:reset:bis}, $P'$ is fragile or pumpable. Otherwise $P'_{[\indm,\indp]}$ is a visible candidate of $P'$ and $||\vu|| \leq \border$ and $\dom{P_{[1,\indp]}} \subset \rect(\border)$ and then $P_{[\indm,\indp]}$ is a visible candidate of $P$ and $||\vu|| \leq \border$ and $\dom{P_{[1,\indp]}} \subset \rect(\boundv)$ (with $\boundv=2\border$).

\end{proof}

Now we can prove the stakes lemma and conclude the article.

\begin{lemma} [Macro 3: stakes lemma]
\label{macro:stakes:bis}
Let $P$ be a path assembly producible by $\mathcal{T}=(T,\sigma,1)$ satisfying macroscopic initial conditions. If $P$ does not satisfy the stakes constraint then it is fragile or pumpable.
\end{lemma}

\begin{proof}
This proof is done in three steps. During the first step, we initialize different tools (seed constraint, fork) to set up a good setting. During the second step, we build the first stake and during the third one we build the second stake. Several incidents may occurs during the construction of the stakes but anyone of these problems leads to the fragility or pumpability of~$P$. 

\vspace{+0.5em}

\noindent \textbf{Initialization:}
 %then either $x_{P_{\ind{4}}}>\freset(\boundv)$ or $y_{P_{\ind{4}}}>\freset(\boundv)$. If $x_{P_{\ind{4}}}>\freset(\boundv)$ then by switching the $x$ and $y$ axis we are in the case $y_{P_{\ind{4}}}>\freset(\boundv)$. 
%Then without loss of generality, we can consider that $y_{P_{\ind{4}}}>\freset(\boundv+1)$. 
see Figures \ref{fig:proof:stakes2} and \ref{fig:proof:stakes2:2} for an illustration of this part of the proof. By the seed lemma (lemma \ref{macro:seed}), we have $\dom{\seed} \subset \rect(\seedbound)$ or $P$ is fragile or pumpable. %Now either $\dom{P_{[1,\ind{4}]}} \subset \rect(\fband(\freset(\boundv)+1))$ and or it fits in a band. In the second case, by lemma \ref{lem:thin} $P$ is fragile or pumpable. 
%We have, $\dom{P_{[1,\ind{4}]}} \subset \rect(\fband(\freset(\boundv)+1))$. 
By lemma \ref{lem:true:init}, either $P$ is fragile or pumpable or there exists $1\leq \indm< \indp \leq \ind{4}$ such that $P_{[\indm,\indp]}$ is a visible candidate segment of direction $\vu$ such that $||\vu||\leq \border$ and $\dom{P_{[1,\indp]}} \subset \rect(\boundv)$. By the definition of macroscopic initial conditions, there exists $1\leq \ind{4} \leq \lastp$ such that $\ind{4}=\min\{i:y_{P_i} = \freset(\boundv+\border)+\border+1)\}$. Note that $\indp \leq \ind{4}$ and remark that by definition of $\ind{4}$, the path assembly $P_{[1,\ind{4}]}$ is an exposed path assembly (with a turn back of $Y_{P_{[1,\ind{4}]}}$). Now either $\dom{P_{[1,\ind{4}]}} \subset \rect(\fband(\freset(\boundv+\border)+\border+1))$ or not. In the second the case, $P_{[1,\ind{4}]}$ fits in a band and by lemma \ref{lem:thin}, $P$ is fragile or pumpable. Otherwise, $\dom{P_{[1,\ind{4}]}} \subset \rect(\fband(\freset(\boundv+\border)+\border+1))$ and by applying lemma \ref{lem:fork} to $P_{[1,\ind{4}]}$, either $P$ is fragile or pumpable or there exists a fork $F$ such that $y_{F_{|F|}}+y_{\vu}\geq \freset(\boundv+\border)+\border+1$ and $\dom{F} \subset \rect(\fband(\freset(\boundv+\border)+\border+1) +\border)$. Since $||\vu||\leq \border$, we have $y_{F_{|F|}} \geq \freset(\boundv+\border)+1$.

\vspace{+0.5em}

\noindent \textbf{Finding the first stake:} By definition of $F$, there exists $\ind{1}$ such that $\ind{1}=\min\{i:y_{F_i}=\freset(\boundv+\border)+1\}$. By definition of a fork, there exists $1 \leq \ind{2} \leq \ind{4}$ such that $F_{\ind{1}}=P_{\ind{2}}$ or $F_{\ind{1}}=P_{\ind{2}}-\vu$. We suppose that $F_{\ind{1}}=P_{\ind{2}}-\vu$ (the other case is symmetric). Remark that $y_{P_{\ind{2}}}\geq \freset(\boundv+\border)+1$ and thus $\pos(P_{\ind{2}}) \notin \rect(\freset(\boundv+\border)+1)$. Now, two cases occur either $P_{[1,\indm]}\cdot F_{[1,\ind{1}]} \cdot (P_{[\ind{2},\lastp]}-\vu)$ is a path assembly producible by $\tiling$ or not. In the second case, since by definition of a fork, the path assembly $P_{[1,\indm]}\cdot F$ is producible by $\tiling$, then $P_{[\ind{2},\lastp]}-\vu$ creates a collision with either $F_{[1,\ind{1}]}$ or $P_{[1,\indm]} \cup \seed$. If $P_{[\ind{2},\lastp]}-\vu$ creates a collision with $F_{[1,\ind{1}]}$ (see Figures \ref{fig:proof:stakes3:3:1} and \ref{fig:proof:stakes3:3:2}), by definition of a fork the path assembly $P_{[1,\indp]}\cdot (F+\vu)$ is producible by $\tiling$ and this path assembly creates a collision with $P_{[\ind{2},\lastp]}$ and thus $P$ is fragile. If $P_{[\ind{2},\lastp]}-\vu$ creates a collision with $P_{[1,\indm]} \cup \seed$ (see Figures \ref{fig:proof:stakes3:2:1} and \ref{fig:proof:stakes3:2:2}), there exists $\ind{2} \leq \ind{3} \leq \lastp$ such that $\pos(P_{\ind{3}})-\vu \in \rect(\boundv)$ and thus $\pos(P_{\ind{3}}) \in \rect(\boundv+\border)$. Then $P$ does not satisfies the reset constraint for $\boundv+\border$ and by the reset lemma \ref{macro:reset}, $P$ is fragile or pumpable. Then, the path assembly $P_{[1,\indm]}\cdot F_{[1,\ind{1}]} \cdot P_{[\ind{2},\lastp]}-\vu$ is producible by $\tiling$ (see Figure \ref{fig:proof:stakes3:1}). Let $\stakem= P_{[1,\indm]} \cdot F_{[1,\ind{1}]}$ be the first stake. Moreover, since $\boundf=\fband(\freset(\boundv+\border)+\border+1)+\border$, then $\dom{\stakem} \subset \rect(\boundf)$.

\vspace{+0.5em}

\noindent \textbf{Finding the second stake:}
see Figure \ref{fig:proof:stakes4} for an illustration of this part of the proof. By definition of index $\ind{4}$ and since $\ind{2}\leq \ind{4}$, there exist two indices $\ind{2} \leq \ind{5} \leq \lastp$ and $\ind{5} \leq \ind{8} \leq \lastp$ such that $\ind{5}=\min\{i \geq \ind{2}: \pos(P_i) \notin \rect(\freset(\boundf+\border))\}$ and $\ind{8}=\min\{i \geq \ind{5}: \pos(P_i) \notin \rect(\fjail(\freset(\boundf+\border+1))\}$. Then either $P_{[\ind{5},\ind{8}]}-\vu$ and $P_{[\ind{5},\ind{8}]}$ intersect or not. In the second case, since $\dom{P_{[1,\ind{5}]}} \subset \rect(\freset(\boundf+\border+1))$ then by the jail lemma (lemma \ref{macro:jailed}), $P$ is fragile or pumpable. In the second case, there exists $\ind{5} \leq \ind{6}\leq \ind{8}$ and $\ind{5} \leq \ind{7} \leq \ind{8}$ such that $\ind{6}=\min\{i \geq \ind{5}:\pos(P_i)-\vu \in \dom{P_{[\ind{5},\ind{8}]}}\}$ and $\pos(P_{\ind{7}})=\pos(P_{\ind{6}})-\vu$. Moreover, $\dom{P_{[\ind{2},\ind{6}]}} \subset \rect(\fjail(\freset(\boundf+\border+1))+1)$. Now, since $\stakem \cdot (P_{[\ind{2},\ind{6}]}-\vu)$ is producible by $\tiling$ then either $P$ is fragile or $P_{\ind{7}}=P_{\ind{6}}-\vu$ and then $P_{\ind{7}}+\vu=P_{\ind{6}}$. Then either $P_{[1,\ind{6}]} \cdot (P_{[\ind{7},\lastp]}+\vu)$ is a path assembly producible by $P$ or not. In the first case $\stakep=P_{[1,\ind{6}]}$ is our second stake and the lemma is true since $\dom{\stakep} \subset \rect(\fjail(\freset(\boundf+\border+1))+\border+1)$ and $\sbound=\fjail(\freset(\boundf+\border+1))+\border+1$ (see Figure \ref{fig:proof:stakes5:1}). Otherwise, $P_{[\ind{7},\lastp]}+\vu$ collides with $P_{[1,\ind{6}]} \cup \seed$. If the collision occurs between $P_{[\ind{7},\lastp]}+\vu$ and $P_{[1,\ind{2}]} \cup \seed$ then there exists $\ind{7} \leq \ind{9} \leq \lastp$ such that $\pos(P_{\ind{9}}+\vu) \in \rect(\boundf)$ and then $\pos(P_{\ind{9}}) \in \rect(\boundf+\border)$ (see Figures \ref{fig:proof:stakes5:2:1} and \ref{fig:proof:stakes5:2:2}). Since $\ind{9}\geq \ind{7} \geq \ind{5}$ and since $\pos(P_{\ind{5}}) \notin \rect(\freset(\boundf+\border))$ then $P$ does not satisfies the reset constraint for $\boundf+\border$ and by the reset lemma \ref{macro:reset}, $P$ is fragile or pumpable. Otherwise there exists a collision between $P_{[\ind{7},\lastp]}+\vu$ and $P_{[\ind{2},\ind{6}]}$ and then there exists a collision between $P_{[\ind{7},\lastp]}$ and $P_{[\ind{2},\ind{6}]}-\vu$ (see Figures \ref{fig:proof:stakes5:3:1} and \ref{fig:proof:stakes5:3:2}). Since $\stakem \cdot (P_{[\ind{2},\ind{6}]}-\vu)$ is producible by $\tiling$ then $P$ is fragile.

%Without loss of generality we suppose that $y_{F_{|F|}}+\vu=TODO$. Then let $i$ such that $P_i=F_{|F|}+\vu$. Then either $P_{[1,\indm]}\cdot F\cdot P_{[i,\lastp]}-\vu$ is path assembly or $P_{[i,\lastp]}-\vu$ collides  with $P_{[1,\indm]}\cdot F$ or the seed. If $P_{[i,\lastp]}-\vu$ collides with $P_{[1,\indm]}$ or the seed then by the reset lemma, $P$ is fragile or pumpable. If $P_{[i,\lastp-\vu]}$ collides with $F$ then either $P$ is not simple which is contradiction or there exist $1\leq k \leq t$ and $i \leq k' \leq \lastp$ such that $P_k$ collides with $P_{k'}-\vu$. In this case, since $P_{[1,\indp]}\cdot F+\vu$ is path assembly producible by TODO then $P_k+\vu$ collides with $P_{k'}$ and thus $P$ is fragile. Now, if $P_{[1,\indm]}\cdot F\cdot P_{[i,\lastp]}-\vu$ is path assembly producible by TODO. Then either $P_{[i,\lastp]}-\vu$ intersect with $P_{[i,\lastp]}$ inside $\rect{}$ by the fourth macroscopic lemma otherwise $P$ is fragile or pumpable. If this intersect is a collision then $P$ is fragile. Otherwise there exists $a>i$ such $P_a=intersect+\vu$ then a reasoning similar to the one done before either $P$ is fragile or we have found the desired stakes.
\end{proof}

\clearpage
\bibliographystyle{plain}
\bibliography{directed}

\appendix

\clearpage
\section{Functions and bounds}
\label{app:bound}
In this appendix, we define precisely all the functions and bounds used along the article. Consider a tiling system $\tiling=(T,\seed,1)$. All the following definitions will depend on $|T|$ and $|\seed|$. We start by $$\fband(x)=4x(2x!)(2|T|+1)^{2x}+2x|\seed|.$$

Let $\fresetstep$ be the function defined as $$\fresetstep(\ell)=|\seed|+2(\ell+3)|T|.$$ Now, consider $\ell \in \mathbb{N}$, then we define $\ell_0=\fresetstep(\ell+1)$ and for $0\leq i \leq 4\ell-1$, let $\ell_{i+1}=\fresetstep(\ell_{i}+1)$. The function $\freset$ is the function $$\freset(\ell)=\ell_{4\ell}.$$

Now, here are the different values of the bounds and the function $\fjail$:

$$\border=\freset(\fband(8|T|+|\seed|)).$$ 

$$\seedbound=8|T|+|\seed|+\border.$$

$$\boundv=2\border.$$

$$\boundf=\fband(\freset(\boundv+\border)+\border+1)+\border.$$

$$\fjail(x)=4(\border^2(2\border+1))(2(2\border^2+\border))!)(2|T|+1)^{2\border^2+\border}+\border x^2+x.$$

$$\sbound=\fjail(\freset(\boundf+\border+1))+\border+1.$$

$$\ebound=\freset(\sbound(\border+1))+1.$$

$$\fbound=\fjail(\ebound+1)+1.$$

$$\text{Note that, }\border<\seedbound<\boundv<\boundf < \sbound < \ebound < \fbound.$$

\clearpage
\section{Figures of section \ref{def:main}: definitions and basic properties}
\label{app:def}
\input{appDef}

\clearpage
\section{Figures of section \ref{sec:macro}: macroscopic reasoning}
\label{app:macro}
\input{appMacro}

\clearpage
\section{Figures of section \ref{sec:dgt}: two dimensional discrete toolbox}
\label{app:tool}
\input{appTool}

\clearpage
\subsection{Illustrations for finding a new split}
\label{app:shrink}
\input{appShrink}

\clearpage
\section{Figures of section \ref{sec:uwl}: modified window lemma and jailed path assembly}
\label{app:uwl}
\input{appUWL}

\clearpage
\subsection{Illustrations for the modified window movie lemma}
\label{app:modi}
\input{appModified}

\clearpage
\section{Figures of section \ref{sec:micro}: microscopic reasoning}
\label{app:micro}
\input{appMicro}

\clearpage
\subsection{Illustrations for building a fork}
\label{app:buildf}
\input{appalgo}

\clearpage
\section{Figures of section \ref{sec:lastpart}: from microscopic reasoning to macroscopic one: proof of the remaining macroscopic lemmas}
\label{app:last}
\input{appLast}

\clearpage
\subsection{Illustrations for building the stakes}
\label{app:stakes}
\input{appStake}

\end{document}